\documentclass[11pt,a4paper]{article}
\usepackage{jheppub,amsmath,  amssymb,slashed,url,bm,textgreek,upgreek}
\usepackage{graphicx}
\usepackage{epstopdf}
\def\t{{ \sf t}} 

\def\tt{{\mathfrak t}}

\def\a{a}

\def\c{{\eurm c}}

\def\be{\begin{equation}}
\def\ee{\end{equation}}

\def\eff{{\mathrm{eff}}}
\def\tilde{\widetilde}

\def\h{\widehat}

\def\D{{\mathcal D}}
\def\S{{\mathcal S}}
\def\SIgma{\Sigma}

\def\V{{\mathcal V}}
\def\sV{{\sf V}}
\def\O{{\mathcal O}}

\def\A{{\mathcal A}}
\def\d{{\mathrm d}}

\def\R{{\mathbb R}}
\def\C{{\mathbb C}}
\def\U{{\mathcal U}}
\def\D{{\mathcal D}}
\def\[{\bigl [}

\def\]{\bigr ]}

\def\T{{\mathcal T}}

\def\Z{{\mathbb Z}}

\def\CC{{\mathcal C}}

\def\t{\widetilde }
\def\h{\widehat}

\def\K{{\mathcal K}}
\def\V{{\mathcal V}}

\def\B{{\mathcal B}}

\def\W{{\mathcal W}}
\def\P{{\mathcal P}}

\def\H{{\mathcal H}}

\def\sW{{ W}}

\def\tilde{\widetilde}

\def\bar{\overline}

\font\teneurm=eurm10 \font\seveneurm=eurm7  \font\fiveeurm=eurm5
\newfam\eurmfam
\textfont\eurmfam=\teneurm \scriptfont\eurmfam=\seveneurm
\scriptscriptfont\eurmfam=\fiveeurm

\font\teneusm=eusm10 \font\seveneusm=eusm7 \font\fiveeusm=eusm5
\newfam\eusmfam
\textfont\eusmfam=\teneusm \scriptfont\eusmfam=\seveneusm
\scriptscriptfont\eusmfam=\fiveeusm

\font\tencmmib=cmmib10 \skewchar\tencmmib='177
\font\sevencmmib=cmmib7 \skewchar\sevencmmib='177
\font\fivecmmib=cmmib5 \skewchar\fivecmmib='177
\newfam\cmmibfam
\textfont\cmmibfam=\tencmmib \scriptfont\cmmibfam=\sevencmmib
\scriptscriptfont\cmmibfam=\fivecmmib

\def\la{\langle}
\def\ra{\rangle}
\def\d{{\mathrm d}}
\def\gen{{\rm gen}}
\def\out{{\rm out}}

\def\th{{\rm th}}
\def\vN{{\rm vN}}
\def\Tr{{\rm Tr}}
\def\i{{\rm i}}
\def\ret{{\rm ret}}
\def\eff{{\rm eff}}
\def\max{{\rm max}}
\def\BH{{\rm BH}}
\def\EH{{\rm EH}}
\def\GHY{{\rm GHY}}
\def\sc{{\rm scr}}
\def\adv{{\rm adv}}
\def\TFD{{\rm TFD}}
\def\a{{\sf a}}
\def\te{\tE}
\def\c{{\sf c}}
\def\E{{\mathcal E}}
\def\RR{\mathcal R}
\def\rad{{\rm rad}}
\def\sstar{{\rm star}}
\def\Page{{\rm Page}}
\def\A{{\sf A}}
\def\B{{\sf B}}
\def\AdS{{\rm AdS}}
\def\tE{t_{\sf E}}
\def\dS{{\rm dS}}
\def\sW{{\mathcal W}}
\def\sE{{\sf E}}
\def\HHI{{\rm HHI}}
\def\sH{{\rm H}}
\def\C{{\sf C}}
\def\D{{\sf D}}
\def\CC{{\sf C}}
\def\CFT{{\sf{CFT}}}

\def\cl{{\rm{cl}}}
\def\ES{{\rm {ES}}}

\title{Introduction to Black Hole Thermodynamics}

 \author{Edward Witten}
\affiliation{School of Natural Sciences, Institute for Advanced Study,\\ 1 Einstein Drive, Princeton, NJ 08540 USA}
\abstract{These notes aim to provide an introduction to the basics of black hole thermodynamics.  After explaining Bekenstein's original proposal that black holes have
entropy, we discuss Hawking's discovery of black hole radiation, its analog for Rindler space in the Unruh effect, the Euclidean approach to black hole thermodynamics,
some basics about von Neumann entropy and its applications, the Ryu-Takayanagi formula, and the nature of a white hole.  
}

\begin{document}\maketitle

\section{Introduction}\label{intro} 

The one thing that almost everyone knows about black holes is that crossing the horizon of a black hole is a one-way trip.
What goes in does not come back out, at least according to classical general relativity.  In other words, entering a black hole
is {\it irreversible}.  

This irreversibility is in obvious tension with  basic physical principles.   For example, if quantum evolution is governed
by a hermitian Hamiltonian $H$, and there is a nonzero matrix element $\la \chi|H|\psi\ra$ for a transition from a state $\psi$ to a state $\chi$,
then there must be a complex conjugate matrix element $\la\psi|H|\chi\ra$ for a transition in the opposite direction.   
 
On the other hand, irreversibility is familiar in everyday life, as explained in a familiar English nursery rhyme. Humpty Dumpty is depicted as an egg sitting up on a wall.\footnote{According to Wikipedia,
the original context for the rhyme is unclear.}    Once the egg falls and breaks, ``all the king's horses
and all the king's men couldn't put Humpty together again.''

Classical physicists saw some tension between such irreversibility and a presumed invariance of the laws of nature under time-reversal.
The tension was largely resolved with the development of thermodynamics and statistical mechanics. 
Irreversible processes are  those in which the entropy increases.
Processes in which the entropy becomes smaller -- for example, a broken egg on the floor spontaneously reassembles and jumps up on the wall -- can happen in
principle but require extreme fine-tuning of initial conditions, so they are exponentially unlikely.   

The basic idea of black hole thermodynamics is that the irreversibility that occurs when an object is absorbed by a black hole is similar to the statistical irreversibility that is familiar in ordinary
physics. 
When a body falls into a black hole, entropy increases.   
A time-reversed scenario in which the black hole spontaneously emits that same body is possible, but requires fine-tuned initial conditions and is prohibitively unlikely, just like a process
in which a broken egg is spontaneously reconstituted.

Since the original work of Bekenstein \cite{bekenstein} and Hawking \cite{hawking} just over half a century ago, as well as the related classical observations
of Christodoulou \cite{Christodoulou} and Bardeen, Carter, and Hawking \cite{BCH}, black hole thermodynamics has raised many challenging questions about
the fundamental nature of quantum mechanics and gravity.   The goal of the present article is to provide a gentle introduction to this fascinating subject.   Hopefully, we will explain
enough to provide an entr\'{e}e to the subject, but many important aspects, both old and new, are omitted.   The literature is much too extensive
to be properly summarized here. Some of the early results not explained here can be found in \cite{wald}, chapters 12 and 14, and in the articles by
Carter, Gibbons, and Hawking in \cite{HI}; and see for example \cite{malda,malda2} for  short introductions to some of the more
contemporary developments.   Some of the background in classical general relativity that helps in understanding the subject more deeply is explained in \cite{wald}, chapters 8 and 9,
 and also in \cite{Witten}, for example. 

 In section 2 of this article, Bekenstein's heuristic proposal concerning black hole entropy is described.  Hawking's celebrated discovery of quantum emission from black holes
is the subject of sections 3 and 4. Section 5 is devoted to Rindler space and the Unruh effect, which provide a simplified framework for understanding some essential aspects of quantum black holes.   The Euclidean approach to black hole thermodynamics is introduced in section 6.    Microscopic von Neumann entropy, as opposed to macroscopic thermodynamic
entropy, is introduced in section 7 and discussed in the context of black holes in general terms in section 8 and more specifically in the framework of the Ryu-Takayanagi formula
in section 9.   Finally, section 10 addresses the question, ``What is a white hole?''

\section{Black Hole Entropy And The Generalized Second Law}\label{gsl}

The Second Law of Thermodynamics says that, in any process that we can observe in practice, the entropy is nondecreasing.   Here, entropy is the usual thermodynamic
entropy.   (In section \ref{twonotions},  we will discuss a related but different notion, the microscopic von Neumann entropy.)   Processes in which the thermodynamic entropy decreases are allowed
by the laws of nature, but are prohibitively unlikely, in practice.

However,  Bekenstein, motivated by a question from his advisor John Wheeler, observed that if we toss a cup of tea into a black hole, the entropy seems to diminish,
 assuming that we assign zero entropy to the black hole.   To avoid concluding that this process violates the Second Law, 
 Bekenstein wanted to assign an entropy to a black hole in such a way that the Second
Law would remain valid when matter falls into the black hole.   For this, he needed to attribute to the black hole an entropy that always increases according to classical general relativity. 

What property of a black hole can only increase?   It is not true that the black hole mass always increases.   A rotating black hole, for instance, can lose mass as its rotation slows down.
But there is a quantity that always increases.  Hawking had just proved the ``area theorem'' \cite{AreaTheorem}, which says that in 
classical general relativity, the area of the horizon of a black hole can only increase.\footnote{For the proof, see for example
\cite{wald}, p. 312, or \cite{Witten}, section 6.3.}  So it was fairly natural for Bekenstein to propose
that the entropy of a black hole should be a multiple of the horizon area.   For example, for a Schwarzschild black hole of mass $M$,
with line element
\be\label{linem}\d s^2=-\left(1-\frac{2 G M}{r}\right)\d t^2+\frac{\d r^2}{1-\frac{2 GM}{r}}+r^2\d\Omega^2\ee
(where $G$ is Newton's constant),
the horizon is at the Schwarzschild radius $r_S=2GM$  and the horizon area is
\be\label{horarea} A=4\pi r_S^2=16\pi G^2 M^2. \ee
Since entropy is dimensionless,\footnote{We measure temperature in energy units, with Boltzmann's constant set to 1.} 
if  black hole entropy is  to be a multiple of the horizon area, the  constant of proportionality will have 
units of inverse area.   From fundamental constants $\hbar, c$ and $G$, one can make the Planck length $\ell_P =(\hbar G/c^3)^{1/2}$
and the Planck area $\ell_P^2=\hbar G/c^3$.   In units with $c=1$, Bekenstein's formula for the entropy of a black hole was
\be\label{bhentropy} S=\frac{A}{4G\hbar},\ee
where the constant $1/4$ was not clear in Bekenstein's work and was determined by Hawking a couple of years later.
The formula with this factor of $1/4$ included is commonly called  the Bekenstein-Hawking entropy of the black hole.

According to this formula, black hole entropy can be extraordinarily large in ordinary terms.    For example,
a black hole with the mass of the Sun has an entropy of roughly $10^{77}$, which is about $10^{18}$ times the entropy of the actual Sun. 
We included the factor of $\hbar^{-1}$ in eqn. (\ref{bhentropy}) to underscore the fact that the formula is quantum mechanical in an essential way, but  henceforth
we will set $\hbar=1$.

Bekenstein's idea was that the entropy of a black hole measures the number of ways that the black hole could have formed.   He
defined a ``generalized entropy'' -- the sum of the black hole entropy $A/4G$ and the ordinary entropy $S_\out$ of matter and radiation
outside the horizon:
\be\label{sgen}S_\gen=\frac{A}{4G}+S_\out.\ee
The generalized entropy was proposed to obey a Generalized Second Law, saying that it is nondecreasing in all processes that we
can observe in practice:
\be\label{practice}\frac{\d S_\gen}{\d t}\geq 0.\ee

Bekenstein considered several tests of the Generalized Second Law.   
For simplicity in the following we consider a Schwarzschild black hole. We want to test whether the generalized second law (\ref{practice}) is valid
when the black hole absorbs matter.  Since Bekenstein considered $S_\out$ to be the usual thermodynamic
entropy, the  statement (\ref{practice}) of the Generalized Second Law assumes that the matter system which is being absorbed by the black hole
is one for which thermodynamics is valid, meaning a system that is close enough to at least local thermodynamic 
equilibrium.\footnote{To formulate a version
of the Generalized Second Law when the matter and radiation outside the black hole are not close to thermal equilibrium, one needs to interpret $S_\out$
as  von Neumann entropy, which we introduce in section \ref{twonotions}, rather than thermodynamic entropy.}     A simple case
to consider is that the black hole is absorbing a beam of black body radiation, say at temperature $T$. 

In $3+1$ dimensions,\footnote{We generally phrase arguments in this article for the standard $3+1$-dimensional case,
as the generalization to $D$ dimensions would not add much.   The discussion of the Ryu-Takayanagi formula is the main exception, for reasons
explained in footnote \ref{exception}.}  the relation between energy $E$, temperature $T$, and entropy $S$ in black body radiation is $E=\frac{3}{4} TS$.
If a black hole of mass $M$ absorbs energy $E\ll M$ from black body radiation, its entropy $A/4G=4\pi GM^2$ increases by $8\pi GME$,
while the  entropy of the radiation decreases by $\Delta S_\out=4E/3T$.   The change in the generalized entropy is
\be\label{likely} \Delta S_\gen =\left(8\pi GM -\frac{4}{3T}\right)E. \ee
In particular, this is positive as long as $\pi r_S>1/3T$, or in other words as long as the typical photon wavelength (which is of order $1/T$) is sufficiently
small compared to the radius $r_S=2GM$  of the black hole.

 However,  $\Delta S_\gen$ becomes negative if $T$ is so small that the typical photon wavelength is much larger than the Schwarzschild radius.   A black hole
 can absorb photons of such great wavelength, though not very efficiently.
The apparent entropy decrease when a black hole absorbs photons of extremely long wavelength does not have a satisfactory resolution in the 
 framework that Bekenstein assumed, which was that whatever falls behind the black hole horizon stays there
 forever.   In thermodynamic terms, since Bekenstein assumed that the black hole does not radiate, one would have
 to assign it a temperature of 0.   Thermodynamics says that at equilibrium, the changes in energy $E$ and entropy $S$
 of a system are related by $\d E= T\d S$ or $\d S=\d E/T$, so a system with $T=0$ should have $\d S=\infty$ if $\d E\not=0$.
 But Bekenstein wanted to attribute a finite, not infinite, entropy to the black hole.

It turned out that the key to understanding the Generalized Second Law for a black hole that is absorbing photons of energy $\ll 1/2GM$
is to take into account the fact that the black hole is strongly emitting such photons.  In the next section, we will explain how Hawking discovered this.

The huge entropy $A/4G$ of a macroscopic black hole appears to imply that the black hole has of order $e^{A/4G}$ quantum states, though
the understanding of these states remains murky to the present day (except in the important special case of a supersymmetric black hole \cite{StromingerVafa}).   
This vast implied degeneracy of a quantum black hole may appear at first
sight to be in tension with the ``no hair theorem'' of classical general relativity \cite{nohair,nohair2}.   The no hair theorem asserts that classically
a black hole, after settling down to a stationary state, is fully characterized by the obvious conserved quantities -- mass, angular momentum, and charge.
Actually, in black hole thermodynamics, the no hair theorem is taken to represent thermalization.   An ordinary thermal system has a sort of no hair behavior.  Pour a cup of
water into a glass.   It will slosh around for a while and will visibly
not be in thermal equilibrium.   After some time, transients will die down and the water in the glass will be in apparent thermal equilibrium,  describable
just by its conserved quantities, though its
detailed microscopic state will not be describable by a truly thermal density matrix.\footnote{The relevant notions will be described in section \ref{twonotions}.  
Pouring  the cup of tea into a glass and letting it settle down is a unitary process but an  irreversible one,
so it increases the thermodynamic entropy without increasing the microscopic von Neumann entropy.  Therefore, even if the initial state of the tea was strictly thermal (described by a thermal
density matrix with equal von Neumann and
thermal entropies),
the final state has a von Neumann entropy less than that of a thermal density matrix at the relevant energy.}   The no hair
theorem is interpreted to say that a black hole does the same.

  \section{Black Hole Evaporation}\label{bhevaporation}
 
 Famously, Hawking discovered in 1974 that at the quantum level, a black hole is not really black.  Hawking reportedly was skeptical about Bekenstein's idea,
 but ended up proving it.   As preparation
 for describing Hawking's work, let us recall the notion of a Penrose diagram.
 
 Penrose diagrams are usually drawn for spherically symmetric spacetimes.   Angular coordinates are suppressed; only the
 time and a radial coordinate are shown.  The main purpose of the Penrose diagram is to exhibit causal relations in a useful
 way.   The diagram is drawn so that radially ingoing or outgoing null geodesics are at a $\pi/4$ angle to the vertical, and any causal
 curve (any null or timelike curve)  is at most at a $\pi/4$ angle from the vertical.  Since the condition for  a curve to be causal
 is invariant under conformal changes of coordinates, one usually makes a conformal mapping such that the whole diagram becomes
 compact and asymptotic regions at infinity are therefore easily visible.

      \begin{figure}
 \begin{center}
   \includegraphics[width=4in]{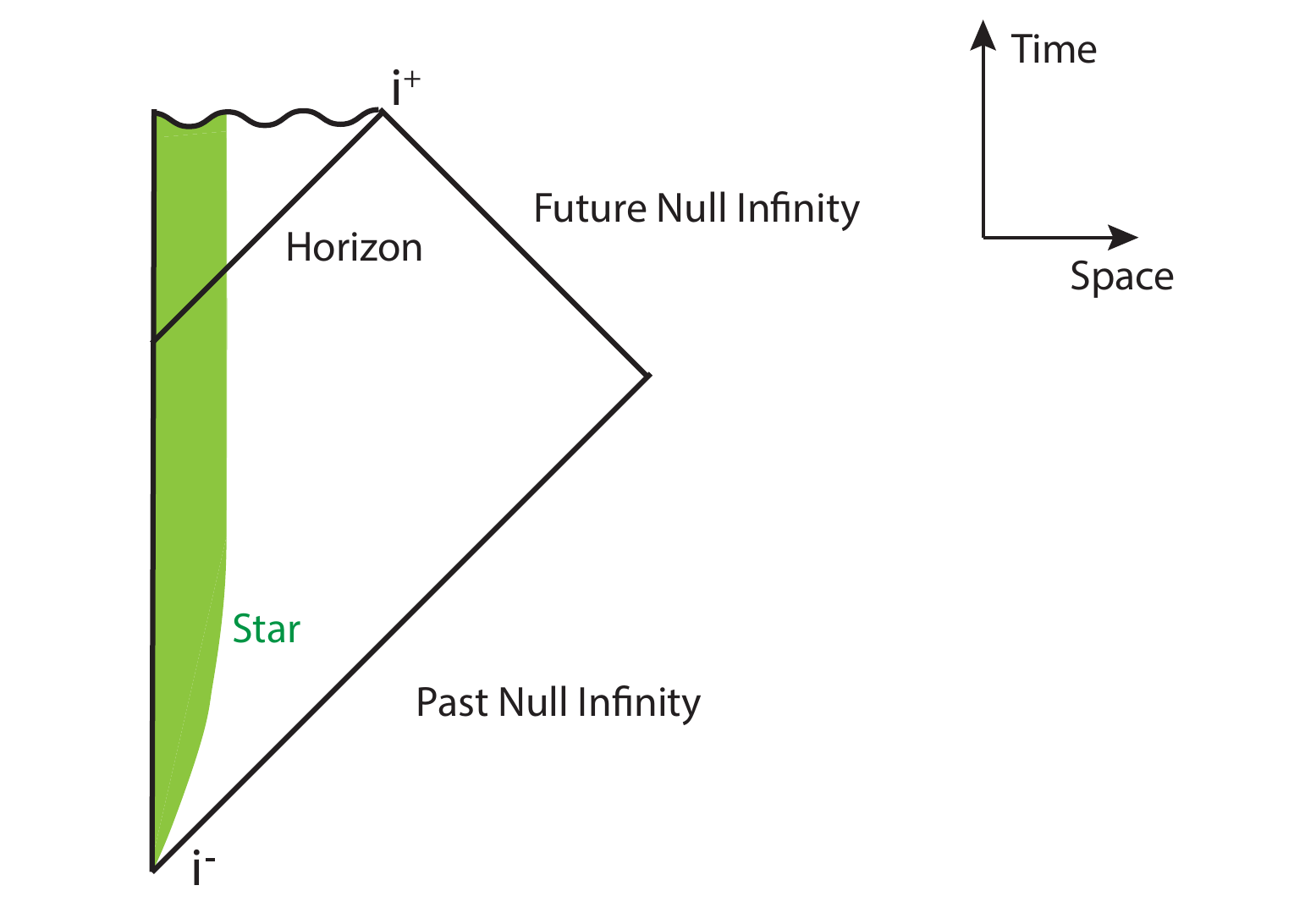}
 \end{center}
\caption{\footnotesize A Penrose diagram describing the collapse of a star to form a black hole.  \label{One}}
\end{figure} 
 
An important example is a Penrose diagram describing spherically symmetric collapse of  a star to a Schwarzschild black hole, as in  fig. \ref{One}.
 The left vertical boundary of the picture is the
origin of polar coordinates at $r=0$.    Shown in green is the worldvolume of the star.
(The fact that the star appears to emanate from a point in the far past is an artifact of the conformal mapping used in drawing the diagram.)   The
star ends its life at the singularity, depicted by the wiggly line at the top of the diagram.   Future and past null infinity are represented by the
diagonal lines on the right boundary, as labeled.  The horizon of the black hole is the diagonal black line inside the picture.    Since causal curves travel at an angle no greater than $\pi/4$ from the vertical, an observer
 outside the horizon can never see beyond the horizon.    The worldline of a massive observer who remains forever outside the horizon (and does not accelerate indefinitely) will end at the point $\i^+$,
 known as future infinity, where the horizon and future null infinity meet.   Anything that can be seen from anywhere in the spacetime outside the horizon can be seen from $\i^+$.  A massive
 observer began life in the far past, at the  point labeled $\i^-$ at the bottom of the figure.
 
 Hawking's discovery of black hole evaporation was based on studying the behavior of a quantum field in a definite classical spacetime background,
 taken to be a Schwarzschild black hole of mass $M$.
 This is potentially a sensible approximation if $M$ is much bigger than the Planck mass $(\hbar c/G)^{1/2}$, 
which is about $10^{-5}$ grams.  Equivalently, it is potentially a sensible approximation if the Schwarzschild radius
of the black hole is much bigger than the Planck length $(\hbar G/c^3)^{1/2}\approx 10^{-33}$ cm. Of course, the framework of quantum field theory in four-dimensional curved spacetime
might break down before reaching the Planck length (due to string theory or Kaluza-Klein theory, for example), but at any rate we expect that the very massive black holes
that are familiar in astrophysics are very far from any such breakdown.   So in particular, for a realistic astrophysical black hole,
Hawking's approximation is expected to be excellent.
 
 We want to analyze what an observer far from the collapsing star will see in the far future, after transients have died down.
 As an idealization, assuming that what are observed are massless fields such as the electromagnetic field, one can think of these observations as being made
 at future null infinity, and more specifically near the upper boundary of future null infinity where it ends at the point $\i^+$.   These conditions correspond to making
 observations at a great distance from the black hole and in the far future.
 
       \begin{figure}
 \begin{center}
   \includegraphics[width=4in]{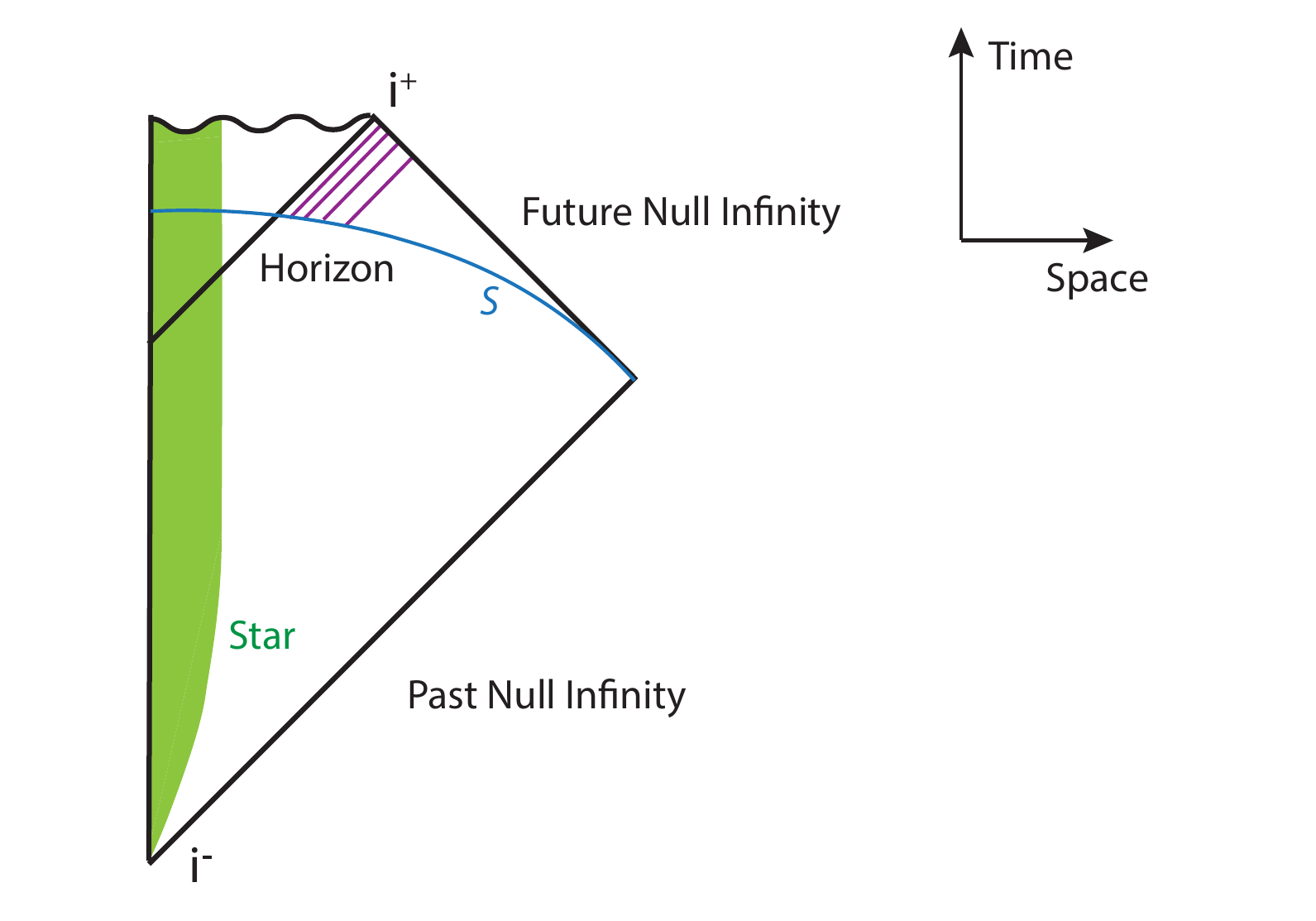}
 \end{center}
\caption{\footnotesize Whatever the distant observer sees in the far future can be traced back to initial conditions on a Cauchy hypersurface, such as the hypersurface $\S$
shown in blue.   This Cauchy hypersurface has been chosen to cross the horizon outside the worldvolume of the collapsing star.   From this Cauchy hypersurface,
signals can propagate to the distant observer at the speed of light.   These signals can propagate along outgoing null geodesics, some of which are indicated
by the purple lines in the figure, which are at a $\pi/4$ angle to the vertical. These outgoing null geodesics, if labeled by the time of a distant observer,
``bunch up'' near the horizon, as shown,
because the redshift diverges there.   \label{Two}}
\end{figure} 
 
 Measurements that an observer will make at, or near, future null infinity can be traced back to initial conditions on a Cauchy hypersurface.
 For this purpose, we can choose any Cauchy hypersurface we want.  It is convenient to choose one that crosses the horizon outside the collapsing star such as the hypersurface
 $\S$ of fig. \ref{Two}.   From any point
 on $\S$, a massless particle might be emitted and propagate to future null infinity at the speed of light.  The diagonal purple lines in the figure represent the trajectories of such
 particles.  
 
 Let $u$ be any coordinate function on $\S$ that vanishes on the horizon and is positive outside, and such that the normal derivative of $u$ is nonzero (and finite) along the horizon.  An outgoing signal
 from the black hole that will eventually be received by a distant observer might propagate through $\S$  at any value of $u$.
 Let $t$ be the time at which the distant observer receives the signal.    The relation between $t$ and $u$ is
 \be\label{doffo} t=4GM \log \frac{1}{u} +C +\O(u), \ee
 where $C$ is a constant that depends on how far away the  observer is, and the precise definition of the function $u$.
   Eqn. (\ref{doffo}) can be justified
 by solving the geodesic equation for an outgoing null geodesic.  At the end of this section, we will explain a convenient choice of the function $u$,
 and a shortcut for doing the calculation.    For now, let us discuss the implications of eqn. (\ref{doffo}).
 
 Eqn. (\ref{doffo}) tells us that as $u\to 0^+$, the time $t$ at which the signal is received by a distant observer diverges, but only logarithmically.   Of course, this divergence
 is related
 to the fact that a signal that originates from behind the horizon -- say at $u<0$ -- will never reach the outside observer.  
 
 We can solve eqn. (\ref{doffo}) to express $u$ in terms of $t$.  Asymptotically for large $t$,
 \be\label{noffo} u = e^{C/4GM}  e^{-t/4GM}. \ee
 At late times, that is if $t$ is large, $u$ is exponentially small.    
 Therefore, late time measurements by the distant observer probe the quantum state at distances exponentially close to the horizon.   In quantum field theory,
 every state looks like the vacuum at short distances, so late time observations by the distant observer are in fact probing the vacuum state at exponentially small
 distances.   The distant observer does not need to wait terribly long before making observations that probe the vacuum at incredibly small distances. For example,
 for a black hole with the mass of the Sun, $4GM$ corresponds to a time of about $2\times 10^{-5}$ seconds, so every time the distant observer waits one second,
 $u$ becomes smaller by a factor $e^{-5\times 10^4}$.  Hence the observer need not wait very long (in human terms) to reach  the ``late time'' regime.

 Moreover,   $\frac{\d u}{\d t}$ is also exponentially small for large $t$,
 which means that a mode that reaches the observer at late times will have undergone an exponentially large redshift on its way.  
  A mode of any given energy $E$ that is observed at a sufficiently late time will have originated from a very high energy
 mode near the horizon.     Roughly
 speaking, a mode of very high energy propagates freely, along a radial null geodesic (such as the geodesics
 represented by the diagonal purple lines of
 fig. \ref{Two}).   If we assume this, we can get a very simple answer for what the distant observer will see.   The assumption is slightly
 oversimplified and a more precise story is explained in section \ref{graybody}.
 
 The distant observer probes the radiation emerging from the black hole by measuring a quantum field $\Psi$. 
 We assume that the distant observer measures $\Psi$ as a function of time $t$ and angular coordinates $\Omega$ at some fixed distance.
   A typical observable
 is a two-point function
 \be\label{dudd}\langle \Psi(\Omega,t)\Psi(\Omega', t')\rangle .\ee
 In a spherically symmetric Schwarzschild background, the field $\Psi$ can be expanded in partial waves.\footnote{Though this is far less obvious, a partial wave
 expansion is also possible in the field of a rotating (Kerr) black hole, using the fact that the usual wave equations in a Kerr geometry are separable.  That leads to a rather similar
 analysis for a rotating black hole.}   The coefficient of each partial wave
 is a $1+1$ dimensional quantum field (the two dimensions being  the distance from the horizon and the time).   
 In the real world, $\Psi$ would probably be a component of the electromagnetic field, which we could expand in vector spherical harmonics.
The ideas needed to understand this case are explained in section \ref{graybody}.   A more general derivation allowing for  arbitrary non-gravitational forces is
 explained in section \ref{euclidean}.
 
 However, we can understand the essence of
 Hawking's discovery by assuming  that a particular partial wave $\psi$ of the field $\Psi$ is, say, a  chiral free fermion in the $1+1$-dimensional sense. (We take $\psi$ to be a {\it chiral} free fermion
 because only the modes that propagate outwards, from the horizon to infinity, are relevant.)   
 A chiral free fermion in $1+1$ dimensions has dimension 1/2, and its two-point function in the vacuum is\footnote{The factor  $(\d u\,\d u')^{1/2}$  is  only a convenient shorthand to
incorporate  the fact that $\psi$ has dimension $1/2$.  Under a change of coordinates from $u$ to some other coordinate such as $t$, as $\psi$ has  dimension $1/2$, it
  transforms by a factor $(\d u/\d t)^{1/2}$. That is important in the derivation of the key result (\ref{nudd}) below.   
 Including the factor $(\d u\,\d u')^{1/2}$ in the formula for the correlation function, with the rule $(\d u)^{1/2}=({\d u/\d t})^{1/2} (\d t)^{1/2}$ for any other function $t$, is a way to build in this
 transformation (and plays no other role).   A more intrinsic description is that because $\psi$ has dimension $1/2$, its two-point function $\la \psi(u)\psi(u')\ra$  is best understood as a half-density 
 rather than a function in each variable $u$ and $u'$. We  make this explicit with the factor $(\d u\, \d u')^{1/2}$.}
 \be\label{indibo} \la \psi(u)\psi(u')\ra = \frac{(\d u\,\d u')^{1/2}}{u-u'}. \ee
In late time measurements of the radiation emitted by a black hole, $u$ and $u'$ are both exponentially small and therefore
 exponentially close to each other.   Since any state looks like the vacuum at sufficiently short distances, in discussing
 what an observer will see at late times,  we can replace $ \la \psi(u)\psi(u')\ra$ by its vacuum expectation value (\ref{indibo}). 
 
\begin{figure}
 \begin{center}
   \includegraphics[width=3.4in]{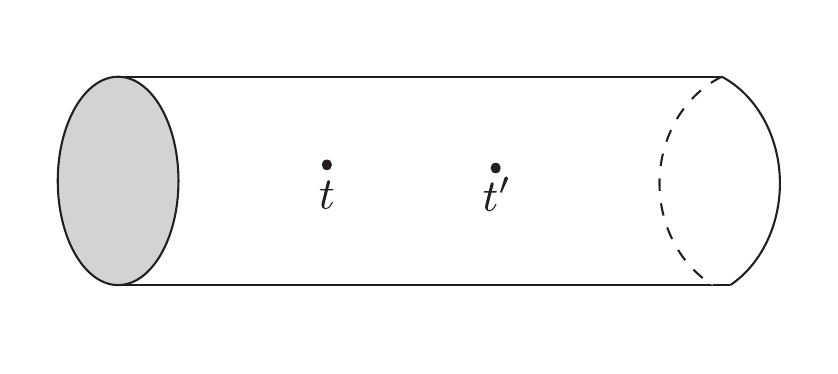}
 \end{center}
\caption{\footnotesize The thermodynamic limit of the thermal two-point function  $\frac{1}{Z} \Tr\, e^{-\beta_\sH  H} \psi(t)\psi(t')$ of a two-dimensional chiral free fermion $\psi$
can be computed by a Euclidean signature path integral on a cylinder of circumference $\beta_\sH$. 
The cylinder is defined by complexifying the time coordinate $t$ to a complex variable $z$, with $t=\rm{Re}\,z$; $z$ is defined to be a periodic variable with
 $z\cong z+\beta_\sH \i$.  The operator $\psi$ is inserted at the points $z=t$ and $z=t'$.
The action is the usual action
$\frac{1}{2\pi}\int \d^2z \psi\partial_{\bar z}\psi $ of a chiral free fermion, and 
 $\psi$ is taken to be antiperiodic in going around the cylinder. 
 In this representation,
antiperiodicity of the thermal correlation function under $t\to t+\beta_\sH \i$ is manifest, and it is also manifest that the only singularity of the two-point function is a simple pole
with residue 1 at $t=t'$.  These properties uniquely determine the answer. \label{Four}}
\end{figure}

 Setting $u=e^{C/4GM} e^{-t/4GM}$, we can turn eqn. (\ref{indibo}) into a formula for the two-point function measured by the
 distant observer at late times:
 \be\label{nudd} \la \psi(t)\psi(t')\ra =\frac{1}{4GM}\frac{(\d t \,\d t')^{1/2}}{e^{(t-t')/8GM}-e^{-(t-t')/8GM}}. \ee
 This is antiperiodic in imaginary time; in fact, it is odd under $t\to t+8\pi GM \i $.   Antiperiodicity with that period
 corresponds to a thermal correlation function\footnote{At temperature $T$,
 fermion correlation functions are antiperiodic under $t\to t+\i/T$, and boson correlation functions are periodic.  For a calculation similar to the one in the text with a bosonic field, one could
 consider a chiral current $J$, of dimension 1, with vacuum expectation value $\la J(u)J(u')\ra =\frac{\d u \, \d u'}{(u-u')^2}$.  Changing variables from $u$ to $t$, one finds
  periodicity of $\la J(t) J(t')\ra$ under $t\to t+8 \pi G M\i$.}   at a temperature $T_\sH=1/8\pi GM$, known as the Hawking temperature of the black hole.  In fact, the right hand side of eqn. (\ref{nudd}) is the two-point function of a chiral free fermion at temperature $T_\sH$.
 In other words it is the thermodynamic limit of  $\frac{1}{Z} \Tr\, e^{-\beta_\sH H} \psi(t)\psi(t')$, where $H$ is the Hamiltonian and $Z$ is the partition function of a chiral free fermion,
 and $\beta_\sH=1/T_\sH$.    
This statement can be verified by a standard textbook calculation, but such a calculation is not really necessary, since
this thermal two-point function is uniquely determined by the following facts: it is antiperiodic under $t\to t+8\pi GM \i$, and
 modulo this antiperiodicity, its only singularity is a simple pole at $t=t'$ with residue 1 (fig. \ref{Four}).

  Since the measurements of a distant observer at late times coincide with what one would expect in a thermal ensemble at temperature $T_\sH=1/8\pi GM$,
  we learn that 
 a black hole, after transients that depend on how it was created die down, radiates thermally at that  temperature.   This explains why Bekenstein had trouble making sense of the interaction of the black hole with photons of energy
  small compared to $1/8\pi GM$.   Such photons are strongly emitted by the black hole, with a large average occupation number in each outgoing mode,
  and in investigating the Generalized Second Law, one has to take into account the entropy increase due to that emission.
 
 We can also now confirm Bekenstein's formula for the entropy of the black hole, and explain how Hawking determined
 the overall constant in this formula.   We use the First Law of thermodynamics  
 \be\label{firstlaw} \d E=T\d S, \ee where the energy $E$ is the black hole mass $M$, and
  for a Schwarzschild black hole  $T=1/8\pi G M$.   Hence  $\d S=8 \pi G M \d M$ so (assuming that $S$ vanishes in the absence
 of a black hole, that is at $M=0$) $S=4\pi G M^2$.   The area of a Schwarzschild black hole is $A=16\pi G^2 M^2$ so the entropy
 is
 \be\label{zindo} S=\frac{A}{4G}.\ee
 This is how Hawking confirmed Bekenstein's ansatz and determined the overall normalization.
 
 If we do not set $\hbar=1$, then the Hawking temperature is actually $T_H=\frac{\hbar}{8\pi GM }$, showing  explicitly that the nonzero temperature is a quantum effect.
  
 In this explanation, we used Hawking's result for the black hole temperature and an assumption that the black hole really is a thermal system to which the First Law will apply to determine
 the entropy and recover the result $S=A/4G$.   Alternatively, if we assume the ansatz $A/4G$ for the entropy and Hawking's result for the temperature, we can read the same
 computation in reverse as a verification that the First Law does hold for Schwarzschild black holes.  Similarly, by a much more detailed analysis, one can show \cite{BCH} that 
 a rotating black with angular momentum $J$  satisfies a more general version of the First Law 
$\d E=T\d S+\Omega \d J$, where $\Omega$ is called the angular potential.
A useful reference  is section 12.5 of \cite{wald}, and a derivation based on a covariant description of gravitational phase space can be found in 
\cite{IW}.    The First Law is an important aspect of the consistency of black hole thermodynamics, and the reader is urged to explore it further, but we will not explain these derivations in
detail in the present article.
However, we briefly return to the First Law in section \ref{morevn}.
 
 One way to justify eqn. (\ref{doffo}) or equivalently (\ref{noffo})  is to introduce the Kruskal-Szekeres coordinates.  A standard definition is
 \begin{align}\label{kscoords} U&=-\left(\frac{r}{2GM}-1\right)^{1/2} e^{r/4GM} e^{-t/4GM} \cr
                                 V&=\left(\frac{r}{2GM}-1\right)^{1/2} e^{r/4GM} e^{t/4GM}.\end{align}
  In terms of these coordinates, the Schwarzschild metric is
  \be\label{ksschw} \d s^2=-\frac{32 G^3 M^3}{r} e^{-r/2GM} \,\d U \, \d V +r^2\d\Omega^2,\ee
  where $r$ is defined implicitly by
  \be\label{impdef} -UV =\left(\frac{r}{2GM}-1\right) e^{r/2GM}. \ee
  The most important application of the Kruskal-Szekeres coordinates is to describe the extension of the Schwarzschild geometry beyond the horizon at $r=2GM$.
   We will return to this in section \ref{thermofield}.   For now, we will just use these coordinates to justify eqn. (\ref{noffo}).   The form (\ref{ksschw}) of the metric
   shows that a radially outgoing or ingoing null geodesic\footnote{A radial  null geodesic is defined as one located at a fixed value of the polar angles, so that $\d\Omega^2=0$
   along such a geodesic. Hence $\d U \,\d V=0$ along such a geodesic and either $\d U=0$ or $\d V=0$.}  must satisfy $\d U=0$ or $\d V=0$, so in other words $U$ or $V$ is constant along such a geodesic.   More specifically, from the formulas
   (\ref{kscoords}), we see that $U$ is constant on an outgoing radial null geodesic and $V$ is constant on an ingoing one.  Eqn. (\ref{kscoords}) shows that $U$ vanishes
   at $r=2GM$ and is negative for $r>2GM$, so for a function that vanishes on the horizon and is positive outside, we can take $u=-U$.  Eqn. (\ref{kscoords}) then gives
   the claimed result $u= C' e^{-t/4GM}$, where $C'=\left(\frac{r}{2GM}-1\right)^{1/2} e^{r/4GM}$ is a constant that depends on the position of the observer and not on the time $t$ at which
   an observation is made.   Actually, it is natural to introduce the retarded time $t_\ret=t-r$ and write the formula for $u$ in the form
   \be\label{zimpdef}u = \left(\frac{r}{2GM}-1\right)^{1/2} e^{-t_\ret/4GM}.\ee   
    For a black hole with the mass of the sun, assuming that $r/2GM$ is large but not exponentially large, $u$ becomes exponentially small as soon as $t_\ret$
    is large compared to $4GM\approx 2\times 10^{-5}$ seconds.               
 
        \begin{figure}
 \begin{center}
   \includegraphics[width=2.5in]{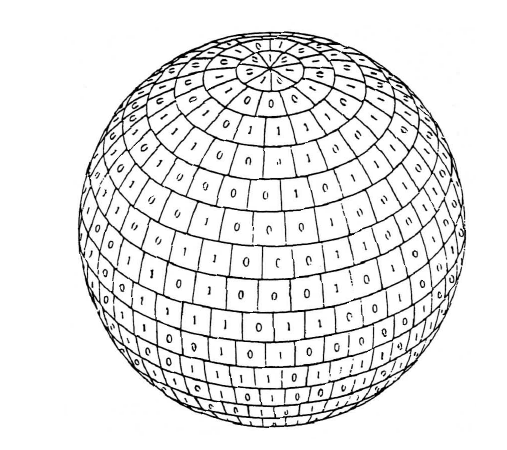}
 \end{center}
\caption{\footnotesize  A visualization by  Wheeler \cite{ItFromBit} of the black hole horizon made out of cells roughly of Planck area, with a bit or
qubit associated to each cell.  \label{Three}}
\end{figure} 
 
 An important detail in this derivation is that it is not necessary to begin the discussion at distances so small, or energies so high, that the laws of nature are unknown.
 We inferred what the distant observer will see at late times by starting with a knowledge of the short distance behavior of the quantum field.   But here, what do we mean by short
 distances?   It is enough that the hypersurface $\S$ is far enough in the past, relative to the observer, so that
the relevant signal originates on  $\S$ at a distance from the horizon that is  very small relative  to the Schwarzschild radius of the black hole; then the details of the black hole geometry do not affect
 the key correlation function $\la\psi(u)\psi(u')\ra$, which will coincide with the expectation value in vacuum.\footnote{The derivation  given here is in the spirit of Hawking's original
 work \cite{hawking}, with an improvement that possibly was first made by Fredenhagen and Haag \cite{FH}.   The improvement is to trace the measurements made by the observer
 not back to initial conditions at past null infinity, but only to initial conditions on a conveniently chosen hypersurface $\S$ that intersects the horizon outside the worldvolume of the collapsing
 star.   In this way, one avoids any discussion of physics at extremely high energies.}
   So for example, in the case of an astrophysical black hole, with a Schwarzschild radius of a few kilometers or more,
 the ``short distance'' scale could be a millimeter: extremely small compared to the size of the black hole, but not nearly small enough to probe the limits of our knowledge of physics.

 We conclude this section with some general remarks about black hole thermodynamics and Hawking radiation.
 Many researchers have thought that, somehow, the entropy $S=A/4G$ means that a black hole
 can be described by some sort of degrees of freedom that live on its horizon -- roughly, with one bit or qubit for every
Planck unit $G$ of area.   In a famous article \cite{ItFromBit},  Wheeler illustrated this idea as in fig. \ref{Three}.  
Even today it remains a challenge to properly justify and understand this picture.

The Hawking temperature $T_\sH=1/8\pi GM$ can be expressed in terms of the Schwarzschild radius $r_S=2GM$ as
\be\label{wiggly} T_\sH=\frac{1}{4\pi r_S}. \ee
Thus the Hawking temperature is of order $1/r_S$, and a typical massless particle emitted by the black hole has a wavelength, measured
at infinity, of order $r_S$ and an energy of order $1/r_S$. 

Energy loss by a radiating  astrophysical black hole is extremely slow.   The total luminosity of a radiating body of surface  area $A$
and temperature $T$ is of order $AT^4$, which in the case of a black hole is a multiple of $1/G^2 M^2$.   Thus in order of magnitude the rate of energy
loss by a radiating astrophysical black hole is
\be\label{enloss} \frac{\d M}{\d t}\sim \frac{1}{G^2 M^2}. \ee
With real world assumptions about the particles emitted by the black hole -- mainly photons and gravitons -- the constant of proportionality in this
relation was computed by Page \cite{Page}.   (This calculation requires understanding the gray body factors, which we introduce in section \ref{graybody}.) 
Following Hawking, energy loss by a radiating black hole is called black hole evaporation.   Eqn. (\ref{enloss}) shows
that evaporation of a black hole with a typical astrophysical mass is a very slow process.   The time for a  solar mass black hole in vacuum to evaporate away
a significant part of its mass is of order $10^{67}$ years.  Of course, in the real world, an astrophysical black hole is not in vacuum and is more
likely to accrete mass than to evaporate.

Since Hawking's approximation of considering a quantum field in a given spacetime background is valid as long as the black hole is
much heavier than the Planck mass, a solar 
mass black hole in vacuum, with initial mass of order $10^{33}$ grams, will shrink to a microscopic size before 
Hawking's analysis breaks down.   We do not really know what happens at that point, but we presume that eventually
the evaporation ends and only stable elementary particles remain.

A fundamental point about Hawking radiation is that the radiation appears to be thermal even though the black hole
could have formed from a pure state.   This has presented a puzzle that drives much of the research in this field and that even today
is only partly resolved.  Hawking's approximations are valid for almost the whole evaporation process and seem to show
that the outgoing state is thermal, ultimately with a very large entropy of order the total number of photons emitted during the evaporation, which is roughly
$M/T_\sH\sim GM^2$ (or about $10^{76}$ for
a solar mass black hole).   But if the formation and evaporation of the black hole are described by the ordinary laws of quantum mechanics,
then if the initial state is pure, the final state should also be pure.

Concretely, the reason that the Hawking radiation seems to be thermal even if the black hole is in a pure state is that the
observations of the distant observer amount to observing the quantum fields only outside the horizon.
Even if a black hole formed from a pure state -- so that we can assume that the state of the whole universe is pure -- the
quantum fields restricted to only part of spacetime are in a mixed state.   That is the essence of the Hawking effect.   We will return to this in section \ref{division}.

 \section{Gray Body Factors}\label{graybody}
 
 In section \ref{bhevaporation}, we assumed, in effect, that a signal emitted from the horizon propagates freely to the distant observer.
 This is oversimplified, since in general, there is a sort of angular momentum barrier around the black hole, as we will see, and an outgoing
 signal might be reflected back towards the horizon.   The derivation of section \ref{bhevaporation} gives a quick way to understand
 the essence of Hawking's discovery,  but here we will give a more precise explanation.
 
 First of all, it is possible in 3+1 dimensions to have a semirealistic model that leads precisely to the analysis in section \ref{bhevaporation}.
 For this, we consider a magnetically charged black hole and a massless electrically charged fermion field $\Psi$ interacting with the black hole.   The partial wave
 of $\Psi$ of lowest possible angular momentum is a massless fermion in the $1+1$-dimensional sense, and its outgoing (chiral) component
 has precisely the properties assumed in section \ref{bhevaporation}.   See \cite{Malda} for a study of such models.   However, more typically,
 as we will see, there is a potential barrier outside the black hole (even for angular momentum zero) and matters are not as simple as assumed in 
 section \ref{bhevaporation}.
 
 \subsection{The Potential Barrier}\label{potential}
 
 For simplicity, we will consider a massless scalar field $\phi$  in the presence of the black hole.    In the real world, it would be more realistic
 to consider the electromagnetic field or the gravitational field.   This would lead to very similar considerations, except that we would have to make a more complicated partial wave
 expansion using vector or tensor spherical harmonics. 
 
 We assume that $\phi$ interacts with gravity only, with minimal coupling via the
 action $-\frac{1}{2}\int \d^4 x \sqrt g  \,g^{\mu\nu}\partial_\mu\phi \partial_\nu\phi.$ 
 In a Schwarzschild background, the action for a mode of angular momentum $l$ is then
 \be\label{zonkyx} I= \int\d t\, \d r \left(\frac{r^2}{2}\frac{1}{1-\frac{2GM}{r}}\left(\frac{\d\phi}{\d t}\right)^2 -\frac{r^2}{2}\left(1-\frac{2GM}{r}\right)\left(\frac{\d \phi}{\d r}\right)^2
 -\frac{l(l+1)}{2}\phi^2\right). \ee  It is convenient to introduce the ``tortoise coordinate'' $r_*=r+2GM\log(r-2GM)$, which satisfies
 $\d  r =\d r_* \left(1-\frac{2GM}{r}\right)$, and ranges over the whole
 real line $-\infty<r_*<\infty$ for $2GM<r<\infty$.   The action becomes 
  \be\label{wonkyx} I= \int\d t\, \d r_* \left(\frac{r^2}{2}\left(\frac{\d\phi}{\d t}\right)^2 -\frac{r^2}{2}\left(\frac{\d \phi}{\d r_*}\right)^2
 -\left(1-\frac{2GM}{r}\right)\frac{l(l+1)}{2}\phi^2\right). \ee
 Setting $\phi=\sigma/r$ and integrating by parts, we get
 \be\label{monky} I=\int \d t\, \d r_*\left( \frac{1}{2}\left(\frac{\d\sigma}{\d t}\right)^2-\frac{1}{2}\left(\frac{\d \sigma}{\d r_*}\right)^2  -\left(1-\frac{2GM}{r}\right)\left(\frac{l(l+1)}{2r^2} +\frac{GM}{r^3}\right)\sigma^2\right). \ee
 In other words, $\sigma$ is effectively a free massless scalar  propagating in  an effective two-dimensional Minkowski space with line element  $-\d t^2+\d r_*^2$ and interacting  with the effective potential
 \be\label{onky} V_\eff=\left(1-\frac{2GM}{r}\right)\left(\frac{l(l+1)}{r^2} +\frac{2GM}{r^3}\right).\ee   This effective potential is positive definite, vanishing near the horizon
and at infinity, with a barrier in between.
  Notably, even if $l=0$, there is a nontrivial effective potential, namely
 \be\label{winky}V_\eff=\frac{2GM}{r^3}-\frac{4(GM)^2}{r^4}.\ee
  The maximum of this potential is at $r_\max=\frac{8}{ 3 }GM$, and the value of the potential at the maximum is $V_\max(l=0)= \frac{27}{1024(GM)^2}$.   
  For $l>0$, the maximum of the potential is greater.   For large $l$, the maximum is approximately at $r=3GM$, and the maximum value of the potential
  is $V_\max(l)\cong \frac{l(l+1)}{27 (GM)^2}$.   
  
  To get from the horizon to infinity, a wave will have to propagate over the potential barrier.   
  Only a wave whose energy 
 is much greater than $\sqrt{V_\max(l)}$   will propagate almost freely\footnote{Such a wave still experiences a phase shift or time delay, but this is not noticeable to the distant
 observer, who does not know when the signal was emitted.}
 from the horizon at $r_*=-\infty$ across the barrier to $r_*=+\infty$.   
 So our previous calculation is good for the $l=0$ mode if the Hawking radiation from the black hole is being observed
 at frequencies much above the Hawking temperature $T_\sH=\frac{1}{8\pi GM}\sim \sqrt{V_\max(l=0)}$. For  $l>0$, the previous calculation is good at frequencies
 much above $l T_H$.   At any given frequency, the analysis of section \ref{bhevaporation}  is completely wrong if $l$ is sufficiently large, because a wave of any given
 energy is far below the barrier if $l$ is sufficiently large.
 
An outgoing mode from the horizon at $r_*=-\infty$ might be scattered back into the black hole by the potential and reabsorbed,  or it might be transmitted across  the barrier
to $r_*=+\infty$.   Heuristically,  our calculation
 in section \ref{bhevaporation} should be modified accordingly; the probability to observe an outgoing particle near  $r_*=\infty$ should be reduced by the
 transmission probability across the barrier.

 Before giving a technical justification of this claim, we will first explain the implications.   Suppose that it is true that the black hole has a temperature $T_\sH$.   Then
 we expect that it can be in equilibrium with a thermal gas at that temperature \cite{Equil1,Equil2}.   
 In equilibrium with such a gas, the black hole is absorbing thermal radiation at temperature $T_\sH$;
 equilibrium can potentially be maintained because the black hole is also emitting thermal radiation at the same temperature.   In the simplest situation,  assume that the
 effective potential vanishes for some partial wave.  (As already noted, this only happens for an electrically charged massless fermion interacting with a magnetically charged black hole.)
 Then the black hole will absorb all of the incident radiation in that partial wave.   Equilibrium is maintained because, as analyzed in section \ref{bhevaporation}, in the absence of a  potential
 barrier, the black hole is also freely emitting thermal radiation at temperature $T_\sH$.    Now consider a more realistic situation with $V_\eff\not=0$.   Not all incident radiation is absorbed.
 In a given partial wave at a given energy, the absorption probability is reduced by a factor equal to the transmission probability from right to left (that is, from $r_*=+\infty$ to $r_*=-\infty$).   To maintain
 equilibrium, the emission probability must be reduced by the same factor.  However, as will be explained in a moment,  the transmission probability through the
 barrier from left to right is the same as the transmission probability from right to left.   So equilibrium of the black hole with thermal radiation at temperature $T_\sH$ is possible if the
 emission from the black hole is reduced relative to our previous result by a factor of the transmission probability from left to right through the barrier.
 
 The statement that the transmission probability from left to right equals that from right to left can be proved as follows.
 A solution of the Klein-Gordon equation for $\sigma$ with frequency $\omega$ has the form  $\sigma(r_*,t)=\lambda(r_*) e^{-\i \omega t}$, with
  \be\label{theq}\left( -\frac{\d^2}{\d r_*^2} +V_\eff(r_*)\right) \lambda(r_*)=\omega^2 \lambda(r^*). \ee
 A solution $\lambda_\omega$  that describes the scattering of a wave incident from the left has the asymptotic behavior
 \be\label{celok} \lambda_\omega(r^*)\sim \begin{cases} e^{\i \omega r_*}+R(\omega) e^{-\i \omega r_*} & r^*\to -\infty \cr
                                                                                            T(\omega) e^{\i \omega r_*} & r^*\to +\infty.\end{cases}\ee
 $T(\omega)$ and $R(\omega)$ are the transmission and reflection amplitudes for a wave  of frequency $\omega$ incident from the left.
 A solution $\t\lambda_\omega$ that describes the scattering of a wave incident from the right similarly has the asymptotic behavior
  \be\label{delok} \t\lambda_\omega(r^*)\sim \begin{cases} \t T(\omega) e^{-\i \omega r_*} & r^*\to -\infty \cr
                                                                                           e^{-\i \omega r_*} + \t R(\omega) e^{\i \omega r_*}  & r^*\to +\infty.\end{cases}\ee
   where $\t T(\omega)$ and $\t R(\omega)$ are transmission and reflection amplitudes
                                                                                           for a wave incident from the right.
   Since $\lambda_\omega$ and $\t\lambda_\omega$ satisfy the same equation (\ref{theq}), the Wronskian                                                                                        
 $\lambda_\omega \frac{ \overset{\leftrightarrow}{\,\d\,}}{\d r_*}\t \lambda_\omega$ is independent of $r_*$.  Comparing the values at $r_*\to \pm\infty$,
 we get the claimed result $T(\omega)=\t T(\omega)$.
 
 \subsection{More Detailed Argument}\label{detailed}
 
       \begin{figure}
 \begin{center}
   \includegraphics[width=4in]{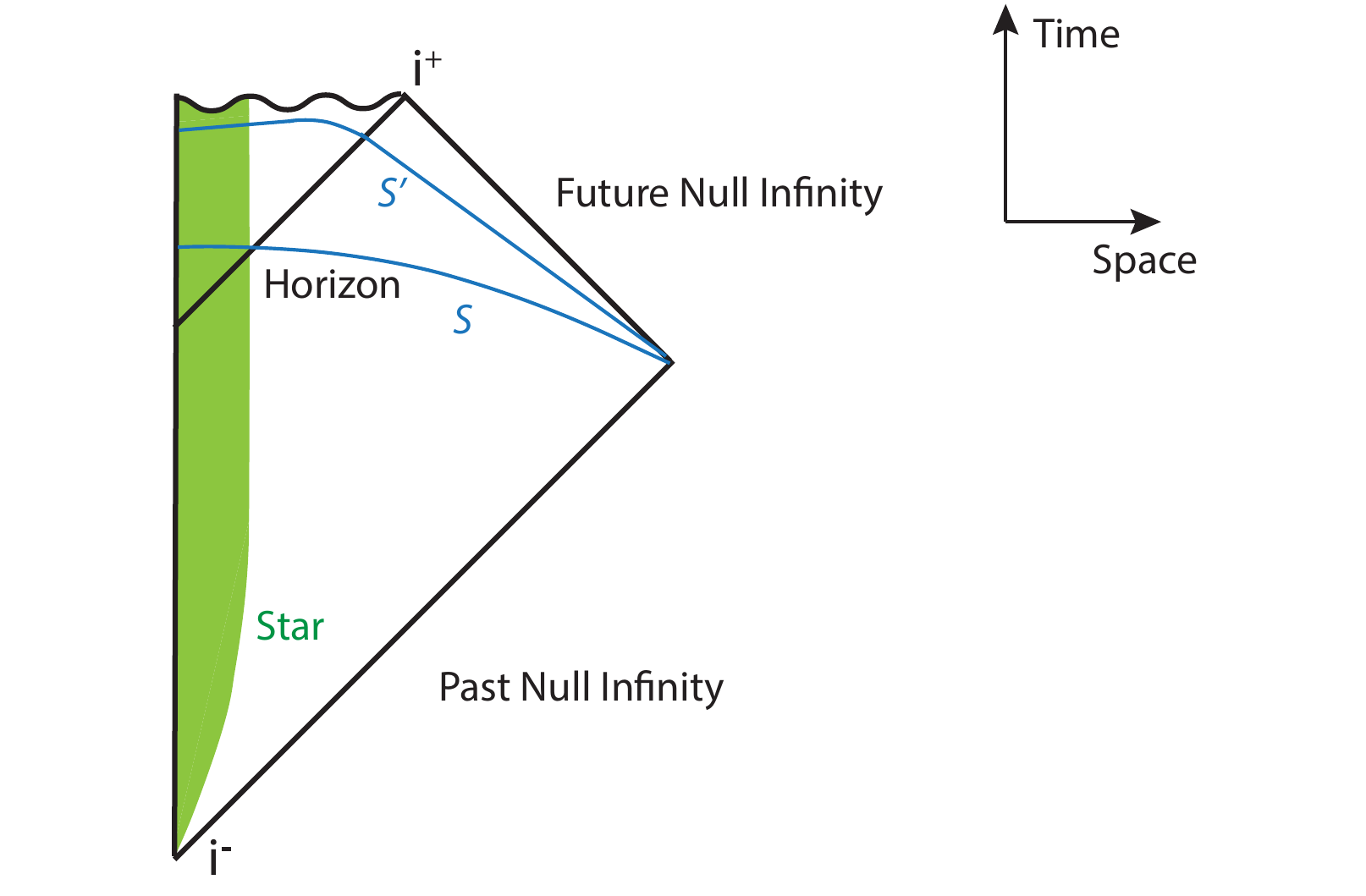}
 \end{center}
\caption{\footnotesize  This picture illustrates a more detailed derivation of the Hawking process.   $\S'$ is a late time Cauchy hypersurface on which a measurement
will be made.     The operator that will be 
measured is $W^\dagger W$, where $W$ is a linear function of a quantum field $\sigma$ and its time derivative $\dot\sigma$ on the surface $\S'$.   
By solving the wave equation backwards in time, starting with ``final data'' on
$\S'$, one can obtain an expression for $W$ in terms of $\sigma$ and $\dot\sigma$ on an earlier
hypersurface $\S$. This leads to a corrected prediction for  black hole radiation that incorporates the interaction of the outgoing radiation with the gravitational field
of the black hole.  \label{Detailedx}}
\end{figure}

 Now, following \cite{hawking,FH}, we will aim for a more technical justification of the
  claim that the thermal radiation rate found in section \ref{bhevaporation} must be multiplied by a factor $|T(\omega)|^2$.
 The late time observer measures, for example, an operator $W^\dagger W$, where $W$ is an operator linear in the field $\sigma$.   Using the field equations, any such
 operator can be expressed in terms of the field $\sigma$ and its time derivative on an arbitrary initial value surface $\S'$:
 \be\label{whatabout} W=\int_{\S'} \d^3 x\sqrt h \left( a(x)\sigma(x) +b(x)\dot \sigma(x)\right),  \ee
 where $a$ and $b$ are functions on $\S'$.
 Here $h$ is the induced metric of the Cauchy hypersurface $\S'$, and $\dot\sigma$ is the derivative of $\sigma$ in the direction normal to $\S'$.
 Though such an expression defines a field operator for any choice of the hypersurface $\S'$ and the functions $a$ and $b$, in order to define
 an operator $W$ that is related in a simple way to the observation that is going to made, it is useful to pick the hypersurface $\S'$ to pass through the detector
 at approximately the time when the measurement will be made, and to choose the functions $a$ and $b$ to be supported near the detector.   In particular,
 the functions $a$ and $b$ are then supported entirely outside the horizon.   It does not matter very much how the  hypersurface $\S'$ behaves away from the detector,
 but it is natural to choose it, as in fig. \ref{Detailedx}, to be everywhere well to the  future of the hypersurface $\S$ of fig. \ref{Two} that was used in the derivation of section \ref{bhevaporation}.
 Assuming that the observer wishes to measure the radiation flux at the angular position of the detector as a function of energy and time,  it is convenient to pick the
 functions $a$ and $b$ to be functions of the distance $r$ from the black hole and to be approximately proportional to $e^{\i \omega r}$, for some $\omega$.   Of course,
 since $a$ and $b$ are supposed to be localized near the detector, they cannot have precisely this exponential form, but they can be supported in a very small range of $\omega$.
 With a little more care with the functions $a$ and $b$,
 we can ensure that $W$ is (very nearly) an annihilation operator for modes of frequency $\omega$, while $W^\dagger$ is a creation operator; thus $W^\dagger W$ is
 a number operator and a measurement of this operator  will reveal the flux of particles at energy $\omega$ at the angular position of the detector.\footnote{To get information about the angular momentum of the emitted particles, one would need to be able to measure interference between events involving absorption of a particle
 at different angular positions.     For example, one could in principle measure $W^\dagger W'+W'^\dagger W$, where $W'$ is
defined like $W$ but for a detector at a different angular position.}
 
 A different representation of the operator $W$ is convenient.   One can view $a$ and $b$ as initial conditions for a solution $f$ of the Klein-Gordon equation $\sum_{\mu=0}^3 D_\mu D^\mu f=0$:
 \be\label{doffox} f|_{\S'}= b(x),~~~ \dot f|_{\S'}=-a(x).  \ee
We can then alternatively write
 \be\label{natabout} W=\int_{\S'} \d \Sigma^\mu \, f \overset{\leftrightarrow}{\partial}_\mu \sigma, \ee
 where $\d \Sigma^\mu$ is the surface element associated to $\S'$.
 The point of this is that since $f$ and $\sigma$ obey the same Klein-Gordon equation $D_\mu D^\mu f=D_\mu D^\mu \sigma=0$, the quantity
$ f \overset{\leftrightarrow}{\partial}_\mu \sigma$ is a conserved current, $D_\mu\left(f \overset{\leftrightarrow}{\partial}{}^\mu \sigma\right)=0$,
and therefore we can define the same operator $W$ by an expression of the same form as eqn. (\ref{natabout}), but with $\S'$ replaced by any other Cauchy hypersurface.
In particular, we can write such a formula with $\S'$ replaced by the hypersurface $\S$ that was used in the derivation in section \ref{bhevaporation}:
\be\label{zatabout}W = \int_{\S} \d \Sigma^\mu \, f \overset{\leftrightarrow}{\partial}_\mu \sigma. \ee

But what does $f$ look like on the hypersurface $\S$?   To find out, we have to start with the initial (or final?) data of eqn. (\ref{doffox}) on the hypersurface $\S'$ and solve the Klein-Gordon
equation backwards in time to find the solution on $\S$.   One general fact is that since the support of $f$ on Cauchy hypersurface $\S'$ was entirely outside the horizon, it follows
that everywhere to the past of $\S'$, $f$ is supported outside the horizon.  In particular then, that is true on $\S$.   More specifically, on the hypersurface $\S'$, $f$ is (very nearly) an incoming wave of frequency 
$\omega$. When we integrate the Klein-Gordon equation backwards in time, $f$ will propagate in towards the black hole until it meets the same potential barrier that we have already
discussed, by which it will be partly reflected and partly transmitted.   The reflected part of the wave will return  -- as we go back in time -- back towards $r=\infty$, and the
transmitted wave will continue to the near horizon region of $\S$.  

 Let $T'$ and $R'$ be the transmission and reflection amplitudes in the time-reversed scattering problem.   Then eqn. (\ref{zatabout}) exhibits $W$ as the sum of $T'$ times a near
 horizon operator and $R'$ times an operator in the Minkowski vacuum near $r=\infty$.     When the
 observer measures $\la W^\dagger W\ra$, the ``long distance'' contributions proportional to $R'$ or $\bar R'$ do not contribute, because the long distance
 operators that appear have vanishing expectation in the Minkowski vacuum
  (the part of $W^\dagger W$ that is proportional to $|R'|^2$ is a number operator whose expectation value in the Minkowski vacuum
 vanishes, and the terms proportional to $T' \bar R'$ or $\bar T' R'$ vanish because an operator linear in $\sigma$ likewise has vanishing expectation value in the Minkowski vacuum).
 So the relevant part of $W^\dagger W$ is just  $|T'|^2$ times the same near horizon operator that we would have if there were no reflection from the barrier.
 
 $T'$ and $R'$ are transmission and reflection amplitudes in a time-reversed version of the scattering problem that we studied earlier.   Time-reversal has the effect
 of complex conjugating the scattering amplitudes, so in particular $T'$ is just the complex conjugate of the amplitude $T$ for transmission through the barrier as defined
 earlier.   So  $W^\dagger W$ is just $|T|^2$ times what it would be if there were no potential barrier, as implicitly assumed in the simple derivation in section
 \ref{bhevaporation}.     Therefore, as claimed, the emission rate from the black
 hole in a given partial wave at frequency $\omega$ is $|T(\omega)|^2$ times the thermal emission rate in the given mode at the Hawking temperature.

These results and extensions of them indicate that in the far future, the quantum fields outside the black hole horizon are in a universal state, known as the Unruh state \cite{U}, 
that does not depend on the details of how the black hole formed.    

 The suppression of the Hawking radiation by what are usually called gray body factors is actually essential for enabling the Hawking process to make sense.
 If the simple derivation of section \ref{bhevaporation} were valid in every partial wave, then we would expect the same thermal emission rate in each partial wave, and, as
 there are infinitely  many partial waves, we would predict an infinite luminosity for the Hawking process.   Instead, because $|T|^2$ vanishes rapidly with increasing $l$,
 the emission is dominated by the first few partial waves and the total luminosity has the order of magnitude claimed in eqn. (\ref{enloss}).
 
 \subsection{Thermodynamic Instability}\label{unstable}
 
 Since thermodynamic equilibrium between a black hole and a thermal gas played a role in motivating this discussion, it probably is time to  point  out that in an asymptotically flat spacetime, 
 once gravitational
 back reaction is taken into account, such equilibrium is actually unstable.   In a sense, there are two reasons for this instability.   The first has to do with the thermodynamics
 of the black hole.   Consider a black hole of mass $M$ that is in equilibrium with a thermal gas at the appropriate Hawking temperature $T=1/8\pi GM$.   Now, consider
 a thermal fluctuation in which the black hole emits a few more particles than it absorbs.  As a result, the black hole mass is reduced, and because the Hawking temperature is
 inversely proportional to the mass, the black hole becomes hotter.   Since the black hole is now hotter than its surroundings, it will now with very high probability emit
 more than it absorbs, and continue to lose mass.   Thus there is a runaway instability that will cause the black hole to disappear.   Conversely, an upward fluctuation in the black hole
 mass will cause the black hole to become cooler and emit less; then it will absorb more than it emits and its mass will grow further, without limit.
 
 This instability reflects the fact  that a black hole in an asymptotically flat spacetime has a negative specific heat.   In general the specific heat of a body of energy $E$ and temperature $T$
 is defined as $C=\frac{\d E}{\d T}$, and thermodynamic stability requires $C\geq 0$.  The black hole with $E=M$ and $T=1/8\pi GM$ has
 \be\label{gofo} C=-8\pi GM^2<0,\ee
 showing the instability.  
 
 The second source of instability is that in fact, in the presence of gravity, it is not possible to have a thermal gas filling an asymptotically flat spacetime.   Consider a portion of space of radius $R$
 filled by a thermal gas of temperature $T$.   The energy density of the gas is of order $T^4$ and its total energy is of order $T^4 R^3$.   The Schwarzschild radius of a body of
 that mass is of order $G T^4 R^3$, so  a  thermal gas filling a region of radius $R$ with $R\gtrsim G T^4 R^3$ will collapse to a black hole.  In other words, a thermal gas
 of temperature $T$ in an asymptotically flat spacetime can at most occupy a region of size
 \be\label{zondy} R\sim \frac{1}{{\sqrt G} T^2}\ee   without collapse to a black hole.  This instability was discussed in \cite{GPY}.
 
 Given this, what is the sense of the argument that we gave concerning equilibrium between a black hole and a thermal gas?   The answer to this question is that, as the derivation
 of the Hawking effect is based on quantum field theory in a fixed spacetime background, the whole analysis is asymptotically valid in the limit $G\to 0$.   As $G\to 0$, 
 the various instabilities that we have mentioned turn off.  
  To see this most clearly, we have to decide what we want to keep fixed as $G\to 0$.
 There is a factor of $G$ in the relation $r_S=2GM$ between the Schwarzschild radius $r_S$ and the mass $M$, so we cannot keep them both fixed as $G\to 0$.   In studying
 black holes, it is more natural to keep $r_S$ fixed as $G\to 0$.  Indeed, in section \ref{bhevaporation}, we worked in a limit $G\to 0$, since we did not consider loop effects due
 to quantum gravity fluctuations, and we worked in a definite spacetime that had a limit as $G\to 0$, which corresponds to keeping fixed $r_S$ rather than $M$ as $G\to 0$.
 It is natural, then, to express the Hawking temperature in terms of $r_S$ rather than $M$:  $T_\sH=\frac{1}{4\pi r_S}.$   Thus in the limit $G\to 0$, the Hawking temperature is fixed
 and the black hole mass $M=r_S/2G$ diverges. 
 
 With this in mind, let us re-examine the upper bound (\ref{zondy}) for the maximum size $R$ of a region of thermal radiation at temperature $T$ that does not collapse to
 a black hole.   
 Setting $T$ to be the Hawking temperature $T_\sH=1/4\pi r_S$,
 we find that the upper bound $R\leq 16\pi^2 r_S^2/\sqrt G$ is of order $1/\sqrt G$ as $G\to 0$ with fixed $r_S$.  
  If the maximum allowed $R$ is much greater than $r_S$  (the upper bound on $R/r_S$ is of order $10^{38}$ for a solar mass black hole), then although the black hole cannot
 be sensibly embedded in a thermal gas that fills all of an asymptotically flat space, it can be embedded in a thermal gas that fills an enormously large volume.
 
 The instability involving a fluctuation in the mass of the black hole also turns off as $G\to 0$.   When a statistical fluctuation occurs and the black hole emits a few more particles than it absorbs, its mass drops by an amount
 of order 1 and its temperature increases  by an amount of order $G$.   This will indeed cause the black hole to continue to emit more than it absorbs, but only by an amount of order $G$,
 and the time scale for the resulting instability to significantly change the black hole mass and the associated geometry is of order $1/G^2$.   
 
 Thus although a black hole in an asymptotically flat spacetime cannot be in perfect equilibrium with a thermal gas filling all of spacetime, one can come exceedingly close to this -- arbitrarily close
 as $G\to 0$.   Actually,  the instabilities that we have considered here can be eliminating by introducing a small negative cosmological constant and replacing
 an asymptotically flat spacetime by one that is asymptotic to Anti de Sitter space.  This regularizes the thermal gas, making it possible to have a thermal gas that fills all space.
 And a black hole of sufficiently large mass in asymptotically Anti de Sitter spacetime has positive specific heat \cite{HawkingPage}.  So in asymptotically Anti de Sitter spacetime,
 one can have perfect equilibrium between a black hole and ambient radiation.  We will consider this case in sections \ref{negative} and  \ref{rbh}, because it provides a setting in which the quantum corrections
to a black hole equilibrium state can be studied seriously.
 It is worth mentioning, however, that in this situation, most of the entropy is in the black hole rather than the radiation; the
 stable state consists of a large black hole and, relatively speaking, not very much radiation.

\section{Thermodynamics of Rindler Space}\label{division}

\subsection{Making The Cut}\label{cut}

The essence of the Hawking effect is that even if the whole universe is in a pure state, the portion of the universe outside the horizon, which is what an outside observer
can see, is in a mixed state, which moreover has thermal properties.

The basic phenomenon actually arises in a simpler context.  There is a similar thermal behavior if one studies the vacuum state in Minkowski space from the
vantage point of an observer who only makes measurements in a Rindler wedge.\footnote{An analysis  along  the following lines was originally made by Unruh and Weiss \cite{UW}. The thermal
nature of Rindler space was first seen by other arguments \cite{U} that we come to later.}   A Rindler wedge is defined
by an inequality such as $x>|t|$ or $x<-|t|$, where in some Lorentz frame $t$ is the time and $x$ is one of the spatial coordinates.

However, we will begin the analysis with a Hamiltonian description on a spatial slice.
A quantum state can be defined on any Cauchy hypersurface $\S$, say the surface $t=0$ in Minkowski space.    For simplicity, consider the theory of a single real scalar field $\phi$.
A  quantum state defined on the hypersurface $\S$ can be regarded as a functional $\Psi(\phi(\vec x))$.  In what follows, we will study the vacuum state $\Omega$.
We  decompose the spatial coordinates $\vec x$ as $(x,\vec y)$,
where $x$ is one of the spatial coordinates and the others are combined into $\vec y$.

  \begin{figure}
 \begin{center}
   \includegraphics[width=4.8in]{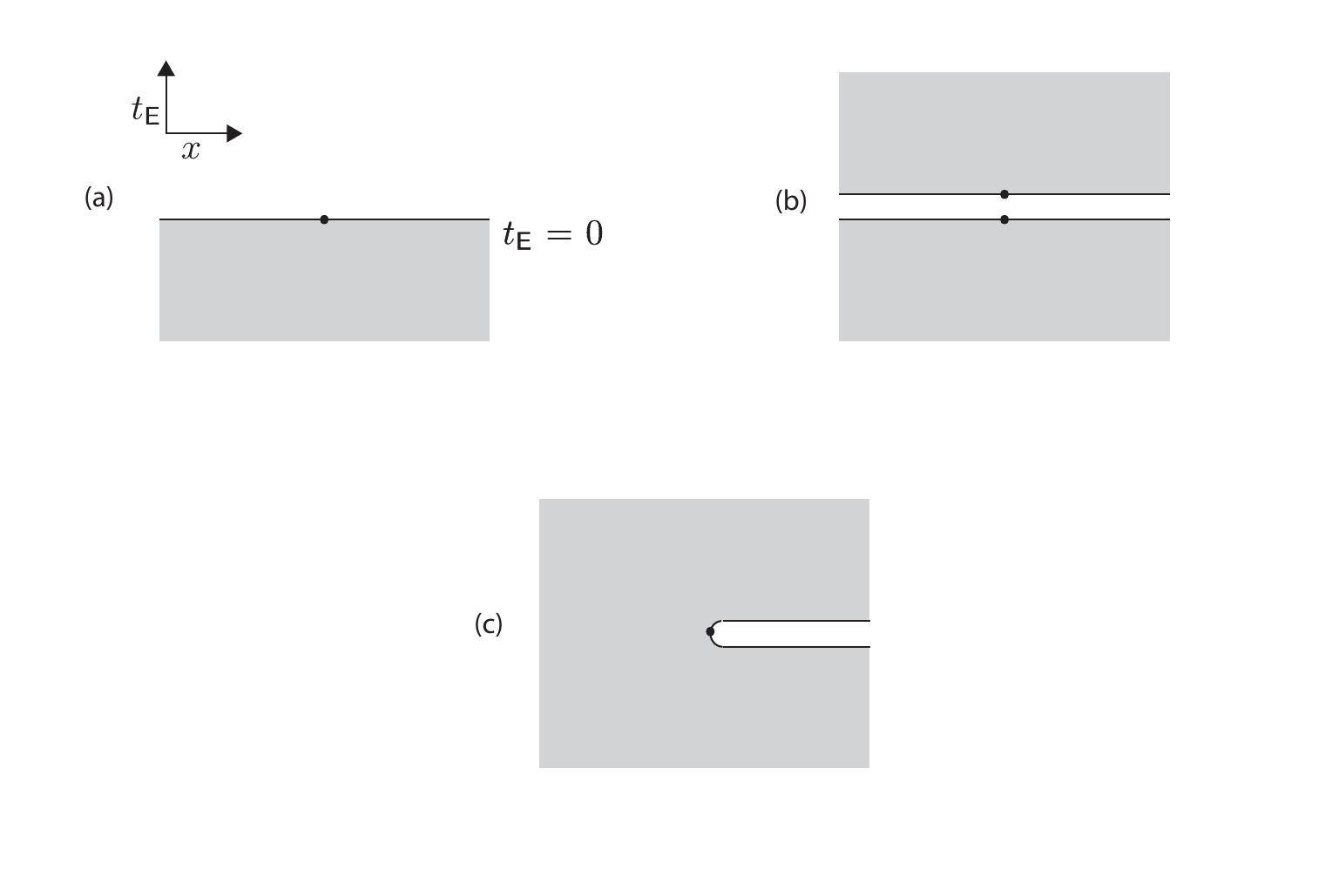}
 \end{center}
\caption{\footnotesize (a) A Euclidean path integral on the half-space $\tE\leq 0$ 
that prepares the vacuum state $|\Omega\ra$. (b) To prepare the pure state density matrix $|\Omega\ra\la\Omega|$ associated to the vacuum, we prepare the ket $|\Omega\ra$
as just described by a path integral on the half-plane $\tE\leq 0$, and
we use a similar path integral on the half-space $\tE\geq 0$ to prepare the bra $\la\Omega|$. Two copies of the $x$ axis, appearing respectively as the boundary of the lower and upper
half-planes,  have been separated for visibility.  (c) To construct the reduced density matrix $\rho$ of the half-line $x>0$, we ``trace out'' the quantum fields
in the region $x<0$. In path integrals, this is accomplished by gluing together the two copies of the negative $x$ axis.   The result is a path integral on Euclidean space with a cut along $\te=0$, $x>0$.   The density matrix $\rho$ is a functional
of the fields just above and just below the cut.  \label{Five}}
\end{figure}

The vacuum state $\Omega$ can be computed by a path integral on a half-space in Euclidean signature.   We set $t=-\i\te$ and integrate over the field $\phi(\te,\vec x)$  restricted to
the half-space $\te<0$, keeping fixed the boundary values $\phi(\vec x)$ at $\te=0$.  This integral, as a function of $\phi(\vec x)$, defines the vacuum wavefunctional $\Omega(\phi(\vec x))$
(fig. \ref{Five}(a)).

The projection operator onto the vacuum state is $\rho= |\Omega\ra \la\Omega|$. This can also be regarded as the density matrix associated to the pure state $\Omega$.   It is straightforward to construct
$|\Omega\ra\la\Omega|$ via path integrals.   First of all, just as we constructed the ket $|\Omega\ra$ by a Euclidean path integral on the lower half space $\te\leq 0$, we can construct the bra $\la\Omega|$ by a similar
Euclidean path integral on the upper half space $\te\geq 0$ (fig. \ref{Five}(b)).    We can thus view the pure state density matrix $\rho$ as a function of pairs of boundary values
\be\label{gurky}\rho(\phi;\phi') =|\Omega(\phi)\ra\la\Omega(\phi')|. \ee
Here $\phi$ is the boundary value of $\phi(\te,\vec x)$ on the upper boundary of the lower half plane, and $\phi'$ is the boundary value of $\phi(\te,\vec x)$ on the lower
boundary of the upper half plane.

Now suppose that we divide the $t=0$ surface $\S$ into the partial Cauchy surfaces $\S_r$ with $x\geq 0$ and $\S_\ell $ with $x\leq 0$.  Corresponding to this, we decompose\footnote{\label{notable}
At many points in discussing Rindler space and the Unruh effect, we make  statements that are oversimplified mathematically.   For example, it is not actually true in continuum quantum field theory that the Hilbert space has a factorization $\H=\H_\ell\otimes \H_r$ (nor is it a direct sum or integral of subspaces
with such a factorization); for a partial explanation of this fact, see the last paragraph of section
\ref{sample}.  The main difficulty here and in other statements involves short distance fluctuations near $x=0$.  Of course, with a lattice cutoff, there is such a factorization, but this spoils Lorentz invariance, which will be important in the derivation.  We also ignore the fact that $\phi_\ell$ and $\phi_r$ coincide
at $t=x=0$.   There is actually a rigorous
approach to the main results that we will obtain about Rindler space.
That approach uses
 Tomita-Takesaki theory, as applied to Rindler space  by Bisognano and Wichmann  \cite{BW} and extended to the black hole context  by Sewell 
  \cite{Sewell}.   For a gentle introduction, see \cite{Notes}, section 5.   The informal arguments
given in the text, however, are highly intuitive and arrive quickly at some important results.} 
 the field $\phi(\vec x)$
as a pair $(\phi_\ell,\phi_r)$ where $\phi_\ell$ is the restriction of $\phi$ to $\S_\ell$ and $\phi_r$ is the restriction of $\phi $ to $\S_r$.   Then we view the ground state wavefunction
as a function $\Omega(\phi_\ell,\phi_r)$.    We introduce a Hilbert space $\H_r$ of functions of $\phi_r$ and a Hilbert space $\H_\ell$ of functions of $\phi_\ell$.   Then formally
$\H=\H_\ell\otimes \H_r$ and in particular
$\Omega\in\H_\ell\otimes \H_r.$

We would like to construct the reduced density matrix of the vacuum state $\Omega$ for an observer who can measure $\phi_r$ only and not $\phi_\ell$.   
For this, we first write eqn. (\ref{gurky}) in more detail, with $\phi=(\phi_\ell,\phi_r)$ and $\phi'=(\phi_\ell',\phi_r')$:
\be\label{nurky}\rho(\phi_\ell,\phi_r;\phi'_\ell,\phi'_r)=|\Omega(\phi_\ell,\phi_r)\ra\la\Omega(\phi'_\ell,\phi'_r)|.\ee
Now to construct a density matrix that is appropriate for observations of $\phi_r $ only, we are supposed to sum over all values of the unobserved variables $\phi_\ell$.
To do this, we set $\phi_\ell=\phi'_\ell$ and integrate over $\phi_\ell$.   This gives the density matrix $\rho_r(\phi_r;\phi_r')$ appropriate for measurements of $\phi_r$:
\be\label{demorc} \rho_r(\phi_r;\phi_r')=\int D\phi_\ell \,|\Omega(\phi_\ell,\phi_r\ra \,\la\Omega(\phi_\ell,\phi_r')|. \ee

How do we represent $\rho_r$ by a path integral?   Before integrating over $\phi_\ell$, we had a pure state density matrix $\rho=|\Omega\ra\,\la\Omega|$ represented as a path
integral over all of Euclidean space but with a ``cut'' on the hyperplane $\S$ defined by $\te=0$ -- and thus with separate boundary values $\phi$, $\phi'$ below and above the cut.
To compute $\rho_r$  as defined in eqn. (\ref{demorc}) requires the following.  In the region $\S_\ell$, we constrain the boundary values above and below the cut to be equal and we integrate over those boundary values.
Geometrically, the effect of this is to glue together the upper and lower half spaces  along $\S_\ell$ and integrate over the value of $\phi$ there.  
We end up with a path integral on all of $\R^4$ except for
a cut along $\S_r$ (fig. \ref{Five}(c)).   The boundary values below and above the cut are $\phi_r$ and $\phi_r'$, and a path integral on  $\R^4$ with this cut and with fixed boundary
values above and below the cut computes the matrix element $\rho_r(\phi_r;\phi_r')$ of the density matrix $\rho_r$ of $\S_r$.

Similarly, a density matrix appropriate for measurements of $\phi_\ell$ only is
obtained by setting $\phi_r=\phi_r'$ in $\rho$ and integrating over $\phi_r$:
\be\label{demorco} \rho_\ell(\phi_\ell;\phi_\ell')=\int D\phi_r\,|\Omega(\phi_\ell,\phi_r\ra \,\la\Omega(\phi_\ell',\phi_r)|. \ee
It can be represented by a path integral on $\R^4$ with a cut along $\te=0$, $x<0$.

  \begin{figure}
 \begin{center}
   \includegraphics[width=3.5in]{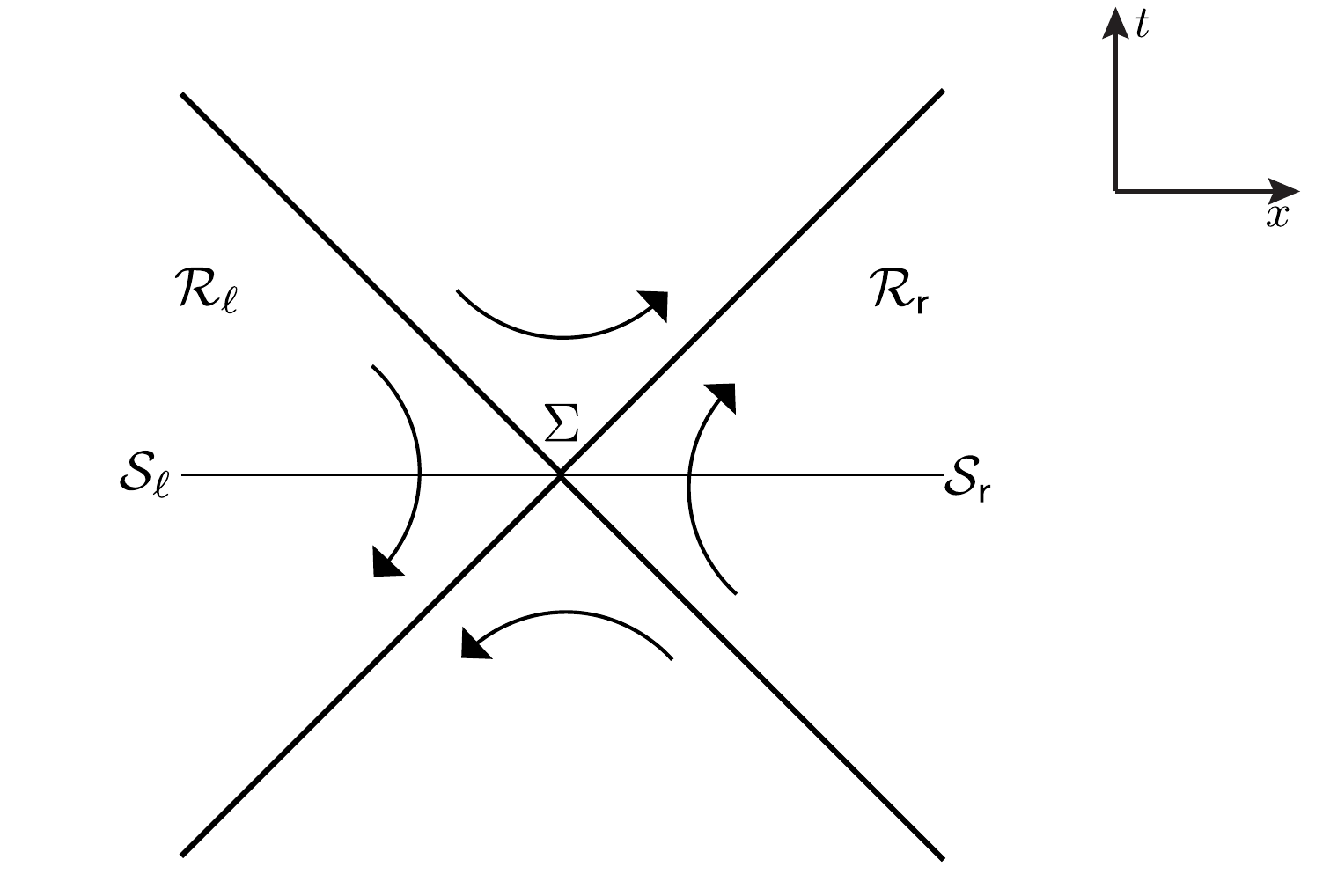}
 \end{center}
\caption{\footnotesize  In Minkowski space, the left and right Rindler wedges $\RR_\ell$ and $\RR_r$ are defined as the domains of dependence of the partial Cauchy hypersurfaces
$\S_\ell$ and $\S_r$.   The diagonal lines mark the boundaries of $\RR_\ell$ and $\RR_r$; they are the past and future horizons of an observer
 who remains forever in $\RR_\ell$ or $\RR_r$ and limit the portion of spacetime that the observer can see or influence.
All past and future horizons meet at the bifurcation surface $\Sigma$,   which also marks the common boundary of $\S_\ell$ and $\S_r$.
The arrows indicate the action of the Lorentz boost generator $K$, which is future-directed timelike in $\RR_r$, past-directed timelike in $\RR_\ell$, and spacelike elsewhere.
   \label{Seven}}
\end{figure}

The relation of all this to Rindler space is as follows.   In Lorentz signature, the  domain of dependence\footnote{\label{dod}In a Lorentz signature spacetime $M$, the domain of dependence
of a set $U$ is the largest set $D(U)$ such that a solution of a standard relativistic wave equation (such as the Klein-Gordon equation) in $D(U)$ is determined by initial data along $U$.
Equivalently, a point $p\in M$ is in $D(U)$ if and only if any causal curve through $p$, if continued far enough into the past and future, eventually meets $U$. This, along with relativistic causality,
implies that a signal observed at $p$ is determined by initial (or final) data on $U$.}
 of the partial Cauchy surface $\S_r$ is the ``right Rindler wedge'' $\RR_r$, defined
by $x>|t|$ (fig. \ref{Seven}).   Whatever is the quantum field theory we are studying, formally its equations of motion determine fields on $\RR_r$ in terms of initial data on $\S_r$.
Therefore any measurement in $\RR_r$ can be viewed as a measurement of $\phi_r$, and the density matrix $\rho_r$ can more covariantly be described as the density matrix
appropriate to measurements in $\RR_r$.   Similarly, the density matrix $\rho_\ell$ is appropriate to measurements in the opposite Rindler wedge $\RR_\ell$, defined by $x<-|t|$.

Following are a few important facts about  the geometry of Rindler space.   First of all, in order to remain for all times in the right Rindler
wedge  $R_r$, an observer must accelerate indefinitely in the future and also in the past, as in eqn. (\ref{unif}) below.   (This will lead to the Unruh effect, as we will discuss.)   The portions of the spacetime visible to such an observer are bounded by the past and future horizons of Rindler space, the diagonal lines in fig. \ref{Seven}.   Similarly, an accelerating observer who remains always in
$R_\ell$ experiences past and future horizons.   The past and future horizons of both observers  intersect at a codimension two surface $\Sigma$ known as the bifurcation surface or
entangling surface,\footnote{In classical general relativity, $\Sigma$ is called the bifurcation surface from which the left and right horizons bifurcate.  Quantum mechanically, it is sometimes 
called the entangling surface of the Rindler wedges $\RR_\ell$ and $\RR_r$, which are entangled across $\Sigma$.}  which
also marks the common boundary of $\S_\ell$ and $\S_r$.
Once we introduce the Penrose diagram of de Sitter space (fig. \ref{desitter}) and the Kruskal-Szekeres extension of Schwarzschild spacetime (fig. \ref{PenroseB}), it will be clear that all this has analogs for de Sitter space and for  a black hole. Indeed, Rindler space is
analogous to the near horizon region of a Schwarzschild black hole of mass $M$ in the limit $M\to\infty$, and to the region near the cosmological horizon in de Sitter space.  That is the reason for its importance.

  We note that $\RR_r$ and $\RR_\ell$ are each invariant
under boosts of the $x-t$ plane.   This symmetry has played no role  up to this point, but now that will change.

\subsection{Boosts and the Unruh Effect}\label{rindler}

The density matrix $\rho_r$ can be understood in another way by emphasizing the rotational symmetry of the $x-\te$ plane.   Actually, it is convenient to first  relate a rotation
in Euclidean signature to a boost in Lorentz signature.   In Lorentz signature, the generator of a boost of the $x-t$ plane is
\be\label{zboost} K=\int_{\S}\d x\d \vec y \, x T_{00}(x,\vec y),\ee
where $T_{00}$ is the energy density.   We can formally write\footnote{In the decomposition $K=K_r-K_\ell$, actually, because of short distance fluctuations near $x=0$, $K_\ell$ and $K_r$ make sense as quadratic
forms (which have matrix elements) but not as operators (which have eigenvectors and eigenvalues and can be measured).  See footnote \ref{notable}.}
\be\label{lboost}K=K_r-K_\ell,\ee
with 
\begin{align}\label{mboost}K_r & =\int_{\S_r}\d x \d\vec y \, xT_{00}(x,\vec y) \cr
                                          K_\ell & =\int_{\S_\ell}\d x \d\vec y \, |x|T_{00}(x,\vec y), \end{align}
 where $K_r$ generates a Lorentz boost of $\phi_r$ and $K_\ell$ generates a Lorentz boost of $\phi_\ell$.
  $K_r$ generates a Lorentz boost of the right Rindler wedge $\RR_r$ and commutes with operators in the spacelike separated wedge
    $\RR_\ell$, while $K_\ell$ generates a Lorentz boost of $\RR_\ell$ and commutes
 with operators in $\RR_r$.  A minus sign was included in eqn. (\ref{lboost}) so that $K_r $ and $K_\ell$ each boost the corresponding Rindler wedge forward in time.   
 There is an additive ambiguity in the definition of $T_{00}$, and we assume this has been fixed so that the vacuum energy density vanishes; otherwise the integrals defining
 $K_\ell$ and $K_r$ diverge.

   \begin{figure}
 \begin{center}
   \includegraphics[width=4.4in]{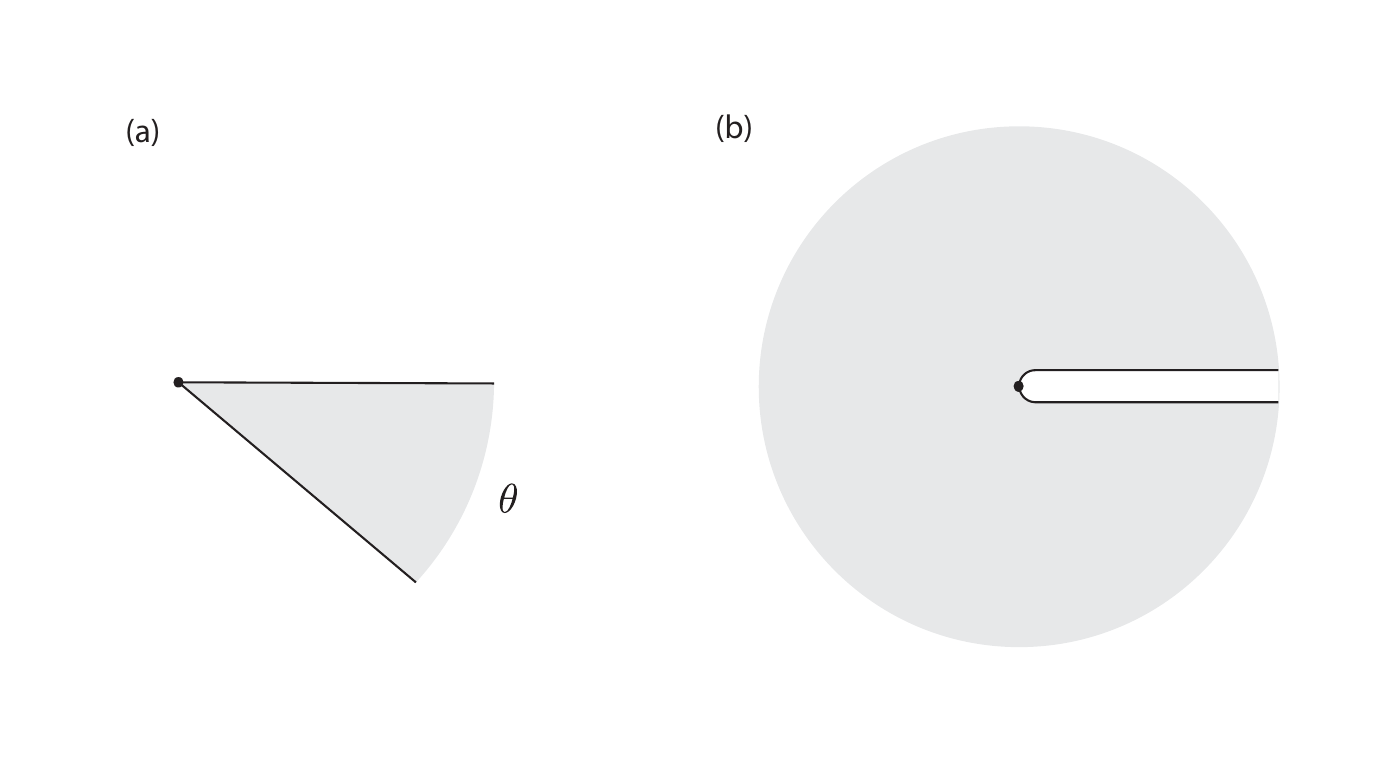}
 \end{center}
\caption{\footnotesize  (a)  A path integral on an angular wedge of angle $\theta$.   This
path integral is generated by the operator $\exp(-\theta K_r)$, where in Minkowski space, $K_r$ generates a boost of $\phi_r$ and commutes with $\phi_\ell$.  
(b)  Setting $\theta=2\pi$ gives the path integral of fig. \ref{Five}(c)) on the whole plane with a cut on the positive $x$ axis.  Similarly, setting $\theta=\pi$ gives the path integral of fig. \ref{Five}(a) that prepares the 
ground state.
   \label{Eight}}
\end{figure} 

 The operator $e^{-\i\eta K_r}$, with real $\eta$, acts on $\phi_r$ by a Lorentz boost with boost parameter $\eta$.   If we set $\eta=-\i\theta$, the Lorentz boost turns into a
 rotation and we get the operator $\exp(-\theta K_r)$ that in Euclidean signature rotates the $x-\te$ plane by an angle $\theta$.   In terms of path integrals,
 this means that to compute a matrix element of $\exp(-\theta K_r)$ acting on $\phi_r$, we need to perform a path integral on a wedge of opening angle $\theta$
 (fig. \ref{Eight}(a)).  If we simply set $\theta=2\pi$ (fig. \ref{Eight}(b)), the wedge of opening angle $\theta$ becomes the cut plane of fig. \ref{Five}(c).  Therefore, we get a formula for the density matrix $\rho_r$:
 \be\label{zilbo}\rho_r=\exp(-2\pi K_r). \ee
 Likewise,
 \be\label{nilbo} \rho_\ell=\exp(-2\pi K_\ell). \ee
 Let us verify that these candidate density matrices are properly normalized to satisfy  $\Tr\,\rho=1$.   Formally, $\Tr\,\exp(-2\pi K_r)$
 (or equally well $\Tr\,\exp(-2\pi K_\ell)$)
 is given by a path integral on the whole plane.   That path integral equals 1, given that the energy-momentum tensor of the theory has been renormalized so that its
 vacuum expectation value vanishes (as was assumed in the definition of $K_r$ and $K_\ell$).
 
The result for $\rho_r$ can be interpreted to mean that if we view $K_r$ as a sort of Hamiltonian of the right Rindler wedge $\RR_r$, then the density matrix $\rho_r$ is thermal
at inverse temperature $2\pi$.   A manifestation of this is the Unruh effect \cite{U} -- an observer in Minkowski space who undergoes constant acceleration (for all time) will
observe thermal correlations.    An example of a uniformly accelerated trajectory is
\be\label{unif} t=L\sinh \frac{\tau}{L},~~ x =L\cosh\frac{\tau}{L},~~\vec y=0, \ee
where $\tau$ is the proper time of the observer, and $1/L$ is the magnitude of the acceleration.   Any uniformly accelerated orbit in Minkowski space has this form for some choice of  Rindler
wedge and some $L$.   The  Lorentz boost generator $K_r$ acts on this orbit as $L\frac{\d}{\d \tau}$, so the observer could interpret $K_r$ as $L$ times the natural Hamiltonian.
Indeed the definition (\ref{mboost}) for $K_r$ shows that for observations near $x=L$, $K_r$ can be approximated as $L H$ where $H=\int_{\S_r}\d x \d\vec y \,T_{00}$ is
the Hamiltonian acting on the right Rindler wedge.

A quick way to become comvinced that the observations of such an observer will be thermal is to continue the orbit to Euclidean signature.  In Euclidean signature,
with  $t=-\i \tE$, $\tau=-\i \tau_E$, the orbit becomes
\be\label{zunif} \tE=L\sin\frac{\tau_E}{L},~~x=L\cos\frac{\tau_E}{L} ,~~\vec y=0.\ee
This orbit is periodic in $\tau_E$ with period $2\pi L$, suggesting that the observer will see thermal correlations at temperature $1/2\pi L$.   This can be justified rigorously
by showing that the correlations measured along the Lorentz signature orbit (\ref{unif}) satisfy the KMS condition that characterizes thermal correlations.   See for example
section 5.4 of \cite{Notes}.   The formula $1/2\pi L$ for the temperature has a natural interpretation.   From the observer's point of view, $K_r$ is interpreted, as noted earlier,
 as $LH$, with $H$ the relevant Hamiltonian, so the formula (\ref{zilbo}) becomes $\rho_r=\exp(-2\pi L H)$, leading naturally to thermal correlations at temperature $1/2\pi L$.
 
 Thus, we have learned that from the standpoint of a uniformly accelerated observer, or for observations restricted to one Rindler wedge, the vacuum state
of Minkowski space appears to be thermal with an appropriate Hamiltonian and temperature. The temperature is inversely proportional to the distance from the edge of 
the Rindler wedge at the bifurcation surface $\Sigma: x=t=0$, so it diverges near $\Sigma$ and vanishes near infinity.    Though the reasoning was slightly heuristic,
the conclusions can actually be justified by verifying the KMS condition for correlators. 
 
 After developing the Euclidean approach to black hole thermodynamics in section \ref{euclidean}, we will be able to give a precisely parallel derivation of the thermal nature
 of the black hole spacetime.    In a way, the main difference is that in the case of the black hole, the temperature measured at infinity does not vanish but equals the Hawking temperature
 of the black hole.
 A noteworthy fact about the derivation that we have given for Rindler space, and the corresponding derivation that we will eventually give for the black hole, is that it requires
 no assumptions whatsoever about the quantum field theory under consideration.    Arbitrary non-gravitational forces may be present. 
  (The assumption that the theory can be characterized by just one scalar field $\phi$ was purely
 for notational convenience.) 
  By contrast, the derivation of black hole evaporation in section \ref{bhevaporation} ignored non-gravitational forces.
                                  
\subsection{The Thermofield Double}\label{another}

For some Hilbert space $\H$, let  $\rho:\H\to \H$ be a density matrix. Recall that in general a density matrix is simply a positive (or non-negative) matrix with trace 1.
  A {\it purification} of $\rho$ is a pure state $\Psi\in\H\otimes\H'$, for some Hilbert space $\H'$, such that $\rho$ is the induced density matrix of the pure
state density matrix $|\Psi\ra\la\Psi|$ on $\H\otimes \H'$:
\be\label{tellmex}\rho=\Tr_{\H'}|\Psi\ra\la\Psi|. \ee

Every density matrix has a canonical purification.   For this, note first that if $\K$ is a vector space and $\V:\K\to\K$ is a linear transformation, then with
respect to a basis we can expand
\be\label{anytrans}\V=\sum_{i,j}v_{ij}|i\ra \la j|.\ee
Given this, we can associate to $\V$ a vector $\Psi_\V$ in a doubled Hilbert space $\K\otimes \K'$: 
\be\label{zvector}  \Psi_\V=\sum_{i,j}v_{ij}|i\ra\otimes |j\ra' .\ee
Here $\K'$ is the complex  conjugate Hilbert space of $\K$, meaning that to each bra $\la j|$ of $\K$ there is canonically associated
a ket $|j\ra'\in \K'$.   Tracing out $\K'$ from the pure state density matrix $|\Psi_V\ra\la\Psi_\V|$, we get
\be\label{nvector}\V \V^\dagger=\Tr_{\K'}|\Psi_\V\ra \la\Psi_\V|.\ee
In particular, if $\Tr\,\V \V^\dagger=1$, so that $\V \V^\dagger$ is a density matrix, then $\Psi_\V$ is a purification of this density matrix.

This enables us to define the canonical purification of a density matrix.   If $\rho$ is any density matrix, then it is
the square of $\rho^{1/2}$.  So $\Psi_{\rho^{1/2}}\in \H\otimes \H'$ is a purification of $\rho$, called the canonical
purification.  

For an important example, consider the thermal density matrix of a system with Hilbert space $\H$ and Hamiltonian $H$ at inverse
temperature $\beta$:
\be\label{dongo} \rho=\frac{1}{Z}\sum_i e^{-\beta E_i}|i\ra \la i|,\ee
where $|i\ra$ are the energy eigenstates with energies $E_i$ and $Z$ is the partition function.   
The canonical purification of a thermal density matrix   is then the state in $\H\otimes \H'$ associated to $\rho^{1/2}$:
\be\label{wongox}\Psi_\TFD=\frac{1}{\sqrt Z}\sum_i e^{-\beta E_i/2}|i\ra\otimes |i\ra'. \ee
This state is also called the {\it thermofield double}.
In general, $\H'$ is the complex conjugate of $\H$, but in  the case of a system with an antilinear time-reversal symmetry whose square is 1, the distinction between $\H$ and $\H'$ is unimportant.
This is often implicitly assumed in  discussions of the thermofield double.
 
The canonical purification of the thermal density matrix $\rho=e^{-2\pi K}$ of Rindler space is $\rho^{1/2}=e^{-\pi K}$.
 This operator is associated to the path integral of fig. \ref{Eight}(a) for the case that the wedge has opening angle $\pi$.   But a wedge of
 opening angle $\pi$ is just a half-plane.   So for this particular value of the angle, the path integral of fig. \ref{Eight}(a) actually reduces to the half-plane path integral
 that we started with in fig. \ref{Five}(a).  The state constructed by the path integral of fig. \ref{Five}(a) is simply the vacuum state $\Omega\in \H_\ell\otimes \H_r$.
 So we learn that the Minkowski space vacuum vector $\Omega$ can be interpreted as the thermofield double state of Rindler space.     In sections \ref{cosmo} and  \ref{thermofield},
 we will describe the analogous statements for de Sitter space and for a Schwarzschild black hole.
 
 To clear up one last detail in this derivation, every quantum field theory has a $\sf{CRT}$ symmetry (charge conjugation $\sf C$ combined with a spatial reflection $\sf R$
 and time-reversal $\sf T$) that exchanges the left and right Rindler wedges and exchanges $\H_\ell$ with $\H_r'$.  (Because $\sf{CRT}$ is antilinear, it exchanges $\H_\ell$
 with $\H_r'$, not $\H_r$.)   So instead of $\H_r\otimes \H_\ell$,  the vacuum vector $\Omega$ can be viewed as a vector in $\H_r\otimes \H_r'$, the expected home of the thermofield double state.
 
 Because it is an important example and
 will be useful in section \ref{anview}, we will work out  the thermofield double state for a bosonic or fermionic harmonic oscillator.  First we consider an ordinary bosonic
 harmonic oscillator with  creation and annihilation
 operators $\a^\dagger$, $\a$ satisfying $[\a,\a^\dagger]=1$ and Hamiltonian $H=\omega \a^\dagger \a$.
 A thermal density matrix at inverse temperature $\beta$ is
 \be\label{zeldo}\rho=\frac{1}{Z}\sum_{n=0}^\infty e^{-n\beta \omega }|n\ra\la n|\ee
 where $|n\ra$ is the $n^{th}$ excited state. The thermofield double state is 
 \be\label{weldo}\Psi_\TFD=\frac{1}{\sqrt Z}\sum_{n=0}^\infty e^{-n\beta\omega/2}|n\ra\otimes |n\ra'. \ee
 Here $|n\ra'$ is the $n^{th}$ excited state of an identical  second harmonic oscillator with creation and annihilation operators $\a'$, $\a'{}^\dagger$.   
Now using $\a^\dagger|n\ra =\sqrt{n+1}|n+1\ra$, etc., we find
\be\label{zoffo}\left(\a^\dagger-e^{\beta \omega/2}\a'\right)\Psi_\TFD = \left(\a-e^{-\beta\omega/2}\a'{}^\dagger\right)\Psi_\TFD =0 . \ee
Moreover, these conditions uniquely determine $\Psi_\TFD$ up to a scalar multiple.
The fermionic analog of these formulas  contains an extra minus sign associated to fermi statistics.
We  consider fermionic creation and annihilation operator $\c,$ $ \c^\dagger$, with $\{\c,\c^\dagger\}=1$, acting on a two-dimensional
Hilbert space with basis consisting of a state $|0\ra$ with $\c|0\ra=0$ and another state $|1\ra=\c^\dagger|0\ra$.   Assuming a Hamiltonian $H=\omega \c^\dagger \c$,
the thermal density matrix is $\rho=\frac{1}{Z}\left(|0\ra\la0|+e^{-\beta \omega}|1\ra\la 1|\right)$.   To construct the thermofield double, we introduce
 a second identical fermionic oscillator with creation and
annihilation operators $\t \c^\dagger $, $\t\c$ that anticommute with $\c,\c^\dagger$.  These operators can be represented in a four-dimensional Hilbert space with a state
$|0,0\ra$ annihilated by both $\c$ and $\t \c$ and additional states $|1,0\ra=\c^\dagger|0,0\ra$, $|0,1\ra=\t\c^\dagger|0,0\ra$, $|1,1\ra =\c^\dagger \t\c^\dagger|0,0\ra$.
The thermofield double state is then
\be\label{wendo}\Psi_\TFD =\frac{1}{Z^{1/2}}\left( |0,0\ra+e^{-\beta\omega/2}|1,1\ra\right), \ee
and satisfies
\begin{align}\label{lendo} \left( \c-e^{-\beta\omega/2} \t\c^\dagger\right)\Psi_\TFD &=0 \cr
                          \left(\c^\dagger+ e^{\beta\omega/2} \t\c\right)\Psi_\TFD&=0.\end{align} 

\subsection{Another View Of The Thermofield Double}\label{anview}

There is  another interesting way, going back to \cite{U}, to show that the vacuum vector $\Omega$ in Minkowski space is the 
 thermofield double state  of the Rindler wedge. This explanation is limited to free field theory, in contrast to the far more general approach that we have
already presented, but  it is illuminating.
  However, for brevity, we will only consider the case of a  chiral free fermion in two spacetime dimensions.
The idea is to show that the vacuum state  obeys conditions that correspond to eqn. (\ref{lendo}).

Consider  two-dimensional Minkowski space with metric $\d s^2=-\d t^2+\d x^2=-2 \d u \d v$, where $v=\frac{1}{\sqrt 2}(t+x)$, $u=\frac{1}{\sqrt 2}(t-x)$ are null coordinates.
The operator $P$ that generates a translation of  $v$ is positive-definite,
annihilating only the vacuum, and satisfies
\be\label{hudo} [P,\O]=-\i \frac{\d}{\d v}\O \ee
for any operator $\O$. 
Consider a hermitian chiral free fermion $\lambda(v)$ satisfying
\be\label{telgo} \{\lambda(v),\lambda(v')\}=\delta(v-v'). \ee
If $\Lambda_\omega=\int_{-\infty}^\infty \d v \,e^{-\i \omega v}\lambda(v)$, then
\be\label{elgo}\{\Lambda_\omega,\Lambda_{\omega'}\} =2\pi \delta(\omega+\omega').\ee
We have
\be\label{ifto} [P,\Lambda_\omega]=\omega \Lambda_\omega,\ee
so operators $\Lambda_\omega$ are creation operators for $\omega>0$ and annihilation operators for $\omega<0$.
To be more precise, these operators are creation and annihilation operators with respect to the Minkowski vacuum
or equivalently they are raising and lowering operators with respect to $P$.

The annihilation operators annihilate the vacuum:
\be\label{hudu} \Lambda_\omega \Omega =0,~~~\omega<0.\ee
More generally, any operator
\be\label{udu}\Lambda'_f=\int_{-\infty}^\infty \d v \,f(v)\lambda(v) \ee
annihilates the vacuum if  the function $f(v)$ is holomorphic and bounded in the upper half $v$-plane.
Indeed, a square-integrable function $f(v)$ is holomorphic and bounded  in the upper half plane if and only if
 \be\label{plunker} f(v)=\int_{-\infty}^0 \d\omega e^{-\i \omega v} g(\omega)\ee
  for some square-integrable function $g(\omega)$;
 the Fourier components of $f(v)$ with $\omega>0$ must be absent, as $e^{-\i\omega v}$ grows exponentially
 in the upper half $v$-plane if $\omega>0$.  But eqn. (\ref{plunker}) implies that $\Lambda'_f$ is a linear combination
 of annihilation operators.

Now let us take the perspective of an observer in  the right Rindler wedge $\RR_r$ defined by $x>|t|$ or $v>0$, $u<0$. 
The Killing vector field that generates a Lorentz boost of the $u-v$ plane is  $k=v\partial_v -u\partial_u$. It generates a symmetry of the right Rindler wedge
 and is future-directed timelike in the wedge.   As before, let $K$ be the hermitian conserved charge associated to $k$.
It is with respect to $K$  that correlations in the Rindler wedge have thermal properties.
So it is natural for an observer in the right Rindler wedge  to decompose the field $\lambda(v)$ in raising and lowering operators with respect to $K$.

  Acting on the chiral fermion field $\lambda(v)$, $k$ reduces to $v\partial_v$,
 and the corresponding conserved charge $K$ acts by
$[K,\lambda(v)]=-\i \left(v\partial_v+\frac{1}{2}\right) \lambda(v)$ (where the $+\frac{1}{2}$ reflects the fact that $\lambda$ has spin $\frac{1}{2}$).  
  An observer in $\RR_r$ measures $\lambda(v)$ only for $v>0$.
So\footnote{The functions $v^{-\i\omega-\frac{1}{2}}$ are delta function normalizable on the half-line $v\geq 0$.}
\be\label{omexo} U_\omega=\int_0^\infty \d v \,v^{-\i\omega-\frac{1}{2}} \lambda(v) \ee
is supported in the wedge $\RR_r$ and 
satisfies
\be\label{Nome} [K,U_\omega]= \omega U_\omega.\ee 
Hence, with respect to $K$, $U_\omega$ is a raising operator, or a creation operator,  if $\omega>0$ and a lowering operator, or an
annihilation operator, if $\omega<0$.
Moreover
\be\label{tellme}U_\omega^\dagger = U_{-\omega}. \ee

However, regardless of $\omega$, $U_\omega$ does not annihilate the Minkowski space vacuum state $\Omega$, since the
function
\be\label{dimo}f(v)=\begin{cases} v^{-\i\omega-\frac{1}{2}} & v>0\cr 0 & v<0 \end{cases}\ee
is not holomorphic in the upper half $v$-plane.
To get an annihilation operator for the Minkowski vacuum that is equivalent to $U_\omega,~\omega<0$ for observations in the right Rindler wedge, we need to modify $f(v)$ to be non-zero for $v<0$ in such a way that
$f(v)$ becomes holomorphic and  bounded in the upper half-plane.   A function that coincides with $f(v)$ for $v>0$
and is holomorphic and bounded in the upper half $v$-plane is $(v+\i\epsilon)^{-\i\omega-\frac{1}{2}}$.  
Here $\epsilon$ is
an infinitesimal positive quantity; a limit $\epsilon\to 0^+$ is understood.    The boundedness in the upper half plane holds for either sign of $\omega$.
 Hence, for all $\omega$, 
\be\label{comexo} V_\omega =\int_{-\infty}^\infty \d v (v+\i\epsilon)^{-\i\omega -\frac{1}{2}} \lambda(v)\ee
is an annihilation operator with respect to the Minkowski vacuum.

We have
\be\label{miliko} \lim_{\epsilon\to 0^+} (v+\i\epsilon)^{-\i\omega-\frac{1}{2}} = \begin{cases}v^{-\i\omega-\frac{1}{2}} & v>0\cr
                                           -\i e^{\pi\omega} \bar v^{\,-\i\omega-\frac{1}{2}} & v<0,\end{cases}\ee
                                           where $\bar v=-v$.   
Now  define  
\be\label{niliko}\t U_\omega=\begin{cases}    \i  \int_0^\infty \d\bar v\,\bar v^{\i\omega-\frac{1}{2}} \lambda(\bar v)    &\omega <0\cr
                                                                       - \i \int_0^\infty \d\bar v\,\bar v^{\i\omega-\frac{1}{2}} \lambda(\bar v)    &\omega >0,\end{cases}\ee
 so that in particular $\t U_\omega^\dagger=\t U_{-\omega}$, in parallel with eqn. (\ref{tellme}).   The prefactors $\i$ and $-\i$ in eqn. (\ref{niliko}) are
 inessential conventions chosen to agree in a sense that will soon be clear
 with the conventions that were used earlier in analyzing a fermionic oscillator.
 More important is that relative to eqn. (\ref{comexo}), we have reversed the sign of $\omega$ in the exponent.   The reason 
is that $v$ increases toward the future in  $\RR_r$, but $\bar v$ increases toward the past in 
 $\RR_\ell$.  Hence the sign reversal is needed if  we want $\t U_\omega$ to look  in $\RR_\ell$ like a creation operator  
  if $\omega>0$ and an annihilation operator if $\omega<0$.
  With these definitions, we see that
  \be\label{zelme}V_\omega=\begin{cases} U_\omega -e^{\pi\omega} \t U_{-\omega}=U_\omega-e^{\pi\omega}\t U_\omega^\dagger & \omega<0\cr                                                                 
    U_\omega +e^{\pi\omega} \t U_{-\omega}=U_\omega+e^{\pi\omega}\t U_\omega^\dagger & \omega>0.  \end{cases}\ee
    We can now confirm that $\Omega$ is the thermofield double state with respect to the right wedge $\RR_r$ at inverse temperature
    $\beta=2\pi$:     the statement that $V_\omega\Omega=0$ for $\omega<0$ matches the first condition in eqn. (\ref{lendo}), and the statement that
    $V_\omega\Omega=0$ for $\omega>0$ matches the second one.

\section{Euclidean Approach To Black Hole Thermodynamics}\label{euclidean}

\subsection{Continuing to Euclidean Signature}\label{disappears}

Gibbons and Hawking \cite{GH}, following earlier work of Hartle and Hawking \cite{HH}, discovered a remarkable alternative approach to black hole thermodynamics based on a continuation to Euclidean signature.   Despite the power of this approach,
we have chosen to present first the Lorentz signature derivation of sections \ref{bhevaporation} and \ref{graybody}, because in that framework the underlying physical principles
are clear.   The Euclidean approach is remarkably powerful and successful, but its foundations are less clear.

The starting point is simply to continue the Schwarzschild metric
\be\label{schwarz}\d s^2=-\left(1-\frac{2 G M}{r}\right)\d t^2+\frac{\d r^2}{1-\frac{2 GM}{r}}+r^2\d\Omega^2\ee
to Euclidean signature by setting $t=-\i \tE$.   We get the Euclidean signature metric
\be\label{eucl}\d s^2=\left(1-\frac{2 G M}{r}\right)\d \tE^2+\frac{\d r^2}{1-\frac{2 GM}{r}}+r^2\d\Omega^2.\ee
Fundamentally, the reason that the simple definition $t=-\i \tE$ leads to a real metric in Euclidean signature is the following.
The Schwarzschild metric is real for real $t$, so if $t$ is regarded as a complex variable, then the Schwarzschild metric is complex conjugated under $t\leftrightarrow \bar t$.
This corresponds in terms of $\tE$ to $\tE \leftrightarrow  -\bar \tE$.   The fact that this operation complex conjugates the metric  does not imply that the metric is real for real  $\tE$.
But the Schwarzschild metric is invariant under the  time-reversal symmmetry $t\leftrightarrow -t$, so it is also complex conjugated by the 
combined action of time-reversal and   complex conjugation  of $t$.  The combined  operation is 
$t\leftrightarrow -\bar t$, or $\tE \leftrightarrow \bar \tE$. The fact that this operation complex conjugates the metric does indeed imply  that the metric is real for real $\tE$.

 Gibbons and Hawking made the remarkable discovery that the Euclidean version of the Schwarzschild metric is perfectly
smooth, complete,  and singularity-free if $\tE$ is interpreted as an angular variable 
and the horizon at $r=2GM$ is interpreted as the origin of polar coordinates.
To see this, let 
\be\label{defrho}{\t r}=4GM\left(1-\frac{2GM}{r}\right)^{1/2}. \ee
The metric is then
\be\label{nefto} \d s^2 = \frac{{\t r}^2\d\tE^2 }{(4GM)^2} + \left(\frac{r}{2GM}\right)^4 \d {\t r}^2 +r^2\d\Omega^2.\ee
Near $r=2GM$, this reduces to
\be\label{wefto} \d s^2 =  \frac{{\t r}^2\d\tE^2 }{(4GM)^2}+\d{\t r}^2 +(2GM)^2\d\Omega^2. \ee
Here we have the product of a two-manifold with metric
\be\label{efto}  \frac{{\t r}^2\d\tE^2 }{(4GM)^2}+\d{\t r}^2 \ee
with a two-sphere of metric $(2GM)^2\d\Omega^2$ and thus radius $2GM$.   
Comparing (\ref{efto}) to the metric of a plane in polar coordinates, namely $\d{\t r}^2+{\t r}^2\d\theta^2$, where $\theta\cong \theta+2\pi$,
we see that the metric (\ref{efto}) is perfectly smooth  if $\tE$ is a periodic variable 
\be\label{pervar}\tE\cong \tE+8\pi GM, \ee
and the circle parametrized by $\tE$ collapses to a point at ${\t r}=0$.   If we assume any other period for $\tE$, then the metric (\ref{efto}) has a conical singularity at the origin.
A metric with this conical singularity does not satisfy Einstein's equations, so if we want to get a complete, smooth metric satisfying Einstein's equations, we have to take
$\tE$ to be periodic precisely with period $\beta_\sH=8\pi GM$.

Notably, the required period of $\tE$ is precisely the inverse of the Hawking temperature $T_\sH=1/8\pi GM$, and equivalently the periodicity
$\tE\to \tE+8\pi GM$ precisely corresponds to the periodicity in imaginary time $t\to t+8\pi GM\i$ that appeared in our derivation in section \ref{bhevaporation}.
Thus the black hole solution is in this sense periodic in imaginary time.   Periodicity in imaginary time is a hallmark of thermal correlations, and the idea of 
Gibbons and Hawking was that the thermal nature of a black hole reflects the fact that the black hole solution is itself periodic in imaginary time.

Since $r$ is positive everywhere in the Euclidean Schwarzschild space, replacing the term $r^2 \d\Omega^2$ 
in the metric
(\ref{eucl})
with $C\d\Omega^2$ (for an arbitrary constant $C$)  does
not change the topology.  After that replacement, the metric is the sum of a metric on $\R^2$ and a metric on $S^2$. Therefore, topologically, the Euclidean Schwarzschild spacetime
is isomorphic to $\R^2\times S^2$.

For $r\to\infty$, the Euclidean Schwarzschild metric  reduces to 
\be\label{flatmetric} \d s^2= \d\tE^2+\d r^2+r^2 \d\Omega^2.\ee
Since $\d r^2+r^2\d\Omega^2$ is just the flat metric on $\R^3$ written in spherical polar coordinates, the metric (\ref{flatmetric}) is the standard product flat metric
on $\R^3\times S^1_{\beta_\sH}$, where $S^1_{\beta_\sH}$ is a circle of circumference $\beta_\sH=\frac{1}{T_\sH}= 8\pi GM$.

In ordinary quantum field theory without gravity, a thermal ensemble on a spatial manifold $\R^3$ at inverse temperature $\beta$ can be studied by a path integral
on $\R^3\times S^1_\beta$.    In gravity, if one expands around the classical solution $\R^3\times S^1_\beta$, then at one-loop order one will find the thermodynamics of
a gas of free gravitons (and other particles if fields other than the gravitational field are present) at inverse temperature $\beta$; higher order corrections will describe interactions, and will also reveal the instabilities described in section \ref{unstable}.  Since this thermal gas arises as a one-loop effect, the resulting entropy and energy densities are of order $G^0$ (as opposed to the Bekenstein-Hawking
entropy of a black hole, which is of order $G^{-1}$).
The Euclidean black hole solution looks like $\R^3\times S^1_{\beta_\sH}$ at big distances, but also contains, of course, a Euclidean version of a black hole in the interior.
The proposal of Gibbons and Hawking was that by expanding around this solution, we would get a description of a black hole interacting with  a gas of thermal radiation.
Since  the relevant value of $\beta$ for the Euclidean black hole solution is precisely the inverse temperature  $\beta_\sH$ of the black hole, the black hole and the radiation have the same
temperature and will be in equilibrium (modulo the usual instabilities).

The partition function, expanded in perturbation theory, will be schematically
\be\label{perfn} Z= e^{-I_\BH} \frac{1}{\sqrt{\det} }\left(1+\cdots\right). \ee
Here $I_\BH$ is the action of the classical black hole solution, $1/\sqrt{ \det}$ schematically represents the one-loop correction, and $\cdots$ represents effects of   two-loop order and higher.
Classical black hole thermodynamics as developed by Bekenstein and Hawking is supposed to appear in $I_\BH$.  The one-loop correction is supposed to give us a thermal
gas of gravitons (and possibly other particles) 
at the Hawking temperature, and their interaction with the black hole.   The higher order corrections will, among other things, describe interactions among
the gravitons, and generate an instability.   

Let us discuss concretely how to extract the black hole thermodynamics from the classical action.   In general, the partition function $Z$ of a thermal system at inverse temperature $\beta$
is interpreted as
$e^{S-\beta E}$, where $S$ is the entropy and $E$ is the energy.   Alternatively, $Z=e^{-\beta F}$, where $F=E-TS$ is the free energy.  Using the first law $\d E=T \d S$, there follows
a well known relation between the entropy and the partition function:
\be\label{Boxy}S=\left(1-\beta\frac{\d}{\d \beta}\right)\log Z.  \ee
In the classical limit, $Z$ is approximated as $e^{-I_\BH}$, so the formula for the entropy is
\be\label{oxy} S=-\left(1-\beta\frac{\d}{\d \beta}\right)I_\BH. \ee

In section \ref{action}, we will compute $I_\BH$ and compare the resulting expression for $S$ to the Bekenstein-Hawking formula.   In section \ref{acomp}, we will make
that comparison in another way.    But even without any of those computations, we can now fill a gap in the logic in section \ref{gsl}.   There, we recalled that Bekenstein was looking
for a nondecreasing quantity that could represent the entropy of a black hole, and, motivated by the Hawking area theorem, which says that the area $A$ of the black hole horizon
is nondecreasing, suggested that the black hole entropy is a constant multiple of the dimensionless quantity
$A/G$.      But from that point of view, any nondecreasing function of $A/G$, such as $(A/G)^2$, would
seem equally plausible.   Why is the entropy precisely a linear function of $A/G$?   Here we simply note that as the Einstein action is proportional to $1/G$, in particular
$I_\BH$ will be a multiple of $1/G$, so if the entropy defined as in eqn. (\ref{oxy}) is going to be a function of $A/G$, it will indeed have to be a constant multiple of $A/G$.
This was Bekenstein's original proposal, though certainly not the original logic.

\subsection{Computing the Action}\label{action}

Computing the Euclidean black hole action $I_\BH$ is not as straightforward as it sounds.  The usual Einstein-Hilbert action on a manifold $M$ of  Euclidean signature is
\be\label{zofo} I_\EH=-\frac{1}{16\pi G} \int_M \d^4x \sqrt g R. \ee
This vanishes in any classical solution of the associated field equations $R_{\mu\nu}=0$, since those  equations imply $R=0$.   

However, this usual form of the action has to be extended by adding the Gibbons-Hawking-York (GHY) boundary term \cite{GH,Y}.  This is necessary because $R$ depends on second
derivatives of the metric.   To understand why that is consequential, let us practice with the simple example of a free particle.   The standard action of a free particle is $I_1=\frac{1}{2}\int\d t \,\dot x^2$.
However, we could derive the same equation of motion from another action such as $I_2=-\frac{1}{2}\int \d t\,x \ddot x.$   Why is $I_1$ better than $I_2$?
 Let us derive the equations of motion from the naive action $I_2$, but on a finite time interval
$[t_1,t_2]$ and with boundary conditions in which the value of $x$ is specified at the endpoints, 
so that the variation of $x$ satisfies 
\be\label{gifme} \delta x(t)=0~{\rm for}~t=t_1,t_2.\ee   Generically, the specified values of $x$ at the endpoints are nonzero, $x(t_1),x(t_2)\not=0$.
We find $\delta I_2=-\int_{t_1}^{t_2}\d t \,\delta x \ddot x-\frac{1}{2}\left[ \delta \dot x \, x\right]_{t_1}^{t_2}.$   Vanishing of the bulk term gives the expected bulk equation of motion
$\ddot x=0$, but vanishing of the second term would give $x(t_1)=x(t_2)=0$.  Thus, the action $I_2$ will only work if we want the specified boundary values of $x$ to vanish.
 That restriction is avoided
for $I_1$: $\delta I_1=-\int_{t_1}^{t_2}\d t \,\delta x \ddot x$, given the boundary condition (\ref{gifme}).
We can also avoid the problem by adding a boundary term to $I_2$ and defining
  \be\label{blogo}I_2'=I_2 +\frac{1}{2} \left[ x\dot x\right]^{t_2}_{t_1} . \ee
The quickest way to verify that this avoids the problem is to observe that after integrating by parts, $I_2'=I_1=\frac{1}{2}\int_{t_1}^{t_2}\d t \dot x^2.$   Thus
adding this particular boundary term has compensated for a non-optimal choice of the bulk action.

The same problem arises for gravity if we study gravity on a manifold $M$ with boundary $\partial M$, with Dirichlet boundary conditions that specify the metric on $\partial M$.
Varying the naive action $I_\EH$ with respect to the metric, one obtains the expected Einstein equations in bulk, but because $I_\EH$ depends on second derivatives of the metric
one also finds an unwanted equation on the boundary which would
make the theory inconsistent (for a generic choice of the assumed boundary metric).    
General relativity is the rare example of a theory in which it is not possible in a covariant fashion to avoid the issue by integrating by parts 
and using a different bulk action.   However, one can avoid the problem by adding a boundary term to the action, namely the GHY boundary term
\be\label{nlogo} I_\GHY=-\frac{1}{8\pi G}\int_{\partial M} \d^3x \sqrt h \,K, \ee
where $K$ is the trace of the extrinsic curvature of $\partial M$, defined in the next paragraph.   The extended action \be\label{exteh} I'_\EH=I_\EH+I_\GHY\ee
 is analogous to
$I_2'$ for the free particle: it leads to a sensible variational problem, with no boundary term in the equations of motion.

To define $K$,  we introduce some conventions.   Local coordinates of  $\partial M$ will be denoted
 $x^i$, $i=1,\dots,3$ and local coordinates of $M$ near $\partial M$ will be denoted as $x^i$ and $y$, where $y=0$ on $\partial M$ and $y>0$ in a neighborhood of $\partial M$. Let
 $h_{ij}$ be the restriction to $\partial M$ of the metric $g$ of $M$.   Let $n$ be the  outward directed
 unit normal to $\partial M$ in $M$.   The extrinsic curvature of $\partial M$ is defined by $K_{ij}=D_i n_j$ and its trace is $K=h^{ij} K_{ij}$.
In Euclidean signature, the definition of $I_\GHY$ is then
\be\label{nlogox} I_\GHY=-\frac{1}{8\pi G}\int_{\partial M} \d^3x \sqrt h K. \ee

Near $\partial M$, it is always possible to put the metric of $M$ in the form
\be\label{thef} \d s^2=\d y^2+h_{ij}(x,y) \d x^i \d x^j.\ee    The outward directed unit normal to $\partial M$ is then
$n=-\partial_y$, 
so  $K_{ij}=-\frac{1}{2}\partial_y h_{ij}(x,y)|_{y=0}$ and $K=-\partial_y\left.\log {\sqrt{\det h}}\right|_{y=0}$.
If therefore we define the volume  of a hypersurface of fixed $y$ by $V(y)=\int_{\partial M}\d^3x\sqrt {\det h(x,y)} $, then  $I_\GHY$ can be expressed in terms
of the derivative of $V(y)$ at $y=0$:
\be\label{GHYev} I_\GHY= \frac{1}{8\pi G}\left.\frac{\d}{\d y}V(y)\right|_{y=0}\ee  

The Euclidean black hole geometry does not have a boundary in the usual sense, but it is not compact, and one needs to consider a sort of large radius limit of the GHY boundary term.
The procedure of Gibbons and Hawking was the following.   They cut off the Euclidean Schwarzschild solution at a large radius $r=\bar r$ and computed $I_\GHY$ with this
cut off in place. Let us call the result $I_\GHY(\bar r;\BH)$.  If one simply takes the limit of $I_\GHY(\bar r;\BH)$ for $\bar r\to\infty$, one finds that it diverges.
The same procedure also would give a divergence in the action for the flat manifold $\R^3\times S^1$.   Since Gibbons and Hawking wanted  the action of $\R^3\times S^1$ to vanish,
they subtracted away the boundary term for the case of $\R^3\times S^1$.  To define this subtraction precisely, 
they cut off $\R^3\times S^1$ in such a way that its boundary geometry is the same as that of the cut off black hole (adjusting the circumference of $S^1$ to make this
possible).   Then what they defined as the action of the Euclidean Schwarzschild solution was the large $\bar r$ limit of the difference of the boundary terms for the black hole
and for $\R^3\times S^1$:
\be\label{tellof} I_\BH = \lim_{\bar r\to\infty}\left(I_\GHY(\bar r; \BH)-I_\GHY(\bar r;\R^3\times S^1)\right). \ee
This trivially vanishes if the black hole is replaced by $\R^3\times S^1$.

With these preliminaries out of the way, the actual calculation is not difficult.
In the black hole solution, the circumference of the circle at $r=\bar r$ is $\beta_\sH\sqrt{1-\frac{2GM}{\bar r}}$, so one considers $\R^3\times S^1$ with that value of the circumference.
With the standard definitions of $r$ in the black hole metric (\ref{eucl}) and the $\R^3\times S^1$ metric (\ref{flatmetric}), a cutoff at $r=\bar r$ means that in each case the
boundary two-sphere has radius $\bar r$.   So with these choices the boundary three-geometries are the same.    To evaluate $I_\GHY$ using eqn. (\ref{GHYev}), 
we need the normal derivative $\frac{\d}{\d y}$ with $y$ normalized so that the metric looks like eqn. (\ref{thef}) near the boundary.
So
\begin{align}\label{gef} \frac{\d}{\d y}=\begin{cases} -\frac{\d}{\d r}  & \R^3\times S^1 \cr
                                                                                 -\left(1-\frac{2GM}{\bar r}\right)^{1/2}\frac{\d}{\d r} & {\rm Black ~hole} .\end{cases}\end{align}
The volume at radius $r$ is $\beta_\sH \left(1-\frac{2GM}{\bar r}\right) ^{1/2}\cdot 4\pi r^2$ for $\R^3\times S^1$, and       $\beta_\sH \left(1-\frac{2GM}{r}\right)^{1/2} \cdot 4\pi r^2$  for the black hole.
So eqn. (\ref{tellof})  together with (\ref{GHYev}) gives
\be\label{nelof} I_\BH=-\lim_{\bar r\to\infty}\beta_\sH \frac{1}{8\pi G}\sqrt{1-2GM/\bar r} \left.\frac{\d}{\d r}\right|_{r=\bar r}\left( 4\pi r^2\sqrt{1-2GM/r}-4\pi r^2 \right)
=\frac{\beta_\sH M}{2}=\frac{\beta_\sH^2}{16\pi G} . \ee
  Using this in eqn. (\ref{oxy}), we finally get the entropy:
  \be\label{zono} S= \frac{\beta_\sH^2}{16\pi G}=\frac{A}{4G}.\ee
Here  we used the usual $A=16\pi G^2 M^2$.                                                                       

\subsection{Another Computation Of The Entropy}\label{acomp}

The computation just described was quite a coup when it appeared, but one may ask a basic question.  Though this computation does give an answer proportional to the
horizon area $A$, the horizon played no particular role in the analysis and one might hope for a more direct explanation of why the  answer is proportional to the horizon area.

There is in fact an alternative derivation that avoids any subtle questions about the boundary contribution, and directly gives an answer proportional to the horizon area. 
(For an early version of this computation, see \cite{CT}.)
In this approach, we just use the fact that, on a manifold $M$ with boundary $\partial M$,  the modified action $I'_\EH$ with the GHY boundary term included has the property that, in expanding around any given metric, its variation is proportional to the Einstein equations, with no additional boundary term,
 as long as we consider only variations of the metric that leave fixed the induced metric on $\partial M$.
In other words, under a metric variation $g_{\mu\nu}\to g_{\mu\nu}+\delta g_{\mu\nu}$ with $\delta g_{\mu\nu}=0$ on the  boundary, the change in $I'_\EH$ to first order  is
\be\label{firstone}\delta I'_\EH=\frac{1}{16\pi G}\int\d^4x \sqrt g g^{\mu\mu'}g^{\nu\nu'}\delta g_{\mu'\nu'}\left(R_{\mu\nu}-\frac{1}{2} g_{\mu\nu}R\right), \ee
with no boundary term.   The specific boundary term $I_\GHY$ was chosen to make this true, but in the alternative computation of the action that we are about to explain,
all one needs to know  about $I_\GHY$ is that eqn. (\ref{firstone}) holds if it is included, and that $I_\GHY$ is defined by a local integral on $\partial M$.  One does not need
any more detailed knowledge about $I_\GHY$.

Since the formula (\ref{oxy}) for the entropy involves a derivative of the action with respect to $\beta$, to compute the entropy at some value of $\beta$, it does not
suffice to compute the action only at that value of $\beta$; one needs to know what happens if $\beta$ is changed.  But since the entropy depends only on the first derivative
of the action with respect to $\beta$, we only need to know what happens if $\beta$ is changed to first order.

In section \ref{action}, we computed the action by studying the Euclidean Schwarzschild solution as a function of $\beta $ or $M$.  When we varied $\beta$, we also varied
$M$ so that the solution remained smooth even at the origin.  

This means that when we varied the metric
with respect to $\beta$, we considered implicitly a metric variation $\delta_\beta g$ chosen so that the Einstein equations remain valid.      On the other hand, let $\delta'_\beta g$
be some other variation of the metric under a change in $\beta$, such that the change in the metric at infinity is the same, but the Einstein equations do not necessarily remain
valid.   Then $\tilde \delta g=\delta_\beta g-\delta'_\beta g$ is a metric variation in which there is no change in the metric at infinity, and in particular no change in the radius 
$\beta$ of the circle at infinity.   But that means, according to eqn. (\ref{firstone}), that as long as we are expanding around a classical solution of Einstein's equations, 
changing the metric by $\tilde \delta g$ does not affect the action to first order.   Therefore, in first order, the change in the action is the
same whether we use $\delta_\beta g$ or $\delta'_\beta g$.   We can make a convenient choice that simplifies the computation of the action.

Instead of varying $\beta$ and $M$ together, as we did in section \ref{action}, we can vary $\beta$ keeping $M$ fixed.   In other words, we leave the Euclidean Schwarzschild
solution (\ref{eucl}) completely unchanged.   All we change is the period $\beta$ of the coordinate $\tE$; instead of setting this to be $\beta_\sH=8\pi GM$, we allow an arbitrary period $\beta$.
(But since we only need the first derivative of the action with respect to $\beta$, it suffices to consider the case that $\beta$ is arbitrarily close to $8\pi GM$.)   
The near horizon metric (\ref{wefto}) is unchanged, as is the associated metric (\ref{efto}) in the two directions orthogonal to the horizon.   The only
difference is that  $\tE$ now has a general period $\beta$ rather than the specific period $8\pi GM$ that makes the metric smooth.   As a result,
the metric (\ref{efto}) has a conical singularity at the ``origin'' ${\t r}=0$.  Rather than the angle subtended by a loop around the origin being $2\pi$, it
is $\frac{\beta}{4 GM}$.   Such a conical singularity produces a delta function in the scalar curvature. 

In general,    if $P$ is a point in a Riemannian two-manifold $N$
that is a conical singularity subtended by an angle $2\pi +\varepsilon$, and $\delta_P$ is a delta function supported at $P$, then
the scalar curvature of $N$ has a delta function contribution 
\be\label{dozo}
R=-2\varepsilon \delta_P+\cdots. \ee
 This statement is actually a consequence of the Gauss-Bonnet theorem.
As a quick check, we will just consider the case $\varepsilon=-\pi$.   We can build a two-sphere by gluing together two identical squares along their boundaries;
this produces a metric that is flat except for four conical singularities with opening angle $\pi$, each corresponding to  $\varepsilon=-\pi$.
  Since this two-sphere is flat except at the four conical singularities, and the Gauss-Bonnet
theorem says that the integral of the scalar curvature for any metric on the two-sphere is $8\pi$, each  of the four conical singularities must produce in the
scalar curvature a delta function with coefficient $2\pi$.   This agrees with eqn. (\ref{dozo}) with $\varepsilon=-\pi$.

In our case, setting $\frac{\beta}{4GM}=2\pi +\varepsilon$, we have $\varepsilon=\frac{\beta-8\pi GM}{4GM}$, so the coefficient of the delta function  at the origin in the scalar curvature
of the two-dimensional metric (\ref{efto}) is $-2\varepsilon=-\frac{\beta-8\pi GM}{2GM}.$ When   $\beta\not=8\pi GM$, 
 the  Euclidean Schwarzschild solution (\ref{eucl})  has a conical singularity on the horizon, subtending  the angle $\beta/4GM$.
  So  its  scalar curvature has a delta function supported on the horizon with  coefficient $-\frac{\beta-8\pi GM}{2GM}$:
\be\label{doto}R=-\frac{\beta-8\pi GM}{2GM}\delta_H. \ee
Here  $\delta_H$ is a delta function supported on the horizon $H$ (that is, the integral of this delta function over the two directions normal to the horizon is 1).

Now we can evaluate the action.   First of all, we do not need to worry about the GHY boundary term.  It is the integral of a local 
expression on the boundary, with an   integrand that is independent of $\tE$.   Therefore, if we vary $\beta$ keeping $M$ fixed,
 the integral that gives the GHY boundary term is simply proportional to $\beta$.   But a multiple of $\beta$ is annihilated by the operator 
 $1-\beta \frac{\partial}{\partial \beta}$ and thus does not contribute in the entropy
formula (\ref{oxy}).
  
  So we can evaluate the entropy just from the bulk part of the action.
Of course, away from the horizon, regardless of the assumed value of $\beta$, the scalar curvature $R$ vanishes (as does the Ricci curvature $R_{\mu\nu}$). Thus,
the only contribution to the bulk action $I_\EH=-\frac{1}{16\pi G}\int \d^4x \sqrt g R$ comes from the delta function on the horizon.   
Hence  $\int\d^4x \sqrt g R= -\frac{\beta-8\pi GM}{2GM} A$, where we get a factor $-(\beta-8\pi GM)/2GM$ by integrating over the directions normal to the horizon and a factor of
the horizon area $A$ from integrating over the horizon.   So the action is
\be\label{nofo}I_\BH =\frac{\beta-8\pi GM}{32\pi G^2 M} A.\ee
And the entropy is
\be\label{wofo}S=-\left.\left(1-\beta\frac{\partial}{\partial\beta}\right)\right|_{\beta=8\pi GM} I_\BH=\frac{A}{4G},\ee
as found previously by other means.

This calculation goes through in much  the same way for other theories with static black holes, such as Einstein-Maxwell theory, in which one can consider
an electrically or magnetically charged black hole.   It is no longer true that the bulk contribution to the black hole action away from the horizon vanishes. But it is given by 
the  integral of a local Lagrangian density  that is independent of $\tE$, so it is proportional to $\beta$ and does not contribute to the entropy.  To analyze a black hole
that is stationary but not static, such as a Kerr black hole, one can give a similar analysis but now using  complex metrics, not just metrics of Euclidean signature \cite{GH}.  

This approach gives a satisfactory explanation of why the entropy of a stationary black hole is always $A/4G$ in the Euclidean approach to Einstein gravity, regardless of
the matter system considered and the charges carried by the black hole.   With the derivation given in section \ref{action}, this seems to require a separate computation in each case.
As an example of such a separate computation, see the derivation of the result (\ref{pilfo}) for the entropy of an AdS-Schwarzschild black hole.   As another example, though we have
here considered Schwarzschild black holes in dimension $D=4$, the generalization to any $D$ is straightforward.  The derivation that we have just given is valid for any\footnote{This statement
assumes
that the  Einstein-Hilbert action is normalized for any $D$ as $-\frac{1}{16\pi G}\int \d^Dx\sqrt g R$.} $D$ and
shows that the entropy is always
$A/4G$, independent of $D$, where now $A$ is the $(D-2)$-volume of the horizon (thus $A$ is only an area in the usual sense if $D=4$).   The Lorentz signature approach and  the derivation of
section \ref{action} can be used to verify this fact but do not give a direct explanation of it.

The derivation of the black hole entropy that we have described is actually the simplest example in which entropy is related to the response of a system to a conical
singularity.   We will see several further examples, centering around the replica trick (section \ref{sample}).

\subsection{Is The Entropy A Counting Of States?}\label{isit}

The Euclidean computation of black hole entropy is quite remarkable, and has been the starting point for many subsequent developments.  But it raises a fundamental
question.  In what sense is the quantity that is computed this way an entropy?   What are the states that are being counted to compute this entropy?  This is certainly one of the central
puzzles concerning black hole thermodynamics.

In the absence of gravity, to compute the partition function of a quantum field theory defined on a spatial manifold $W$, say at inverse temperature $\beta$, one simply
performs a path integral on $W\times S^1_\beta$.   This path integral manifestly has a state-counting interpretation: the states are defined on $W$ and they are propagating around the
circle $S^1_\beta$.  

The gravitational path integral on the Euclidean Schwarzschild spacetime  does not have that sort of interpretation, because the topology is wrong.  Near infinity the spacetime
looks like $\R^3\times S^1_\beta$, but in the interior, at the horizon,  the circle shrinks to a point. As a result, as noted earlier,
 the topology is actually $\R^2\times S^2$, not $W\times S^1$ for some $W$.
So the path integral on the Euclidean Schwarzschild spacetime  does not have any evident, or known, interpretation as a sum over states at inverse temperature $\beta$.

That is why in these notes, we have presented first what was essentially  Hawking's original derivation in Lorentz signature \cite{hawking} (with a later refinement \cite{FH}).
It is true that the Lorentz signature derivation also does not give much insight about what are the quantum states that account for the huge Bekenstein-Hawking entropy of
a black hole.   But at least in the Lorentz signature approach, 
 the principles involved in deducing  thermal emission from the black hole are clear cut.   The fact that the results deduced in Lorentz signature agree with what is 
found in the Euclidean derivation is one of the strongest reasons to believe that the Euclidean approach is correct.   

An entropy $S=A/4G$ suggests a number of states of order $e^{A/4G}$.   For a macroscopic black hole, this is a remarkably big number, especially given that
at the classical level a black hole is quite featureless, as summarized in the no hair theorem \cite{nohair,nohair2}.  Neither the Lorentzian approach nor the Euclidean approach
really explains this vast degeneracy.   Nonetheless the huge inferred number of quantum black hole states suggests
that somehow a black hole should be understood quantum mechanically as a complex dynamical system of some kind whose description requires a Hilbert space of this
enormous size.   That is actually the modern viewpoint, as explained for example in \cite{malda}.

\subsection{Negative Cosmological Constant}\label{negative}

There are very interesting and completely different lessons to be learned by studying horizons and black holes in a world with a positive or negative cosmological constant $\Lambda$.
Here we consider the case $\Lambda<0$, originally analyzed by Hawking and Page \cite{hp},   and in section \ref{cosmo} we take $\Lambda>0$.

In $3+1$ dimensions, the maximally symmetric solution of Einstein's equations with $\Lambda<0$ 
is Anti de Sitter space, which can be described by the metric
\be\label{adsmet}\d s^2= -\sV \d t^2+\frac{1}{\sV} \d r^2+r^2\d\Omega^2,\ee
where
\be\label{vis} \sV=1+\frac{r^2}{b^2}, \ee
\be\label{bdef} b=\sqrt{-\frac{3}{\Lambda}},\ee
 $\d\Omega^2$ is the metric on a unit two-sphere, and $0\leq r<\infty$, $-\infty<t<\infty$.  
The symmetry group is $SO(2,3)$.   One generator of this group is the time translation symmetry $\frac{\partial}{\partial t}$.  
We call the corresponding conserved charge the Hamiltonian $H$,  and its eigenvalue the energy $E$.
As usual, an ensemble in which one sums over all states weighted by $e^{-\beta H}$ can be defined by rotating to imaginary time
 with $t=-\i \tE$ and taking  $\tE$ to be a periodic variable with period $\beta$. 
We arrive at the metric 
\be\label{adsth}\d s^2= \sV \d \tE^2+\frac{1}{\sV} \d r^2+r^2\d\Omega^2.\ee
Though this metric is well-behaved  if $\tE$ is considered to be real-valued, we usually will consider it for the case that $\tE$ is a periodic variable, with some period $\beta$.
In that case, we call (\ref{adsth})  the metric of $\AdS_\beta$ (thermal AdS space with inverse temperature $\beta$), and we call the corresponding ensemble a thermal ensemble
with temperature $T=1/\beta$.
 As we explain at the end of this discussion, the modern interpretation is that $T$ is the temperature measured
 in a dual conformal field theory on the boundary \cite{MaldaDual,GKP,EW}.    However, for now we just note that $T$ is far from being the temperature that would be measured
 locally by an observer living in this spacetime.   The circumference of the circle parametrized by $\tE$ at a given value of $r$ is
 $\beta\sqrt{\sV(r)}$, which for very large $r$ grows as $\beta r/b$.   The locally measured temperature is the inverse of this and vanishes for large $r$  as $b/\beta r$.
Hence a particle of any given energy is very unlikely to be found at very large $r$. 
 
In the $\AdS_\beta$ metric (\ref{adsth}), the function $\sV(r)$ is positive-definite, so the circle parametrized by $\tE$ never shrinks to a point, and the topology of
$\AdS_\beta$ is $\R^3\times S^1$, the same as it would be if $\sV(r)$ were identically equal to 1. 
 
 The AdS-Schwarzschild solution describing a spherically symmetric black hole of mass $M$ has the same form as eqn. (\ref{adsmet})  or (in Euclidean signature) eqn. (\ref{adsth}), but now
 with
 \be\label{ergy} \sV(r)=1-\frac{2GM}{r}+\frac{r^2}{b^2}. \ee   It is no longer true that $\sV(r)$ is positive-definite; it is negative for sufficiently small positive $r$.  
  The black hole horizon is at the largest zero of $\sV(r)$, which we denote as $r_+$.   Thus 
 \be\label{invox} 1-\frac{2GM }{r_+}+\frac{r_+^2}{b^2}=0 \ee  or
 \be\label{ninvo}M=\frac{r_+}{2G}\left(1+\frac{r_+^2}{b^2}\right).\ee  The parameter $M$  is the ADM energy (the eigenvalue of  the Hamiltonian $H$) measured at infinity.
 Clearly, if $b$ is extremely large compared to $GM$, or equivalently if the size of the black hole is much less than the cosmological scale $\frac{1}{\sqrt{-\Lambda}}$, this solution can be well approximated  out to $r\gg GM$ by the usual Schwarzschild solution in asymptotically flat spacetime.  
 However, if $GM\gtrsim b$ or equivalently the black hole radius is comparable to the cosmological scale or bigger, the solution is quite different from the asymptotically flat case.

 For a black hole in an asymptotically Anti de Sitter spacetime, there is no close analog of the derivation of the Hawking effect that we explained in Lorentz signature in
 section \ref{bhevaporation}.  That is because a particle emitted from the black hole will always be reflected back by a sort of potential barrier near $r=\infty$; see \cite{BF},\cite{EW}.   However, it is straightforward to imitate the Euclidean derivation.
   Expanding $r=r_++{\t r}^2$ near $r=r_+$ and following the same steps as in section \ref{disappears}, we find that the Euclidean 
   AdS-Schwarzschild metric is completely smooth and complete
 if $\tE$ is a periodic variable with period
 \be\label{dunco}\beta=\frac{4\pi r_+ b^2}{b^2+3r_+^2}. \ee
 Once again, the modern interpretation  is that $T=1/\beta$ is the temperature measured in the boundary conformal field theory.
 
 The classical thermodynamics of the AdS black hole can be analyzed by either of the two methods in sections \ref{action} and \ref{acomp}.
 To imitate the derivation of section \ref{action}, we have to compute the action of the Euclidean black hole, subtracting the action of $\AdS_\beta$ adjusted to have the
 same asymptotic form of the metric.  In contrast to the case of a black hole in asymptotically flat spacetime, the GHY boundary term makes no contribution to the difference
 in action between the two solutions, because the function $\sV(r)$ is independent of  $M$ up to a term of relative order $1/r^3$, too small to be important.
Everything therefore comes from the bulk Einstein-Hilbert action
 \be\label{difo} I_\EH=-\frac{1}{16\pi G}\int\d^4x\sqrt g \left(R-2\Lambda\right).  \ee
 The Einstein equations imply that $R=4\Lambda$, so the action of a classical solution is
formally  $ \frac{3}{8\pi G b^2}V,$ 
 where $V$ is the volume of the spacetime and we used eqn. (\ref{bdef}).   As the black hole and $\AdS_\beta$ solutions both have infinite volume, 
 we regularize by truncating these spacetimes at $r=\bar r$ for some very large $\bar r$,
 and then  we subtract from the volume of the truncated black hole solution the volume of a thermal AdS space also cut off at $r=\bar r$ and with a suitably adjusted period of $\tE$ so
 that the two solutions have the same boundary metric.
 In the black hole solution, at $r=\bar r$, the circumference of the circle parametrized by $\tE$  is $\beta\sqrt{1-\frac{2GM}{\bar r}+\frac{\bar r^2}{b^2}}$.
 The thermal AdS solution with the same circumference is $\AdS_{\beta'}$ with
 \be\label{dino}\beta'=\beta  \frac{\sqrt{1-\frac{2GM}{\bar r}+\frac{\bar r^2}{b^2}}}    {\sqrt{1+\frac{\bar r^2}{b^2}}}   
 =\beta\left(1-\frac{GMb^2}{\bar r^3}+\cdots\right),\ee
 where the omitted terms vanish fast enough to be irrelevant.  So finally the action is 
 \be\label{nifo} I=\frac{3}{8\pi Gb^2}\Delta V, \ee where $\Delta V$ is the difference between the black hole and thermal AdS  volumes computed with these cutoff parameters.
 It is straightforward to compute that the volume of the black hole cut off at $r=\bar r$ is
 \be\label{wifo} V(\BH) =\frac{4\pi}{3}\beta\left(\bar r^3-r_+^3\right). \ee
 The volume of $\AdS_{\beta'}$ also cut off at $r=\bar r$ is
 \be\label{nifox}V(\AdS_{\beta'})=\frac{4\pi}{3}\beta\left(\bar r^3-GMb^2\right)=\frac{4\pi}{3}\beta\left(\bar r^3-\frac{b^2 r_+}{2}-\frac{r_+^3}{2}\right). \ee
 The regularized action of the black hole solution is the difference, or
 \be\label{tifo} I=\frac{\pi r_+^2(b^2-r_+^2)}{G(b^2+3r_+^2)}.\ee
 Relating $I$ to the partition function by $I=-\log Z$ and defining the entropy by $S=\left(1-\beta\frac{\d}{\d \beta}\right)\log Z$,
 where the derivative can be evaluated using (\ref{dunco}), we get
 \be\label{pilfo} S=\frac{\pi r_+^2}{G}=\frac{A}{4G}.\ee
 
 Alternatively, we can follow the logic of section \ref{acomp}. This tells us immediately, without any computation, that the entropy in the classical limit is $A/4G$.

In contrast to an ordinary Schwarzschild black hole in asymptotically flat spacetime, an AdS-Schwarzschild black hole cannot have an arbitrary temperature.
On the contrary, eqn. (\ref{dunco}) shows that $\beta$ has a maximum, and therefore $T=1/\beta$ has a minimum, at $r_+=r_0=\frac{b}{\sqrt 3}$, with
$\beta=\beta_0=\frac{2\pi b}{\sqrt 3}$.   For $\beta<\beta_0$, there are two black hole solutions, corresponding to the two values of $r_+$ that satisfy the quadratic equation
(\ref{dunco}).   To decide which solution, if either, is stable, we recall that thermodynamic stability requires that the specific heat $\frac{\partial M}{\partial T}$
should be positive.  Equivalently, $\frac{\partial M}{\partial \beta}$ should be negative.   Since eqn. (\ref{ninvo}) shows that $\frac{\partial M}{\partial r_+}$ is always
positive, the condition for stability is 
\be\label{tenz} \frac{\partial \beta}{\partial r_+}<0.\ee
From eqn. (\ref{dunco}),  this is so precisely if $r_+>r_0$.   In other words, the ``small'' black hole solutions with $r_+<r_0$ are thermodynamically
unstable, but the ``large'' ones with $r_+>r_0$ are stable.  The instability for small $r_+$ should not be a surprise, because for $r_+\ll b$, the solution goes over
to the usual Schwarzschild solution in asymptotically flat spacetime, whose thermodynamic instability was discussed in section \ref{unstable}. 

Since there is no black hole solution for $T<1/\beta_0$, how do we describe thermodynamics in Anti de Sitter space at low temperatures?   The answer is straightforward: at low
temperatures, we consider an ordinary thermal ensemble of particles in AdS space, described by perturbation theory around the classical solution $\AdS_\beta$.  
 For $\beta>\beta_0$,
this is presumed to be the dominant or perhaps only  classical solution  that contributes to a thermal ensemble at the given value of
$\beta$.  For $\beta<\beta_0$, there are two classical solutions to consider, namely $\AdS_\beta$  and the black hole with $r_+>r_0$.
We expect that in the limit $G\to 0$, the thermal ensemble is dominated by the solution of lower action.   The difference between the actions of the two solutions
is precisely what we computed in eqn. (\ref{tifo}).   The black hole solution dominates if and only if this difference of actions is negative, which is so  precisely if $r_+>b$.
Therefore, for $r_+>b$, the black hole is stable; a thermal gas without a black hole, described by the $\AdS_\beta$ solution, is stable locally at this temperature but should be able to  tunnel to a state
that contains a black hole.\footnote{An explicit description of this tunneling is unknown.}   For $r_+<b$, the thermal gas is stable.
For $\frac{b}{\sqrt 3}<r_+<b$, the action of the black hole solution is bigger, so although the black hole is stable locally, it can potentially  tunnel to
thermal $\AdS_\beta$.  The phase transition between thermal AdS and the black hole  at $r_+=b$ is called the Hawking-Page transition.   Of course, this is only a true sharp phase transition
in the limit $G\to 0$. 

When $\AdS_\beta$ is the dominant solution, the entropy is of order $G^0$ for small $G$, as noted earlier.   When the black hole dominates, the entropy is of order $G^{-1}$.

Hawking and Page made several further observations about this problem.  First of all, because the locally measured temperature vanishes at infinity, a particle of
given energy is confined to a bounded volume of space, even though the total volume of space at, say, time $t=0$ is infinite.   At a given temperature $T$, typical
particles have energy $E\lesssim T$ and therefore are confined in a bounded volume.   Thus a thermal gas in an asymptotically Anti de Sitter spacetime
has a bounded total energy, total entropy, and total number of particles.  
Related to this, in Anti de Sitter space, there is a well-defined microcanonical ensemble, in which one fixes the total energy rather than the temperature.   By contrast, in an asymptotically
flat spacetime, the space of states of bounded energy is of infinite dimension and there is not a well-defined microcanonical ensemble.    For more on the microcanonical ensemble in
Anti de Sitter space, see \cite{hp}.

From the preceding formulas, one can find the following scaling relations at large $r_+$ or high temperature:
\begin{align}\label{scaling} S&\sim T^2\cr E=M&\sim T^3 \cr S&\sim E^{2/3}.\end{align}   
The third relation, which follows from the first two, has the following consequence, as noted by Hawking and Page.    The number of states at energy $E$, which is expected to be roughly
$e^{S(E)}$, grows less than exponentially with $E$, and therefore the partition function $\Tr\,e^{-\beta H}$ converges for all $\beta>0$.   This contrasts with a black
hole in asymptotically flat spacetime, where $S\sim E^2$ and therefore the partition function diverges; that is another manifestation of the thermodynamic instability
of such a black hole.  

 \begin{figure}
 \begin{center}
   \includegraphics[width=4.8in]{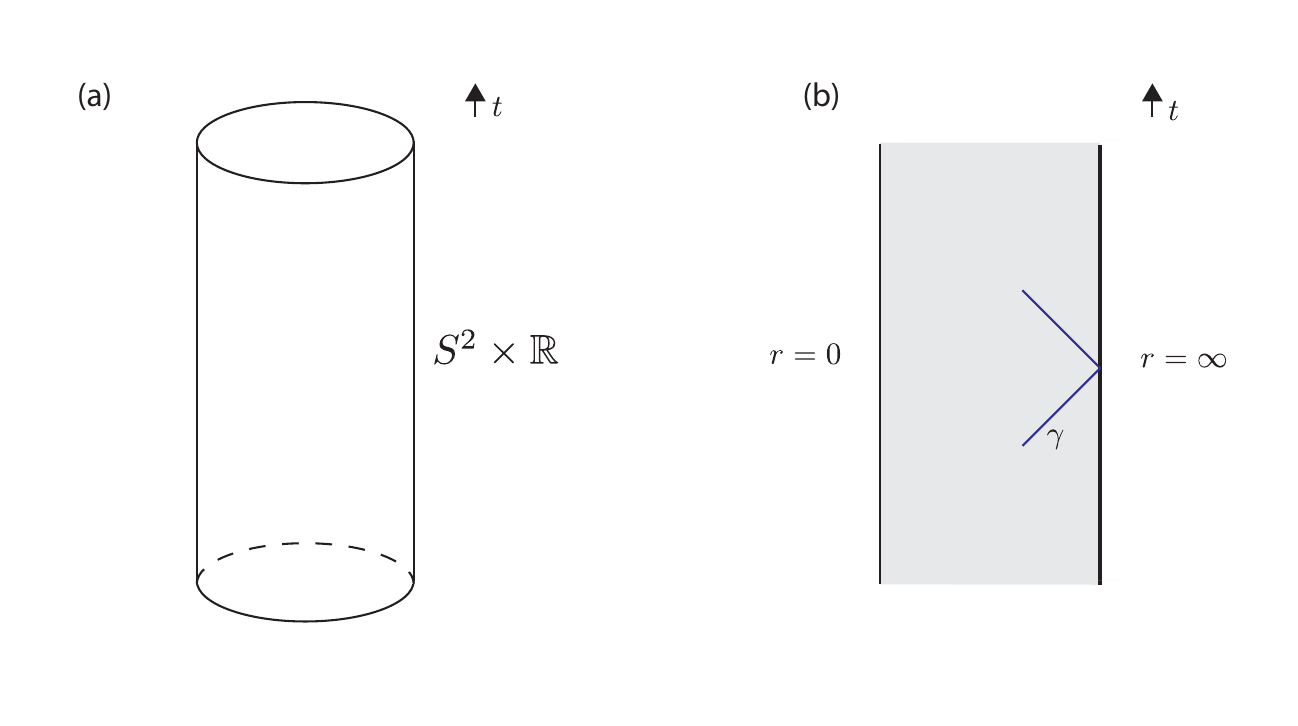}
 \end{center}
\caption{\footnotesize (a) Topologically, four-dimensional Anti de Sitter space is $B\times \R$ where $B$ is a three-dimensional ball, with conformal boundary $S^2$, and $\R$ parametrizes time.
  (b) Penrose diagram of Anti de Sitter space. As usual, time runs vertically in the diagram. 
   Plotted from left to right is the distance from the origin in $B$; $r=0$ is the left edge of the diagram and the conformal
boundary at $r=\infty$ is the right edge. 
The conformal boundary is infinitely far away if approached in a spacelike direction.
 But a massless particle, propagating
at a $\frac{\pi}{4}$ angle to the vertical on a worldline such as $\gamma$, can reach the conformal boundary and return in a finite time.  Hence boundary conditions on the conformal
boundary affect physics in the interior, making possible AdS/CFT duality.   \label{AdS}}
\end{figure} 

The  scaling laws in eqn. (\ref{scaling}) have a more specific interpretation.   They are characteristic of a $2+1$-dimensional conformal field theory
formulated in finite volume.   Indeed, this is related to the modern interpretation of  AdS black hole thermodynamics, which involves a dual conformal
field theory defined on the conformal boundary of spacetime.  The basic idea of how this duality arises can be briefly described as follows
(for illustration of some relevant facts, see fig. \ref{AdS}).
  Let $z=1/r$, so that $r=\infty$
corresponds to $z=0$.   Near $z=0$, the AdS metric can be approximated as
\be\label{mellow} \frac{\d z^2}{z^2}+\frac{1}{z^2} \left(-\d t^2+\d\Omega^2\right). \ee
What lies at $z=0$ is a copy of $\R\times S^2$, where $S^2$ is a two-sphere and $\R$ is parametrized by $t$.   
$\R\times S^2$  is sometimes called the Einstein static universe (in three spacetime dimensions).
More generally, a spacetime $M$ is said to be asymptotically locally
Anti de Sitter if one can pick a coordinate $z>0$ that vanishes at spatial infinity such that the
metric near $z=0$ takes what is known as the Fefferman-Graham form \cite{FG}:
\be\label{ellow}\frac{\d z^2}{z^2}+\frac{1}{z^2} \sum_{i,j} h_{ij} \d x^i \d x^j+\cdots, \ee
where $x^i$ are local coordinates on a  manifold $N$ of one dimension less,  $h_{ij}$ is a metric on $N$, and omitted terms are regular at $z=0$. Here the manifold $N$ that lies at $z=0$ is not part of $M$, as the distance to $z=0$ diverges:  $\int_0^\epsilon \frac{\d z}{z}=\infty$.  Rather $N$ is said to be the conformal
boundary of $M$, or the virtual boundary of $M$ at (spatial) infinity.\footnote{\label{dodads}It is sometimes convenient to complete $M$ by adjoining the virtual boundary $N$.   The completed space
$\bar M$ is a manifold with boundary  $N$. $\bar M$  does not have a natural Riemannian metric, but it does have a natural conformal structure, because if one multiplies
the metric in (\ref{mellow}) by a Weyl factor $z^2$, then it extends over $\bar M$. 
  An example of why this is useful is that it enables one to define the domain of dependence $D(U)$ of
a set $U\subset M$.   Denoting as $\bar U$ the closure of $U$ in $\bar M$, $D(U)$ is defined as the domain of dependence $D(\bar U)$ of $\bar U$ in $\bar M$, by the usual definition of footnote
\ref{dod}.  (The boundary conditions along $N$ that are used in defining the AdS/CFT correspondence ensure that the two definitions in footnote \ref{dod} give the same
result for  $D(\bar U)$.)}
  Equation (\ref{ellow}) may appear to show that $N$ is endowed with a natural Riemannian metric $h_{ij}$,
but this is misleading.   The Fefferman-Graham form (\ref{mellow}) of the metric is invariant modulo less singular terms under $z\to e^{\varphi} z$    where $\varphi$ is an arbitrary real-valued
function on\footnote{To be more precise, a coordinate transformation $(z,x^i)\to (e^\varphi z,x^i -\frac{z^2}{2} h^{ij}\partial_j\varphi)$ preserves the asymptotically locally AdS
form (\ref{mellow}) of the metric, while transforming the metric of $N$ as described in the text.}
$N$. 
This transforms the metric on $N$ by a Weyl transformation $h\to e^{2\varphi}h$.   So $N$ has a conformal structure (a Riemannian metric up to Weyl transformation)
but not a Riemannian structure. To illustrate this, we can compare the symmetry group of AdS space to that of its conformal boundary.
The isometry group of AdS space in four  spacetime dimensions
is\footnote{More precisely, assuming the time is taken to range over $-\infty<t<\infty$,  the AdS  isometry group is a cover of $SO(2,3)$ in which, roughly, the $SO(2)$ factor
is ``unwrapped.'' The same cover is the conformal group of the boundary.}  
$SO(2,3)$.  This group is also the conformal group (the group of diffeomorphisms that preserve the metric up to a Weyl transformation)
of the Einstein static universe $\R\times S^2$.   In comparing the symmetries, we have to allow conformal transformations of the boundary, not just isometries, so we have to take
into account the fact that intrinsically the boundary has only a conformal structure.

Though the distance to $z=0$ along a path of fixed $x^i$ is infinite, 
 it turns out that a light ray in an asymptotically AdS manifold can reach $z=0$ and return back in a finite time, as illustrated in fig. \ref{AdS}(b);  see for example \cite{EW}.   
 As a result, manipulating the boundary conditions along $N$
can affect physics in $M$.  By manipulating the boundary conditions along $N$, while also imposing suitable initial and final conditions in the far past and future, one can define
correlation functions on $N$ that satisfy all the general properties of correlation functions in a conformal field theory (CFT).  These ``boundary correlation functions''
are the closest analog in asymptotically AdS spacetime of the usual $S$-matrix elements that can be defined in an asymptotically flat world.

In many important examples of what the bulk gravitational theory might be, there are very strong reasons to believe
 that what can be defined  on the conformal boundary $N$ is not just a nice set of correlation functions that obey axioms of conformal field theory
but a full-fledged CFT whose content is precisely equivalent to the content of the  bulk gravitational theory in one dimension more.   In many instances, quite a few of which were described
in the original paper \cite{MaldaDual}, considerations of string theory and $M$-theory motivate precise duality conjectures relating a specific conformal field theory in $d$ dimensions
to a specific gravitational theory on an asymptotically AdS spacetime of one dimension more.
  By now, this AdS/CFT duality has been explored and  tested from
many points of view, and the evidence for it is extensive.   For  reviews, see for example \cite{VH,MaldaNew}.   

To be precise, in the AdS/CFT correspondence, one specifies the manifold $N$ on which one wishes to define the boundary CFT, together with the conformal structure of $N$.
Then in the dual gravitational description, one must allow all bulk manifolds $M$ that are asymptotic to $N$.  This in particular will entail a sum over topologies.  For example, suppose that we wish to study a thermal ensemble
in the dual CFT at inverse temperature $\beta$, on a spatial manifold $S^2$ of unit radius.   For this, we define the CFT on the Euclidean signature spacetime $N=S^2\times S^1_\beta$.
The AdS/CFT duality then instructs us that to give a dual gravitational description of this thermal ensemble, we should sum over
Euclidean manifolds that are asymptotically locally AdS with conformal boundary $N$.  For small $G$, one expects the gravitational description of the thermal ensemble to be dominated
by a classical solution of least action that has $N$ as its conformal boundary.  There are two candidates, namely $\AdS_\beta$ and the AdS black hole.   Which of them dominates
depends on $\beta$, as we have seen.  Clearly this reasoning leads back to the analysis of AdS black hole thermodynamics originally given by Hawking and Page in \cite{hp} and summarized
in the preceding paragraphs.   
The thermal ensemble of the boundary CFT on a compact manifold $S^2$ is well-defined at all temperatures,         
and satisfies the scaling relations\footnote{To be precise, this is true in a CFT that when formulated on the compact manifold $S^2$ has a discrete spectrum at all energies.
A generic CFT has that property, but there are special CFT's whose spectrum on $S^2$ is discrete only below some energy $E_0$ and continuous above that.
In such a case, the dual bulk description involves additional degrees of freedom (such as branes with special properties) such that the derivation of the scaling relations
(\ref{scaling}) is not valid.   For example, see \cite{SW}.}
 (\ref{scaling}) at high temperatures.   So those relations are a simple example of a prediction of the AdS/CFT duality.

From what we have said so far, the dual CFT provides an attractive  setting in which to understand AdS black hole thermodynamics.   But the applications of this duality go far beyond that.
For example, simple constructions in the boundary CFT can be used to create much more general density matrices than a thermal one, and the duality then gives a framework to
study these more general density matrices via gravity.  That will be important in discussing the Ryu-Takayanagi formula in section \ref{rt}.   At a more abstract level,
the boundary CFT is an ordinary quantum field theory governed by ordinary quantum mechanical laws.   Transferring this observation to the bulk,
 AdS/CFT duality provides a powerful reason to believe
that processes involving black holes can be described by unitary quantum mechanical evolution, though our understanding of quantum gravity does not yet
enable us to exhibit this fact directly. 

\subsection{Positive Cosmological Constant}\label{cosmo}

  \begin{figure}
 \begin{center}
   \includegraphics[width=3.8in]{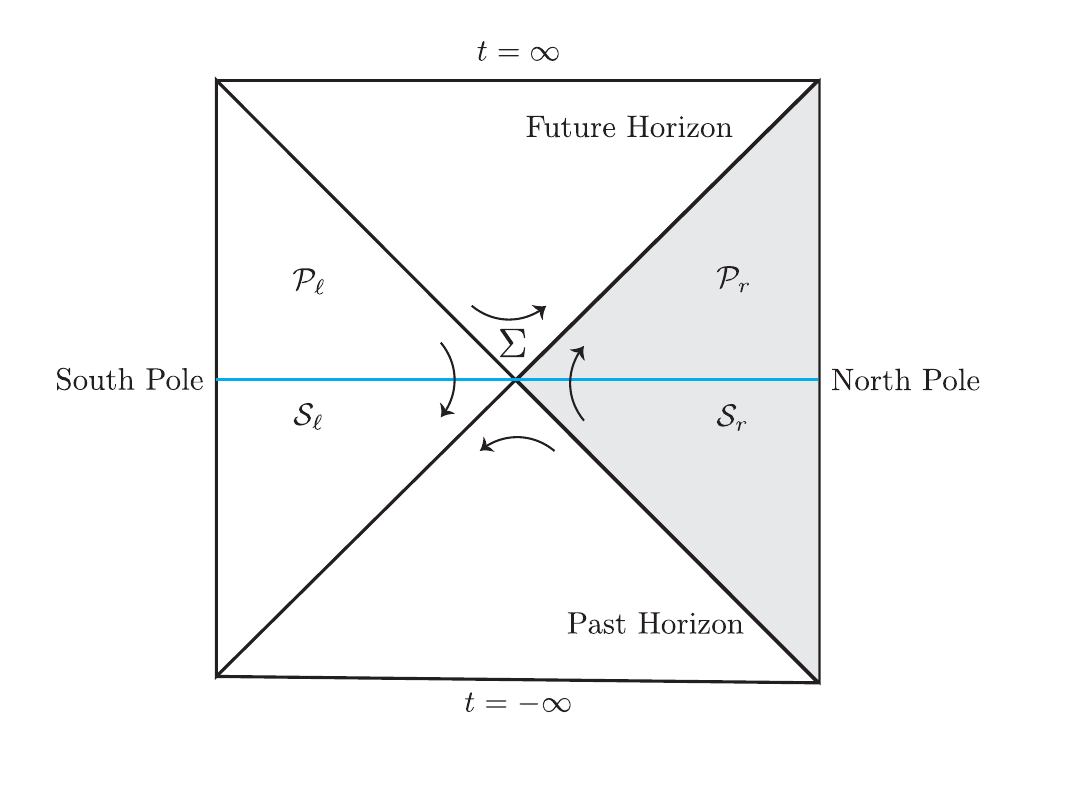}
 \end{center}
\caption{\footnotesize  The Penrose diagram of de Sitter space.  The worldline of a geodesic  observer who remains at rest at the north pole of the sphere is
the right edge of the diagram.   Shaded is the portion of the spacetime that the observer can see and influence, namely  the right static patch $\P_r$. Spacelike separated from $\P_r$
is a complementary left static patch $\P_\ell$.   $\P_r$ and $\P_\ell$ are  bounded, as shown, by past and future horizons that
intersect at the bifurcation surface $\Sigma$.   A Cauchy hypersurface $\S$ runs horizontally through $\Sigma$; its portions in $\P_r  $ and $\P_\ell$ are labeled
$\S_r$ and $\S_\ell$, respectively.  $\P_r$ is the domain of dependence of $\S_r$ and $\P_\ell$ is the domain of dependence of $\S_\ell$.  The arrows indicate the action of the
symmetry generator $H$, which is future-directed timelike in $\P_r$, past-directed timelike in $\P_\ell$, and spacelike elsewhere.   \label{desitter}}
\end{figure}

Apart from black hole horizons, there can also be horizons in cosmology.   A cosmological horizon bounds the portion of spacetime that a given observer can see or influence.
To be more precise, a future cosmological horizon bounds the region of spacetime that the observer can see, and a past cosmological horizon bounds the region that the observer can
influence.   Gibbons and Hawking proposed \cite{GH2}  that also in the case of a cosmological horizon, the quantity $\frac{A}{4G}$, where $A$ is the horizon area, can be interpreted as an entropy.
One motivation for this  proposal was that the area theorem \cite{AreaTheorem} applies equally to black hole and cosmological horizons, showing that $A$ is monotonically non-decreasing, and 
suggesting that $\frac{A}{4G}$ has some sort of thermodynamic interpretation for cosmological horizons, as for black hole horizons.  

An important difference between a cosmological horizon and a black hole horizon is that a cosmological horizon is very observer-dependent and describes the region of spacetime accessible
to some particular observer.  In the case of a black hole horizon in an asymptotically flat or asymptotically AdS spacetime, we usually consider an observer at a great distance and the precise
choice of observer does not affect the definition of the horizon.

A prototype of a spacetime with cosmological horizons is de Sitter space, which is the maximally symmetric solution of Einstein's equations with cosmological constant $\Lambda>0$.
In this particular case, unlike generic cosmologies, a temperature can be associated to the cosmological horizon, as we will see.
We will describe three different presentations of the de Sitter metric, which will exhibit various important aspects: exponential growth, topology, symmetries, and horizons.

To begin with, the  de Sitter geometry can be described by the line element
\be\label{nilbac}\d s^2=\varrho^2\left(-\d t^2+\cosh^2 t\,\d\Omega_3^2\right),\ee
where  $-\infty<t<\infty$, $\d\Omega_3^2$ is the metric of a three-sphere of unit radius, and
\be\label{ilbar}\varrho=\sqrt{\frac {3}{\Lambda}}. \ee
The  symmetry group of de Sitter space  is  $SO(4,1)$, though only an $SO(4)$ subgroup (rotating the three-sphere) is manifest in these coordinates. 
 The spatial section of de Sitter space at any constant time $t$ is a three-sphere of radius $\varrho \cosh t$.   
These spatial sections
expand exponentially towards either the future or the past; the part of the spacetime with $t\gg 1$ is believed to be an approximation to the present
universe.   These spatial sections are compact, so de Sitter space is a closed universe.

Any observer in de Sitter space  will experience past and future cosmological horizons.  For instance, we can consider an observer whose worldline is a geodesic.   An example
of a geodesic is the one that remains for all $t$ at the north pole of the three-sphere.   The metric along this worldline is just $\varrho^2\d t^2$, so the observer's proper time $\tau$
satisfies
\be\label{sattime}\d\tau=\varrho\,\d t. \ee
 Any other geodesic  is related to this one by
an $SO(4,1)$ transformation.    As usual, a Penrose diagram helps in visualizing the causal relations in de Sitter space.   In the Penrose diagram, we depict only two
coordinates in the spacetime -- the time and the ``latitude'' on the three-sphere.   The other polar angles on the sphere are suppressed.   A conformal mapping is made so that in the Penrose
diagram, causal curves travel at an  angle at most $\pi/4$ from the vertical.   The Penrose diagram of de Sitter space is depicted in fig. \ref{desitter}.  It is a square with $t=-\infty$
at the bottom and $t=+\infty$ at the top; the ``latitude'' on the sphere is plotted horizontally, and the left and right edges of the diagram represent the south and north poles of the sphere, respectively.
That means, in particular, that the right edge of the Penrose diagram
 is the worldline of an observer who remains at the north pole at all times (the left  edge is a geodesic that remains always
at the south pole).    The future horizon of the observer is the diagonal that slopes from the lower left to the upper right; the past horizon is the opposite diagonal.
The intersection of the two diagonals is called the bifurcation surface $\Sigma$.   It is a two-sphere of radius $\varrho$ and therefore area $A=4\pi\varrho^2$.   Any point on the past
or future horizon represents a two-sphere of the same area.

Another way to describe de Sitter space, making manifest its $SO(4,1)$ symmetry, is to represent it as a hyperboloid in a five-dimensional Minkowski space with metric
$-\d X_1^2+\sum_{j=2}^5 \d X_j^2$:
\be\label{dsmet} -X_1^2+\sum_{j=2}^5 X_j^2=\varrho^2. \ee
The relation to the  description in eqn. (\ref{nilbac}) is as follows.  Let 
\be\label{mildo} X_1=\varrho \sinh t,\ee
and for $j>1$, let $X_j=\varrho\cosh t\, Z_j$ with $\sum_{j=2}^5 Z_j^2=1$, so that the $Z_j$ parametrize
a unit three-sphere.
Then the metric $-\d X_1^2+\sum_{j=2}^5 \d X_j^2$ coincides with eqn. (\ref{nilbac}), with $\d\Omega_3^2=\sum_{j=2}^5 \d Z_j^2$.
In this representation, assuming that the north pole on the three-sphere  is taken to be at $Z_2=1$, with other $Z_j$ vanishing, 
 the observer's
orbit, parametrized by the observer's proper time $\tau$, takes the form
\be\label{zoggo}X_1=\varrho\sinh \frac{\tau}{\varrho},~~X_2=\varrho\cosh\frac{\tau}{\varrho},~~X_3=X_4=X_5=0.\ee
An important role will be played by the $SO(4,1)$ generator  $H$ that generates a Lorentz boost of the $X_1-X_2$ plane while acting trivially on $X_3,X_4,X_5$.
$H$ acts as $\varrho\frac{\d}{\d \tau}$ along the observer's  orbit.
The fixed point set of $H$ is the two-sphere $X_1=X_2=0$.   This is the bifurcation surface, a two-sphere $\Sigma$ of radius $\varrho$.

The portion of the spacetime that the observer can both see and influence, which  is bounded by the past and future horizons, is what we will call the right static patch $\P_r$.  The 
region spacelike from $\P_r$ will be called  the left static patch $\P_\ell$.  
Another change of coordinates gives a useful description of $\P_r$.  Let $X_1=\sqrt{\varrho^2-r^2}\sinh\frac{\tau}{\varrho}$, $X_2=\sqrt{\varrho^2-r^2}\cosh\frac{\tau}{\varrho}$,
and $X_j=r Y_j$ for $j=3,4,5$ with  $\sum_{j=3}^5 Y_j^2=1$.   The de Sitter metric becomes
\be\label{staticpatch}\d s^2=-\left(1-\frac{r^2}{\varrho^2}\right)\d \tau^2 +\frac{\d r^2}{1-\frac{r^2}{\varrho^2}}+r^2\d\Omega^2.\ee
The portion of de Sitter space that this coordinate system describes is precisely the right static patch $\P_r$. The observer worldline is at $r=0$, and  the coordinate $\tau$ is the observer's proper time.  The observer's horizon is in this description at  $r=\varrho$, with any  $\tau$, and is once
again a two-sphere of radius $\varrho$. (The $r,\tau$ coordinate system actually breaks down at the horizon and does not describe the portion of the spacetime beyond the horizon.)
 This way to write the metric of de Sitter space hopefully exhibits an analogy between the de Sitter cosmological horizon at $r=\varrho$ and the Schwarzschild horizon
 at $r=2GM$.   The difference is that in the case of a black hole, we usually consider an observer living ``outside'' the  horizon, while in the static patch metric (\ref{staticpatch}),
 the observer lives at $r=0$ and is surrounded by the cosmological horizon.
  The boost generator $H$ of the $X_1-X_2$ plane corresponds in this coordinate system to a symmetry that
 advances the observer's proper time
 \be\label{iko}H \sim \varrho \frac{\partial}{\partial\tau}. \ee
So this symmetry generator is $\varrho$ times the natural Hamiltonian, from the observer's point of view.   The static patch is time-independent if $\tau$ is viewed as the time
coordinate.  This is the reason for the name
``static patch.''

The fact that the time translation symmetry of the observer is a Lorentz boost of the $X_1-X_2$ plane is hopefully reminiscent of Rindler space, where a Lorentz boost
of the $x-t$ plane in Minkowski space is interpreted as the time translation symmetry of a uniformly  accelerating observer.\footnote{Here and in section \ref{thermofield},
 we will explain that the derivation of the thermal nature of Rindler   space given in 
\cite{UW} and presented in section \ref{division} has  close analogs for de Sitter space and for a Schwarzschild black hole.
This  was originally argued in  \cite{TJ}. For another perspective on the analogy between the three cases, see \cite{KW}. The framework of Tomita-Takesaki theory applies in a similar way
to all three cases \cite{Sewell}.}     Comparing figs. \ref{Seven} and \ref{desitter}, hopefully an analogy between the two cases  is evident.
The two spacelike separated static patches $\P_r$ and $\P_\ell$ correspond to the two Rindler wedges $\RR_r$ and $\RR_\ell$.      The boundaries of
$\P_r$ and $\P_\ell$ -- or of $\RR_r$ and $\RR_\ell$ -- are past and future horizons, which meet at a bifurcation surface $\Sigma$.    To the future of the future horizons
is a future wedge that an observer in $\P_r$ or $\P_\ell$ -- or in $\RR_r$ or $\RR_\ell$ -- cannot see.  Similarly to the past of the past horizons is a past wedge that these observers
cannot influence.   The boost symmetry that is future directed timelike
in $\P_r$ or $\RR_r$ is past directed timelike in $\P_\ell$ or $\RR_\ell$ and is spacelike in the past and future wedges.  A difference between the two cases is that in de Sitter space, an 
observer whose worldline is a geodesic can remain forever in $\P_r$ or $\P_\ell$, but in Rindler
space, to remain forever in $\RR_r$ or $\RR_\ell$, the observer must accelerate indefinitely in both the past and the future. 

De Sitter space can be continued to Euclidean signature by setting $t=-\i \tE$ in eqn. (\ref{nilbac}) or $X_1=-\i X_{1,\sE}$ in  eqn. (\ref{dsmet}).
The resulting Euclidean signature manifold is simply a four-sphere $S^4$ of radius $\varrho$, now with symmetry $SO(5)$:
\be\label{oggo} X_{1,\sE}^2+\sum_{j=2}^5 X_j^2=\varrho^2. \ee
In this description, $H$ acts by a rotation of the $X_{1,\sE}-X_2$ plane.   Analytic continuation from a Lorentz boost generator in Lorentz signature to  a rotation generator 
in Euclidean signature was  important in the analysis of Rindler space and plays a similar role in analyzing de Sitter space.

From the explicit description (\ref{zoggo}) of the observer's orbit, we see that after continuing the proper time $\tau$ to complex values, this orbit is periodic
in $\tau$ with period $\i\beta_\dS$, where $\beta_\dS= 2\pi\varrho$.   Therefore, in any $H$-invariant state that can be defined by analytic continuation from Euclidean signature, the observer will see
thermal correlations at temperature $T_\dS=1/\beta_\dS$, which is known as the de Sitter temperature.  Such a result was first found in \cite{FHN}, \cite{GH2}, from a different
but related point of view.

Actually, in quantum field theory in a background de Sitter space,\footnote{The linearization of gravity around de Sitter 
space can be viewed as a quantum field theory in a background de Sitter space.   But if gravity is treated beyond leading order, the de Sitter symmetry generators
become constraint operators and a quite different language is needed.}
 there is a natural de Sitter invariant state $\Psi_\dS$, sometimes called the Bunch-Davies state \cite{CT2,SS2,BD,Mo,Al}, which can be defined by
analytic continuation from Euclidean signature.   Literally, one starts with correlation functions on $S^4$,
\be\label{bumblebee}\bigl\la \phi(X_{1,\sE},\vec Z)\phi(X_{1,\sE}{}', \vec Z')\cdots \bigr\ra_{S^4} \ee
(where $\phi$ is a generic quantum field and again  $\vec Z=(X_2,\cdots,X_5)$).
These are unnormalized correlation functions, defined without dividing by a normalization constant.   The natural normalization constant  would be the $S^4$ partition function $Z$.
Then one defines Lorentz signature correlation functions by  analytic continuation to imaginary values of $X_{1,\sE}$, which correspond to real values of $X_1$.  Those
analytically continued correlation functions are by definition the expectation values of the same products of operators in the state $\Psi'_\dS$:
\be\label{tumblebee}\bigl\la \Psi'_\dS|\phi(X_1,\vec Z)\phi(X_1', \vec Z')\cdots |\Psi'_\dS\bigr\ra. \ee   We call this state $\Psi'_\dS$ rather than $\Psi_\dS$ because it is unnormalized;
tautologically, $\la\Psi'_\dS|\Psi'_\dS\ra=\la 1\ra_\dS=Z$.   So the normalized de Sitter state is
\be\label{umbletree} \Psi_\dS=\frac{\Psi'_\dS}{\sqrt Z}.\ee
Since the Euclidean correlation functions are manifestly $SO(5)$-invariant, their continuation to Lorentz signature is $SO(4,1)$-invariant, and in particular $H$-invariant.
Accordingly, in this state, the observer will see thermal correlations at the de Sitter temperature.

Analytic continuation of correlators from Euclidean signature has given a very quick and efficient way to describe the state $\Psi'_\dS$.  This
characterization of the state is also illuminating, as it makes manifest the invariance of the state under the de Sitter symmetry $SO(4,1)$.   However, it is important to also understand
how to describe the state as a functional of the quantum fields on a suitable spacelike hypersurface.  Apart from other applications that will become clear, this will fill 
 a serious gap in the definition of $\Psi'_\dS$.  We described
this state by a set of correlation functions, but to show that there really is a quantum state with those correlation functions, we need to verify a positivity condition
$\la \Psi'_\dS|\O^\dagger \O|\Psi'_\dS\ra\geq 0$, where $\O$ is a general linear combination of products $\phi(x_1)\phi(x_2)\cdots \phi(x_s)$.  This is obvious from  an alternative
definition of $\Psi'_\dS$ that we explain next.

  \begin{figure}
 \begin{center}
   \includegraphics[width=4.8in]{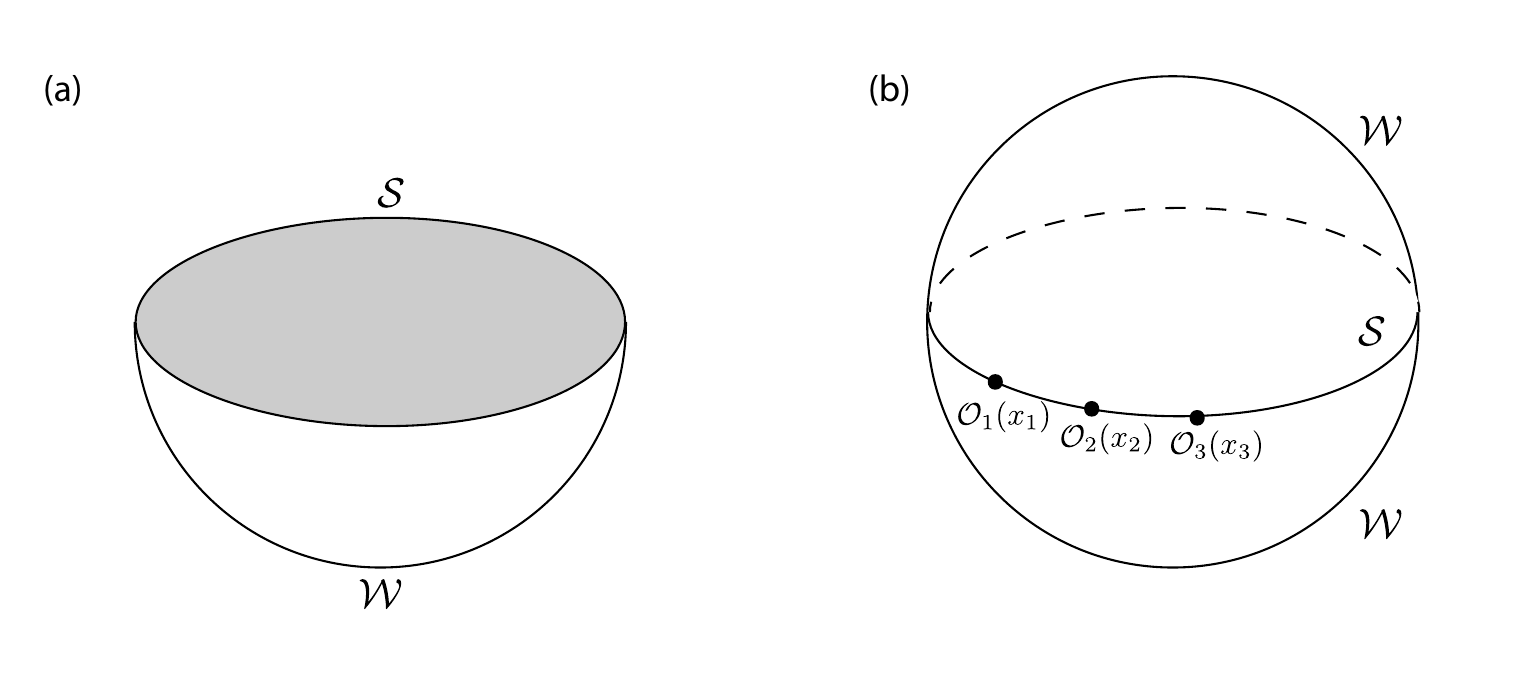}
 \end{center}
\caption{\footnotesize  (a) If $\S$ is an initial value surface in spacetime, a state $\Psi_\sW(\phi_\S)$ of a quantum field $\phi_\S$ on $\S$ can be defined by
a path integral over a Euclidean manifold $\sW$ of boundary $\S$.  In our application, $\S$ is the equator of $S^4$, and $\sW$ is the lower hemisphere.
(b) An expectation $ \la\Psi_\sW|\O_1(x_1)\cdots \O_s(x_s)|\Psi_\sW\ra $ is then computed as follows: a path integral on the lower hemisphere of $S^4$
computes the ket $|\Psi_\sW\ra$, and a path integral on the upper hemisphere of $S^4$ similarly computes the bra $\la \Psi_\sW|$.   To compute the matrix
element, the two hemispheres are glued together with insertion of $\O_1(x_1)\cdots \O_s(x_s) $ on the equator.   The upshot is that 
$ \la\Psi_\sW|\O_1(x_1)\cdots \O_s(x_s)|\Psi_\sW\ra $ is computed by a path integral on $S^4$ with the indicated operator insertions.
  \label{hemisphere}}
\end{figure} 

With this aim, we first note that as the relation $t=-\i \tE$ implies that $t=0$ coincides with $\tE=0$, it follows that the Euclidean hypersurface $\tE=0$ coincides
with the Lorentz signature hypersurface $t=0$, so a state defined by Euclidean methods at $\tE=0$ can be viewed as a state defined in Lorentz signature at $t=0$.
In Euclidean signature, one frequently useful way to define a quantum state on a hypersurface $\S$ is to do a path integral on a manifold $\sW$ of boundary $\S$
as a function of the boundary values on $\S$.   We will call this state $\Psi_\sW(\phi_\S)$, where $\phi_\S$ represents a generic quantum field on $\S$.
The general definition of $\Psi_\sW(\phi_\S)$ is (fig. \ref{hemisphere}(a))
\be\label{tolox}\Psi_\sW(\phi_\S)=\int_{\left.\phi_\sW\right|_\S=\phi_\S}  D\phi_\sW\, e^{-I(\phi_\sW)},\ee
where $\phi_\sW$ schematically represents a quantum field on $\sW$, $I(\phi_\sW)$ is the corresponding action, and the integral is over all fields $\phi_\sW$ on $\sW$ that
coincide with $\phi_\S$ on $\S$. The state $\Psi_\sW$ is unnormalized; we do not divide this path integral by any normalization factor.
 To implement the definition (\ref{tolox})  in the case at hand, we note from eqn. (\ref{oggo}) that in the  description of Euclidean de Sitter space
as a four-sphere, $\tE=0$ corresponds to $X_1=0$.   For a manifold whose boundary is  the hypersurface $\S$ defined by $X_1=0$, we can take the ``lower'' hemisphere
$\sW$ defined by $X_1\leq 0$.   

With these choices,  the state $\Psi_\W$ actually coincides with $\Psi'_\dS$.   To see this, let us first compare expectation values in the two states of a general product of local
operators $\O_1(x_1)\cdots \O_s(x_s)$, where $x_1,\cdots, x_s$ are  points\footnote{The points must be distinct in order for the matrix elements that we will discuss
to be well-defined.}   in $\S$.   To compute
\be\label{herbo}  \la\Psi'_\dS|\O_1(x_1)\cdots \O_s(x_s)|\Psi'_\dS\ra, \ee
with general points $x_1,\dots, x_s$ in de Sitter space of Lorentz signature, in general we are supposed to start with correlation functions on $S^4$ 
and then analytically continue from real $\tE$ to  real $t$.   However, in the particular case that the operators are all inserted in $\S$, no analytic continuation is necessary
because $\S$ already lies at real $t$, indeed at $t=0$.   So the definition of the state $\Psi'_\dS$ tells us that the correlation function (\ref{herbo}) is to be computed by
a path integral on $S^4$ with the operators inserted on the equator $\S$.

Now let us compare this to the corresponding expectation value
\be\label{herbox}  \la\Psi_\sW|\O_1(x_1)\cdots \O_s(x_s)|\Psi_\sW\ra \ee
in the state $\Psi_\sW$.   To compute this expectation value, we represent the ket $|\Psi_\sW\ra$ by a path integral on the lower hemisphere
$X_1\leq 0$, and  the bra $\la\Psi_\sW|$ by a path integral on the upper hemisphere $X_1\geq 0$.   An inner product $\la\Psi_\sW|\Psi_\sW\ra$ would be computed by a path integral on the sphere obtained by gluing together the two hemispheres.
To compute the desired expectation value  of $\O_1(x_1)\cdots \O_s(x_s)$, we glue the hemispheres together in the same way but now  with insertion
of $\O_1(x_1)\cdots \O_s(x_s)$ on their common boundary $\S$ (fig. \ref{hemisphere}(b)).  But this just builds up a path integral on $S^4$ with the desired operator insertions on $\S$ -- the
same recipe that computes the corresponding expectation value in the state $\Psi'_\dS$.   So we conclude that
\be\label{nerbo} \la\Psi'_\dS|\O_1(x_1)\cdots \O_s(x_s)|\Psi'_\dS\ra = \la\Psi_\sW|\O_1(x_1)\cdots \O_s(x_s)|\Psi_\sW\ra \ee
for arbitrary local operators $\O_1,\cdots, \O_s$ inserted  on the equator $\S$.

Is the restriction to insertions on $\S$ essential?   The answer to this question is ``no,'' basically because if $\O(x)$ is a local operator, then its derivatives of any finite order
are also local operators.   In free field theory, we can argue as follows.  Let $\phi$ be a free field that obeys, for example, a second order equation of motion.
Write $\dot \phi$ for $\frac{\partial \phi}{\partial \tE}$. 
Then by its equation of motion,  $\phi$ is determined  throughout $S^4$   in terms of the values of $\phi$ and $\dot\phi$ on the initial value surface $\S$, 
along with  the singularities at points where other operators are inserted, which are determined by universal operator product relations.
Hence  the equality (\ref{nerbo}) for points $x_i\in\S$, with the $\O$'s taken independently to be $\phi$ or $\dot\phi$, immediately implies a similar 
equality
\be\label{nerbox} \la\Psi'_\dS|\phi(x_1)\cdots \phi_s(x_s)|\Psi'_\dS\ra = \la\Psi_\sW|\phi(x_1)\cdots \phi(x_s)|\Psi_\sW\ra \ee
for any $x_1,\cdots,x_s$ in de Sitter space.   In free field theory, this suffices to show that $\Psi_\dS=\Psi_\sW$.

To reach the same conclusion in a non-free theory, we  use the fact that if $\O$ is a local operator, then so are its $t$ derivatives of any order.  
The basic idea is to try to expand a local operator $\O(t,x)$ that depends on both $t$ and a point $x\in \S$ as a sum of local operators at $t=0$.  For a local operator $\O(t,x)$
that depends on both the time $t$ and a point $x\in \S$, 
let $\O^{[k]}(x)=\left.\frac{\partial^k\O(t,x)}{\partial t^k}\right|_{t=0}$.   Then inside any matrix element in which this series converges, we can expand
\be\label{zdonk}\O(t,x)=\sum_{k=0}^\infty \frac{t^k}{k!}\O^{[k]}(x). \ee
We want to generalize eqn. (\ref{nerbo}) to a statement about operators $\O_i(t_i,x_i)$, that depend on time variables $t_i$ as well as the points $x_i\in\S$.
We make the expansion (\ref{zdonk}) for each of the operators involved.   Term by term we have an equality
\be\label{nerboxx} \la\Psi'_\dS|\O_1^{[k_1]}(x_1)\cdots \O^{[k_s]}_s(x_s)|\Psi'_\dS\ra = \la\Psi_\sW|\O^{[k_1]}_1(x_1)\cdots \O^{[k_s]}_s(x_s)|\Psi_\sW\ra\ee
since all operators involved are local operators  inserted on $\S$ and (\ref{nerbo}) holds for any such operators.
For distinct points $x_i$ and $x_j$ in $\S$, the points $(t_i,x_i)$ and $(t_j,x_j)$ are spacelike separated if $|t_i|$ and $|t_j|$ are small enough.   
Hence the left and right hand sides of the relation 
\be\label{werbo} \la\Psi'_\dS|\O_1(t_1,x_1)\cdots \O_s(t_s,x_s)|\Psi'_\dS\ra = \la\Psi_\sW|\O_1(t_1,x_1)\cdots \O_s(t_s,x_s)|\Psi_\sW\ra \ee
have power series expansions around $t_1=\cdots=t_s=0$  that are convergent for sufficiently small $|t_i|$.   So the term by term equality (\ref{nerbo}) implies
an equality for sufficiently small $|t_i|$, and hence for all $t_i$ by analyticity.   This is enough to show that $\Psi'_\dS=\Psi_\sW$.   Among other things, this confirms that $\Psi'_\dS$
is a quantum state; the original definition of $\Psi'_\dS$ by specifying a set of correlation functions did not imply that it has the positivity properties of a quantum state.  It was indeed not obvious
from the original definition that $\la \Psi'|\O^\dagger \O|\Psi'\ra\geq 0$ for all $\O$, where $\O$ is a general linear combination of products $\phi(x_1)\phi(x_2)\cdots \phi(x_s)$.

The equality $\Psi'_\dS=\Psi_{\sW}$ has a more simple analog in Minkowski space that was important in section \ref{rindler}.   The Minkowski space analog of $\Psi'_\dS$ is simply
the vacuum state $\Omega$.  The Minkowski space analog of $\Psi_\sW$ is the state prepared by a path integral on a half-space as in fig. \ref{Five}(a).   In the case of Minkowski
space, the usual argument that these states coincide is as follows.      A path integral over an infinite Euclidean
time (in this case, the Euclidean time elapsed in the half-space) projects onto the ground state, the state of minimum energy.   In Minkowski space, the vacuum state $\Omega$
is the unique state of minimum energy.  Therefore, a path integral on the Euclidean half-space prepares the vacuum state.   The reason that we cannot make a similar argument
in de Sitter space is that the state $\Psi'_\dS$ does not minimize anything.   Indeed, though a quantum field in a background de Sitter space has an $SO(4,1)$ symmetry group,
none of the $SO(4,1)$ generators are bounded below, because none of the corresponding Killing vector fields are everywhere future directed causal.   For example, the
Killing vector field  that is related to the conserved charge $H$, and that maps the observer worldline forwards in time in a sense described earlier,
 is future directed timelike in the static patch causally accessible to the observer,
but not elsewhere.
 
   \begin{figure}
 \begin{center}
   \includegraphics[width=4.1in]{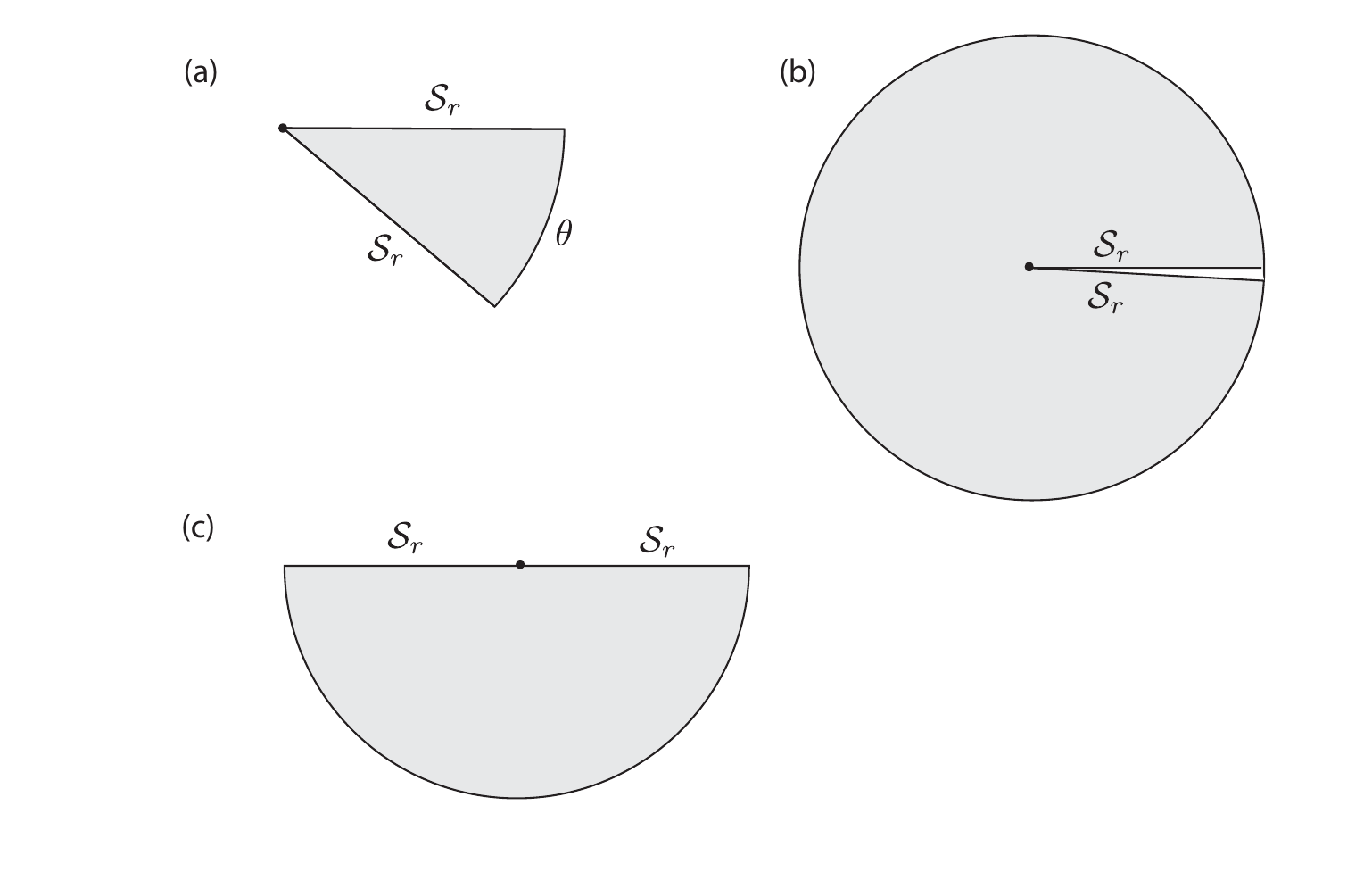}
 \end{center}
\caption{\footnotesize (a) 
 The operator $\exp(-\theta\varrho H_r)$ can be represented by a path integral on a wedge in $S^4$ obtained by rotating the three-dimensional hemisphere
 $\S_r$ through a Euclidean angle $\theta$, as in fig. \ref{Eight}(a) in the analysis of  Rindler space.
   (b) Setting $\theta=2\pi$, we get the density matrix appropriate for observations in $\S_r$ or more generally in the static patch $\P_r$, apart from a normalization factor.   (c)
Setting $\theta=\pi$, we recover the definition of the state $\Psi'_\dS$. We learn that the normalized state $\Psi_\dS$ is the thermofield double state of the two entangled regions $\S_\ell$ and $\S_r$,
or, after continuing to Lorentz signature, of their domains of dependence, namely the static patches
$\P_\ell$ and $\P_r$.
  \label{splittingx}}
\end{figure}

 That is why the proof that $\Psi'_\dS=\Psi_{\sW}$ required a much more elaborate argument than is needed to justify the corresponding statement in Minkowski space.
 But once we know that $\Psi'_\dS=\Psi_{\sW}$, it is straightforward to imitate all of the arguments and conclusions of the analysis of Rindler space in 
 section \ref{division}. That should not be a great surprise, given the analogy between de Sitter space and Rindler space that was explained earlier.   We will express the
 following results in an informal language.\footnote{As in footnote \ref{notable}, the Hilbert space decomposition and splitting of $H$ assumed shortly
 are  not truly correct in continuum quantum field theory.
 There is an alternative rigorous formulation in terms of Tomita-Takesaki theory.}
   Split the 
  the surface $\S$ as $\S_\ell\cup \S_r$, where $\S_r$ is the part of $\S$ to the right of the horizon and $\S_\ell$ is the part to the left (fig. \ref{desitter}).   Thus $\S_\ell$ and $\S_r$
  are the ``left'' and ``right'' hemispheres in $\S$.
  Correspondingly, split the Hilbert space of the theory as
   \be\label{welfo}\H=\H_\ell\otimes \H_r,\ee
 where $\H_\ell$ and $\H_r$ describe degrees of freedom on $\S_\ell$ and $\S_r$, respectively.
 The symmetry generator $H$ can be defined as an integral over $\S$,
 \be\label{tunigo}H=\int_\S \d \Sigma^\mu T_{\mu\nu} V^\nu,\ee
 where $T_{\mu\nu}$ is the stress tensor, $V^\nu$ is the Killing vector field on de Sitter space associated to the conserved charge $H$, 
 and $\d\Sigma^\mu$ is the appropriate integration form on $\S$.
 Similarly to the splitting of the boost generator $K$ in eqn. (\ref{lboost}), split $H$
 as 
 \be\label{zongo}H=H_r-H_\ell,\ee
 where $H_r$ and $H_\ell$ are defined by integrals similar to eqn. (\ref{tunigo}) restricted to $\S_r$ or $\S_\ell$.
So formally $H_r$ acts on $\H_r$ and $H_\ell$ acts on $\H_\ell$.

Given the state $\Psi_\dS\in\H$, we want to understand the corresponding reduced density matrices $\sigma_r$ and $\sigma_\ell$ that act on $\H_r$
and $\H_\ell$, respectively.   In Lorentz signature,
similarly to what happens in  Rindler space, $\P_r$ is the domain of dependence of $\S_r$ and $\P_\ell$ is the domain of dependence of $\S_\ell$.    
So $\sigma_r$ is the appropriate density matrix for observations in the static patch $\P_r$ and likewise
$\sigma_\ell$ is the appropriate density matrix for observations in $\P_\ell$. In other words,  if $\O$ is any operator in the region $\P_r$, then  $ \la\Psi_\dS|\O|\Psi_\dS\ra=\Tr\,\O\sigma_r$,
with a similar formula in $\P_\ell$.
It will be convenient to introduce unnormalized density matrices 
\be\label{unnorm}\sigma_\ell=\frac{\sigma'_\ell}{Z},~~\sigma_r=\frac{\sigma'_r}{Z}.\ee
Thus for any operator $\O$ in $\P_r$, 
$\la\Psi'_\dS|\O|\Psi'_\dS\ra=\Tr\,\O\sigma'_r,$
and similarly for $\P_\ell$.

The main claims are as follows.  
Similarly to what happens in Rindler space, one has 
 \be\label{wingo} \sigma'_\ell=\exp(-2\pi \varrho H_\ell),~~~~~\sigma'_r=\exp(-2\pi \varrho H_r). \ee
 Hence
  \be\label{wingot} \sigma_\ell=\frac{1}{Z}\exp(-2\pi \varrho H_\ell),~~~~~\sigma_r=\frac{1}{Z}\exp(-2\pi \varrho H_r). \ee
 Thus observations in $\P_r$ or $\P_\ell$ are thermal with Hamiltonian $H_r $ or $H_\ell$ at the de Sitter temperature
 $T_\dS=1/2\pi \varrho$.
    And also in analogy with Rindler space,
 the pure state $\Psi_\dS\in \H_\ell\otimes \H_r$ can be viewed as the thermofield double state $\Psi_\TFD$ of the entangled pair $\P_\ell$ and $\P_r$.

 Once one knows that $\Psi'_\dS=\Psi_{\sW}$, the proofs of these statements precisely parallel the corresponding arguments concerning Rindler space.  As in fig. \ref{splittingx}(a)
 (and as in fig. \ref{Eight} in the case of Rindler space),
 one interprets the operator $\exp(-\theta \varrho H_r)$ (or $\exp(-\theta \varrho H_\ell)$) in terms of a path integral on a wedge defined by rotating $\S_r$ (or $\S_\ell$) inside $S^4$
 through a Euclidean
 angle $\theta$.   Setting $\theta=2\pi$ as in fig. \ref{splittingx}(b), this path integral computes $\sigma'_r$ (or $\sigma'_\ell$), leading to the statements (\ref{wingo}) and hence (\ref{wingot}).
 Since the thermal density matrix of $\P_r$ is  $\frac{1}{Z}\exp(-2\pi\varrho H_r)$, the thermofield double state is  $\Psi_\TFD=\frac{1}{\sqrt Z}\exp(-\pi \varrho H_r)$.   But $\exp(-\pi\varrho H_r)$, which rotates $\S_r$ through an angle $\pi$ (fig. \ref{splittingx}(c)), is represented by the path integral
 on the lower hemisphere of $S^4$ that was the definition of $\Psi_\sW$.  From  $\Psi'_\dS=\Psi_\sW=\exp(-\pi\varrho H_r)$, it follows  that 
 \be\label{zuffy} \Psi_\dS=\frac{1}{\sqrt Z}\exp(-\pi\varrho H_r)=\Psi_\TFD.\ee
 
All of this has been quantum field theory in a fixed background de Sitter spacetime.   Now we will consider gravity.
Gibbons and Hawking proposed that not only does quantum field theory in de Sitter space have the thermal interpretation that we have explained, but in the context of gravity,
$\frac{A}{4G}$, where as usual $A$ is the horizon area, is an entropy, which they said ``measures the lack of information of the observer about the regions which he cannot see.'' 
In many respects, de Sitter entropy is  less well understood than black hole entropy.  But pragmatically, we can ask how to compute the   de Sitter entropy by Euclidean path integrals.   In the case of a black hole in an asymptotically flat or asymptotically AdS
spacetime, the Euclidean path integral  is interpreted as $e^{S-\beta E}=e^{-\beta F}$, where $E$ is the energy and $F$ is the free energy.   The energy that
appears in this formula is the ADM energy measured at spatial infinity. As de Sitter space is a closed universe, there is no spatial infinity so there is no conserved
charge analogous to the energy that could appear in such a formula.\footnote{At $G=0$, which corresponds to studying quantum fields in a fixed background spacetime, one has conserved charges associated to the $SO(4,1)$ symmetry of de Sitter space.   None of these charges is bounded below, since none of the corresponding Killing vector fields is everywhere future-directed  timelike.
 As soon as $G>0$, these conserved charges
become constraints that annihilate physical states.  They do not remain as operators in the quantum theory.}   It is proposed therefore that the Euclidean path integral for gravity and possibly other fields in de Sitter space should be interpreted simply
 as $e^S$.
In the semiclassical limit, this would be approximated as $e^{-I}$, where $I$ is the action, so the semiclassical formula is just $S=-I$.
  Since the Einstein equations with a cosmological constant imply (in four dimensions) that $R=4\Lambda$, the
form (\ref{difo}) of the Einstein-Hilbert action with a cosmological constant implies that 
\be\label{invo} I=-\frac{\Lambda V}{8\pi G},\ee
where $V$ is the volume of the Euclidean solution.  
As a four-sphere of radius $\varrho$ has volume $\frac{8\pi^2}{3}\varrho^4$, we find, with $\Lambda=\frac{3}{\varrho^2}$,  that the action is $-\frac{\pi \varrho^2}{G}$.  The horizon
is a two-sphere of radius $\varrho$ and area $4\pi\varrho^2$, so this  result for the action implies the expected
\be\label{melme}S=-I=\frac{A}{4G}.\ee
It is also possible to adapt the reasoning of section \ref{acomp} and prove directly  that $I=-\frac{A}{4G}$ without computing either $I$ or $A$.  

Eqn. (\ref{invo}) is applicable for the action of any solution of Einstein's equations with a cosmological constant and shows that for $\Lambda>0$, the action is a negative
multiple of the volume.  So one might expect the dominant solution to be the one of largest volume. In any dimension, among classical solutions of Einstein's equations with $\Lambda>0$, a  sphere has maximum volume.   This is proved in
\cite{bc}, p. 254; there is also a nice explanation in the Wikipedia article  ``Bishop-Gromov Inequality.''
Hence  it has been proposed that for small $G$, with $\Lambda>0$, the sum over topologies in  the Euclidean path integral of gravity is dominated by the Euclidean version of de Sitter space.

\subsection{The Thermofield Double State of a Black Hole}\label{thermofield}

For quantum fields propagating in the  Schwarzschild black hole spacetime, 
there is a close analog of the state $\Psi_\dS$,  sometimes called the Hartle-Hawking-Israel
state  $\Psi_\HHI$ \cite{HH,I}.   This state describes
a black hole in equilibrium with radiation at the Hawking temperature.  In constructing this state and describing its basic properties, it does not matter much
whether the cosmological constant is zero or negative.\footnote{There is no such construction with positive cosmological constant.  There exists a spherically
symmetric dS-Schwarzschild solution with $\Lambda>0$, but because the black hole and cosmological horizons have different temperatures, an equilibrium state
does not exist.}  For definiteness, we consider the case $\Lambda=0$ and make some brief remarks about the case $\Lambda<0$ at the end.

We recall the Schwarzschild solution
\be\label{tonzo}\d s^2=-\left(1-\frac{2GM}{r}\right)\d t^2+\frac{\d r^2}{1-\frac{2GM}{r}}    + r^2\left(\d\theta^2+\sin^2\theta \d\varphi^2\right) \ee
and its Euclidean continuation with $t=-\i\tE$:
 \be\label{tonzox}\d s^2=\left(1-\frac{2GM}{r}\right)\d \tE^2+\frac{\d r^2}{1-\frac{2GM}{r}}+ r^2\left(\d\theta^2+\sin^2\theta \d\varphi^2\right). \ee

As in the case of the de Sitter invariant state $\Psi_{\dS}$, we will consider two different ways to characterize the state $\Psi_\HHI$.
The equivalence of the two can be proved the same way that we proved $\Psi'_\dS=\Psi_\sW$ in section \ref{cosmo}.
First, we can define $\Psi_\HHI$ by saying that correlation functions in this state are analytic continuations from Euclidean signature.
In other words, with $\phi$ a generic quantum field, we begin by considering correlation functions defined by path integrals on the Euclidean Schwarzschild spacetime:
\be\label{winkly} \bigl\la \phi(t_{{\sf E},1},r_1,\theta_1,\varphi_1)\phi(t_{{\sf E},2},r_2,\theta_2,\varphi_2)\cdots \phi(t_{{\sf E},n},r_n,\theta_n,\varphi_n)\bigr\ra_{\ES}. \ee
Then we analytically continue to imaginary values of $t_{{\sf E},1},t_{{\sf E},2},\cdots,t_{{\sf E},n}$, which correspond to real values of $t_1,t_2,\cdots,t_n$,
to define the corresponding correlation functions in a state that we call  $\Psi'_\HHI$:
\be\label{inky}\la\Psi'_\HHI|\phi(t_1,r_1,\theta_1,\varphi_1)\phi(t_2,r_2,\theta_2,\varphi_2)\cdots \phi(t_n,r_n,\theta_n,\varphi_n)|\Psi'_\HHI\ra. \ee
This suffices to characterize the state $\Psi'_\HHI$.   The correlation functions in eqn. (\ref{winkly}) are unnormalized, so the state $\Psi'_\HHI$ is unnormalized.
By definition, $\la\Psi'_\HHI|\Psi'_\HHI\ra=\la 1\ra_{\ES}=Z$, where $Z$ is the path integral on the Euclidean Schwarzschild spacetime  with no operator insertion.
So the normalized version of the state that we have defined is
\be\label{normver}\Psi_\HHI=\frac{1}{\sqrt Z}\Psi'_\HHI.\ee  

From this construction, it is immediate that $\Psi_\HHI$ has a simple interpretation far from the black hole.  For $r\gg 2GM$, the Euclidean Schwarzschild spacetime is
asymptotically $ \R^3\times S^1$, with the circumference of $S^1$ being the inverse of the Hawking temperature $T_\sH$.  Since a path integral on $\R^3\times S^1$ with that value of the
circumference describes a thermal ensemble at temperature $T_\sH$, we conclude that far from the black hole the correlation functions in the state $\Psi_\HHI$ reduce to thermal
correlators at that temperature.   Actually, since in eqn. (\ref{inky}) we have analytically continued from Euclidean time to real time, these correlation functions, at $r\gg 2GM$,
are real time correlators in a thermal ensemble.   

 The state $\Psi_\HHI$ is time-independent, since the correlators (\ref{inky}) are certainly invariant under time translations.  So it describes an equilibrium state of thermal radiation
 at the Hawking temperature interacting with a black hole.    These statements apply in the limit $G\to 0$.   Beyond that limit, one will encounter instabilities described in section
 \ref{unstable}. In the case $\Lambda<0$, $\Psi_\HHI$ can be  defined precisely as we have just done, and the instabilities are avoided if the black hole mass is big enough.
 
The state $\Psi_\HHI$ has a surprising property that is somewhat hidden in the way we have presented it and is more obvious in an alternative description of $\Psi_\HHI$ as a functional
of quantum fields on an initial value surface $\S$.   To find this alternative construction, one can  imitate 
 the procedure that we followed in section \ref{cosmo} in the case of de Sitter space, but care is needed.
 In analyzing de Sitter space, we observed that the relation $t=-\i \tE$ implies that the Lorentz signature hypersurface $t=0$ coincides with the Euclidean signature hypersurface $\tE=0$.
This suggested to define the quantum state on the $\tE=0$ hypersurface, which in the case of de Sitter space can be immediately interpreted as a Cauchy hypersurface in Lorentz
signature.

If we do precisely this in the case of a black hole, we run into a snag.  The problem can be seen if we recall that the near horizon geometry of the Euclidean black hole,
after changing variables from $r$ to a suitable variable $\t r$ that vanishes on the horizon is (as in eqn. (\ref{wefto}))
\be\label{uto}\d s^2=\frac{\t r^2 \d\tE^2}{(4GM)^2}+\d \t r^2.\ee
plus the metric of a two-sphere of radius $2GM$.   Eqn. (\ref{uto}) describes $\R^2$ in polar coordinates if $\tE$ is regarded as a polar angle of period $8\pi GM$.
Thus the condition $\tE=0$ defines a ray in $\R^2$, ending at the horizon, which is the origin of polar coordinates at $\t r=0$.  Hence an attempt to define a Cauchy hypersurface
by a condition $\tE=0$ will fail: this condition defines a partial Cauchy hypersurface, a manifold with a boundary at the horizon.

Instead, we should slightly modify the approach and define $\S$ to be the fixed point set of the time-reversal symmetry $t\leftrightarrow -t$. 
As we noted in section \ref{disappears}, this symmetry is the reason that  the Schwarzschild solution has a continuation to Euclidean signature.
  In Lorentz signature, the fixed point set of $t\leftrightarrow -t$ is simply the hypersurface
$t=0$.  In Euclidean signature, the symmetry $t\leftrightarrow -t$ becomes $\tE\leftrightarrow -\tE$.  Since $\tE$ is a periodic variable of period $\beta_\sH=8\pi GM$, the transformation $\tE\leftrightarrow -\tE$ actually
has fixed points both at $\tE=0$ and at $\tE=\frac{1}{2}\beta_\sH$.  In the copy of $\R^2$ described in eqn. (\ref{uto}), the two rays $\tE=0$ and $\tE=\frac{1}{2}\beta_\sH$ fit together
into  a copy of $\R$ that actually is a geodesic.   This is no coincidence; the fixed point set of a symmetry of a manifold is  always a submanifold, never a submanifold with boundary.  The problem
with defining a submanifold by $\tE=0$ was that $\tE$ fails to be a good coordinate at the horizon.

   \begin{figure}
 \begin{center}
   \includegraphics[width=4.1in]{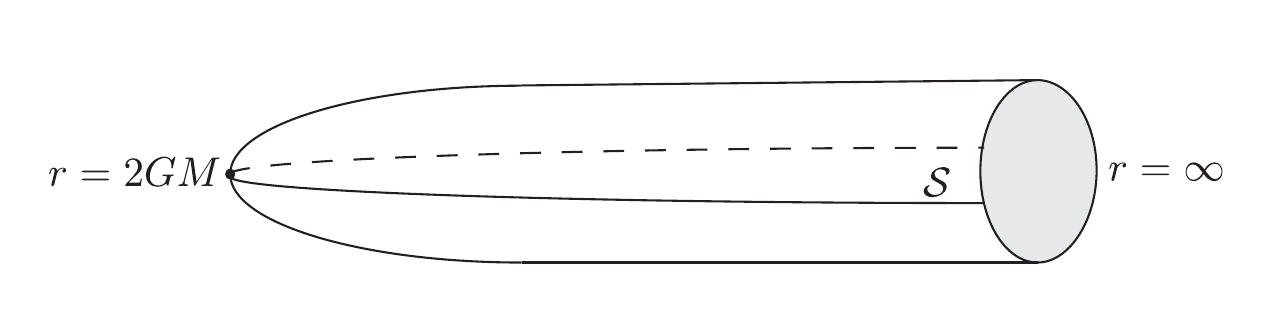}
 \end{center}
\caption{\footnotesize  
This two-dimensional surface of revolution is the restriction of the Euclidean Schwarzschild spacetime   to fixed values of the polar angles $\theta,\phi$. The ``tip''
  of the ``cigar'' represents the black hole horizon at $r=2GM$.   The time-reversal symmetry $\tE\leftrightarrow -\tE$ acts as a reflection of the surface, leaving fixed the tip. 
 The  fixed point set $\S$ of time-reversal  is a copy of $\R$ that comes in from $r=\infty$, bends around the tip of the cigar,  and returns to $r=\infty$.  \label{cigar}} 
\end{figure}

We can visualize the essential aspects of the Euclidean black hole metric   eqn. (\ref{tonzox})  if we restrict to some fixed value of the polar angles $\theta,\varphi$.
The restriction is a two-manifold $U$, parametrized by $r$ and $\tE$, with line element
\be\label{revsurf} \d s^2=\left(1-\frac{2GM}{r}\right)\d \tE^2+\frac{\d r^2}{1-\frac{2GM}{r}}.\ee
  $U$  is a surface of revolution because $\tE$ is an angular variable and the metric is invariant under
 constant shifts of $\tE$.   Since the asymptotic
circumference of the circles  parametrized by $\tE$ is a constant $\beta_\sH$, $U$ is roughly a semi-infinite cigar (fig. \ref{cigar}).   In this picture, we can visualize the codimension one fixed point set  $\S$
of the symmetry $\tE\leftrightarrow -\tE$.   Starting in the asymptotic region at large $r$ on the ``near'' side of $U$ at $\tE=0$, $\S$ continues inward all the way to the horizon
at $r=2GM$, which is the ``tip'' of the cigar,
and then returns back to large $r$  on  the ``far'' side of $U$ at $\tE=\frac{1}{2}\beta_\sH$.

Now that we have defined a hypersurface $\S$ in the Euclidean Schwarzschild spacetime, we can describe  the state $\Psi'_\HHI$
by a path integral on a manifold  $\sW$ with boundary $\S$.   We can define $\sW$ by $0\leq \tE\leq \frac{1}{2} \beta_\sH$, for example, corresponding to the bottom (or top)
half of the surface of revolution in fig. \ref{cigar}.  Then we define $\Psi'_\HHI$ by a path integral on $\sW$ with fixed boundary values on $\S$:
\be\label{omigo} \Psi'_\HHI(\phi_\S)=\int_{\left.\phi_\sW\right|_{\S}=\phi_\S}D \phi_\sW \,e^{-I(\phi_\sW)}. \ee
That the two definitions of $\Psi'_\HHI$ are equivalent follows from precisely the same argument that we used in section \ref{cosmo} to show that $\Psi'_\dS=\Psi_\sW$.

   \begin{figure}
 \begin{center}
   \includegraphics[width=5.1in]{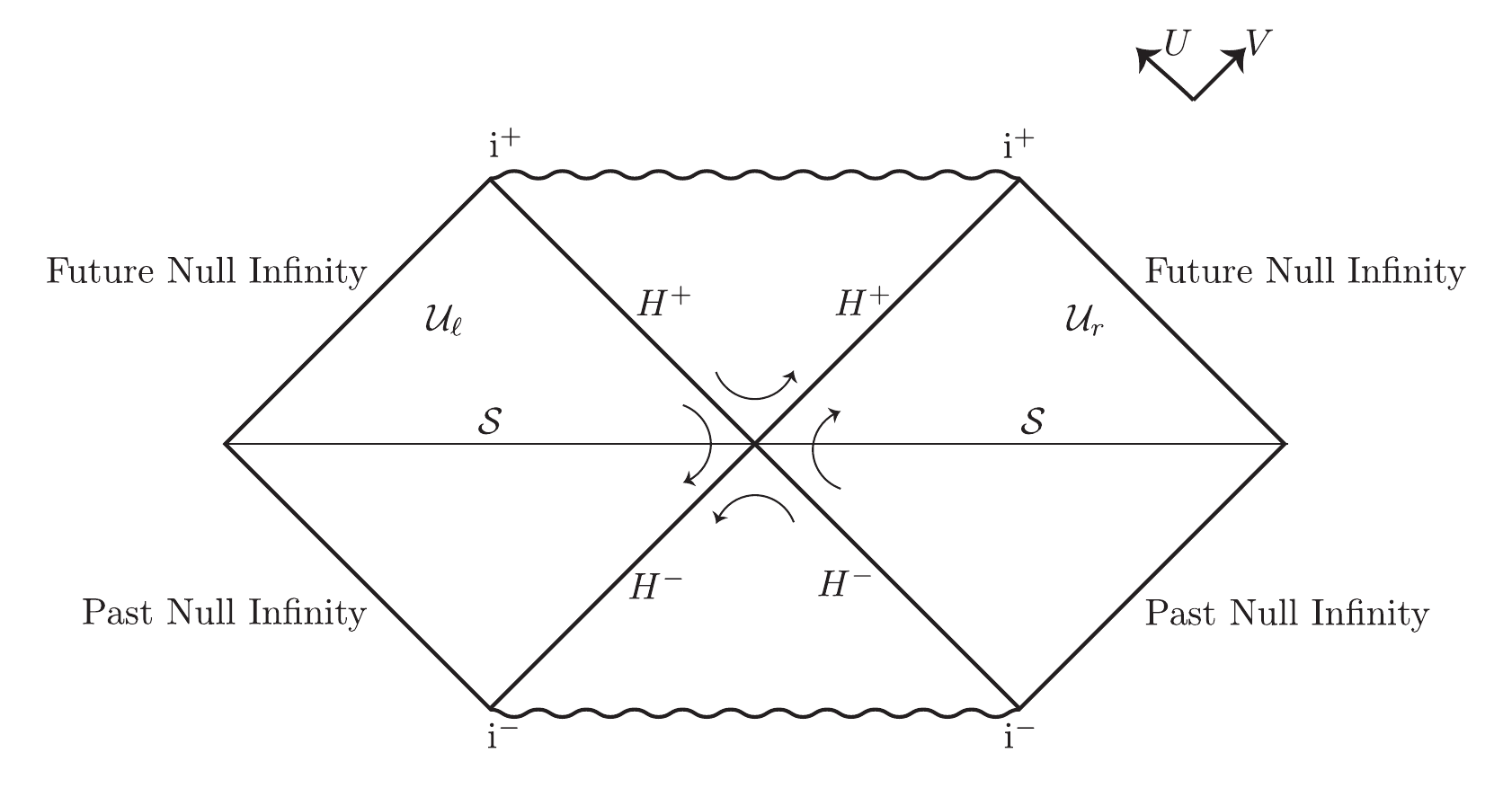}
 \end{center}
\caption{\footnotesize   The Penrose diagram of the extended asymptotically flat Schwarzschild spacetime. $\U_\ell$ and $\U_r$ are the two asymptotically 
flat regions at infinity.  They are bounded by 
past and future horizons labeled  $H^+$ and $H^-$.
$\mathcal S$ is the initial value surface that in the Euclidean picture is depicted in fig. \ref{cigar}. It passes from the asymptotically flat region on the left, through the ``wormhole''
and into the asymptotically flat region on the right.   The arrows indicate the action of the Killing vector field $h$
that generates time translations outside the horizon in the original Schwarzschild solution.   It is future-directed timelike in the asymptotically flat region on the right,
past-directed timelike in the asymptotically flat region on the left, and spacelike in other regions.
  \label{PenroseB}}
\end{figure} 

Since the state $\Psi_\HHI$ can be described as a functional of quantum fields on $\S$, it has a Lorentz signature interpretation as a quantum state of fields in the spacetime
to which $\S$ evolves.   Here we run into a remarkable phenomenon, which will not  surprise readers who are already familiar with the analytic
continuation of the Schwarzschild solution beyond the horizon.   In Euclidean signature, the hypersurface $\S$ has two ``ends,'' both of them at $r\to\infty$ but with $\tE$ equal to 0  or $\frac{1}{2}\beta_\sH$.   These ends are completely equivalent because of the symmetries of the Euclidean Schwarzschild solution.
Correspondingly, the Lorentz signature spacetime for which $\S$ supplies initial data has two equivalent asymptotically flat regions at infinity, joined by a sort of ``wormhole'' and
separated by a common horizon.  

 This spacetime is a real analytic  extension of the original Schwarzschild spacetime and was
 discovered well before the dawn of black hole thermodynamics.   The full analytic extension of the Schwarzschild spacetime is most efficiently seen using the Kruskal-Szekeres
coordinates, which we used for a more limited purpose in section \ref{bhevaporation}:
 \begin{align}\label{KScoords} U&=-\left(\frac{r}{2GM}-1\right)^{1/2} e^{r/4GM} e^{-t/4GM} \cr
                                 V&=\left(\frac{r}{2GM}-1\right)^{1/2} e^{r/4GM} e^{t/4GM}.\end{align}
 In Schwarzschild coordinates used in eqn.  (\ref{tonzo}), the metric is singular at $r=2GM$.   The exterior of the horizon at  $r>2GM$  corresponds in the coordinates (\ref{KScoords})
 to $U<0$, $V>0$.   In terms of $U$ and $V$, the Schwarzschild metric becomes
   \be\label{newmet} \d s^2=-\frac{32 G^3 M^3}{r} e^{-r/2GM} \,\d U \, \d V +r^2\d\Omega^2,\ee
                   with $r$ defined implicitly as a function of $U$ and $V$ by
                     \be\label{nimpdef} -UV =\left(\frac{r}{2GM}-1\right) e^{r/2GM}. \ee
   In these coordinates, the metric remains perfectly smooth at $r=2GM$, so the spacetime smoothly continues past the horizon.   However, curvature invariants in the Schwarzschild
   geometry diverge at $r=0$, so we limit these formulas to $0<r<\infty$, which corresponds to $UV<1$.    The extended Schwarzschild geometry has a symmetry $U\leftrightarrow -V$,
   so in addition to the ``original'' asymptotically flat black hole exterior at $U<0$, $V>0$, there is an isomorphic ``second'' copy at $U>0$, $V<0$.      A Penrose diagram is shown
   in fig. \ref{PenroseB}.              The spacelike hypersurface $\S$ is defined by $U=-V$, which generalizes the more naive condition $t=0$.    $\S$ extends from one
   asymptotically flat region at $U,-V\gg 0$ to the other one at $U, -V\ll 0$, passing through a\footnote{This wormhole is sometimes called
   the Einstein-Rosen bridge \cite{EinsteinRosen}, as those authors understood this aspect of the surface $\S$, though not the full picture exhibited by the Kruskal-Szekeres coordinates.} 
``wormhole'' near $U=V=0$.      
   The diagonals in the Penrose diagram with $U$ or $V$ vanishing represent past and future horizons for
   observers to the left or right of the horizon.  Details are hopefully clear from the Penrose diagram.
   
   What about the ``time translation'' symmetry of the extended Schwarzschild spacetime generated by  the vector field $h=\frac{\partial}{\partial t}$, which is associated
   to conservation of energy?  
   In Kruskal-Szekeres coordinates, we find that
   \be\label{ksboost}h= \frac{1}{4GM}\left(V\partial_V-U\partial_U\right). \ee
   Apart from the factor of $\frac{1}{4GM}$, this is a Lorentz boost generator $V\partial_V-U\partial_U$.
   Near the bifurcation surface at $U=V=0$, $U$ and $V$ are good inertial coordinates and $h$ looks like the generator of a Lorentz boost. The formula (\ref{ksboost}) tells us
   that the Killing vector field that generates time translations at infinity looks near the bifurcation surface like  $\kappa=\frac{1}{4GM}$ times a Lorentz boost generator $V\partial_V-U\partial_U$. The
   coefficient $\kappa$ in this relationship
   is called the surface gravity of the black hole, and the derivation in section \ref{eucl} shows that the Hawking temperature is $T_H=\frac{\kappa}{2\pi}$.
   
   The Penrose diagram of fig. \ref{PenroseB} has an obvious similarity to the corresponding diagrams of fig. \ref{Seven} for Rindler space and fig. \ref{desitter} for de Sitter space.
   The fundamental reason for this is that what to an observer at infinity in Schwarzschild spacetime is time translation symmetry looks like a Lorentz boost near the horizon, as a result of which 
   the near horizon region of the Schwarzschild spacetime can be modeled by Rindler space.   
      As in Rindler space and de Sitter space, the Penrose diagram of the extended Schwarzschild solution is divided into four ``wedges'' 
 by diagonal lines $U=0$ and $V=0$ that represent past and future horizons  and meet at the bifurcation surface $\Sigma$ defined by $U=V=0$.
   The right wedge with $V>0,U<0$, which we will call $\U_r$, is the region exterior to the black
   hole horizon for an observer in one of the two asymptotically flat ``ends'' of the spacetime.    Spacelike separated from it is the left wedge $\U_\ell$ with $V<0,U>0$, which is the region exterior
   to the black hole horizon for an observer in the opposite asymptotically flat end of the spacetime.  There is also a future wedge $U,V>0$, which an observer in $\U_\ell$ or $\U_r$ cannot
   see, and a past wedge $U,V<0$, which an observer in $\U_\ell$ or $\U_r$ cannot influence.      The vector field $h$ is future-directed timelike in $\U_r$, past-directed timelike in $\U_\ell$,
   and spacelike in the past and future wedges.   In particular, the past and future wedges have no timelike Killing vector field and there is no sense in which they can be considered
   time-independent. 
     
  We decompose
   the initial value surface $\S$ as $\S_\ell\cup \S_r$, where $\S_r$ is the portion at $U<0,\,V>0$, and $\S_\ell$ is the portion at $U>0,\,V<0$.  The domain of dependence 
   of $\S_r$ is the whole outside-the-horizon wedge $\U_r$, and the  domain of dependence of $\S_\ell$ is the whole outside-the-horizon  wedge 
 $\U_\ell$.  Associated to this, we formally split\footnote{The usual remark of footnote \ref{notable} applies again.  The splitting claimed in the text are
   not strictly valid in continuum quantum field theory. Tomita-Takesaki theory supplies a rigorous approach.   (In the context of quantum gravity, as opposed to quantum field
   theory in a fixed background spacetime, the splitting $H=H_r-H_\ell$ is rigorous if $H_r$ and $H_\ell$ are interpreted as the ADM energies at infinity.)}  the Hilbert space as
   \be\label{uddu}\H =\H_\ell\otimes \H_r,\ee
   where $\H_\ell$ and $\H_r$ describe degrees of freedom on $\S_\ell$ or $\S_r$, respectively.  Similarly, we split the conserved charge  $H$ associated to the Killing vector field $h$
   as
   \be\label{wuddu} H=H_r-H_\ell, \ee
   where $H_r$ acts on fields on $\S_r$, and $H_\ell$ acts on fields on $\S_\ell$.   $H_r$ is the natural Hamiltonian that generates time translations in $\U_r$, and
   similarly $H_\ell$ is the natural Hamiltonian in $\U_\ell$.
   
    As in the analysis of Rindler space and of de Sitter space, there are two key statements about this
   setup.  First, the density matrices that describe observations in the state $\Psi_\HHI$ in the regions $\U_r$ and $\U_\ell$ are respectively
   \be\label{pokko}\sigma_r=\frac{1}{Z}\exp(-\beta_\sH H_r),~~\sigma_\ell=\frac{1}{Z}\exp(-\beta_\sH H_\ell),\ee
   where $Z$ is a normalization factor ensuring that $\Tr\,\sigma_r=\Tr\,\sigma_\ell=1$.
This statement confirms that an observer in either asymptotically flat region at large $r$ will see thermal equilibrium at the Hawking temperature.   Second,
   the state $\Psi_\HHI$ can be viewed as the thermofield double state of the two entangled regions $\U_\ell$ and $\U_r$.   Proofs of these
   statements are precisely as in the discussions of Rindler space and of de Sitter space.   One represents the operator $\exp\left(-\theta\frac{\beta_\sH}{2\pi}H_r\right)$
   by a path integral on a wedge obtained by rotating $\S_r$ through an angle $\theta$ inside the Euclidean Schwarzschild spacetime.   Then by setting $\theta=2\pi$,
   one arrives at the claim (\ref{pokko}) concerning $\sigma_r$.  Of course $\sigma_\ell$ is analyzed similarly.  By setting $\theta=\pi$, one deduces that $\Psi_\HHI$ 
   can be interpreted as the thermofield double state that purifies the thermal density matrices $\sigma_r$ and $\sigma_\ell$.
   
   The ``wormhole'' near $U=V=0$ provides a kind of geometrical connection between the asymptotically flat regions $\U_\ell$ and $\U_r$.  However, $\U_\ell$ and $\U_r$
   are spacelike separated and there is no possibility of communicating from one region to the other.  An observer who enters the ``wormhole'' hoping to reach the other side
   will instead end up at the black hole singularity at $UV=1$.   The fact that one cannot transmit information from one side of the wormhole to the other is actually 
   a special case of ``topological censorship'' \cite{topocensor,topotwo} (see for example \cite{Witten} for more on this).\footnote{On the other hand, it is possible for observers entering
   the wormhole from opposite sides to meet in the black hole interior. This is a puzzling fact with no obvious counterpart for an ordinary bipartite quantum system.}    Quantum mechanically, $\U_\ell$ and $\U_r$ are entangled
   in the thermofield double state.   In general, entanglement between two quantum systems $\A$ and $\B$ establishes a sort of correlation between them, but  this
   correlation cannot be exploited to transmit information.   Indeed, nothing that one can do to system $\A$ will change the density matrix of system $\B$, or will transmit information to one
   who only has access to system $\B$.   It has been proposed that quantum entanglement in general should be understood as some sort of generalization of the
   wormhole that connects the two ends of the Schwarzschild geometry \cite{MaldaSuss}.

      \begin{figure}
 \begin{center}
   \includegraphics[width=3.1in]{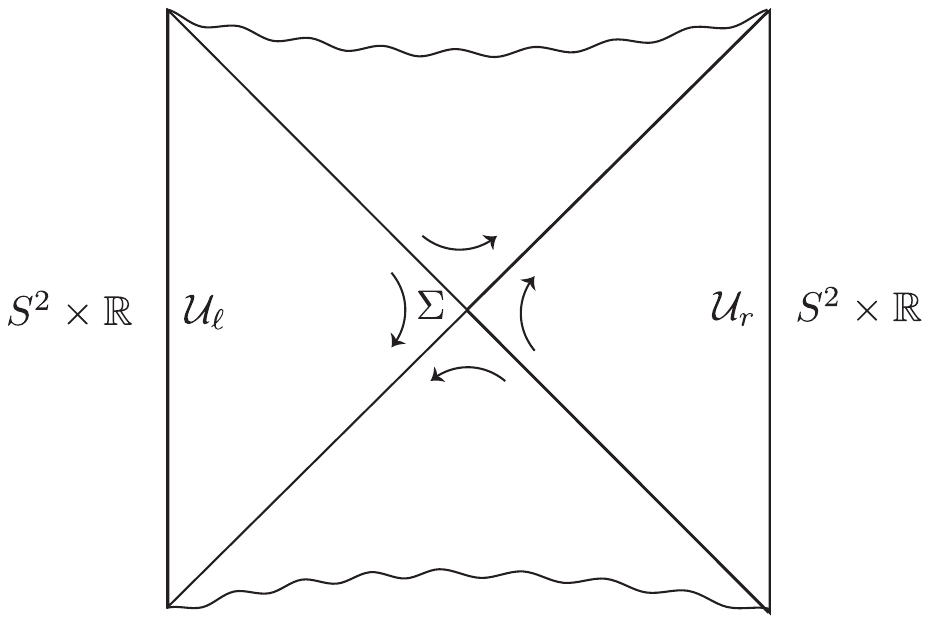}
 \end{center}
\caption{\footnotesize   The Penrose diagram of an AdS-Schwarzschild spacetime.   As in the case of an asymptotically flat Schwarzschild spacetime, there are two distinct
outside-the-horizon regions $\U_\ell$ and $\U_r$, connected by a ``wormhole.'' The conformal boundary of $\U_\ell$ is a copy of $S^2\times\R$, represented by the left
edge of the diagram; the conformal boundary of $\U_r$ is another copy of $S^2\times \R$, represented by the right edge.  The diagonal lines represent the past and future horizons of an observer
living in $\U_\ell$ or $\U_r$.  The arrows represent the action of the Killing vector field $h$ that generates time translations in the region $\U_r$;
it is future-directed timelike  in $\U_r$, past-directed timelike in $\U_\ell$, and spacelike in other regions.  Near the bifurcation surface $\Sigma$, it looks like the generator of a Lorentz
boost, as in Rindler space.  
  \label{PenroseC}}
\end{figure}

All this, including the definition and properties of the state $\Psi_\HHI$ and the extension of the spacetime beyond the horizon to reveal a second asymptotic region, works
 in precisely the same way for the AdS-Schwarzschild solution, whose thermodynamics we studied in section \ref{negative}. 
The main difference is that 
 the boundary of the Penrose diagram outside the horizon
is timelike (fig. \ref{PenroseC}), rather than null as in the asymptotically flat case.  
 This timelike boundary represents the conformal boundary of the extended AdS-Schwarzschild spacetime.
It consists of two copies of the Einstein static universe $\R\times S^2$, one for each of the two asymptotically AdS regions of the spacetime.
In the  AdS-Schwarzschild  case, if the black hole is sufficiently massive, the state $\Psi_\HHI$ is not affected by any thermodynamic instability.  This state  is believed to be well-defined
to the extent that the bulk gravitational theory is well-defined.  In particular, in those cases in which a boundary CFT dual is known, and therefore it is believed that the bulk
gravitational theory is completely well-defined, it is believed that the state $\Psi_\HHI$ is completely and nonperturbatively well-defined.   Indeed, it is simply the dual, in AdS/CFT duality,
of the thermofield double state of a pair of entangled CFT's living at the two timelike boundaries of the Penrose diagram.  This thermofield double state is well-defined for any 
thermal quantum system.

 \section{Two Notions of Entropy}\label{twonotions}
 
 \subsection{Von Neumann Entropy}\label{vn}
 
 So far in this article, ``entropy'' has referred to thermodynamic entropy $S_\th$, which in ordinary physics obeys the Second Law
 \be\label{secondlaw}\frac{\d S_\th}{\d t}\geq 0,\ee
 with $t$ the time.   However, a more microscopic point of view about entropy has turned out to be important in understanding black holes.   Though we could start
 with classical physics, for brevity we will consider the quantum case.  A quantum mechanical system with Hilbert space $\H$ in a general,
 possibly ``mixed'' state has a density matrix $\rho$,  which is a positive operator on $\H$ with $\Tr\,\rho=1$.
The von Neumann entropy of $\rho$  is defined by
 \be\label{vonN} S_\vN(\rho)=-\Tr\,\rho\log\rho.\ee
$S_\vN(\rho)$  is invariant under unitary transformations $\rho\to U\rho U^{-1}$.   In particular, it is invariant under 
unitary quantum mechanical evolution with a Hamiltonian $H$; in such evolution, $\rho$ evolves according to $\frac{\d\rho}{\d t}=-\i [H,\rho]$, implying that its eigenvalues
 do not change and
 \be\label{zolk}\frac{\d S_\vN}{\d t}=0. \ee   This remains so even if the Hamiltonian is time-dependent.
  
 The relation between $S_\th$ and $S_\vN$  is that $S_\th$ is the largest that $S_\vN$ can be given the macroscopic
 state of a system:
 \be\label{olk} S_\vN\leq S_\th .\ee
 To explain this inequality, we will work in a microcanonical ensemble. Consider a quantum system that has $N$ states with observed values of the macroscopic
 observables (such as energy, charge, and so on).  The thermodynamic entropy is defined as the logarithm of the number
 of microstates of a system that are consistent with the macroscopic observables.   So if there are $N$ such states, the 
 thermodynamic entropy is $S_\th =\log N$.   What about the von Neumann entropy?   
 A general density matrix $\rho$
 of such a system can be diagonalized
 \be\label{delgo}\rho=\sum_{i=1}^N p_i |i\ra \la i|,\ee
 with orthonormal states $|i\ra$, and nonnegative constants $p_i$ that obey $\sum_i p_i=1$, to ensure $\Tr\,\rho=1$.
 One can interpret the formula to mean that the given system is in state $|i\ra$ with probability $p_i$.  
 The von Neumann
 entropy of a system with this diagonal density matrix is
 \be\label{welgo} S_\vN=-\Tr\,\rho\log\rho=-\sum_{i=1}^N p_i\log p_i.\ee
 A pure state is the case that 
one of the $p_i$ is 1 and the others vanish; then $S_\vN=0$.   Otherwise $S_\vN>0$, as $-x\log x>0$ for $0<x<1$.
 A simple exercise with Lagrange multipliers shows that the maximum of $-\sum_{i=1}^N p_i\log p_i$ under the 
 constraint $\sum_i p_i=1$ is achieved when all $p_i$ are equal to $1/N$, in which case the system is said to be in a maximally mixed state and
 $S_\vN=S_\th=\log N$.
 This accounts for the inequality (\ref{olk}), and shows that the inequality is saturated if and only if all microstates
 consistent with macroscopic observations are equally probable.   To summarize,
 \be\label{delmog} 0\leq S_\vN\leq \log N,\ee
 with the two inequalities saturated respectively for a pure state or a maximally mixed state.
  
The generalized entropy $S_\gen=\frac{A}{4G}+S_\out$  of a black hole is not constant in time and instead is proposed to obey a Generalized Second Law. So it is  a
 version of thermodynamic entropy.    However, microscopic von Neumann entropy is also important
 in understanding black hole thermodynamics.
 
 It can be very hard to know if a system that appears to be in thermal equilibrium is truly in a thermal state with $S_\vN
 =S_\th$, or is really in a state in which $S_\vN\ll S_\th$ with only one or a relatively small number of the $p_i$ being nonzero.   That is because,
 since the states $|i\ra$ in this discussion are by hypothesis  microstates consistent with the macroscopic properties of the system, they are hard to distinguish.
 A strong version of this statement is the Eigenstate Thermalization Hypothesis (ETH), which says that for a generic macroscopic quantum system, generic microstates
 (usually taken to be energy eigenstates) cannot be distinguished from a thermal ensemble by any reasonably simple measurement \cite{ETH,ETH2}.  A simple measurement
 is, for example, a measurement of an $n$-point correlation function, where $n$ is a fixed number independent of the  number of atoms in the system under study (which, near the
 thermodynamic limit,  is presumed
 to be much larger than $n$).
 
 One can view von Neumann entropy as the natural microscopic definition of entropy, if one is able to probe all of the microscopic details of a system.
 It has been called the fine-grained entropy.   By contrast, if one ignores the differences between states that cannot be distinguished in any reasonably simple
 way, one arrives at the thermodynamic entropy, a coarse-grained notion of entropy.

 Let us consider a black hole that forms in the real world by the collapse of a star.  
Write $S_{\th}(\sstar)$ for the thermodynamic entropy of a star prior to collapse, and $S_\vN(\sstar)$ for its von Neumann entropy.
They obey the usual inequality 
\be\label{usual}S_\vN(\sstar)\leq S_\th(\sstar).\ee   What happens when the star collapses to a black hole?   We will assume that
this collapse is governed by ordinary quantum mechanical laws and that the collapsing star is isolated.  Since (as in eqn. (\ref{zolk})), the von Neumann entropy of an
isolated system is unchanged under quantum mechanical evolution, the von Neumann entropy is unchanged when the star collapses to a black hole,
and remains equal to $S_\vN(\sstar)$:
\be\label{holsome}S_\vN(\BH)=S_\vN(\sstar).\ee   For the thermodynamic entropy of the black hole, we take the Bekenstein-Hawking formula 
$S_\th(\BH)=A/4G$.   As we noted in section \ref{gsl}, this is vastly greater than the thermodynamic entropy $S_{\th}(\sstar)$ of the star prior to collapse:
\be\label{omy}S_\th(\sstar)\ll S_\th(\BH) .\ee
Putting these facts together, a black hole in the real world forms
in a state with von Neumann entropy much less than its thermodynamic entropy:
\be\label{zomy} S_\vN(\BH)\ll S_\th(\BH). \ee 
Both for an  ordinary quantum system and for a black hole, it can be very difficult to distinguish a state of low von Neumann entropy from a state of true
thermal equilibrium with $S_\vN=S_\th$. In the case of a black hole, the difficulty is expressed at the classical level by the no hair theorem, which says that after transients die down, a black hole
that formed from collapse is described by a simple solution of Einstein's equations that depends only on its mass, charge, and angular momentum.

\subsection{More On  von Neumann Entropy}\label{morevn}

A few facts about von Neumann entropy will be useful.\footnote{For somewhat more detailed introductions, see for example \cite{Preskill,WittenQI}.}
Consider two quantum systems $\A,\B$ with Hilbert spaces $\H_\A$, $\H_\B$.   The composite or ``bipartite'' system $\A\B$ then has Hilbert space
$\H_{\A\B}=\H_\A\otimes \H_\B$.   Let $\Psi_{\A\B}\in \H_{\A\B}$ be any pure state and let 
\be\label{zingo}\rho_\A=\Tr_{\H_\B}|\Psi_{\A\B}\ra\la \Psi_{\A\B}|, ~~\rho_\B=\Tr_{\H_\A}|\Psi_{\A\B}\ra\la \Psi_{\A\B}| \ee
be the corresponding density matrices.   The entropies $S(\rho_\A)=-\Tr_{\H_\A}\,\rho_\A\log\rho_\A$, $S(\rho_\B)=-\Tr_{\H_\B}\rho_\B\log\rho_\B$ are in this
context called entanglement entropies, because they result purely from the way the subsystems $\A,\B$ are ``entangled'' in the underlying pure state
$\Psi_{\A\B}$.     We usually write, for example, just $S_\A$ rather than $S(\rho_\A)$ if it is clear what density matrix is intended.

  A generic pure state $\Psi_{\A\B}$ of the combined system is not a tensor product $\psi_\A\otimes \chi_\B$ of pure states of the two subsystems; rather,
its canonical form up to the action of the unitaries $U,V$ on $\H_A$ and $\H_B$ is
\be\label{tolfo}\Psi_{AB}=\sum_i\sqrt{p_i} \psi_\A^i\otimes \zeta_\B^i \ee
where we can assume that the states $\psi_\A^i\in\H_\A$, $\zeta_\B^i\in\H_\B$ are normalized to $\la\psi_\A^i|\psi_\A^j\ra=\la\zeta_\B^i|\zeta_\B^j\ra=\delta^{ij}$.
Moreover, assuming that $\Psi_{\A\B}$ is normalized to $\la\Psi_{\A\B}|\Psi_{\A\B}\ra=1$, we have $\sum_i p_i=1$.
With $\Psi_{\A\B}$ as in eqn. (\ref{tolfo}), the density matrices $\rho_\A$ and $\rho_\B$ are
\be\label{plimy} \rho_\A=\sum_i p_i |\psi_\A^i\ra\la \psi_\A^i|~~~~   \rho_\B=\sum_i p_i |\zeta_\B^i\ra\la \zeta_\B^i| . \ee

Since $\rho_\A$ and $\rho_\B$ have the same eigenvalues, they certainly have the same von Neumann entropies, which in this context are entanglement entropies. 
From eqn. (\ref{plimy}), it is clear that this entanglement entropy vanishes if and only if there is only one term in the sum in eqn. (\ref{tolfo}),
that is, if and only  if $\Psi_{\A\B}=\psi_\A\otimes \zeta_\B$ is a tensor product of pure states of the two subsystems.  When that is not the case, the two subsystems
are said to be entangled.

Beyond having the same von Neumann entropy, 
$\rho_\A$ and $\rho_\B$  have the same values for any invariants that depend only on their eigenvalues.   The most important examples are the R\'{e}nyi entropies. 
The R\'{e}nyi entropy of order $\alpha$ of a density matrix $\rho$ is 
defined for $\alpha>0$, $\alpha\not=1$ by
\be\label{renyi}R_\alpha(\rho)=\frac{1}{1-\alpha}\log \Tr\,\rho^\alpha .\ee
It follows from this definition  that the R\'enyi entropies of any density matrix are non-negative:
\be\label{renyipos} R_\alpha(\rho)\geq 0, \ee
with equality only for  a pure state.  To show this, one just observes that  eigenvalues of $\rho$ are valued in the interval $[0,1]$,
and hence $\Tr\,\rho^\alpha$ is a decreasing function of $\alpha$; since $\Tr\,\rho=1$, $\Tr\,\rho^\alpha$ is bounded above by 1 if $\alpha>1$ and bounded below by 1 if $\alpha<1$,
leading to the inequality (\ref{renyipos}).    Another important fact is that $R_\alpha(\rho)$ is non-increasing\footnote{To prove this, let $p_i$ be the eigenvalues of $\rho$
and set $z_i =p_i^\alpha/\sum_j p_j^\alpha$.  A straightforward computation shows that $\frac{\d R_\alpha(\rho)}{\d \alpha}=-\frac{1}{(1-\alpha)^2} \sum_i z_i\log(z_i/p_i)$.
The collections of numbers $\{z_i\}$ and $\{p_i\}$ are both probability distributions (sets of nonnegative real numbers whose sum is 1), and we show in analyzing eqn. (\ref{lommy})
below that $\sum_i z_i \log(z_i/p_i)$ is non-negative for any two probability distributions.}
 as a function of $\alpha$:
\be\label{enyipos}\frac{\d R_\alpha(\rho)}{\d\alpha}\leq 0. \ee

We also see that the von Neumann entropy is a limit of R\'enyi entropies,
\be\label{renyilim}\lim_{\alpha\to 1} R_\alpha(\rho)=S_\vN(\rho).\ee
So the definition of $R_\alpha$ is extended by defining $R_1(\rho)=S_\vN(\rho)$.  (Similarly the definition of $R_\alpha(\rho)$ is extended for
$\alpha=0,\infty$ by taking limits.)   The definition (\ref{renyi}) makes it clear that $R_\alpha(\rho)$ only depends on the eigenvalues of $\rho$,
and therefore if the bipartite system $\A\B$ is in a pure state, $R_\alpha(\rho_\A)=R_\alpha(\rho_\B)$.  

As a special case of this, we observed in section \ref{another} that every density matrix $\rho_\A$ for a quantum system $\A$ can be purified by a pure state $\Psi_{\A\B}$ on
some bipartite system $\A\B$.   We now see that $\A$ and its purifying system $\B$  have the same von Neumann entropy, and likewise the same R\'{e}nyi entropies.
This is true regardless of the choice of $\B$, which could be, but need not be, another copy of $\A$.

Now imagine an observer $O_\A$ with the ability to manipulate system $\A$ (but no access to system $\B$), and reciprocally an observer $O_\B$ with access only to system $\B$.
$O_\A$ can act on a state $\Psi_{\A\B}$ by an operator $U_\A\otimes 1_\B$, where $U_\A$ is a unitary operator on system $\A$ and $1_\B$ is the identity operator on system $\B$.
Similarly $O_\B$ can act by $1_\A\otimes U_\B$, where $U_\B$ is a unitary on system $\B$.   Jointly, the two observers can map $\Psi_{\A\B}$ to $(U_\A\otimes U_\B)\Psi_{\A\B}$.
This will transform the density matrices $\rho_\A$ and $\rho_\B$ of the two subsystems by
\be\label{mizo} \rho_\A\to U_\A\rho_\A U_\A^{-1}, ~~~ \rho_\B\to U_\B\rho_\B U_\B^{-1}.\ee
Conjugation by a unitary operator does not change the eigenvalues of the density matrices, so it does not change the entanglement entropies $S_\A=S_\B$ of the
two systems, or  their R\'enyi entropies.   Before leaving this question,
we should contemplate a more general scenario in which
 the observer $O_\A$ brings into the picture an experimental apparatus with Hilbert space $\K_\A$, and the observer $O_\B$
likewise brings into the picture an experimental apparatus with Hilbert space $\K_\B$.   Before any experiments begin, $\K_\A$ and $\K_\B$ are initialized in the pure states
$\chi_\A$ and $\chi_\B$ respectively.\footnote{What is important is not that these states are pure but that there is no entanglement of $\K_\A$ or $\K_\B$ with each other or with the original
system $\A\B$.   Assuming this,  there is actually no essential loss of generality in assuming that the experimental equipment is initialized in a pure state; if $\K_\A$ and $\K_\B$ are initialized in mixed states, we can extend  $\K_\A$ and $\K_\B$ by adjoining the purifying systems and thereby reduce to the case that $\K_\A$ and $\K_\B$ are initialized in pure states. It does not matter whether the
observers manipulate the purifying systems or not.}
   The overall system is thus initialized in the pure state $\chi_\A\otimes \Psi_{\A\B}\otimes \chi_\B$.  
 The observer $O_\A$ thus has access to the composite system $\t \A= \A\K_\A$, and the observer $O_\B$ has access to the composite
system $\t \B=\B\K_\B$.   Before the observers take any action, the entanglement entropy of the composite systems $\t \A$ and $\t \B$ is the same as the entanglement entropy of systems $\A$ and $\B$.
Observer $O_\A$ can act on the combined system $\t \A\t \B$ with a operator $U_{\t \A}\otimes 1_{\t \B}$, where $U_{\t \A}$ is a unitary on system $\t \A$ and $1_{\t \B}$ is the identity on system
$\t \B$; similarly $O_\B$ can act with $1_{\t \A}\otimes U_{\t \B}$.   The result is to transform the state $\chi_\A\otimes \Psi_{\A\B}\otimes \chi_\B $ to 
$(U_{\t \A} \otimes  U_{\t \B})(\chi_\A\otimes \Psi_{\A\B}\otimes \chi_\B) $, thus conjugating the density matrices $\rho_{\t \A}$ and $\rho_{\t \B}$: 
\be\label{vizo} \rho_{\t \A}\to U_{\t \A}\rho_{\t \A}U_{\t \A}^{-1},~~~~~\rho_{\t \B}\to U_{\t \B}\rho_{\t \B}U_{\t \B}^{-1}. \ee
This leaves unchanged the entropies $S_{\t \A}$ and $S_{\t \B}$, which remain equal to each other and to the original $S_\A$ and $S_\B$.  So in this more general sense, the two observers
acting separately  cannot modify the entanglement entropy between the subsystems that they control.

Another important concept is the {\it relative entropy} between two density matrices $\rho$ and $\sigma$ on the same Hilbert space $\H$.   It is defined
as
\be\label{tommy} S(\rho||\sigma)=\Tr\,\rho\log\rho -\Tr\,\rho\log\sigma. \ee
This quantity has applications to the question of how hard it is to distinguish $\rho$ and $\sigma$ by an experiment; see for example \cite{WittenQI} for more on that.
For our purposes here, all we need to know is that $S(\rho||\sigma)$ vanishes if $\rho=\sigma$, as is obvious, and is positive otherwise:
 \be\label{pommy}S(\rho||\sigma)>0~~{\rm if}~~\rho\not=\sigma.\ee  This is called {\it positivity of relative entropy}.
   To prove this statement, we can first diagonalize $\sigma$ in some basis, and then we let
$\rho_D$ be the matrix whose diagonal elements in that basis coincide with those of $\rho$, while its off-diagonal elements vanish.  $\rho_D$ is a positive
matrix with trace 1, so it is a density matrix.
Using $\Tr\,\rho\log\sigma=\Tr\,\rho_D\log \sigma$, we have
\be\label{ommy}S(\rho||\sigma) = S(\rho_D||\sigma) + S(\rho_D)-S(\rho),\ee
so to prove positivity of $S(\rho||\sigma)$, it suffices to prove positivity of $S(\rho_D||\sigma)$ and of $S(\rho_D)-S(\rho)$.
For the first statement, we use the fact that $\rho_D$ and $\sigma$ are diagonal in the same basis, say with diagonal matrix
elements $p_i$ and $q_i$, respectively.  So
\be\label{lommy} S(\rho_D||\sigma)=\sum_i p_i (\log p_i - \log q_i).\ee
The right hand side is the classical relative entropy (or Kullback-Liebler divergence) between two probability distributions $\{p_i\}$ and $\{q_i\}$.
One way to prove that this is positive is to use the fact that the classical relative entropy is a convex function on the space of probability distributions. Concretely, for $0\leq t\leq 1$, define the probability distribution $p_i(t) =(1-t)q_i +t p_i$, which coincides with $\{q_i\}$ at $t=0$ and with $\{p_i\}$ at $t=1$.
Then let $f(t)=\sum_i p_i(t)(\log p_i(t) - \log q_i)$.   For $0\leq t\leq 1$, we have
\be\label{bulsing}\ddot f(t) =\sum_i \frac{(p_i-q_i)^2}{p_i(t)}\geq 0.\ee
Since also $f(0)=\dot f(0)=0$, it follows that $f(1)\geq 0$, with equality only if $\{p_i\}=\{q_i\}$ or in other words $\rho_D=\sigma$.
   But $f(1)=S(\rho_D||\sigma)$.  To finish the proof of eqn. (\ref{pommy}), we also
need to show that $S(\rho_D)\geq S(\rho)$, with equality only if $\rho_D=\rho$.   This can be proved by a somewhat similar use of concavity.
For $0\leq t\leq 1$, define the density matrix  $\rho(t)=t\rho +(1-t)\rho_D$.   Using
\be\label{bilgox}\log\rho(t)=\int_0^\infty \d s\left(\frac{1}{s+1}-\frac{1}{s+\rho(t)}\right),\ee
we compute
\be\label{nilgox}\left.\frac{\d}{\d t} S(\rho(t))\right|_{t=0}=-\Tr\, (\rho-\rho_D)\log \rho_D=0, \ee
where the vanishing holds because $\rho-\rho_D$ is strictly off-diagonal in the basis in which $\rho_D$ is diagonal.   We also compute
\be\label{ilgox}\frac{\d^2}{\d t^2} S(\rho(t))=-\int_0^\infty\d s\,\Tr\,\dot\rho \frac{1}{s+\rho(t)}\dot\rho\frac{1}{s+\rho(t)}.\ee
The integrand is positive if $\rho\not=\rho_D$, as it is $\Tr\,B^2$ where $B$ is the self-adjoint operator $(s+\rho(t))^{-1/2}\dot\rho
(s+\rho(t))^{-1/2}$, so $\frac{\d^2}{\d t^2}S(\rho(t))<0$.   Combining these facts, we have $S(\rho(1))<S(\rho(0))$, or $S(\rho_D)>S(\rho)$, completing the proof of the inequality (\ref{pommy}).

An important special case is the following.   Let $\rho_{\A\B}$ be any density matrix on a bipartite system $\A\B$ with Hilbert space $\H_{\A\B}=\H_\A\otimes \H_\B$.
Define the reduced density matrices $\rho_\A=\Tr_{\H_\B}\,\rho_{\A\B}$, $\rho_\B=\Tr_{\H_\A}\, \rho_{\A\B}$.  Then we can define a density matrix $\sigma_{\A\B}=\rho_\A\otimes \rho_\B$ for
the combined system.   Here $\sigma_{\A\B}$ is equivalent to $\rho_{\A\B}$ for measurements of only system $\A$ or only system $\B$, but ignores the information contained in
$\rho_{\A\B}$ concerning correlations between the two systems.   Using $\log\,\sigma_{\A\B}=\log\rho_\A\otimes 1+1\otimes \log\rho_\B$
we have $S(\rho_{\A\B}||\sigma_{\A\B})=\Tr\,\rho_{\A\B}\log\rho_{\A\B}-\Tr\,\rho_{\A\B}\log\sigma_{\A\B}=-S(\rho_{\A\B}) -\Tr\,\rho_{\A\B}(\log\rho_\A\otimes 1+1\otimes \log\rho_\B)
=-S(\rho_{\A\B}) -\Tr\,\rho_\A\log\rho_\A-\Tr\,\rho_\B\log\rho_\B=-S_{\A\B}+S_\A+S_\B$, so the positivity of relative entropy implies that
\be\label{zildo}S_\A+S_\B\geq S_{\A\B}.\ee
This is called {\it subadditivity} of entropy.  An equivalent statement is that  the {\it mutual information} $I(\A:\B)=S_\A+S_\B-S_{\A\B}$ is non-negative.

Subadditivity of entropy has an interesting corollary when it is combined with the existence of purifications.   As usual, we  can purify the density matrix $\rho_{\A\B}$; we 
 introduce a third system $\C$ and a pure state $\Psi_{\A\B\C}\in \H_\A\otimes \H_\B\otimes \H_\C$ such that $\rho_{\A\B}=\Tr_{\H_\C}\,|\Psi_{\A\B\C}\ra\la\Psi_{\A\B\C}|$.
Then $S_{\A\B}=S_\C$, and likewise, since the system $\B\C$ is purifying system $\A$, $S_\A=S_{\B\C}$.   So the inequality (\ref{zildo}) becomes
$S_{\B\C}+S_\B\geq S_\C$.   Exchanging the roles of $\C$ and $\A$, we get 
\be\label{pildo} S_{\A\B}+S_\B\geq S_\A.\ee  For example, this inequality is saturated if the combined system $\A\B$ is
pure, in which case as we have discussed $S_\A=S_\B$ and $S_{\A\B}=0$.
The two statements  (\ref{zildo}) and (\ref{pildo}) combine 
to give the {\it Araki-Lieb inequality}:
\be\label{tildo}S_\B\geq |S_{\A\B}-S_\A|. \ee
Similarly, subadditivity plus the fact that $\A\B\C$ is pure implies that $S_\A+S_\C\geq S_{\A\C}=S_\B$
and $S_\B+S_\C\geq S_{\B\C}=S_\A$, so
\be\label{wildo} S_\A+S_\C\geq S_\B \geq S_\A-S_\C. \ee

Von Neumann entropies also satisfy a deeper inequality known as {\it monotonicity of relative entropy}.   
This inequality  says that tracing over or forgetting a subsystem can only
reduce the relative entropy.  In more detail, if $\rho_{\A\B}$ and $\sigma_{\A\B}$ are two density matrices of the system $\A\B$, and $\rho_\A=\Tr_{\H_\B}\,\rho_{\A\B}$, $\sigma_\A
=\Tr_{\H_\B}\,\sigma_{\A\B}$ are the reduced density matrices on system $\A$, then
\be\label{miso} S(\rho_{\A\B}||\sigma_{\A\B})\geq S(\rho_\A||\sigma_\A).\ee
This was first proved in \cite{LR}; a relatively accessible proof was given in \cite{PN}, and can also be found, for example, in sections 3.6 and 4.3 of \cite{Notes}.
Monotonicity of relative entropy has an important application in proving a version of the Generalized Second Law \cite{Wall}, but unfortunately that argument will not be explained
here.  Monotonicity of relative entropy has  a corollary that is known as {\it strong subadditivity} of entropy.\footnote{Conversely, from strong subadditivity one can deduce
monotonicity of relative entropy by a rather short argument.}   For this, let $\rho_{\A\B\C}$ be a density matrix on a tripartite system
$\A\B\C$ with Hilbert space $\H_{\A\B\C}=\H_\A\otimes \H_\B\otimes \H_\C$ and define the reduced density matrices such as $\rho_{\A\B}=\Tr_{\H_\C}\,\rho_{\A\B\C}$,
$\rho_\C=\Tr_{\H_\A\otimes \H_\B}\,\rho_{\A\B\C}$, etc.   Let $\sigma_{\A\B\C}=\rho_{\A}\otimes \rho_{\B\C}$.    Then the calculation that led to eqn. (\ref{zildo}) shows
that $S(\rho_{\A\B\C}||\sigma_{\A\B\C})=S_{\A}+S_{\B\C}-S_{\A\B\C}$.   Tracing out $\C$ from $\sigma_{\A\B\C}$, we get $\sigma_{\A\B}=\Tr_{\H_\C}\,\sigma_{\A\B\C}=\rho_\A\otimes \rho_\B$.
So $S(\rho_{\A\B}||\sigma_{\A\B})=S_\A+S_\B-S_{\A\B}$.   The inequality $S(\rho_{\A\B\C}||\sigma_{\A\B\C})\geq S(\rho_{\A\B}||\sigma_{\A\B})$ therefore becomes an inequality
that is known as  strong subadditivity:
\be\label{ziggo}S_{\A\B}+S_{\B\C}\geq S_{\A\B\C}+S_\B. \ee
This inequality also has an interesting variant that can be proved by using the fact that any density matrix  has a purification.  Let $\D$ be a quantum system such that the
state of $\A\B\C\D$ is pure.  Then $S_{\A\B}=S_{\C\D}$, $S_{\A\B\C}=S_\D$, so the inequality of strong subadditivity becomes
\be\label{liggo}S_{\C\D}+S_{\B\C} \geq S_\B +S_\D. \ee     The difference $S_{\C\D}-S_\D$ is called conditional entropy $S(\C|\D)$, and classically it is always non-negative, roughly
because one's lack of knowledge of the combined system $\C\D$ is at least as great as one's lack of knowledge of system $\D$ by itself.   
Quantum mechanically, 
it is possible to have $S_{\C\D}-S_\D<0$ if $\C$ and $\D$ are entangled (for example if $\C\D$ is in an entangled pure state, so $S_{\C\D}=0$, $S_\D>0$), and similarly
 it is possible to have $S_{\B\C}<S_\B$ if $\C$ and $\B$ are entangled.
But the sum $(S_{\C\D}-S_\D)+(S_{\C\B}-S_\B)$ is always nonnegative.  This is a statement of {\it monogamy of entanglement}; entanglement between $\C$ and one quantum system $\D$
limits how much entanglement there can be between $\C$ and another quantum system $\B$.

As a last illustration  of von Neumann entropy, we will reconsider the First Law of thermodynamics.   A thermal density matrix for a state with Hamiltonian $H$,
inverse temperature $\beta$, and partition function $Z$ is
is $\rho=\frac{1}{Z}e^{-\beta H}$.   The average energy in this state is
\be\label{omigox} E=\Tr\,H\rho, \ee
and the entropy is $S=-\Tr\,\rho\log \rho$.   To explore the First Law,
we consider an arbitrary first order deformation of $\rho$.  We get
\be\label{tomigox}\d E=\Tr\,H\d\rho \ee
and 
\be\label{omigix}\d S=-\Tr \,\d\rho\log \rho=\beta \Tr\,H \d\rho .\ee
Here we used $\log\rho=-\beta H-\log Z$ and\footnote{In proving that $\Tr\,\rho \,\d\log \rho=0$,  $\rho$ and $\d\rho$ cannot be assumed to commute.
Via  (\ref{bilgox}),  we get $\Tr\,\rho\,\d\log\rho=\int_0^\infty\d s \Tr\,\rho \frac{1}{s+\rho}\d\rho \frac{1}{s+\rho}=\int_0^\infty \d s
\Tr\, \rho \frac{1}{(s+\rho)^2}\d\rho= \Tr\,\d\rho=\d\Tr\,\rho=0$ (since $\Tr\,\rho=1$).} $\Tr\,\rho\,\d\log\rho=0$.
So with $T=1/\beta$, we have the First Law $\d E=T\d S$.  The derivation shows that this relation is valid, in first order, for an {\it arbitrary} deformation of $\rho$ (not necessarily preserving
thermal equilibrium). A standard fact is that 
for a deformation of $\rho$ in which  thermal equilibrium is maintained through all stages (with a varying temperature or changes in other thermodynamic variables), the First Law can be integrated to give information about a non-infinitesimal deformation of $\rho$.   But the derivation just given  shows that the statement holds in first order for
arbitrary perturbations.   This statement has a limit in classical General Relativity (\cite{IW}, Theorem 6.1):  with $S$ defined as $A/4G$, the First Law, in its more general version
$\d E=T\d S+\Omega\d J$, holds for an arbitrary first order deformation of a stationary, possibly rotating, black hole.

\subsection{The Page Curve}\label{Page}

Suppose that a star in a quantum mechanical pure state collapses to form a back hole.   The collapse is expected to be a unitary process, so the black hole forms in a pure state.
Then the black hole begins to decay by emitting Hawking radiation.    According to Hawking's analysis, the black hole is emitting purely thermal radiation, modulated by gray body
factors.   This means that the von Neumann entropy of the radiation equals its thermodynamic entropy
\be\label{inono}S_\vN(\rad)=S_\th(\rad),\ee
assuming that in defining the thermodynamic entropy $S_\th(\rad)$, one takes into account a knowledge of its spectrum, including gray body factors.   In particular, $S_\vN(\rad)$
steadily increases, according to Hawking's analysis.

Assuming that black hole evaporation is a unitary process, the combined system consisting of the black hole and the radiation it emits remains pure as the evaporation proceeds.
Therefore the black hole and the radiation have equal von Neumann entropies:
\be\label{equalvn} S_\vN(\BH)=S_\vN(\rad). \ee
Now we must remember the fundamental inequality (\ref{olk}) between the von Neumann and thermodynamic entropies of any system:
\be\label{fundid}S_\vN\leq S_\th. \ee
For the thermodynamic entropy of the black hole, we take the Bekenstein-Hawking entropy
\be\label{bhtake} S_\th(\BH)=\frac{A}{4G}.\ee
Therefore, we expect that at all times,
\be\label{tilox} S_\vN(\rad)=S_\vN(\BH)\leq S_\th(\BH)=\frac{A}{4G}.\ee

    \begin{figure}
 \begin{center}
   \includegraphics[width=3.1in]{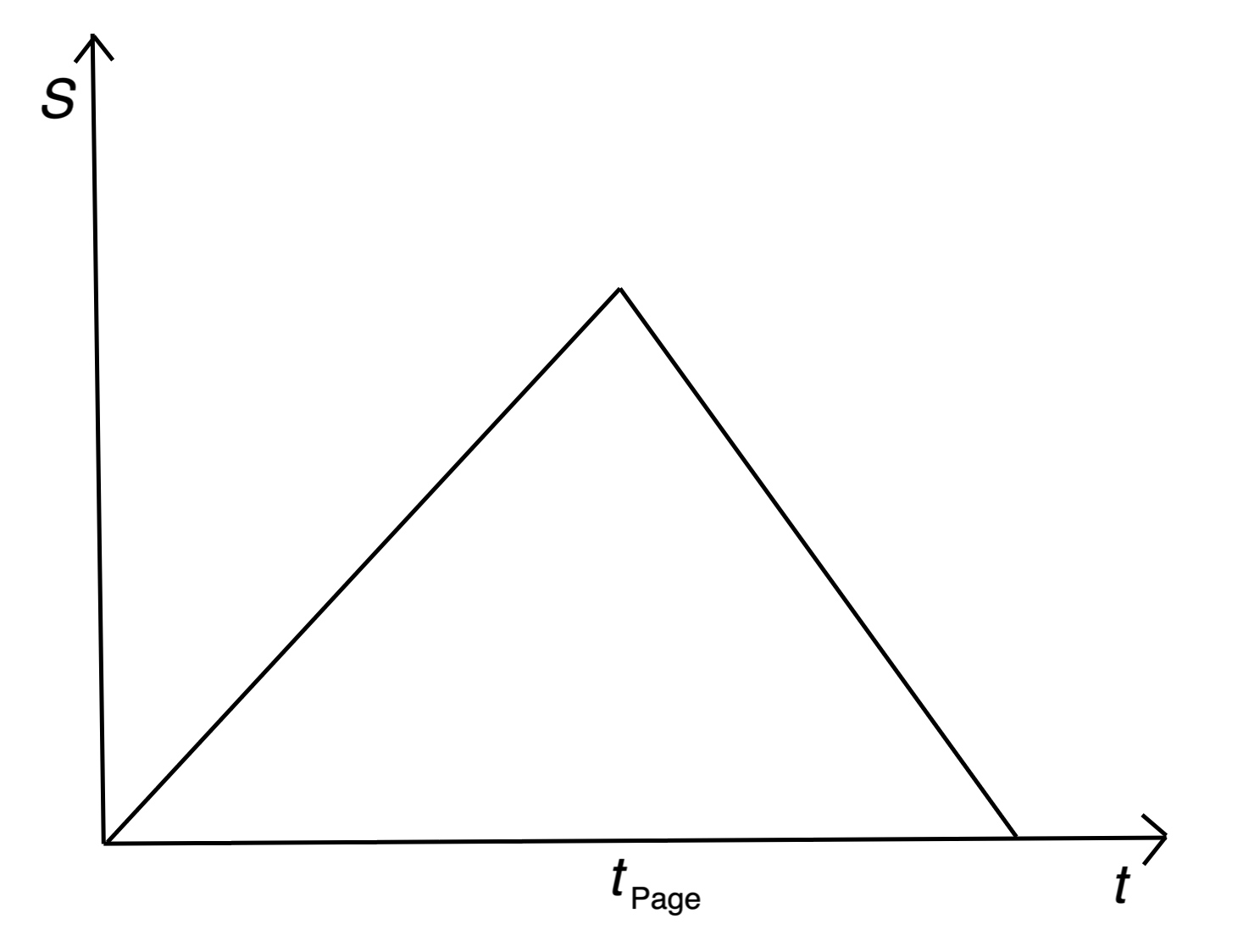}
 \end{center}
\caption{\footnotesize  A schematic illustration of the Page curve. Plotted is the entropy $S$ of an evaporating black hole, assumed to form in a pure state, as a function
of the time $t$.   The entropy increases steadily until the Page time and thereafter decreases steadily.   The turnabout at the Page time is believed to become sharp
in the limit $G\to 0$.   \label{pagecurve}}
\end{figure} 

At early times, there is no problem with this inequality: $S_\vN(\rad)\ll A/4G$.   However, it was observed by Page \cite{PageCurvePaper} that as $S_\vN(\rad)$ continues to increase in accord with
(\ref{inono}), while the area $A$ decreases, eventually the inequality (\ref{tilox}) will be saturated and then, if (\ref{inono}) remains valid, it will be violated. What happens then?

First of all, Page proposed, based on assuming that black hole dynamics is sufficiently complex and generic that it can be modeled by a random unitary process, that the equality (\ref{inono})
remains valid for as long as it is consistent with the inequality (\ref{tilox}).    The time at which this fails, meaning that the thermodynamic entropy of the radiation equals
the thermodynamic  entropy of the remaining black hole, is called the Page time $t_\Page$.   What happens beyond this time?  Page proposed that once
the inequality $S_\vN(\BH)\leq S_\th(\BH)$ is (nearly) saturated, it continues to be (nearly) saturated for all times, until the black hole becomes so small that thermodynamic reasoning fails.   
The idea behind this proposal is the following.   When $S_\vN(\BH)=S_\th(\BH)$, this means that the black hole is in perfect thermal equilibrium, with a truly thermal density matrix
to good approximation.   Once perfect thermal equilibrium
is achieved, one expects it to be maintained by any adiabatic process.  As long as the black hole is macroscopic, its evaporation  is a very slow, adiabatic process that
one expects would preserve thermal equilibrium.  

In short, Page proposed  that of the two inequalities
\begin{align} \label{twoineq}S_\vN(\rad)&\leq S_\th(\rad)\cr  S_\vN(\BH)&\leq S_\th(\BH), \end{align}
to very high accuracy, the first is saturated until the Page time and the second is saturated after the Page time.   Equivalently, at all times,
\be\label{eqbal} S_\vN(\BH)=S_\vN(\rad)={\rm min}(S_\th(\BH),S_\th(\rad)). \ee
We can also summarize this prediction by saying that, of the two inequalities in eqn. (\ref{twoineq}), whichever one is more restrictive, subject to the condition
$S_\vN(\BH)=S_\vN(\rad)$, is (nearly) saturated at any given time.   The curve of von Neumann entropy as a function of time that follows from this reasoning
is called the Page curve, schematically illustrated in fig. \ref{pagecurve}.

We would like to slightly extend this reasoning to the case that the star that collapsed to form a black hole was not initially in a pure state.   After all, even if the star formed 
in a pure state, by the time the star collapses to form a black hole, it has become entangled with the radiation that it
has emitted and is no longer in a pure state.   However, the von Neumann entropy of the star that is collapsing to form a black hole satisfies the usual inequality
\be\label{melmo} S_\vN(\sstar)\leq S_\th(\sstar). \ee
In the following analysis, it will be convenient to use the fact that every density matrix has a purification.   So there is some quantum system $\C$ such that
the combination of the star and $\C$ is in a pure state.   This system and the star have equal von Neumann entropies:
\be\label{elmo} S_\vN(\sstar)=S_\vN(\C). \ee
If the star was in a pure state at birth, $\C$ might be simply the radiation that the star has emitted during its lifetime prior to collapsing to a black hole, but whether that is so will not be relevant.

Now we consider the collapse of the star to form a black hole.   As explained in section \ref{vn}, because 
 the thermal entropy of the star is much less than the thermal entropy of the black hole to which it collapses,
the  black hole is born in a state of low entropy,
\be\label{pilo}S_\vN(\BH)\ll S_\th(\BH). \ee
Moreover,  $S_\vN(\C)=S_\vN(\BH)$ when the black hole is born, as $\C$ and the star were in a pure state, and the collapse of the
star to a black hole was unitary.

We want to generalize the previous discussion of the Page curve of a black hole that is born in a pure state to this more realistic case of a black hole born in a state of low entropy
compared to its thermal entropy.
As before, the black hole starts to radiate, so the thermal and von Neumann entropies of the radiation increase.   Hawking's analysis indicates that they  are equal at least initially.
But, as in the case that the black hole formed in a pure state, eventually this will lead to a contradiction.   To analyze this situation, we note that since initially the star and $\C$ were in a pure
state, it follows that after the star collapses and the black hole begins to radiate, assuming this evolution is unitary, the tripartite system consisting of the black hole, the radiation, and $\C$ will be
in a pure state.   Therefore, we have from eqn. (\ref{wildo})
\be\label{niloc} S_\vN(\rad)+S_\vN(\C) \geq S_\vN(\BH)\geq S_\vN(\rad)-S_\vN(\C). \ee   Moreover, as $\C$ does not participate in the evolution at all, $S_\vN(\C)$ is independent of time.
Since $S_\vN(\C)$ is extremely small compared to typical values of $S_\vN(\rad)$ and $S_\vN(\BH)$ during the subsequent evolution,
eqn. (\ref{niloc}) tells us that $S_\vN(\rad)$ and $S_\vN(\BH)$ will be very nearly equal throughout the evolution, though not precisely equal as in the case
of a black hole that forms in a pure state.  Therefore one expects that as in the case of a black hole that forms in a pure state, $S_\vN(\rad)$ and $S_\vN(\BH)$ will go up together  in tandem,
and then go back down together in tandem.   With a suitable assumption of randomness of the evolution, one can generalize the analysis
in \cite{PageCurvePaper} and make slightly more precise statements.

The form of the Page curve, as sketched in fig. \ref{pagecurve}, suggests some sort of phase transition at the Page time -- a transition that becomes sharp in the limit $G\to 0$.
  Something like this
has indeed been found \cite{P,AEMM}, as we briefly indicate in section \ref{pageagain}.    

\section{Black Holes and Von Neumann Entropy}\label{bhvn}

\subsection{Bekenstein-Hawking Entropy as Entanglement Entropy}\label{bhent}

 \begin{figure}
 \begin{center}
   \includegraphics[width=2.1in]{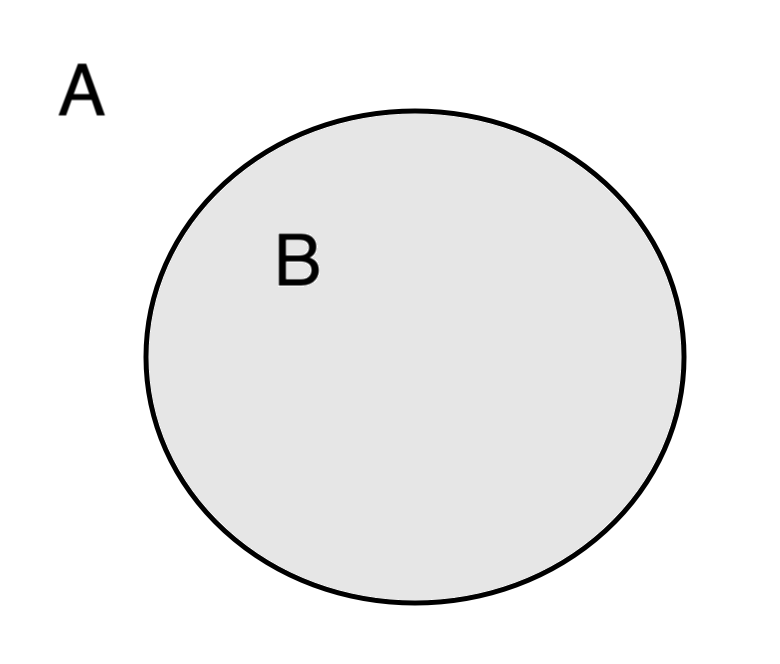}
 \end{center}
\caption{\footnotesize  Dividing space into two regions $\A$, $\B$.   \label{sorkin}}
\end{figure}

An early proposal to relate black hole entropy to entanglement entropy was made by Sorkin \cite{SorkinA} in 1983 and and extended by Bombelli, Koul, Lee, and Sorkin
a few years later  \cite{SorkinB}, in work that initially attracted little attention.
For other early suggestions, see  \cite{Thooft,Srednicki}.
The idea was to interpret the Bekenstein-Hawking entropy of a black hole in terms of the entanglement entropy of quantum fields on opposite sides of the horizon.  
 
  For a simpler problem, in  some quantum field theory in Minkowski space, divide space at time $t=0$ (in some Lorentz frame)  into two complementary regions $\A$ and $\B$
(fig. \ref{sorkin}).  Let $\Omega $ be the vacuum state of the theory, 
and let $\rho_\A$ and $\rho_\B$ be the density matrices appropriate to measurements in the state $\Omega$ in
regions $\A$ and $\B$, respectively.    We constructed such a density matrix in a particular case (Rindler space) in section \ref{cut}, though
in that analysis we discussed temperature only and not entropy.
We  will discuss
some further examples in section \ref{sample}.    
The idea now is to consider the 
 von Neumann entropies $S_\vN(\rho_\A)$ and $S_\vN(\rho_\B)$.  These are equal to each other,
and can be regarded as entanglement entropies, since the overall state $\Omega$ is pure.    One finds that $S_\vN(\rho_\A)$ and $S_\vN(\rho_\B)$ are ultraviolet
divergent.   The divergence comes from short distance modes supported near the boundary $\Sigma$ between the two regions,
and accordingly the coefficient of the leading divergence is proportional to the area $A$ of  that boundary.  The computation in section \ref{sample} will
exhibit the divergence in the entanglement entropy in an illustrative case.  

The idea in \cite{SorkinA}-\cite{Srednicki} was to interpret black hole entropy as  entanglement entropy between the regions behind and outside the horizon,
with somehow  gravity cutting off the ultraviolet divergence, leaving an entanglement entropy that is still proportional to the area $A$, but with a finite  coefficient
$1/4G$.   This idea makes a great deal of intuitive sense, as it
matches two ideas:

(1) $\frac{A}{4G}$ is the irreducible entropy of the system for someone who has access only to the region outside the horizon.

(2)  The entropy of a black hole is proportional to the horizon area because  the horizon supports roughly one bit or qubit per Planck unit of area, as in Wheeler's picture (fig. \ref{Three}); here
the modes in question are the short distance modes that dominate the entanglement entropy.

This idea has not evolved into a precise proposal, even today, but further developments have certainly shown that in black hole thermodynamics, it is important
to consider the microscopic von Neumann entropy, and not only the thermodynamic entropy that Bekenstein analyzed originally.   A few, but realistically only a few, of
these further developments will be described in the rest of this article.

A decade after the initial proposal, Susskind and Uglum \cite{SU} made the following very simple observation.   The generalized entropy of a black hole
\be\label{nubbo} S_\gen=\frac{A}{4G}+S_\out ,\ee
is better defined than either term is separately,
if $S_\out $ is  understood as the von Neumann entropy  of fields outside the horizon.
We have already remarked that $S_\out$ has an ultraviolet divergence proportional to the horizon area.   In four dimensions, this is a quadratic divergence
\be\label{lubbo} S_\out = f \Lambda^2 A+\cdots,\ee 
where $f$ is a constant and $\Lambda$ is an ultraviolet cutoff.   On the other hand, in Hawking's original calculation of black hole evaporation   (and in our sketch of this
calculation in section \ref{bhevaporation}), loop effects that renormalize Newton's constant were not taken into account.   This means that we should think of the ``$G$'' that
appears in that analysis as a bare Newton constant $G_0$, and thus we should have written the generalized entropy as
\be\label{wubbo}S_\gen=\frac{A}{4G_0}+S_\out.\ee
 At one-loop order, the renormalization of Newton's constant has the general form
\be\label{dumbox}\frac{1}{4G_0}=\frac{1}{4G}-f'\Lambda^2,\ee
again with a quadratic divergence and a  constant $f'$.     Susskind and Uglum observed that, at least at one-loop order, $f'=f$; therefore the ultraviolet divergence cancels in $S_\gen$,
when it is expressed in terms of the physical parameter $G$.    $S_\gen$  is thus
better-defined than either of the two terms on the right hand side of eqn. (\ref{wubbo}).  These arguments were later extended; see section  \ref{rbh}.

There are two reasons that it is important in this derivation to interpret $S_\out$ as the von Neumann entropy of the quantum fields outside the horizon, not thermodynamic entropy.   First, von Neumann entropy
and not  thermodynamic entropy has the ultraviolet divergence that is needed to cancel the ultraviolet divergence in the renormalization of Newton's constant.   Second, interpreting
$S_\out$ as von Neumann entropy greatly increases the scope of the formula.  With this interpretation, $S_\gen$ is defined for an arbitrary state of the quantum fields, not 
necessarily a state for which thermodynamics is valid.

This analysis could possibly be interpreted to suggest that the part of $\frac{A}{4G}$ that results from renormalization of $\frac{1}{G}$ reflects  entanglement entropy
of the quantum fields, but  the bare contribution $\frac{A}{4G_0}$ does not.  Is it possible to interpret all of the black hole entropy as entanglement entropy of quantum fields?
 That question has motivated the suggestion \cite{Jacobson} that $\frac{1}{G_0}=0$, which would mean that there
is no Einstein-Hilbert term in the classical action and the usual gravitational action arises entirely from loop effects.   This idea has been called ``induced gravity,'' originally proposed
with a different motivation \cite{Sakharov}.   (In induced gravity,
it  is usually assumed that the gravitational field is present to begin with, and only the gravitational action, not the gravitational field, is induced.)
From a standard point of view, the absence from the classical action of a bare Einstein-Hilbert term is in principle possible, but there is no obvious reason to make this
assumption, since the absence of a classical Einstein-Hilbert term would not reflect any symmetry.  And precisely for that reason, the absence of an Einstein-Hilbert term
in the classical action does not seem to be a well-defined statement, as it depends on the renormalization scheme that one is planning to use.

In a sense string theory may give an improved version of something similar to induced gravity.  
Let us recall that in ordinary quantum field theory, the classical action and the quantum corrections are quite different things.   The classical action is not induced from anything;
it is postulated to define the theory.   Then the quantum corrections are deduced from the classical action together with Feynman diagrams and the like.   
The quantum corrections, being non-local, can describe
correlations between particles in different places and in particular between particles or fields on opposite sides of the horizon.   So they can contribute to the entanglement entropy.
The classical action, as the integral of a local expression, does not provide such correlations so it does not contribute to entanglement entropy.

Now compare to string theory.   In perturbative string theory, at least, the starting point is a two-dimensional field theory.   What in spacetime is interpreted as the classical action
comes from the path integral of the two-dimensional field theory on a Riemann surface of genus zero, and the quantum corrections similarly come from path integrals on
Riemann surfaces of genus $g>0$.
In a sense, the classical action is ``induced'' from the two-dimensional field theory; it has a similar origin to the quantum corrections.  If the genus $g>0$ contributions  can contribute
to the entanglement entropy across the horizon, it is logical that the genus zero contribution can as well.   The two-dimensional path integral on a surface of genus zero is not really
local from a spacetime point of view; it just becomes local in an asymptotic expansion at low energies.   So it is not obvious why there could not be a genus
zero contribution to the entanglement entropy.   It was indeed suggested by Susskind and Uglum \cite{SU} that there
is a classical contribution to the entanglement entropy across a horizon, coming from genus zero string worldsheets that are partly outside the horizon and partly behind it.
Unfortunately, even thirty years later,  this line of thought has not been backed up by a real calculation.

\subsection{A Sample Computation}\label{sample}

Most calculations of von Neumann entropy in quantum field theory are based on a simple device known as the {\it replica trick}, which in this context was introduced by Callan and Wilczek
\cite{CW,HLW}. (A somewhat similar replica trick was  used earlier in spin glass theory \cite{AE}.)   The idea is simple.   In many situations, it is practical to use path integrals
to construct the density matrix $\rho$ of a region. In that case, as we will see, it is also comparatively easy to describe the positive integer powers $\rho^n$
of the density matrix.   However, to describe $\log\rho$ is usually difficult, and this usually makes a direct calculation of $S(\rho)=-\Tr\,\rho\log\rho$ difficult.
Instead one computes $\Tr\,\rho^n$ for positive integer $n$.   This computation is done, as we will explain, 
 by considering $n$ copies or replicas of the original system, whence the name ``replica trick.''
After computing $\Tr\,\rho^n$ for positive integer $n$, one analytically continues it to a holomorphic function on the half-plane ${\rm Re}\,n\geq 1$ and then one computes
the entropy from the formula
\be\label{izzo}S(\rho)=-\left.\frac{\d}{\d n}\right|_{n=1}\Tr\,\rho^n. \ee

Of course, one question here is whether the analytic continuation away from integer values of $n$ exists and is unique.   In ordinary quantum mechanics, one can address
this question as follows.  As $\rho$ is a positive operator with trace 1, it follows that in the half-plane
${\rm Re}\,n\geq 1$, the function $f(n)=\Tr\,\rho^n$ is holomorphic and bounded by $|f(n)|\leq 1$.  So the analytic continuation exists.  
The continuation is unique by Carlson's theorem.
According to this  theorem (see for example
\cite{Boas}, p. 153), a holomorphic function in the half-plane ${\rm Re}\,n\geq 1$ that coincides with $f(n)$ at positive integers is unique, under condition
of being bounded or even under a much weaker condition that allows some exponential growth.  So in ordinary quantum mechanics, there is no difficulty with the existence and uniqueness of the analytic continuation away from positive integer
values of $n$.   In quantum field theory, these considerations are not really applicable, as the formalism of density matrices is not  rigorous in quantum field theory.\footnote{The factorization
of Hilbert space that is usually used in defining density matrices is not valid in continuum quantum field theory, as previously remarked in
 footnote \ref{notable}. See also a remark at the end of this section.}  But in practice, the use of the replica trick to compute entropies in quantum field theory has been very effective.   
 
 The basic strategy to compute $\Tr\,\rho^n$ is as follows. Suppose that $\rho$ is a density matrix on a Hilbert space $\H$.
   Then $\H^{\otimes n}=\H\otimes \H\otimes \cdots\H$ is the Hilbert space of a composite system consisting of $n$ copies or replicas of the original system.
   On $\H^{\otimes n}$, one defines a product density matrix 
    $\rho^{[n]}=\rho\otimes \rho\otimes \cdots\otimes \rho$ that describes $n$ replicas all in the same state $\rho$.  
         Whatever method is available to describe $\rho$ can be repeated $n$
    times to describe $\rho^{[n]}$. Explicitly
        \be\label{zilbox}\rho^{[n]}{}^{i_1\cdots i_n}_{j_1\cdots j_n} =\rho^{i_1}_{j_1}\rho^{i_2}_{j_2}\cdots \rho^{i_n}_{j_n}. \ee
  To compute $\Tr\,\rho^n$, arrange the $n$ replicas in cyclic order and contract the ``bra'' state  (or lower index) of the $i^{th}$ replica with the ``ket'' state  (or upper index) of the $i+1^{th}$:
    \be\label{ilbo} \Tr\,\rho^n =\rho^{i_1}_{i_2}\rho^{i_2}_{i_3}\rho^{i_3}_{i_4}\cdots \rho^{i_n}_{i_1}.\ee
    A slightly more abstract description is as follows.     If $P$ is the operator that cyclically permutes the $n$ replicas, then
   \be\label{usefulone}\Tr_{\H}\,\rho^n=\Tr_{\H^{[n]} }\,P\rho^{[n]}.\ee
   
   Once $\Tr\,\rho^n$ is computed, this immediately gives the R\'enyi entropies  $R_n(\rho)=\frac{1}{1-n}\log \Tr\,\rho^n$ of integer order $n>1$.  Other R\'enyi entropies and
   the von Neumann entropy are then computed by analytic continuation.

We will explain how to carry out this program in
 a simple example, analyzed in \cite{CW,HLW,CC}, that  is highly illustrative and which  also has important applications, for instance in \cite{P,AEMM}.
 To understand this example, the reader will need a basic knowledge of two-dimensional conformal field theory (CFT) and twist fields.  Unfortunately it will not be practical to fully
 explain the background here.
 
 \begin{figure}
 \begin{center}
   \includegraphics[width=3.6in]{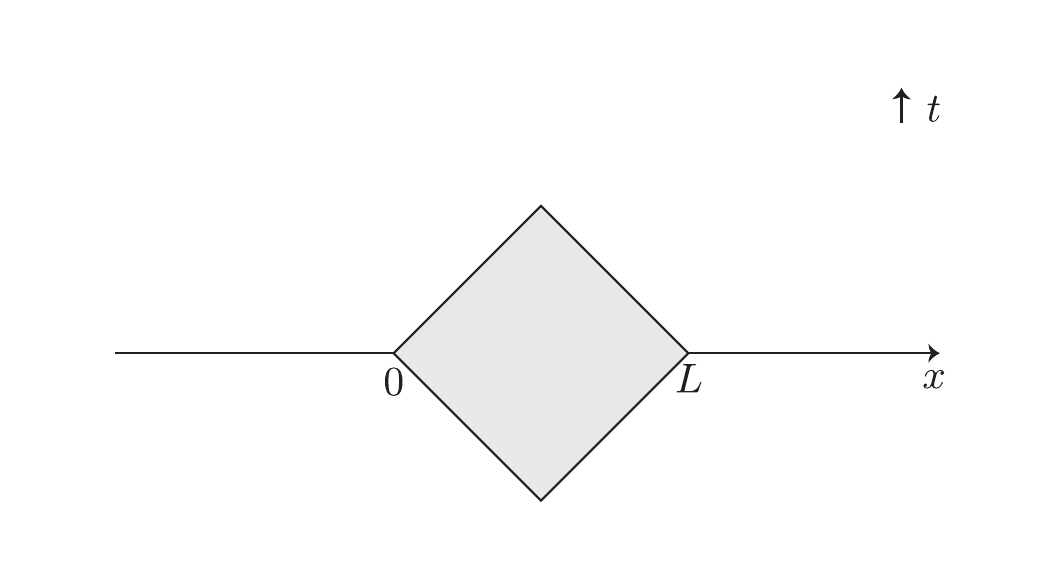}
 \end{center}
\caption{\footnotesize The shaded region is the domain of dependence of the 
 interval $I=[0,L]$ at $t=0$  in two-dimensional Minkowski space.    \label{domain}}
\end{figure}

 \begin{figure}
 \begin{center}
   \includegraphics[width=6in]{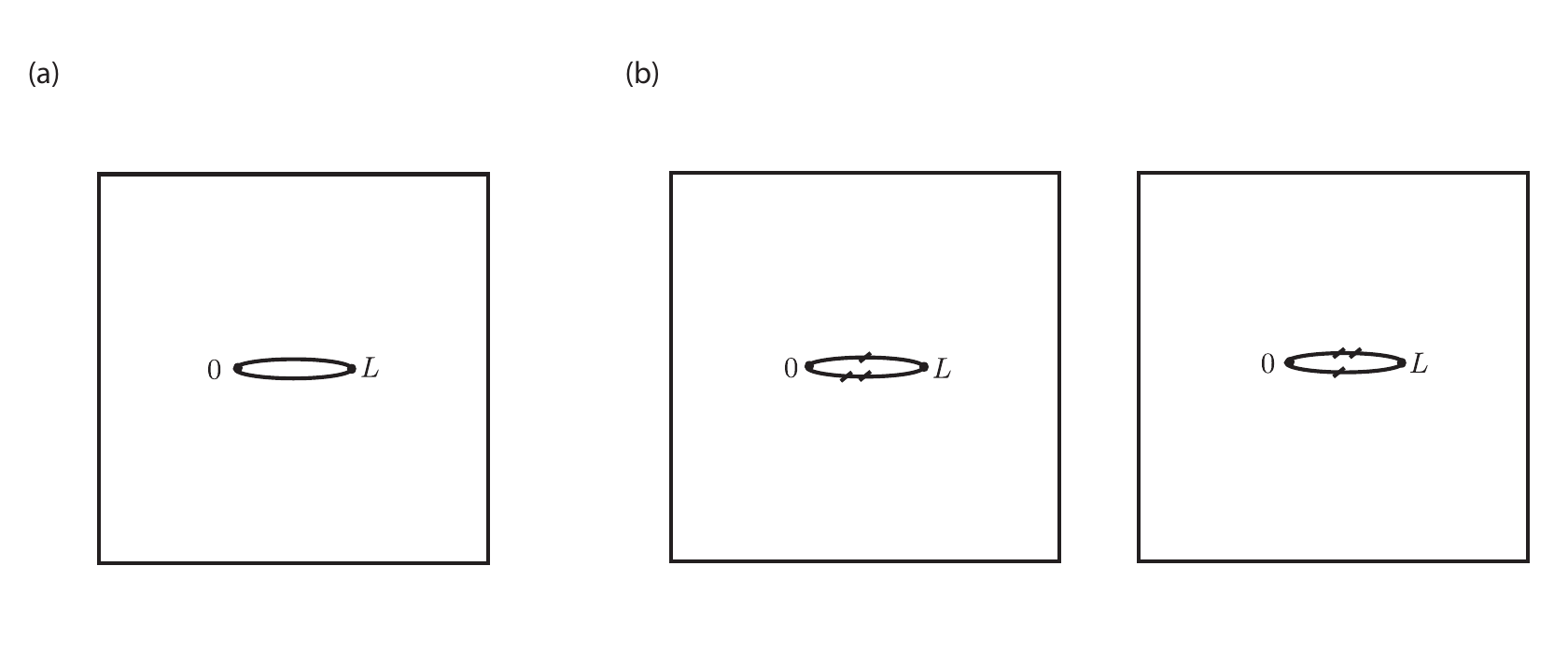}
 \end{center}
\caption{\footnotesize (a) The density matrix $\rho$  of the interval $I=[0,L]$ is represented by a path integral on a Euclidean
$\R^2$ with a cut along $I$. (The boundaries
above and below the cut have been separated for visibility.)
 Concretely the density matrix is a function $\rho(\phi;\phi')$ where $\phi$ and $\phi'$ are respectively the boundary values along $I$ below
and above the cut.   (b) To compute $\Tr\,\rho^2$, we start with two copies of the path integral that computes $\rho$
and glue the boundary above the cut in each one to the boundary below the cut in the other, as marked.
 The resulting surface is a two-fold cover of $\R^2$ branched at the endpoints of $I$, and the path integral
on this manifold computes $\Tr\,\rho^2$.  $\Tr\,\rho^n$ is computed similarly starting with $n$ copies and gluing them in a cyclic arrangement.    \label{density}}
\end{figure} 

In two-dimensional Minkowski space with metric $\d s^2=-\d t^2+\d x^2$, we denote the initial value surface $t=0$ as $\S$, and we let $I\subset \S$ be the interval $0\leq x\leq L$.
In this spacetime, we consider a CFT with holomorphic and antiholomorphic central charge $c$, and vacuum vector $\Omega$.  
We call this theory $\T$ and generically denote the fields in this theory as $\phi$.
We want to compute the entropy of the density matrix $\rho$ that describes measurements in the state $\Omega$ in the region $I$.   Equivalently, just as in the discussion of Rindler
space in section \ref{cut}, this
is the density matrix that describes measurements in the domain of dependence of $I$ (fig. \ref{domain}), since fields in the domain of dependence are determined by fields along $I$.
In computing the entropy of the density matrix $\rho$, we are going to run into the characteristic ultraviolet divergence of all such computations.   In the particular example
that we are studying here, if there were no ultraviolet divergence, then conformal invariance, or even just global scale invariance, would imply that the answer is independent of $L$.
Instead, we will run into an ultraviolet divergence, which makes possible a logarithmic dependence on $L$.   A drawback of this example is that, since we are in two spacetime dimensions,
the boundary of the region $I$ consists of a pair of points, each of ``area'' 1,  so in this calculation we will not see clearly that the coefficient of the ultraviolet divergence is a multiple of
the area.   (A rather similar calculation in dimension $D>2$  does make that clear.) 

The first step, of  course, is to construct the density matrix $\rho$.   This can be done precisely in the way that we constructed the density matrix for Rindler space in section 
\ref{cut}.   After continuing to Euclidean signature by $t=-\i \tE$, the  projection operator  $|\Omega\ra\la\Omega|$ onto the vacuum is represented, as in fig. \ref{Five}(b),
 by a path integral on $\R^2$
with a cut on the line $\tE=0$.   The bra $\la\Omega|$ is represented by a path integral on the region $\tE>0$ as a function of the boundary values of the fields  above the
cut; the ket $|\Omega\ra$ is similarly represented by a path integral on the region $\tE<0$ as a function of the boundary values of the fields below the cut.  Now suppose we want
to construct 
a density matrix appropriate for measurements only on a portion $I$ of the $x$ axis, whose complement we will call $I^c$.  To do this, we glue together the upper and lower
half spaces along $I^c$, leaving a cut only on $I$.  The logic is the same as in section \ref{cut}: to ``trace out'' the fields on $I^c$ from the pure state density matrix 
$|\Omega\ra\la\Omega|$, we set the boundary values along $I^c$ to be equal above and below the cut and then integrate over them; this has the effect of erasing the
cut along $I^c$.

The resulting construction of the density matrix 
is depicted in fig. \ref{density}(a).  As in the discussion of Rindler space, the density matrix can be viewed as a function $\rho(\phi;\phi')$, where $\phi'$ denotes
fields on the boundary above the cut and $\phi$ denotes fields on the boundary below the cut.      Now let us  compute  $\Tr\,\rho^n$ for a positive integer $n$.   
For this purpose, as described earlier,  we take $n$ disjoint copies of the $z$-plane, each with a cut on the interval $[0,L]$ of the real axis, to represent the $n$-fold tensor
product $\rho^{[n]}=\rho\otimes\rho\otimes \cdots\otimes\rho$.   Then after arranging the replicas in cyclic order, we contract the bra state in the $i^{th}$ replica with the
ket state in the $i+1^{th}$.   Geometrically this contraction is accomplished by gluing together the boundary above the cut in the $i^{th}$ copy of the $z$-plane to
the boundary below the cut in the $i+1^{th}$.     This gluing recipe is illustrated for $n=2$ in fig. \ref{density}(b).   The gluing constructs a Riemann surface $C_n$ that is an $n$-fold
cover of the complex $z$-plane, with branch points at $z=0$ and $z=L$, and no other branch points on the complex $z$-plane or at $z=\infty$.  The monodromy around
the branch points is a cyclic permutation of the $n$ sheets at $z=0$ and an inverse cyclic permutation at $z=L$.
$C_n$ can be described by the equation
\be\label{meliflo} y^n=\frac{z}{z-L}, \ee
which describes a cover of the complex $z$-plane with precisely the right branch points and monodromy. 

The cyclic arrangement of the $n$ replicas and the rule for gluing each one to the next are invariant under a cyclic permutation of the replicas.  This cyclic permutation generates
a symmetry group $\Z_n$ that we will call a replica symmetry.
In the algebraic description (\ref{meliflo}) of the Riemann surface $C_n$, the generator of the replica symmetry is $y\to e^{2\pi \i /n}y$.

The upshot of all this is that $\Tr\,\rho^n$ is given by the path integral of the CFT under study on $C_n$.   However, there is a simpler approach.   Away from the branch points,
 $C_n$ is just $n$ copies of the original complex $z$-plane $C$.  Instead of studying one copy of theory $\T$ on an $n$-fold cover of $C$, it is equivalent, away from the
branch points, to study $n$ copies of theory $\T$ on $C$.   We will denote as $\T^n$ the CFT that consists of $n$ copies of theory $\T$.
We would like to compute $\Tr\,\rho^n$ by studying theory $\T^n$ on $C$.  In doing this, what are we supposed to say about the branch points of the covering map $C_n\to C$?
This question actually has a simple answer.    Theory $\T^n$ is invariant under the group of permutations of the $n$ copies.   Under broad conditions,\footnote{The discrete
symmetry must be one that could be gauged; its 't Hooft anomaly must vanish. That condition is satisfied in the present example.} to a discrete symmetry $\gamma$ of a CFT, one can associate
a ``twist field'' $\Theta_\gamma$, with the property that in going around a point $p$ at which the operator $\Theta_\gamma(p)$ is inserted, the fields undergo the automorphism
$\gamma$.   Such twist fields were first constructed in \cite{DFMS} and have relatively simple properties; in particular, 
the twist field of lowest possible dimension
for a given $\gamma$ is a conformal primary.  In our problem,  
at $z=0$, we want a lowest energy conformal primary\footnote{In a lattice regularization, as introduced shortly, one would meet here a non-universal linear combination of all possible
twist fields for given $\gamma$.  But in the limit that the lattice scale $\varepsilon$ becomes small, the dominant contribution will come from the twist field for $\gamma$ of lowest
possible dimension.   That is why the relevant twist field $\Theta_{(n)}$ is the one of lowest dimension, which in particular is a conformal primary.  There
are many other conformal primary twist fields of higher dimension for the same $\gamma$.}  twist
field $\Theta_{(n)}$ that cyclically permutes the $n$ sheets; at $z=L$, we need the conjugate twist field $\bar\Theta_{(n)}$ that cyclically permutes the $n$ sheets in the opposite direction.  

Instead of studying the path integral of theory $\T$ on $C_n$, it is equivalent, and more transparent, to study the path integral of theory $\T^n$ on $C$, with a pair of twist field
insertions.  The resulting formula for $\Tr\,\rho^n$ is a two-point function of twist fields inserted at $z=0$ and $z=L$:
\be\label{polo} \Tr\,\rho^n=\la \Theta_{(n)}(0) \bar\Theta_{(n)}(L)\ra. \ee

It turns out that $\Theta_{(n)}$ and its conjugate $\bar\Theta_{(n)}$ are primary fields of dimension
\be\label{orto} \Delta_n=\frac{c}{12}\left(n-\frac{1}{n}\right).\ee
Before explaining how to obtain this result, we will explain how to use it to compute the entropy.

In general,  if $\O$ is a CFT primary field of scaling dimension $\Delta$ and $\bar \O$ is its conjugate, 
the two point function is 
\be\label{tellmet}\la \O(0)\bar \O(L)\ra = w L^{-2\Delta}, \ee
with a constant $w$ that depends on how the operator $\O$ is normalized.   Applying this in the present context, we may seem to have a contradiction.  The quantity $\Tr\,\rho^n$ that we are trying
to compute is a dimensionless function of $n$, of course, but the two-point function $\la \Theta_{(n)}(0)\bar\Theta_{(n)}(L)\ra$ will be proportional to $L^{-2\Delta_n}$.
Apart from $L$, there appears to be no other dimensionful quantity in the problem, so how can we possibly get a dimensionally correct answer for $\Tr\,\rho^n$?

The answer to this question involves the fact that in continuum relativistic quantum field theory, the density matrix formalism is not strictly applicable.   Of course,
with a suitable lattice regularization, density matrices do exist.   A lattice regularization breaks Lorentz invariance, but in the present discussion, unlike our previous analysis
of Rindler space, that is not a problem. A lattice regularization introduces another dimensionful parameter -- the lattice scale $\varepsilon$  -- making it possible
to write a dimensionally correct formula. 
Near the continuum limit -- that is, for small $\varepsilon$ -- the resulting formula  will scale  with $L$ as $L^{-2\Delta_n}$, the expected CFT behavior for the two-point function
of an operator of dimension $\Delta_n$.       The dimensionally correct version of the formula for $\Tr\,\rho^n$ is
\be\label{inot} \Tr\,\rho^n =w(n) \left(\frac{\varepsilon}{L}\right)^{2\Delta_n} =w(n) \left(\frac{\varepsilon}{L}\right)^{\frac{c}{6}\left(n-\frac{1}{n}\right)}. \ee
Here $w(n)$ is a dimensionless function of $n$; it is non-universal and  depends on the specific cutoff used.   Note that $w(1)=1$, since $\Tr\,\rho=1$.

Analytic continuation of this result is immediate; the only singularity is at $n=0$.  It is now  straightforward to compute $S(\rho)=-\Tr\,\rho\log\rho$:
\be\label{plimbo} S(\rho)=-\left.\frac{\d}{\d n}\right|_{n=1}\left( w(n) \left(\frac{\varepsilon}{L}\right)^{\frac{c}{6}\left(n-\frac{1}{n}\right) }\right)= \frac{c}{3}\log\frac{L}{\varepsilon}   -w'(1).\ee 
The logarithmic term $\frac{c}{3}\log \frac{L}{\varepsilon}$ is universal and has many applications, 
but the constant $-w'(1)$ is not universal; it depends on the cutoff-dependent function $w(n)$ in eqn. (\ref{inot}).   It is noteworthy that the universal result depends on the cutoff
 $\varepsilon$, illustrating the fact that in quantum field theory, such entanglement entropies are ultraviolet divergent.
  In two dimensions, the dependence on $\varepsilon$ is only logarithmic; an analogous computation in $D$ dimensions
gives a leading divergence proportional to $\frac{A}{\varepsilon^{D-2}}$, where $A$ is the area (in the $(D-2)$-dimensional sense) of the boundary of the region considered.

 \begin{figure}
 \begin{center}
   \includegraphics[width=4.9in]{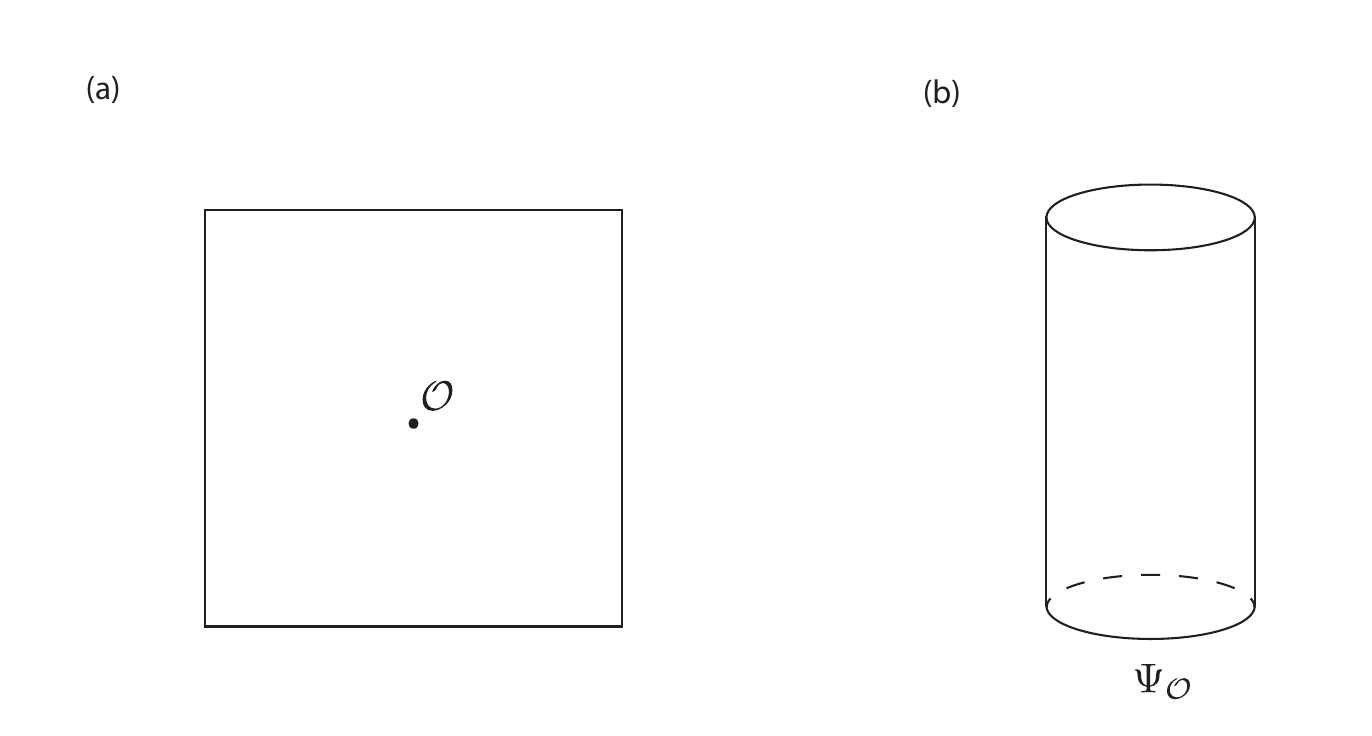}
 \end{center}
\caption{\footnotesize The state-operator correspondence of conformal field theory:  a  local operator $\O$  of scaling dimension $\Delta$ corresponds to a state
$\Psi_\O$ of energy $E=\Delta-\frac{c}{12}$, provided the theory is quantized on a circle of circumference $2\pi$.   To prove this, one starts with the picture in (a)
with the operator $\O$ inserted at a point $p$ on $\R^2$; then one makes a conformal mapping from $\R^2$ with the point $p$ removed to an infinite cylinder of circumference $2\pi$.
The operator $\O$ is transformed into a state $\Psi_\O$ that is inserted at one end of the cylinder  as in (b).  Taking into account the scaling dimension of $\O$ and the conformal
anomaly, one learns that the state $\Psi_\O$ has energy $\Delta-\frac{c}{12}$. \label{stateoperator}}
\end{figure} 

It remains to explain the formula (\ref{orto}) for the dimension of the operator $\Theta_{(n)}$.   For this purpose, we use the state-operator correspondence of conformal
field theory (fig. \ref{stateoperator}).   In a conformal field theory with central charge $c$, an
 operator of dimension $\Delta$ corresponds to a state of energy $E=\Delta-\frac{c}{12}$, if the theory is quantized on a circle
of standard circumference $2\pi$.     But if the circumference is $2\pi R$, then the energy is
\be\label{manygo}E_R=\frac{1}{R}\left(\Delta-\frac{c}{12}\right). \ee

The operator of lowest dimension is the identity operator, with $\Delta=0$.  The corresponding state is the ground state; on a circle of circumference $2\pi$, its energy is
$-\frac{c}{12}$.  This negative 
ground state energy of a CFT quantized on a circle is a generalization of the usual Casimir energy of a free field.  

Theory $\T^n$ is the product of $n$ copies of theory $\T$, so if theory $\T$ has central charge $c$, then $\T^n$ has central charge $nc$.
Hence if  we quantize theory $\T^n$ on a circle of circumference $2\pi$, as in fig. \ref{stateoperator}(b), then an operator $\O$ of dimension $\Delta$ corresponds  to a state
 $\Psi_\O$ of energy $\Delta-\frac{nc}{12}$. In general, a quantum state  of theory $\T^n$ on the cylinder can be viewed as a state of theory $\T$ on an $n$-sheeted cover
 of the cylinder.     If $\O$ is a twist field associated to a  permutation $\gamma$ of the $n$ copies of $\T$, then the state $\Psi_\O$ is what is called a twisted sector state \cite{orbifold}, which means in the case of theory $\T^n$ that
 the $n$ sheets of the cover are permuted by $\gamma$ in going around the cylinder.   The specific twist field $\Theta_{(n)}$ is associated to a cyclic permutation of the
 $n$ factors, so if  $\O=\Theta_{(n)}$, then the $n$ sheets are cyclically permuted in going around the cylinder.   That means that the $n$-fold cover of the cylinder is a connected
 manifold, a cylinder of circumference $2\pi n$.   So instead of viewing $\Psi_\O$ as a state in theory $\T^n$ on a cylinder of circumference $2\pi$, we can view it as a state of
 theory $\T$ on a circumference $2\pi n$.  As such, since $\Theta_n$ is the primary operator of lowest dimension for the given permutation,
  $\Psi_\O$ is the  primary state of lowest energy, namely the
 ground state. So its energy is given by eqn. (\ref{manygo}) with $\Delta=0$ and $R=n$ and is therefore $-\frac{c}{12 n}$.  
 The upshot then is that the energy of the state $\Psi_{\Theta_{(n)}}$ is $\Delta_n -\frac{nc}{12} $ if computed one way, and $-\frac{c}{12n}$ if computed another way.
 Comparing these formuls, we get the claimed result (\ref{orto}) for $\Delta_n$.
 
 The ultraviolet divergent part of the entanglement entropy that we computed is $\frac{c}{3}\log \frac{1}{\varepsilon}$.   This comes from short distance modes
 near the ends of the interval.  As the interval has two ends which make equal contributions, the short distance modes near one end of the interval contribute
 $\frac{c}{6}\log \frac{1}{\varepsilon}$.  This is also the divergent part of the entanglement entropy in any two-dimensional quantum field theory that, while not being conformally invariant,
 is asymptotic in the ultraviolet to a conformal field theory of central charge $c$.
 Though we computed the entanglement entropy for the vacuum state, this divergent contribution is universal, independent of the state,
 because all states look like the vacuum at short distances.  
 
  Related to the ultraviolet divergence that we found in the entanglement entropy is a conformal anomaly.   Suppose that we make a Weyl transformation of
 the metric on $\R^2$ by $g\to e^{2v} g$, for a real-valued function $v$.   This multiplies all  locally measured lengths by $e^v$, so  the short distance cutoff $\varepsilon$
 is replaced by $e^v \varepsilon$.   In particular, after a general  spatially dependent Weyl tansformation, the short distance cutoff is  spatially dependent.   We found
 that the entanglement entropy of the interval has a divergent contribution $\frac{c}{6}\log \frac{1}{\varepsilon}$ at each end.  So under $\varepsilon\to e^v\varepsilon$,
 the entanglement entropy is shifted at each end by $-\frac{c}{6}v$.   The total ``conformal anomaly'' in the entanglement entropy is a sum of contributions at the endpoints $z=0,L$:
 \be\label{sanom} \Delta S =-\frac{c}{6}\left( v(0)+v(L)\right). \ee
 
  The universal divergence in the entanglement entropy shows that it is not true that the Hilbert space $\H$
 of a quantum field theory on the real line has a factorization $\H_I\otimes\H_{I^c}$ as the tensor product of a Hilbert space $\H_I$ of modes supported on an interval $I$ and
 a Hilbert space $\H_{I^c}$ of modes supported on the complement $I^c$ of $I$; nor is $\H$ the direct sum or integral of subspaces with such a factorization.
 If $\H$ did have such a form, there would exist states of finite (or even zero) entanglement entropy.  The absence of such a factorization is related to the fact that the algebra of
 operators in a region such as $I$ or $I^c$ is a von Neumann algebra of Type III, for which density operators cannot be defined.

\subsection{The Bekenstein Bound}\label{Bek}

The Bekenstein bound is an interesting and relatively simple example  in which understanding ``entropy'' to be microscopic von Neumann entropy -- rather than thermodynamic
entropy -- made it possible to unravel a longstanding puzzle.

Some years after his original proposal concerning black hole entropy, Bekenstein \cite{Bek2} revisited the original work, and made the following proposal.
Suppose that a black hole of mass $M$ and therefore of radius $r_S=2GM$ absorbs a body of size $R$, energy $E$, and 
entropy $S$.    Assuming that $E\ll M$, the black hole entropy changes, as we actually computed
in section \ref{gsl}, by approximately $8\pi GME$.   On the other hand, the entropy $S$ of the infalling body disappears.
So a process in which the black hole absorbs the given body satisfies the Generalized Second Law if and only if
\be\label{tongo} 8\pi GME >S. \ee
If one naively says that a black hole of radius $2GM$ and therefore diameter  $4GM$ can only absorb a body of size $R< 4GM$, then an inequality
\be\label{wongo} 2\pi RE >S \ee
will suffice to ensure that the Generalized Second Law is not violated. 

Bekenstein observed that the inequality (\ref{wongo}), which became known as the Bekenstein bound,
 does not depend on Newton's constant and makes no mention of gravity or black holes,
but is just a statement about the matter system that is possibly falling into the black hole.   This motivated Bekenstein to propose that the inequality
is a universal inequality about relativistic quantum systems.  Since the reasoning that led to the Bekenstein bound is rather heuristic,  one might prefer to state
the Bekenstein bound as the assertion that there is some constant $k$ such that
\be\label{longo} k RE>S\ee
for all matter systems.

Plenty of criticisms could be made of this proposal.  
 For one thing, it is not really true that a black hole cannot absorb a larger body.   In the real world, astronomers
observe ``tidal disruption events''  (TDE's) in which a black hole absorbs a potentially much larger star (though in  most observed TDE's, the black hole is at least as big as the star).  
However, a real star has relatively low entropy, and a TDE in the real world actually does satisfy the Generalized Second Law, though not some of the assumptions
in heuristic arguments that motivate the Bekenstein bound.
 
 There were at least two other important objections.  
 First, a simple argument appears to show that the Bekenstein bound cannot possibly be
 true as a universal statement about relativistic quantum field theories.    Consider a free field theory with $N$ scalar fields  all of the same mass $m$.   Consider a box of some size
 $R$ and place one particle inside the box (in a maximally mixed state in which each of the $N$ possible particle types is equally likely).   The mass of the resulting system does not depend on $N$, but its entropy receives a contribution $\log N$ because there are $N$ choices of
 which species of particle to place inside the box.   So it seems clear that the Bekenstein bound is violated if $N$ is sufficiently large.
 
 A quite different objection involves the question of whether the Bekenstein bound is interesting in a case in which its meaning is clear.
 First let us consider a case in which the statement of the bound does have a clear meaning.   Consider a box of size $R$ containing black body radiation of temperature $T$.  (Massless particles
 in the box present a stronger challenge to the Bekenstein bound than massive ones as they have less energy for given entropy.)    Since the $S$ in the Bekenstein bound was presumed to 
 be thermodynamic entropy, we assume
 that $T$ is large enough so that thermodynamics is applicable.   For this, we need
 \be\label{donko} RT\gg 1. \ee
The total entropy of radiation of temperature $T$ filling a region of size $R$ and volume of order $R^3$ is of order $R^3T^3$; the total energy
is of order $R^3T^4$.     In order of magnitude, therefore, the ratio $RE/S$, which is supposed to
 be bounded below by a constant, is actually of order $RT\gg 1$.   Thus the Bekenstein bound is satisfied for such a system, but is not tight enough to be very interesting.  
 
 In what situation is the Bekenstein bound actually interesting?   To try to do better, we can take $RT\sim 1$.
 Two things go wrong. Frst, then the box only contains a few particles and thermodynamics is not applicable.   Second, we really should take account of the mass of the box
 and  since this mass contributes to the energy but not the entropy, that  causes the Bekenstein bound to be trivial even if $RT\sim 1$.
 
 To do better, we should get rid of the box, and consider one or a few massless particles without the box.   But then what does the Bekenstein bound mean?
 What is the entropy of a state consisting of a single particle without a box?   Thermodynamics is certainly not applicable.    And for that matter, in relativistic quantum mechanics, the ``size'' $R$
 of a state consisting of only one or a few particles is  somewhat murky, as particles cannot really be localized.   
 In other words, it seemed that in the situation in which it was interesting, the Bekenstein
 bound was ill-defined.  
 
 However, Casini \cite{Casini}, partly inspired by earlier work \cite{MMR}, showed that with a suitable reinterpretation 
 of the terms, a version of the Bekenstein bound actually is valid as a universal statement in quantum field
 theory.    The idea was to exploit the positivity of relative entropy, for measurements in a Rindler wedge.   Let $\Omega$ be the ground state of some quantum field theory,
 and let $\Psi$ be some other state in which we want to test the Bekenstein bound.   In section \ref{rindler}, we determined the density matrix $\sigma$ of the state $\Omega$
 restricted to the partial Cauchy hypersurface $\S_r$ defined by $t=0,\, x>0$, or equivalently to the right Rindler wedge $\RR_r$ defined by $x>|t|$:
 \be\label{zolbo} \sigma=\exp(-2\pi K_R),~~~~~K_R=\int_{x\geq 0}\d x \,\d\vec y \, x T_{00}(x,\vec y).\ee
 We do not know much about the corresponding density matrix of a general state $\Psi$; let us just call this density matrix $\rho$.   Positivity of relative entropy says that
 \be\label{olbo}S(\rho||\sigma)\geq 0. \ee
 The familiar definition is\footnote{Relative entropy
 for measurements in a spacetime region in quantum field theory has a rigorous definition due to Araki \cite{Araki}, using Tomita-Takesaki theory.
 For an introduction, see \cite{Notes}.   However, that definition does not lend itself well to the following analysis.  For a recent attempt to circumvent this difficulty, see \cite{KLS}.}
 \be\label{dumbo}S(\rho||\sigma)=\Tr\,\rho\log\rho -\Tr\,\rho\log\sigma.\ee    Here, although $S(\rho||\sigma)$ is actually unambiguous and ultraviolet finite,
  as the following analysis
 will essentially show, the two terms separately do not have that property.   However, one can add and subtract $\Tr\,\sigma\log\sigma$ in such a way as to write $S(\rho||\sigma)$ as
 the sum of two terms that are each unambiguous and ultraviolet finite:
 \be\label{tolbo}S(\rho||\sigma) =\left(\Tr\,\rho\log\rho-\Tr\,\sigma\log\sigma\right)+\left(-\Tr \,\rho\log\sigma+\Tr\,\sigma\log\sigma\right). \ee
 
 Let us first discuss the first term $\Tr\,\rho\log\rho-\Tr\,\sigma\log\sigma$.  Formally, this is a difference of entropies: $-\Tr\,\sigma\log\sigma$ is $S_\vN(\sigma)$, the
 von Neumann entropy of the density matrix $\sigma$, and similarly $\Tr\,\rho\log\rho$ is $-S_\vN(\rho)$, the negative of the von Neumann entropy of $\rho$.
 In a Hamiltonian approach, these von Neumann entropies measure entanglement in the states $\Psi$ or $\Omega$  between modes in the partial Cauchy hypersurface $\S_r$ 
 and modes in the complementary partial Cauchy hypersurface $\S_\ell$; in a covariant description, they measures entanglement between modes in the Rindler
 wedge $\RR_r$ and modes in the complementary Rindler wedge $\RR_\ell$.    
The von Neumann entropy in this problem is ultraviolet divergent because of entanglement between
 short wavelength modes close to but on opposite sides of the common boundary of $\S_\ell$ and $\S_r$ at $x=0$.  That common boundary, which is often called the
 entangling surface,  
  is also the ``edge'' or corner at which $R_\ell$ and $R_r$ meet; it was denoted as $\Sigma$ in  fig. \ref{Seven}.
   We illustrated the ultraviolet  divergence in the entanglement entropy in a concrete example in section \ref{sample}.
Because every state looks like the vacuum at short distances, the ultraviolet divergence is independent of the state
 and therefore the difference $\Delta S= S_\vN(\rho)-S_\vN(\sigma)$ is ultraviolet finite.   Now we can understand one
  contribution on the right hand side of eqn. (\ref{tolbo}): it is  precisely $-\Delta S$.
 
 From eqn. (\ref{zolbo}), we have $\log \sigma = -2\pi K_R$ and therefore the other contribution in eqn. (\ref{tolbo}) is $2\pi\left( \Tr\,\rho K_R-\Tr\,\sigma K_R\right)$.   
 The definition of the density matrix $\rho$ is that for any operator $\O$ supported on $\S_r$ (or in the wedge $\RR_r$),  $\Tr\,\rho\O=\la\Psi|\O|\Psi\ra$.
 Similarly, $\sigma$ has the property that for any such $\O$, $\Tr\,\sigma\O=\la\Omega|\O|\Omega\ra$.   Applying this principle with $\O=K_R$ and using the
 definition of $K_R$, we get 
\be\label{doofus}  -\Tr \,\rho\log\sigma+\Tr\,\sigma\log\sigma=2\pi \int_{\S_r} \d x\d\vec y \,x   \left(\la\Psi| T_{00}(x,\vec y)|\Psi\ra -\la\Omega|T_{00}(x,\vec y)|\Omega\ra\right). \ee
In quantum field theory, the definition of the operator $T_{00}(\vec x)$ is not straightforward; it is subject to an additive renormalization.  However, this
renormalization only involves an additive $c$-number, which cancels in the difference $\la\Psi| T_{00}(x,\vec y)|\Psi\ra -\la\Omega|T_{00}(x,\vec y)|\Omega\ra$.
So that difference, and therefore eqn. (\ref{doofus}), is well-defined, not affected by the renormalization ambiguity.  One usually picks a renormalization scheme in 
which $\la\Omega|T_{00}(x,\vec y)|\Omega\ra=0$; with such a choice the quantity in eqn. (\ref{doofus}) is $2\pi \E$, with
\be\label{oofus} \E=\int_{\S_r}\d x\,\d\vec y\, x\la\Psi|T_{00}(x,\vec y)|\Psi\ra .\ee
The inequality of positivity of relative entropy thus becomes 
\be\label{bumble} 2\pi \E\geq \Delta S, \ee
with equality only if $\Psi=\Omega$.  

Casini proposed this inequality as a rigorous version of the Bekenstein bound.   The quantity $S$ on the right hand side of the original Bekenstein bound (\ref{wongo}) 
is replaced here with $\Delta S$, the difference of von Neumann entropies between a general state $\Psi$
and the vacuum state $\Omega$.   In case $\Psi$ differs from $\Omega$ by the presence of a matter system that is well localized away from the entangling surface $\Sigma$,
$\Delta S$ will be approximately the von Neumann entropy of this matter system.   However, $\Delta S$ is well-defined for an arbitrary state $\Psi$, even if $\Psi$ differs from
the vacuum by, say, the presence of a single particle, which -- to the extent such localization makes sense relativistically -- may be located partly in $\RR_r$ and partly in $\RR_\ell$.
As for $\E$, Casini interpreted this as a substitute for the product $RE$ in the original Bekenstein bound.   (Thus, Casini defined a rigorous substitute for the product $RE$ 
but not for $R$ and $E$ separately.)   To explain the motivation for this interpretation of $\E$, suppose that $\Psi$ describes an object or matter state of some kind  that can be understood
semiclassically and that has size $R$ and energy $E$.   To use the inequality (\ref{bumble}) to make  a statement about the size, energy, and entropy of this object, we place it in $\S_r$,
that is at $x>0$.     On the other hand, to make the inequality (\ref{bumble}) as sharp as possible, we want to make $\E$ as small as possible under the constraint that the object is 
supported at $x>0$.   So, as the object under study has size $R$, we place it in the region $0<x<R$.   But then, for a system of energy $E$ localized in that range of $x$,
the order of magnitude of $\E$ is $\E\sim ER$.   Thus in a situation in which the terms in the original Bekenstein bound have a clear meaning, the inequality (\ref{bumble}) has a similar
import to the Bekenstein  bound.   But it has the virtue of being rigorously true for an arbitrary quantum state.

To underscore that the inequality (\ref{bumble}) goes far beyond any semiclassical  reasoning, we may point out that there exist states $\Psi$ such that the left and right hand
sides of the inequality (\ref{bumble}) are both negative, and still, of course, the inequality is satisfied.   To make $\E$ negative, let $\Upsilon$ be any state such that the
matrix element $\la\Upsilon|K_R|\Omega\ra\not=0$. (Such states exist since $K_R|\Omega\ra\not=0$.)     Then a suitable linear combination of $\Omega$ and $\Upsilon$
has negative $\E$.   The inequality then implies that this state has negative $\Delta S$.   In free field theory, another way to construct a state with negative $\Delta S$ is roughly the following.
Take any finite set of modes in $R_r$ that are entangled with a corresponding finite set of modes in $R_\ell$, and disentangle those particular modes without affecting the state of other
modes.  This reduces the entanglement entropy between the two Rindler wedges so it produces a state with $\Delta S<0$.

What was wrong with the attempt to disprove the Bekenstein bound by considering a theory with $N$ free fields of mass $m$ for very large $N$?  This was originally understood in \cite{MMR},
with additional analysis in  \cite{Casini}.
Suppose we place a particle  (in a mixed state of entropy $\log N$) near  $x=R$.  Naively this adds $\log N$ to the entropy and can violate the Bekenstein bound.
 It turns out that if $N$ is large enough to cause a problem,
vacuum fluctuations are  important.  
   To try to violate the Bekenstein bound, we take  $\log N>2\pi mR$ or $N>e^{2\pi mR}$.  The probability of a vacuum fluctuation in which a particle-antiparticle
pair appears and separates a distance $L$ is roughly $e^{-mL}$.   The number of such pairs with the particle near  $x= R$ and the antiparticle at $x<0$ is roughly
$N e^{-mR}$.    So if $N>e^{2\pi m R}$, the number of such pairs is overwhelmingly large, and the additional entropy due to an added particle cannot be computed
without taking into account the particles that are  present due to vacuum fluctuations.  Detailed calculation \cite{MMR,Casini} verifies consistency with the Bekenstein bound.

\subsection{R\'{e}nyi Entropy and Generalized Entropy of a Black Hole}\label{rbh}

Here we will analyze the claim cited in section \ref{bhent} that the generalized entropy $S_\gen=\frac{A}{4G}+S_\out$ is a well-defined quantity,
not subject to ultraviolet divergences, in any theory of gravity (possibly interacting with other fields) that has been satisfactorily defined or renormalized.    We will do this analysis
for a pair of black holes  entangled in the thermofield double state $\Psi_\HHI$.  In such a case, each of the two black holes is in perfect thermal equilibrium, so the distinction between
thermodynamic entropy and microscopic von Neumann entropy is not important.   As a statement about von Neumann entropy, the
 argument extends readily beyond the state $\Psi_\HHI$ to other states prepared by Euclidean 
path integrals, such as we will discuss in section \ref{just}. 

Actually, it is convenient to first consider a more general problem of showing that R\'{e}nyi entropies are well-defined in the thermofield double state.   
We recall that the R\'{e}nyi entropy of order $\alpha$ of a density matrix
$\rho$ is defined by
\be\label{mozo} R_\alpha(\rho)=\frac{1}{1-\alpha}\log \Tr\,\rho^\alpha. \ee
The von Neumann entropy is the limit of the R\'{e}nyi entropy for $\alpha\to 1$, so what we learn about R\'enyi entropy applies also to von Neumann entropy.

In making the argument, we will consider an AdS-Schwarzschild black hole with $\Lambda<0$.  
Taking $\Lambda<0$ provides an infrared regulator that eliminates the thermal instabilities 
described in section \ref{unstable}  (in the case of a sufficiently massive black hole), so we are not limited to lowest order of perturbation theory.
Indeed, if the gravitational theory considered in asymptotically AdS spacetime has a known CFT dual on the boundary, then we expect the gravitational theory to be
nonperturbatively well-defined and the main conclusion about the R\'enyi entropies to be likewise valid non-perturbatively.   Even if a dual CFT is not known, and may not exist,\footnote{A generic theory of gravity, such as Einstein gravity with or without a cosmological constant and  with 
no other fields or with only finitely many other fields, may well lack a sensible ultraviolet completion.  The cases
in which dual CFT's are known are derived from string/M-theory and correspond to theories of gravity that do have ultraviolet completions.}
the calculation of the entropies will make sense to the extent that the theory does.

Another way to avoid infrared instabilities, while keeping to $\Lambda=0$, is to take the limit that the black hole mass goes to infinity.  Then the near horizon region outside  the black
hole  converges to Rindler space.   This is a  useful framework, studied for example in \cite{CW}.    But here we will work in the AdS-Schwarzschild setting, as this makes the
role of gravity more transparent.

We will see that, for investigating the finiteness of the entropy, a theory of quantum fields in a fixed spacetime background is fundamentally different from a theory
in which gravity is dynamical.  
We start by considering an ordinary quantum field in a fixed gravitational background, which we take to be the maximal extension of the AdS-Schwarzschild spacetime,
with Penrose diagram depicted in fig. \ref{PenroseC}.  As before, we denote the asymptotically AdS regions to the right and left of the horizon as $\U_r$ and $\U_\ell$.
As explained in section \ref{thermofield}, the density matrix $\sigma_r$ that describes observations in the region $\U_r$ in the thermofield double state $\Psi_\HHI$
is \be\label{hyto}\sigma_r=\frac{1}{Z}e^{-\beta_\sH H_r}, \ee
where $\beta_\sH$ is the relevant Hawking temperature and $H_r$ is the Hamiltonian that acts on the  Hilbert space $\H_r$ of region $\U_r$.  
To compute the R\'enyi entropy $R_\alpha(\sigma_r)$, we need to compute
\be\label{zoldo} \Tr\,\sigma_r^\alpha =\frac{\Tr\, e^{-\alpha\beta_\sH H_r}}{Z^\alpha}. \ee
Now recall that the operator $e^{-\beta_\sH H_r}$ rotates the partial initial value surface $\S_r$ discussed in section \ref{thermofield} through an angle $2\pi$
inside the Euclidean AdS-Schwarzschild spacetime.   Therefore, the operator $e^{-\alpha\beta_\sH H_r}$ rotates $\S_r$ through an angle $2\pi\alpha$, producing a
singularity at the horizon with a deficit angle $2\pi(1-\alpha)$.   The quantity $\Tr\,e^{-\alpha\beta_\sH H_r}$ is simply the partition function of the field $\phi$
in this singular spacetime.   As discussed in section \ref{bhent}, that partition function will be ultraviolet divergent because of the singularity.    
There is no way to eliminate the divergence by redefining $H_r$.   Since commutators of $H_r$ with $\phi$ are required to generate time translations, the only allowed
redefinition of $H_r$ is an additive constant.  But because of the normalization condition $\Tr\,\sigma_r=1$, an additive constant in $H_r$ would be compensated by a rescaling
of $Z$, with no effect on $\sigma_r$. 

Now let us compare this with what happens if the gravitational field is taken to be dynamical, and not just a $c$-number background.   Then to compute
$\Tr\,e^{-\alpha \beta_\sH H}$, we are supposed to do a path integral over asymptotically AdS metrics whose conformal boundary is $S^2\times S^1_{\alpha\beta_\sH}$,
where $S^2$ is a unit two-sphere and $S^1_{\alpha\beta_\sH}$ is a circle of circumference $\alpha\beta_\sH$.   This is the recipe of the AdS/CFT correspondence, but it
is actually also the procedure followed by Hawking and Page (though not described in precisely the same language) long before AdS/CFT duality was formulated \cite{hp}.   For small $G$,
one expects the path integral to be dominated by a classical solution of minimum action with the appropriate asymptotic behavior.  As explained in section \ref{negative},
  the  classical solution of minimum action is  believed to be thermal $\AdS_{\alpha\beta_\sH}$ or a black
hole of appropriate mass, depending on the value of $\alpha\beta_\sH$.   Both of these solutions are smooth; no singularity appears in the computation.   Assuming any necessary renormalizations have been performed to make the gravitational path integral well-defined,
the expansion around the dominant classical solution is manifestly going to give a finite answer.   Indeed, if  the assumed gravitational theory is such that a boundary CFT exists,
what the gravitational path integral  computes is equivalent to the manifestly well-defined quantity $\Tr_{\sf {CFT}}\, e^{-\alpha \beta_\sH H}$, where $H$ is the Hamiltonian of the 
boundary CFT quantized on a unit two-sphere, and $\Tr_{\sf{CFT}} $ denotes a trace in the CFT Hilbert space.

In general, in any quantum field theory that requires regularization of divergent quantities, the finiteness of physical amplitudes depends on cancellations between
divergent loops and counterterms. 
In low energy quantum field theory with gravity, defining the gravitational path integral involves canceling the divergences in loop diagrams against
classical counterterms.   In particular, the well-definedness of the gravitational path integral that computes $\Tr\,e^{-\alpha\beta_\sH H}$ will depend on such cancellations
between loop effects and counterterms.   This is quite analogous to the claim of Susskind and Uglum that the generalized entropy is finite because of a cancellation
between a loop divergence in $S_\out$ and a renormalization counterterm that affects the value of $G$ in the classical  $\frac{A}{4G}$ contribution to black hole entropy.
 
 Indeed, the original one-loop analysis of the generalized entropy \cite{SU}  is essentially  a more explicit
special case of what we have just discussed.   Here we roughly follow \cite{CW}.   The argument is most easily expressed in terms of the effective action $I_\eff$ of the theory.
At one-loop order, this is a sum
\be\label{zingox}I_\eff = I_\cl+ \frac{1}{2}\log \det\, D .  \ee
Here $I_\cl$ is the classical action, defined in terms of a bare Newton constant $G_0$, and  $\det\,D$ is the one-loop determinant of the matter fields (one-loop gravitational fluctuations
are included, and for simplicity the matter fields are assumed to be bosonic; fermion determinants contribute with the opposite sign).  
If the theory has been successfully renormalized, then $I_\eff$ is well-defined and finite; concretely, the cutoff dependence of $G_0$ is defined to cancel divergences in
$\frac{1}{2}\log \det\, D$.   
 The quantum-corrected 
black hole solution is an extremum of the quantum effective action $I_\eff$, not of the classical action $I_\cl$.     The definition of $I_\eff$ is that the partition function is
$Z=e^{-I_\eff}$, so as in eqns. (\ref{Boxy}),(\ref{oxy}), the black hole entropy at inverse temperature $\beta$ is 
\be\label{bodo}S(\beta)=\left(1-\beta \frac{\d} {\d\beta}\right)\log Z(\beta)=-\left(1-\beta \frac{\d} {\d\beta}\right)I_\eff(\beta)\ee
or in more detail
\be\label{nodo}S(\beta)
=-\left(1-\beta \frac{\d} {\d\beta}\right)\left( I_\cl +\frac{1}{2}\log \det\, D\right).\ee
Here $I_\eff(\beta)$ is the effective action of a black hole that has inverse temperature $\beta$, and has a mass $M(\beta)$ that can be found by extremizing the effective
action for a given $\beta$ and evaluating the ADM mass at infinity.   The function $M(\beta)$ will receive quantum corrections and will not coincide with the classical result.

In eqn. (\ref{nodo}) as just described, to evaluate the derivative with respect to $\beta$, 
we should vary $\beta$ and vary $M$ as a function of $\beta$ so that the quantum-corrected black hole solution remains smooth at the horizon,
and then evaluate the right hand side of eqn. (\ref{nodo}).
But just as in section \ref{another}, since in eqn. (\ref{nodo}) we only need the first derivative with respect to $\beta$, and $I_\eff(\beta)$ is obtained by evaluating the effective
action at an extremum, we will get the same result if we vary $\beta$ keeping $M$ fixed, producing a conical singularity at the horizon.  
When we do the calculation that way, the contribution to eqn.  (\ref{nodo}) from $ -\frac{1}{2}\log \det\, D$ is just the replica trick calculation of the matter entropy
$S_\out$ outside the horizon.   The contribution of the conical singularity at the horizon is $A/4 G_0$.  Of course, $G_0$ is ultraviolet divergent as in eqn.
(\ref{dumbox}).   But  the form (\ref{bodo}) makes it obvious that these divergences cancel, since ultraviolet divergences have been canceled in defining $I_\eff(\beta)$.

An interesting detail about this is that the area $A$ receives quantum corrections:
\be\label{ziff} A=A_\cl +G b_1+\O(G^2),\ee
where $A_\cl$ is the classical area and the one-loop correction is $Gb_1$.   The contribution $A/4G_0$ to the generalized entropy 
from the $I_\cl$ term in eqn. (\ref{nodo}), is then, using eqn. (\ref{dumbox})
for $G_0$,
\be\label{melbox}   \frac{1}{4}\left(\frac{1}{G}- f'\Lambda^2\right)\left(A_\cl +G b_1\right).  \ee
We see that in order $G^0$, the correction to the area contributes only the finite term $b_1/4$ to the entropy.  But in order $G$, there is a divergent term
$-f'\Lambda^2b_1/4$.   Therefore, in two-loop order and higher, the correction to the horizon area plays a role in the cancellation of divergences
that makes the generalized entropy finite.

\section{The Ryu-Takayanagi Formula}\label{rt}

At this point, it is hopefully clear that von Neumann entropy as well as thermodynamic entropy is important in black hole physics.
As there is a Bekenstein-Hawking area formula for the thermodynamic entropy, one may wonder if there is also a similar area formula
for von Neumann entropy.   In fact,  in the context of AdS/CFT duality, this question was answered before it had been widely asked by the Ryu-Takayanagi (RT) formula \cite{RT},
with several important later refinements \cite{HRT,LM,BDHM,FLM,NEW}.  

In section \ref{hm}, we will describe a heuristic motivation for the RT formula.   In section \ref{original}, we explain the original setting in which the RT formula
was formulated, and some interesting tests of it in that context.
In section \ref{just}, we sketch a proof of the RT formula, following \cite{LM}.   In section \ref{pageagain}, we briefly summarize a few of the further developments.

\subsection{Heuristic Motivation}\label{hm}

 \begin{figure}
 \begin{center}
   \includegraphics[width=6.6in]{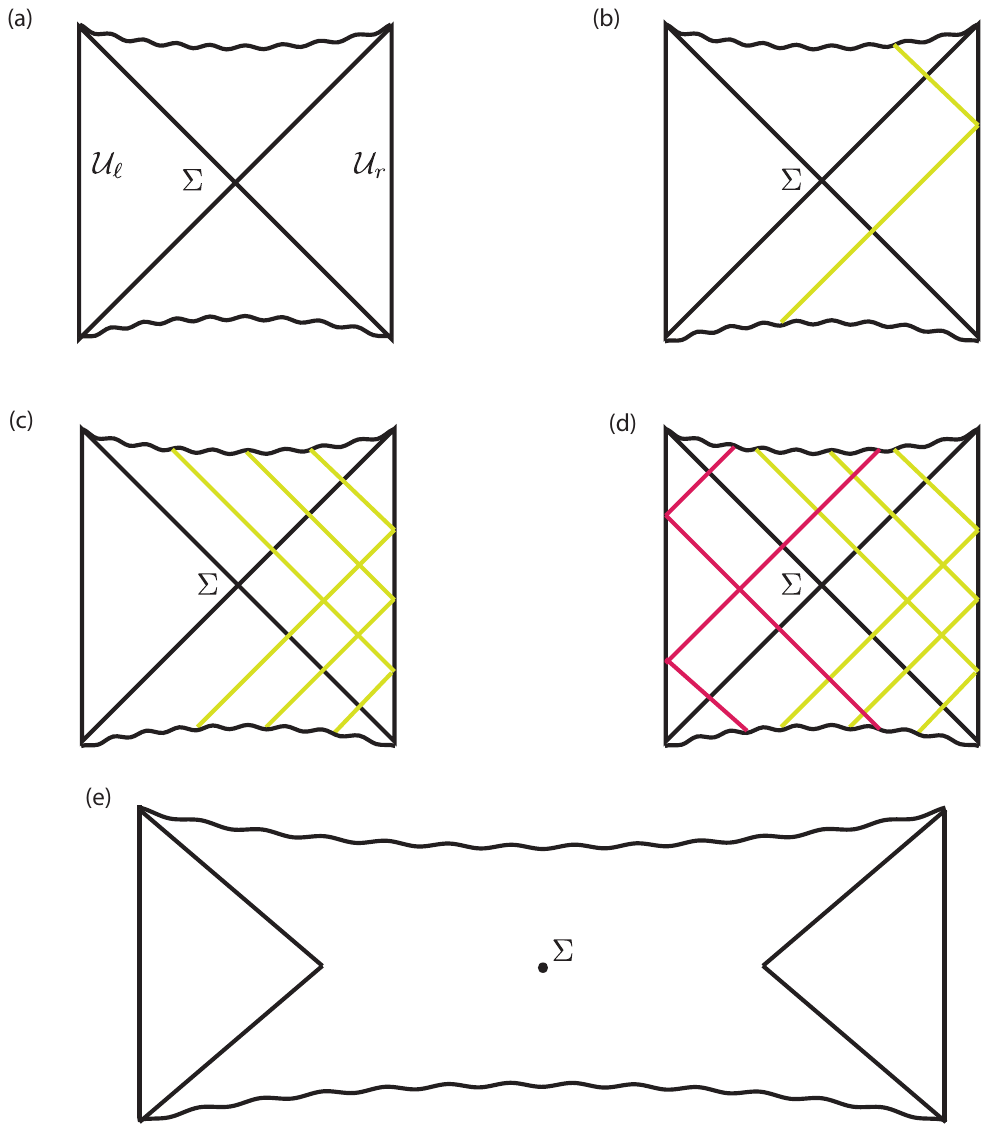}
 \end{center} 
\caption{\footnotesize (a) Two black holes, in different asymptotically AdS regions, entangled  in the thermofield double state $\Psi_\HHI$.  Shown are 
the exterior regions $\U_\ell$ and $\U_r$, and the bifurcation surface $\Sigma$.  
(b) An observer $O_r$ with access to the right conformal boundary creates a new
 state by acting with a unitary operator that modifies the state near a given point on the conformal boundary.   Depicted is the spacetime history of this state assuming that
 it is allowed to evolve to the past and future via the unperturbed Hamiltonian.  Back reaction on the geometry (which causes the horizon to move outward)
 is ignored here and in (c) and (d).  (c) The observer $O_r$ makes multiple such
  perturbations at different times.   (d) An observer $O_\ell$ with access to the left conformal boundary makes similar perturbations of the state. (e)  A possible outcome of these manipulations, with back reaction
 taken into account.
 The original bifurcation surface $\Sigma$ lies far behind the horizons of $O_r$ and $O_\ell$, deep  inside the ``wormhole'' that connects the two sides.    $\Sigma$
  is not a horizon any more, but it is an extremal surface and its area is unchanged. \label{TFDmany}}
\end{figure} 

We start with a heuristic justification of the RT formula based on considering a pair of black holes, 
 entangled in a general state.    To begin with, consider  two black holes
in thermal equilibrium, entangled in the thermofield double state $\Psi_\TFD$.       Shown in the Penrose diagram of
fig. \ref{TFDmany}(a)  are the diagonal lines that represent past
and future horizons for an observer in the left or right exterior region $\U_\ell$ or $\U_r$, and the bifurcation surface $\Sigma$ at which the past and future horizons meet.  As usual
$\U_\ell $ and $\U_r$ are spacelike separated from $\Sigma$ and are respectively to its  left or right.   The conformal boundary of the spacetime consists of two copies of the Einstein static universe
$\S^2\times \R$, appearing as the left and right boundaries of the Penrose diagram.   

An observer $O_r$ with access to the right boundary, by manipulating the boundary conditions in a way described in the AdS/CFT correspondence \cite{EW}, 
 can create a new state that differs from the original one in region $\U_r$
(and in the past and future wedges as well, as we will see), but not in $\U_\ell$; conversely an observer $O_\ell$ with access to the left boundary can disturb the spacetime in region $\U_\ell$
but not in $\U_r$.   
However, whatever can be done by these boundary observers will not disturb the entanglement entropy between the two sides (see the analysis of eqns. (\ref{mizo}) and (\ref{vizo})).
We will use this as a clue to suggest what sort of area formula could represent the von Neumann entropy.

What can $O_r$ do to disturb the state?   At some chosen time $t$, $O_r$ can modify the state by injecting a particle in from the boundary (fig. \ref{TFDmany}(b)).
(Concretely, this is done by manipulating the boundary conditions in a way that is familiar in the AdS/CFT correspondence \cite{EW}; note that here we do not need to assume
the existence of a full-fledged CFT dual of the bulk gravitational theory under study.)  
As explained in \cite{SS}, the most convenient way to look at the outcome of injecting a particle is the following.  The action of the observer creates from $\Psi_\TFD$ a new state $\Psi'$ whose
evolution we can usefully study in the unperturbed dynamics; in other words, once the new state $\Psi'$ is selected, we evolve it forwards and backwards in time using the unperturbed
Hamiltonian, as if the observer is not there.   Followed forwards in time, a particle injected into the spacetime by the observer $O_r$ will most likely eventually fall across the future
horizon; however, if we follow the same trajectory backwards in time, the particle will be reflected from the conformal boundary of the spacetime, return inward, and most likely eventually fall across
the past horizon.   The spacetime evolution of the perturbed state is depicted in fig. \ref{TFDmany}(b).  This is a schematic picture in which the modification of the geometry by the injected particle is
not shown (there is no claim, for example,  that the resulting spacetime can be obtained by gluing together  pieces of AdS Schwarzschild solutions with different masses).

Of course $O_r$ can make such manipulations repeatedly, creating a state in which many particles emerge from the past horizon, are reflected from the right conformal boundary, and eventually
fall back behind the future horizon, as in  \ref{TFDmany}(c).      This figure has not been drawn realistically to show the back-reaction on the geometry of the infalling particles (and interactions
among these particles have also been ignored).
Even if the injected  particles have modest energy, the resulting back-reaction can be large if the particles  are injected into the spacetime at widely separate times  and out of
time order \cite{SS,SSagain}.  The consequences of  this
back-reaction are  very interesting and surprisingly complicated, with an intimate relation to chaos in black hole physics \cite{SS}, as partly anticipated in \cite{DT}.
For our present purposes of motivating the RT formula, the only aspect of this that we need to know is that injecting particles from the right boundary into the bulk increases the ADM mass of the spacetime
as measured on the right boundary, causing the horizon to move outward.   Therefore, in fig. \ref{TFDmany}(c), the surface labeled $\Sigma$ no longer lies on the past or 
future horizon from the point of
view of $O_r$, though as nothing has been done to the part of the spacetime that is to the left of $\Sigma$, 
this surface is still the intersection of past and future horizons from the
point of view of $O_\ell$.

However, the observer $O_\ell$ is, of course, similarly free to modify the state by injecting particles from the left boundary.   This will produce the still more complicated spacetime of
 fig. \ref{TFDmany}(d),
which again has been drawn ignoring back-reaction and particle interactions.    For our purposes, what is important about this picture is the following.   First of all,  the 
observers $O_\ell$ and $O_r$ have been able to modify the geometry everywhere except at the bifurcation surface $\Sigma$.  Second, after these manipulations, the past and future horizons
of $O_r$ have moved to the right, the past and future horizons of $O_\ell$ have moved to the left, and the surface $\Sigma$ is behind the past and future horizons of both observers.
A schematic depiction  of this  is shown in fig. \ref{TFDmany}(e).  The past and future horizons of $O_\ell$ are far to the left of the past and future horizons
of $O_r$.   Between the horizons of $O_\ell$ and $O_r$ is a long ``wormhole,'' out of sight of each observer and beyond their influence.   The former bifurcation surface $\Sigma$ is still present, with its area unchanged, somewhere in the wormhole.

Now, since whatever the two observers have done has not changed the entanglement entropy between the two sides, if we are going to find an area formula for that entanglement
entropy, it must be the area of a surface whose area the two observers are unable to change.  But since the only place where the actions of $O_\ell$ and $O_r$ do not disturb
the geometry is the former bifurcation surface $\Sigma$, we conclude that if the area of any surface in the spacetime is going to represent the entanglement entropy of interest, 
this is most plausibly going to be $\Sigma$.

But how do we characterize the surface $\Sigma$?  Before the manipulations by $O_r$ and $O_\ell$, this surface was the bifurcation surface where the two horizons meet.  
After those manipulations, $\Sigma$
is no longer on the  horizon of either observer.   If this surface is going to represent an entropy, we need another way to characterize it.

With this aim, we go back to the original spacetime of fig. \ref{TFDmany}(a), before any manipulations by the two observers.   
The extended Schwarzschild solution has a time translation symmetry and an $SO(3)$ symmetry that rotates the polar angles (which are not shown in the Penrose diagram).
The Penrose diagram exhibits an additional  $\Z_2\times \Z_2$
symmetry, where one $\Z_2$ is a spatial reflection that exchanges the left and right ends of the figure, and the other $\Z_2$ is a time-reversal symmetry that exchanges the past and future.
  The bifurcation surface  $\Sigma$  is invariant  under $\Z_2\times \Z_2$, and this 
 ensures that $\Sigma$  is an {\it extremal surface}, meaning a surface whose
area is invariant to first order if it is slightly displaced in any direction.   This extremal property can also be deduced from the fact that $\Sigma$ is the fixed point set of the time-translation
symmetry.\footnote{Because time-translation symmetry rescales $U$ and $V$, it implies that the horizon area is independent of $U$ at $V=0$ and independent of $V$ at $U=0$.
(We see this explicitly in eqn.  (\ref{nimpdef}): if $U$ or $V$ vanishes, then $r=2GM$.)   Hence at $U=V=0$, the $U$ and $V$ derivatives of the horizon area both vanish,
so the bifurcation surface at $U=V=0$ is an extremal surface.}

Rather than invoking the symmetries, we can also demonstrate that $\Sigma$ is an extremal surface with formulas.   The area of the two-sphere
represented by a point in the Penrose diagram is $4\pi r^2$, where the relation between $r$ and the Kruskal-Szekeres null coordinates $U,V$ is\footnote{This formula is for an asymptotically
flat Schwarzschild black hole, but the analogous formula for an AdS-Schwarzschild black hole is similar.}
\be\label{dotme} \left(\frac{r}{2GM}-1\right) e^{r/2GM}=-UV. \ee
The bifurcation surface $\Sigma$ is at $U=V=0$, and from the formula (\ref{dotme}), it is evident that $\left.\frac{\partial r}{\partial U}\right|_{U=V=0}=\left.\frac{\partial r}{\partial V}\right|_{U=V=0}=0$.
Since the area is $4\pi r^2$, the fact that $r$ is stationary at the horizon again shows that $\Sigma$ is an extremal surface.

$\Sigma$ has a few additional important properties.   Let $T$ and $X$ be timelike and spacelike coordinates such that $U=T-X$, $V=T+X$, so $-UV=-T^2+X^2$.
From eqn. (\ref{dotme}) we see that 
\be\label{icono}\frac{\partial^2 r}{\partial X^2}>0. \ee
Thus the area of $\Sigma$ is a local minimum if $\Sigma$ is moved in a spatial direction.\footnote{It is also true that the area of $\Sigma$ is a local maximum if $\Sigma$
is moved in a timelike direction.   This condition need not be stated separately, as for Einstein gravity coupled to a reasonable matter system (satisfying the null energy condition),
it follows from (\ref{icono}) together with the Raychaudhuri equation. For example, see \cite{Twice}.} 
Another important fact is that $\Sigma$ is homologous to either the left or right conformal boundary of the spacetime.  In fact, the homology can be made through any Cauchy hypersurface
that passes through $\Sigma$, such as the surface $t=0$ that is the fixed point set of the time-reversal symmetry.

Finally, we can formulate the RT formula, 
 or more precisely its time-dependent HRT generalization \cite{HRT}, for the sort of general two-sided geometry that we have been considering here.  According to the RT or HRT formula,
 the von Neumann entropy of a density matrix that describes observations made at the left boundary or at the right boundary, or equivalently the entanglement entropy
between the two sides, is (to lowest order in $G$)
\be\label{nubo} S_\vN=\frac{A(\SIgma)}{4G},\ \ee
where $A(\Sigma)$ is the area of an extremal surface $\Sigma$ that is a local minimum of the area in spatial directions, and is homologous to either conformal boundary.
If there is more than one such surface, then they are called candidate RT (or HRT) surfaces, and the entropy is given by $A(\Sigma)/4G$, where $\Sigma$ is the candidate RT
surface that has the least area among all such surfaces.\footnote{If the candidate RT surface of minimum area is not unique, then of course it does not
matter which one we pick in evaluating the RT formula for the leading order entropy.  In this case, it is believed that there are quantum corrections to the entropy
of order $G^{-1/2}$, rather than the usual $\O(1)$ \cite{MWW}.}

In particular, by the sort of manipulations sketched in fig. \ref{TFDmany}, the observers $O_\ell$ and $O_r$ may be able to create a spacetime with multiple candidate RT
surfaces.   But the RT/HRT formula says that as these manipulations cannot change $S_\vN$, any such additional candidate RT surfaces have an area at least as large as that of the original
bifurcation surface  $\Sigma$.  This is a nontrivial prediction about classical general relativity.

In section \ref{original}, we will discuss more examples of the RT formula at work, in the original setting in which it was proposed.

\subsection{The Original Setting}\label{original}

The RT formula \cite{RT} was originally 
formulated as a statement about the entropy of a region in the boundary conformal field theory.   We recall  that according to this duality, a particular $d$-dimensional CFT, formulated on a $d$-manifold $Y$, has
the same content as a $D=d+1$-dimensional gravitational theory\footnote{\label{exception} In this article, we have generally discussed black holes in the familiar  $D=4$ world.
Extension  to other values of $D$ would generally not have added much.
However, in analyzing the RT formula, we will consider arbitrary $D$ for several reasons.   One reason is that in string theory, there are important examples with various values of $D$.
Another reason is that  the case $D=3$, $d=2$ is illuminating and easy to visualize.  Yet another reason is that if one wants to use the RT formula to learn about gravity,
one might consider  $D=4$, $d=3$ to be the most natural case, but if one wants to use it to study entanglement entropy in quantum field theory, then one might be particularly
interested in $D=5$, $d=4$.  The cases $D=4,d=3$ and $D=3,d=2$ are also important in some condensed matter applications. So in short it seems artificial to limit
a discussion of the RT formula to a particular value of $D$. }   formulated on an asymptotically locally AdS manifold $X$ whose conformal boundary is $Y$.   In principle,
for given $Y$, one must take into account all possible choices of $X$, but in the sort of simple applications of the RT formula that we will discuss, there is a particular $X$
that dominates for small $G$.   In fact (as in \cite{RT}), the main example that we will discuss here is that $Y$ is the $d$-dimensional Einstein static universe $S^{d-1}\times \R$,
and $X$ is $\AdS_D$, Anti de Sitter space of dimension $D=d+1$.   
As in  the original formulation of the RT formula, we will assume a time-reversal symmetry $t\to -t$ that leaves fixed the initial value surfaces $\S\subset X$ and $\S'\subset Y$, respectively 
of dimension $d$ and $d-1$. Here $\S'$ is the conformal boundary of $\S$.
(To be more precise, we formulate the CFT on a manifold $Y$ with a time-reversal symmetry, and we assume -- as is true in simple examples, including those
we will discuss -- that the relevant $X$ also possesses this symmetry.)   Now, let $\A$ be some region in $\S'$.  We assume that the boundary of $\A$ is an embedded
$d-2$-dimensional manifold $\partial\A\subset \S'$.
For some given state $\Psi$ of the quantum fields on $Y$, let $\rho_\A$ be the density matrix appropriate for measurements in the region $\A$. 
The original version of the RT formula answered the following question: What is the von Neumann entropy $S_\vN(\rho_\A)$?   The proposed
answer was that this entropy is given by a formula analogous to the Bekenstein-Hawking entropy formula
\be\label{rtform} S_\vN(\rho_\A)=\frac{A(\Sigma_\A)}{4G}+\cdots,\ee 
where $\Sigma_\A$ is a particular surface known as the RT surface of $\A$, and 
 the omitted terms are $\O(1)$ for small $G$.     The definition of $\Sigma_\A$ is as follows:
 among all $d-1$-dimensional surfaces in $\S$ whose boundary coincides with $\partial \A$ and which are also homologous to $\A$, $\Sigma_\A$ is the one whose area $A(\Sigma_\A)$ is smallest.   To be more precise, this ``area'' is really a renormalized area, as discussed at the end of this section, 
and as $\Sigma_\A$ has dimension $d-1$,
it is really an area in the $d-1$-dimensional sense (thus it is only an ordinary area if $D=4$, $d=3$).  The statement that $\Sigma_\A$ is homologous to $\A$ means that
$\Sigma_\A$ and $\A$ together are the boundary of a region in $\S$.   
There is no explicit dependence on the quantum state $\Psi$ in  the formula (\ref{rtform}), so if true this formula
 implies that for small $G$, the leading  contribution to the entropy depends on $\Psi$
only to the extent that the geometry depends on $\Psi$.    When it is clear what density matrix is intended, we abbreviate $S_\vN(\rho_\A)$ as $S_\A$.

 \begin{figure}
 \begin{center}
   \includegraphics[width=3.6in]{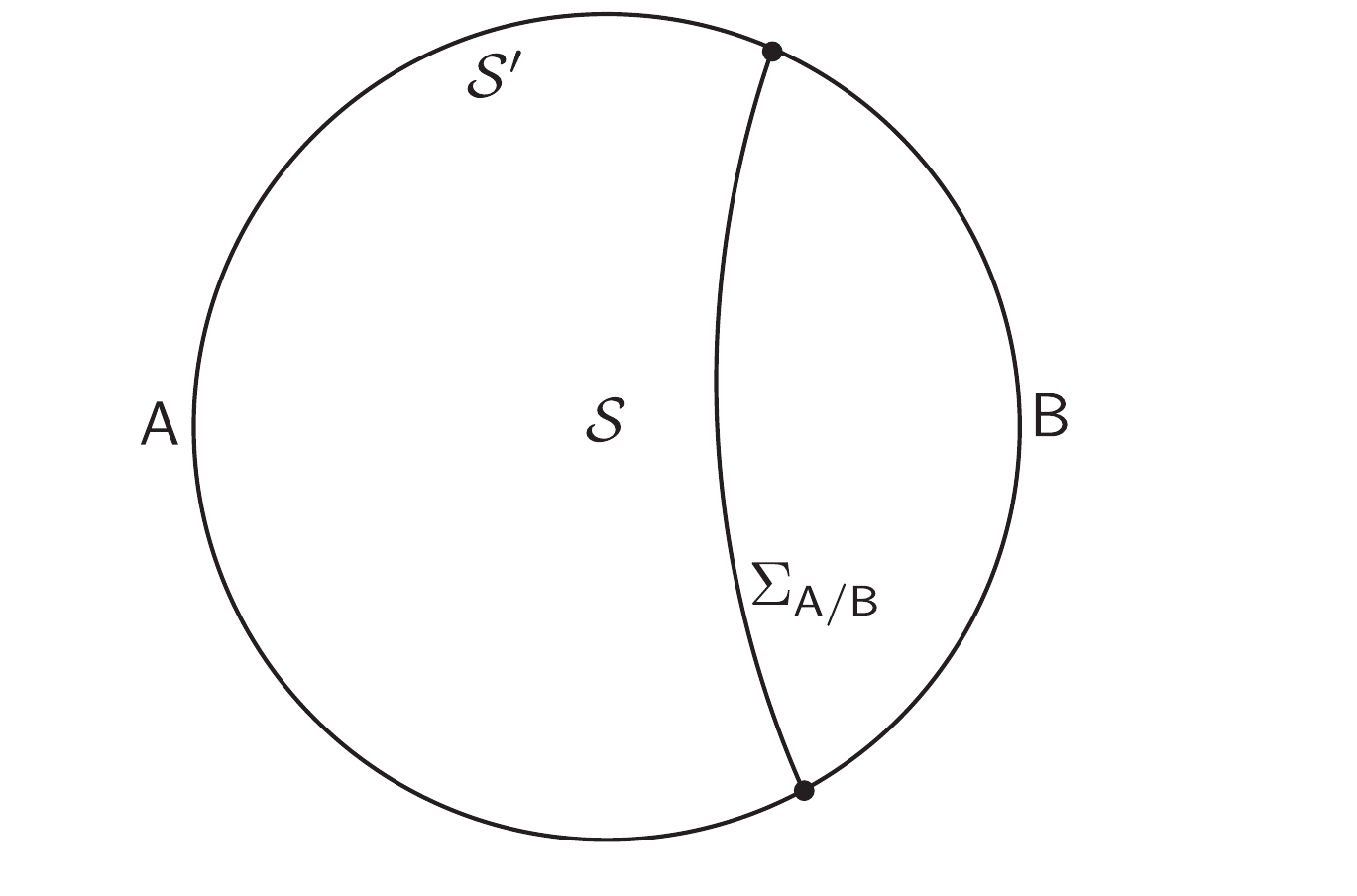}
 \end{center}
\caption{\footnotesize The disc $\S$, with conformal boundary $\S'$, is a Cauchy hypersurface in $\AdS_3$.  $\S'$ is decomposed as the union
of two complementary  intervals $\A, \B$, as shown.  $\A$ and $\B$ have the same RT surface, denoted $\Sigma_{\A/\B}$.   It  is the  curve in $\S$ of minimal (renormalized) length whose ends are the endpoints of $\A$
(or $\B$).  As a curve of minimal length, it is a geodesic.   \label{RTpic2}}
\end{figure}

If there are several surfaces  $\Sigma_{\A,\alpha}\subset \S$ that obey the necessary topological conditions to be the RT surface of $\A$
($\partial \Sigma_{\A,\alpha}=\partial\A$, and $\Sigma_{\A,\alpha}$ is homologous to $\A$) and that locally minimize
the area, then they are called candidate RT surfaces.   The true RT surface is the candidate RT surface of minimal area.  

The case $d=2$ is particularly easy to visualize.   As a first example, we will take the full spacetime to be $\AdS_3$, with conformal boundary the Einstein static universe
$S^1\times \R$, and the state to be, for example,
the CFT ground state (or any other state that produces negligible back-reaction on the geometry).      Then a  time-reversal symmetric  initial value surface 
is the disc\footnote{$\S$ is a disc topologically, but its metric is that of Euclidean AdS$_2$.} $\S$ of fig. \ref{RTpic2}; its conformal boundary is a circle $\S'$.  To illustrate the RT formula, we decompose $\S'$ as the union of complementary closed intervals $\A$, $\B$ (by which
we mean intervals that share common endpoints but are otherwise disjoint).  For $d=2$, $\Sigma_\A$ has dimension
 $d-1=1$, so it is an embedded one-manifold and its ``area'' is really a
(renormalized) length.  An embedded one-manifold of minimal length is a geodesic.  So in fact, $\Sigma_\A$ is just a geodesic in $\S$  that connects the endpoints of $\A$, as shown in the figure.
In this particular example, the geodesic in $\S$ that connects the endpoints of $\A\subset \S'$ is unique.  Since it is unique, it is 
 the RT surface $\Sigma_\A$. The union of   $\A$ and $\Sigma_\A$ is the boundary
of a region in $\S$, so the homology constraint is satisfied.

In this simple example, as  the combined system $\A\B$
is in a pure state -- namely the ground state of the CFT -- we expect $S_\A=S_\B$.   From the point of view of the RT formula, this is true because $\A$ and $\B$
have the same RT surface.   Indeed, $\A$ and $\B$ have the same endpoints, so the geodesic $\Sigma_\A$ connecting the endpoints of $\A$ is trivially the same as the geodesic
$\Sigma_\B$ connecting the endpoints of $\B$.   We denote this common RT surface of $\A$ and $\B$ as $\Sigma_{\A/\B}$.

 \begin{figure}
 \begin{center}
   \includegraphics[width=5.6in]{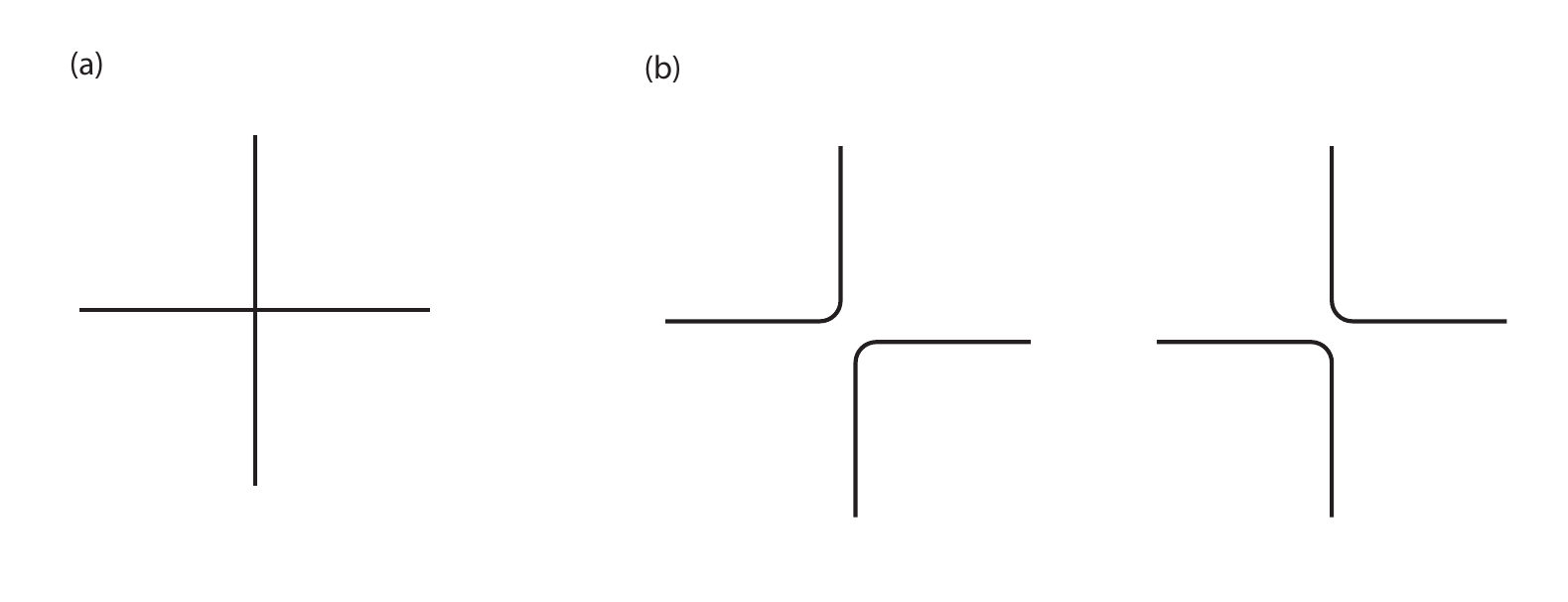}
 \end{center}
\caption{\footnotesize (a) A geodesic that contains a crossing.  (b) Such a crossing can be resolved in two ways.   This does not affect the endpoints of the geodesic,
and one of the resolutions will satisfy the homology constraint.  The resolution shortens the length of the curve.  So a geodesic with a self-crossing is never the RT surface.  \label{resolution}}
\end{figure} 

In this particular example, there was only one geodesic connecting the endpoints of $\A$, so it is inevitably the RT surface $\Sigma_\A$.  In a more complicated example, 
there might be several candidates joining the endpoints of $\A$.  In a negatively curved two-manifold such as $\S$, any geodesic locally minimizes the length, so any such geodesic that satisfies the homology constraint is a candidate RT surface.   In looking for the true RT surface, we can restrict ourselves to embedded geodesics, because a geodesic 
that is not embedded can always be shortened by resolving the crossing (fig. \ref{resolution}), so is never the RT surface.
(An RT surface is not required to be connected, so we do not have to worry about whether resolving the crossing affects  whether the geodesic is connected.)
Of these candidates, the true RT surface $\Sigma_\A$ is the one of least renormalized length.  

 \begin{figure}
 \begin{center}
   \includegraphics[width=5.1in]{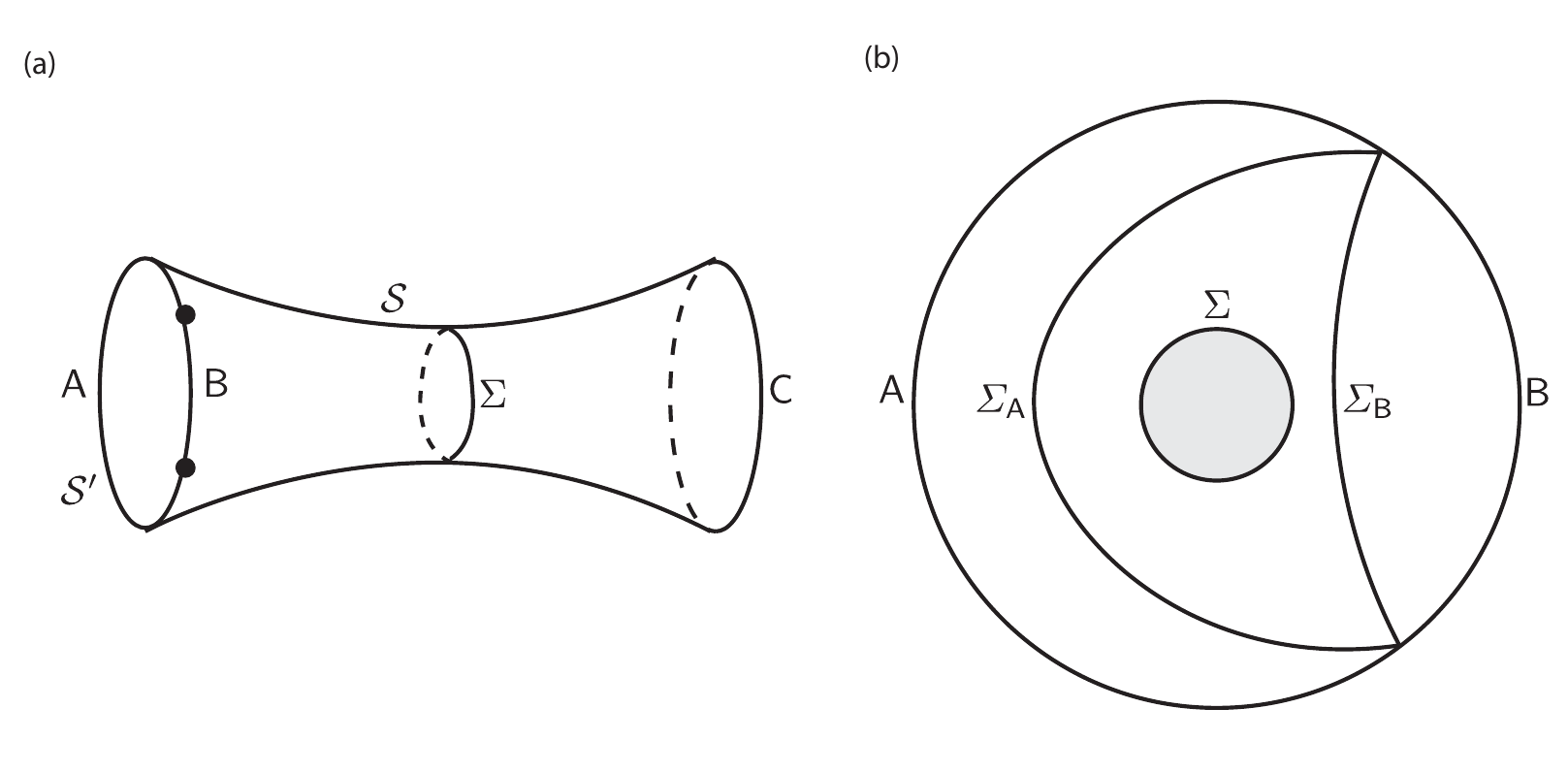}
 \end{center}
\caption{\footnotesize (a) This figure represents a time zero slice $\S$ of a two-sided  AdS-Schwarzschild black hole in dimension $D=3$ (that is, a BTZ black hole).   The conformal boundary now has two components.
One of those components, labeled $\S'$, is decomposed as the union of complementary regions $\A,$ $\B$ as before.   (b)   In applying the RT formula to regions $\A$ and $\B$,
only the part of (a) to the left of the bifurcation surface is relevant, since geodesics with endpoints on $\S'$ will not reach $\Sigma$.   The relevant part of fig. (a) is drawn
here  in a convenient view.  It is a disc with a grayed-out hole in the center, representing the black hole.   The boundary of the hole is the bifurcation surface $\Sigma$.   There are three relevant geodesics in this figure: two of them
are embedded geodesics $\varSigma_\A$ and $\varSigma_\B$
that connect
the endpoints of $\A$ and $\B$ and are homologous respectively to $\A$ and $\B$, and the third is the closed embedded geodesic $\Sigma$.    \label{BTZpic}}
\end{figure} 
It is instructive to see what happens if the combined system $\A\B$ is not in a pure state.  
A small entropy of the combined system will not matter; to get something interesting we should put the combined system in a state with entropy of order $1/G$, which is most naturally
accomplished by introducing  a black hole.
 So  we consider two copies of the Einstein static universe $S^1\times \R$ entangled
in the thermofield double state $\Psi_\HHI$, at some temperature.   The bulk spacetime is then the maximally extended Schwarzschild solution
(in three dimensions, the BTZ black hole with zero angular momentum \cite{BTZ}).
A $t=0$ slice of the black hole, which we call $\S$,  is an Einstein-Rosen bridge or
``wormhole'' connecting the two components of  the conformal boundary
(fig. \ref{BTZpic}(a)).  $\S$ is a surface of constant negative curvature with  $\Sigma$, the bifurcation surface at which various horizons meet, at its center.
The conformal boundary of $\S$ is the disjoint union of two circles, one for each asymptotically AdS world outside the black hole horizon.
We focus on one component of the conformal boundary, which we call $\S'$.  
We divide  $\S'$ as the union of two complementary intervals $\A$ and $\B$.   It is a simple fact of differential geometry that a geodesic connecting two points on the same component of 
the conformal boundary of $\S$ never reaches $\Sigma$; it remains everywhere on one side of $\Sigma$.  (This can be proved using the constant negative curvature of $\S$.)  Hence to understand candidate RT surfaces of $\A$ or $\B$,
we can restrict attention to the left half of fig. \ref{BTZpic}(a), which is drawn in a convenient way in fig. \ref{BTZpic}(b).  Depicted in this figure is an annulus; the outer boundary
of the annulus is the circle $\S'=\A\cup \B$, and the inner boundary is the surface $\Sigma$, which is the intersection of $\S$ with the horizon.   
The ``hole'' at the center of the annulus represents the black hole.

In fig. \ref{BTZpic}(b), any two points on the boundary are joined by precisely two (connected) embedded geodesics, labeled in the figure as $\varSigma_\A$ and $\varSigma_\B$. (There
are also infinitely many self-intersecting geodesics with the same endpoints that wrap around the hole any number of times.) We
call these $\varSigma_\A$ and $\varSigma_\B$ as we do not yet know if they are the true RT surfaces $\Sigma_\A$ and $\Sigma_\B$.  $\varSigma_\A$  goes around the black hole to the left and 
$\varSigma_\B$  goes around it to the right.  There is just one more embedded geodesic, namely the horizon $\Sigma$.
Since these are the only three embedded geodesics, a candidate RT surface must be built from them.   The candidate RT surfaces for $\A$ are $\varSigma_\A$ and the
disjoint union $\varSigma_B\cup \Sigma$, which is also  homologous to $\A$.   Similarly the candidate RT surfaces for $\B$ are $\varSigma_\B$ and $\varSigma_\A\cup \Sigma$.
Note that $\varSigma_\B$ is not a candidate RT surface for $\A$ as it is not homologous to $\A$; similarly $\varSigma_\A$ is not a candidate for $\B$.

If $\A$ and $\B$ are of approximately equal size, then
$\varSigma_\A$ and $\varSigma_\B$ are indeed the true RT surfaces, because $\varSigma_\A$ and $\varSigma_\B$ have almost the same renormalized length,
so the renormalized length of $\varSigma_\A$ is less than that of the other candidate $\varSigma_\B\cup \Sigma$.
   Note that if $\A$ and $\B$ are not of the same size, then $\varSigma_\A$ and $\varSigma_\B$ are of unequal
lengths and therefore,   
 the RT formula implies that $S_\A\not= S_\B$.   It is no surprise that $S_\A$ and $S_\B$ can be unequal, as the presence of the black hole means that
the combined system $\A\B$ is in a thermal state, not a pure state.

Now let  $\C$ be  the second boundary  of $\S$ at the right of fig. \ref{BTZpic}(a).  Since $\C$ has no boundary, its RT surface will be an embedded closed geodesic that is homologous to $\C$.
 There is only one candidate, namely the horizon $\Sigma$.   So $\Sigma=\Sigma_\C$ is the RT surface of $\C$
and therefore $S_\C$  is just the Bekenstein-Hawking entropy of the black hole.
 In fact, the RT formula was constructed to incorporate this expectation.

In this situation, $\Psi_\HHI$ is a pure state of the combined system $\A\B\C$.  Hence $S_\A=S_{\B\C}$,
and subadditivity of entropy $S_\B+S_\C\geq S_{\B\C}$ tells us that
\be\label{junco} S_A\leq S_\B+S_\C. \ee 
If $\Sigma_\A$ and $\Sigma_\B$ are the RT surfaces of $\A$ and $\B$, then the inequality (\ref{junco}) implies that 
\be\label{numbox}A(\Sigma_\A)\leq A(\Sigma_\B)+A(\Sigma_\C)=A(\Sigma_\B)+A(\Sigma).\ee
This inequality is satisfied if $\A$ and $\B$ are of approximately equal size, but it is violated if  $\A$ is much bigger than $\B$.   
In that case, the true RT surfaces of $\A$ and $\B$ are  $\Sigma_\A=\varSigma_\B\cup \Sigma$ and $\Sigma_\B=\varSigma_B$.
Clearly the inequality (\ref{numbox}) is then saturated.

Going back to fig. \ref{RTpic2}, instead of discussing the von Neumann entropy of $\A$ or $\B$, let us discuss the von Neumann entropy of the whole boundary $\S'$.
Since the whole system is in a pure state, the entropy of $\S'$ should vanish.
   To get that answer from the RT formula,
the RT surface $\Sigma_\S'$ must be empty.   Indeed, the empty surface is an allowed RT surface in this example because (1) $\S'$ has no boundary, so the condition that
the boundary of $\Sigma_{\S'}$ coincides with the boundary of $\S'$ allows $\Sigma_{\S'}$  to be empty, and (2) $\S'$ is the boundary of a two-manifold in $\S$ (namely $\S$ itself),
so it is homologous to the empty set, and therefore the empty set satisfies the homology condition that an RT surface is supposed to satisfy.
   By contrast in fig. \ref{BTZpic}(a), $\C$ has no boundary, but it is not homologous to the empty set, so the empty surface is not
a candidate RT surface for $\C$.   This is as expected as in that example, the quantum state restricted to $\C$ is thermal, with a positive von Neumann entropy.

These examples have illustrated a few important points: the RT surface of a region must be homologous to that region; the empty set is allowed as a possible RT surface;
and as parameters are varied, there can be a phase transition in the location of the RT surface.   We will describe another two illustrative examples.  

 \begin{figure}
 \begin{center}
   \includegraphics[width=5.1in]{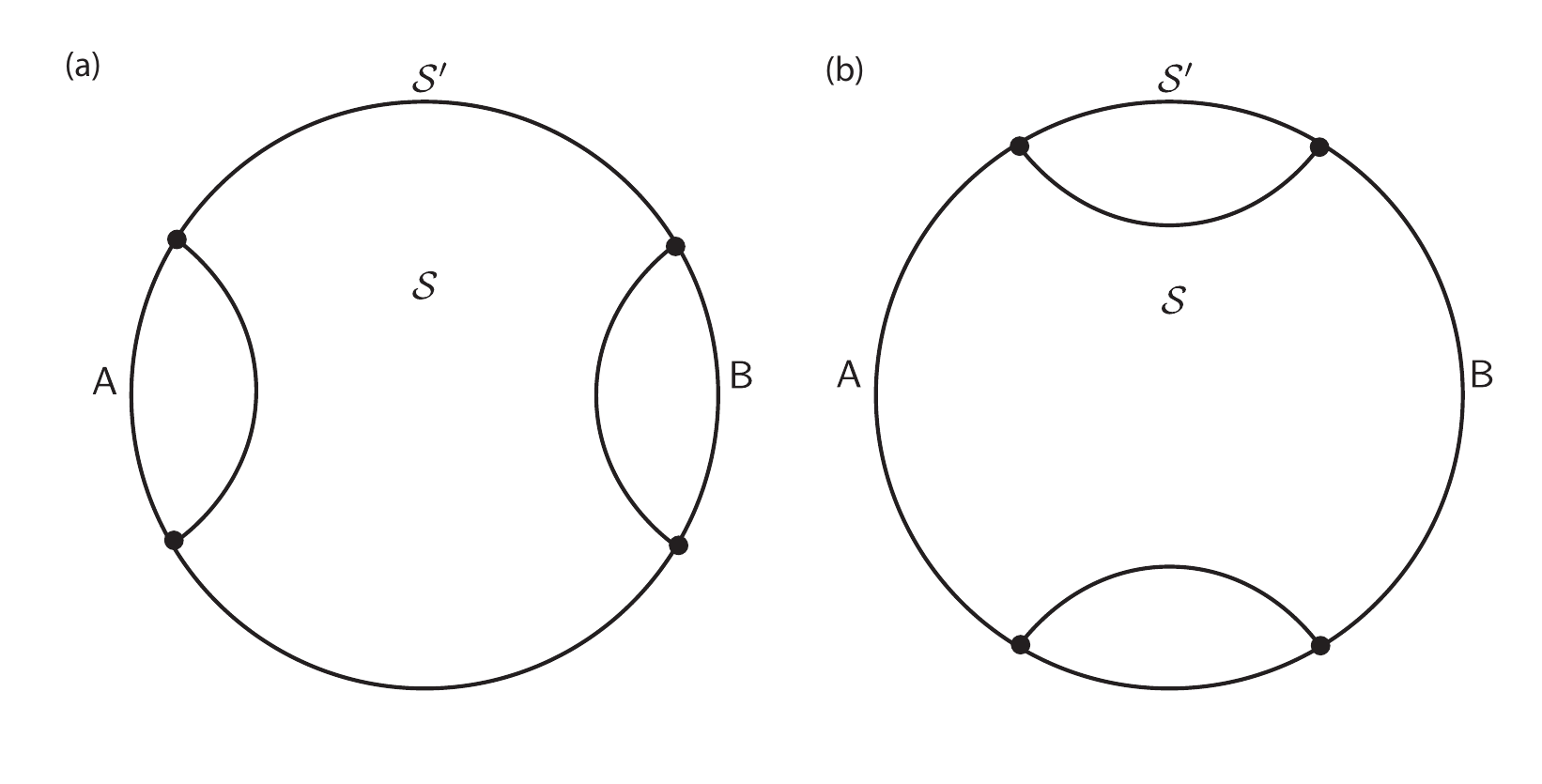}
 \end{center}
\caption{\footnotesize (a) $\S$ is an initial value surface in $\AdS_3$.  $\A$ and $\B$ are disjoint intervals in the conformal boundary $\S'$.
If $\A$ and $\B$ are sufficiently small compared to their separation, then the RT surface of the union $\A\cup \B$ is just the union of the separate RT surfaces of $\A$ and $\B$.
The mutual information $I(\A:\B)$ vanishes in leading order $1/G$.
(b) If instead $\A$ and $\B$ are sufficiently large compared to their separation, a quite different candidate RT surface has a smaller renormalized length.
In this case, the mutual information $I(\A:\B)$ is nonvanishing in leading order.    \label{TwoIntervals}}
\end{figure} 

 \begin{figure}
 \begin{center}
   \includegraphics[width=3.1in]{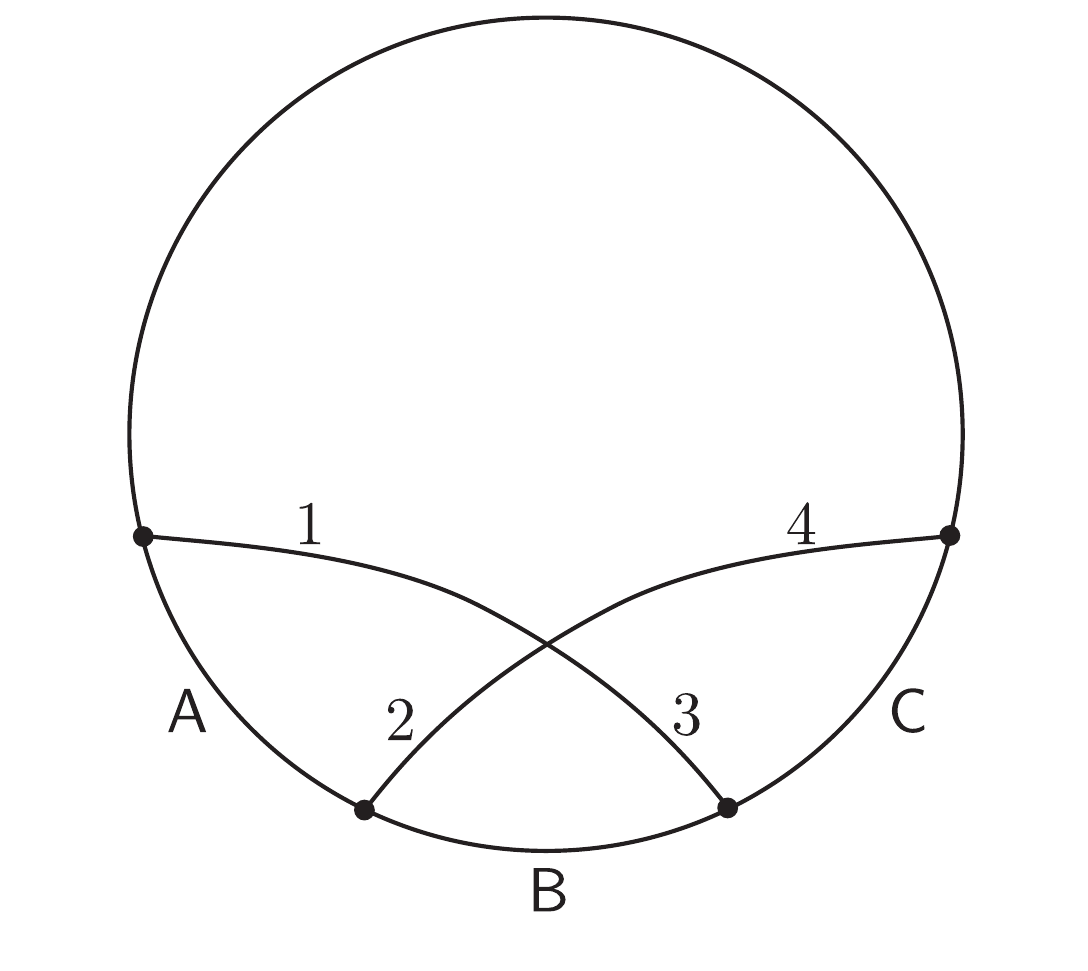}
 \end{center}
\caption{\footnotesize Three regions $\A,\B,\C$ in the conformal boundary of an initial value surface in $\AdS_3$.   The RT surface of $\A\B$ is the union of the segments labeled 1 and 3;
the RT surface of  $\B\C$ is the union of segments labeled 2 and 4.   The union of 2 and 3 is a surface whose endpoints comprise the boundary of $\B$, so its  renormalized length
exceeds that of the RT surface of $\B$; similarly the union of 1 and 4 has a renormalized length that exceeds that of the RT surface of the combined region $\A\B\C$.  So
the RT formula predicts the inquality $S_{\A\B}+S_{\B\C}\geq S_\B+S_{\A\B\C}$ of strong subadditivity.  \label{SSA}}
\end{figure} 

In fig. \ref{TwoIntervals}, we return to the case that the bulk manifold is AdS$_3$ with $t=0$ slice $\S$,  but now we consider two {\it disjoint} intervals $\A$, $\B$ in the conformal
boundary $S'$ of $\S$ \cite{Headrick}.  We already know that the RT surface $\Sigma_\A$ of  $\A$ is the geodesic connecting the endpoints of $\A$, and $\Sigma_\B$ is described similarly.
But what is the RT surface of $\A\B$, the disjoint union of the two intervals $\A$ and $\B$?  One candidate RT surface is the disjoint union $\Sigma_\A\cup\Sigma_\B$; this is
locally area-minimizing, and it is homologous to the disjoint union of $\A$ and $\B$.  That is actually the true RT surface if the intervals $\A$ and $\B$ are small enough, as in fig.
\ref{TwoIntervals}(a).
When that is the case, the inequality of subadditivity of entropy  is saturated in order $1/G$:
\be\label{melox} S_\A+S_\B=S_{\A\B}+\O(1). \ee
Equivalently, in this regime, the mutual information $I(\A:\B)=S_\A+S_\B-S_{\A\B}$ vanishes in order $1/G$.   However, if $\A$ and $\B$ are sufficiently large, a different candidate RT surface is shorter
(fig. \ref{TwoIntervals}(b)), and $I(\A:\B)$ is nonzero in order $1/G$.   This gives another interesting example of a phase transition in the location of the RT surface.

For a final example, we note that the rather deep inequality of strong subadditivity of entropy has a simple geometrical explanation in the context of the RT formula, at least in the time-symmetric
case.
This is briefly explained in fig. \ref{SSA}.  For a fuller discussion in a general $D$-dimensional context, see \cite{TakHead}.  

If the spacetime is not assumed to be invariant under time-reversal,  then one must define the RT surface not as a minimal area surface in space that satisfies a homology
constraint, but as an extremal surface in spacetime that satisfies such a constraint  and has the least area among all such surfaces 
\cite{HRT}.  It is still possible to deduce strong subadditivity in this more general context \cite{Wall2}.
The necessary argument is  much more involved and depends on subtle properties of the Einstein equations.   

Up to this point, we have ignored the fact that these entropies and areas are all divergent.   Indeed, one of the original arguments for the RT formula \cite{RT}
was that it correctly reproduces the expected ultraviolet divergence of the entanglement entropy of a region in quantum field theory. We recall that for a general region $\A$ in a $d$-dimensional
quantum field theory defined on a manifold $Y$, the entropy $S_\A$ has a leading ultraviolet divergence proportional to the area of $\partial\A$:
\be\label{divent}S_\A=\begin{cases} k \frac{A(\partial \A)}{\varepsilon^{d-2}}+\dots & d>2\cr
                                                    k\log\frac{1}{\varepsilon}+\cdots & d=2,\end{cases} \ee
where $\varepsilon$ is a short distance cutoff, $k$ is a constant that depends on the theory, the omitted terms are less singular as $\varepsilon\to 0$, and 
the behavior for $d=2$ is familiar from  section \ref{sample}.
This formula is valid in any quantum field theory, conformally invariant or not.   In the conformally invariant case, the formula for $S_\A$ is not invariant under Weyl transformations of
the metric on $Y$.   Indeed, there is no Weyl-invariant notion of the ``area'' $A(\partial \A)$ of the $d-2$-manifold $\partial \A$.   To define a ``constant'' cutoff $\varepsilon$, one has to fix a Weyl 
frame on $Y$ -- that is, one has to endow $Y$ with a Riemannian metric $h$, not just a conformal class of such metrics -- and then the area is well-defined.
All this is in perfect parallel with  the behavior of the RT formula.
We recall that the metric of an asymptotically locally AdS manifold $X$ of dimension $D=d+1$  near its conformal boundary $Y$
has the form
\be\label{confb}\d s^2=\frac{1}{z^2}\left(\d z^2+\sum_{i,j=1}^d  g_{ij}(x)\d x^i\d x^j\right) ,\ee
where $z>0$ in $X$ and $z=0$ on the conformal boundary $Y$.   The function $z$ is only uniquely defined up to multiplying by a positive function on $Y$.   A particular choice of
the function $z$ endows $Y$ with the Riemannian metric $g$ (and not just a conformal class of metrics) and also makes it possible to define a cutoff on the areas by restricting
to $z\geq \varepsilon$ for some $\varepsilon>0$.  
The metric on $Y$ can be restricted to  $\partial\A$, giving a  Riemannian metric that we will call $h$.     If $\Sigma_\A\subset X$ is an extremal surface that is asymptotic at infinity to $\partial \A$, then the metric of $\Sigma_\A$ looks near its conformal boundary   like
\be\label{nonfb} \d s^2 =\frac{1}{z^2}\left(\d z^2+\sum_{i,j=1}^{d-2} h_{ij}(y)\d y^i\d y^j\right),  \ee where $y^1,\dots,y^{d-2}$ are local coordinates on $\partial\SIgma_\A$. 
The leading behavior, for small $\varepsilon$, of the contribution of the  region $z\geq \varepsilon$ to the ``area''  of $\Sigma_\A$ (in the $D-2$-dimensional sense) is then
\be\label{fonb} A(\Sigma_\A)\sim \int_{\varepsilon} \frac{\d z}{z^{d-1}} \int_{\partial\A}\d^{d-2} y \sqrt{\det h}\sim \begin{cases} \frac{A(\partial\A)}{(d-2)\varepsilon^{d-2}} & d>2 \cr
  A(\partial\A)\log\frac{1}{\varepsilon} & d=2. \end{cases}\ee
 All this   matches the CFT answer (\ref{divent}).  
 
  A noteworthy fact,  surprising but characteristic of the AdS/CFT correspondence, is that although the
 same leading divergence arises in the boundary field theory or the bulk gravitational theory, the interpretation is different. 
 In the boundary  field theory, the divergence in the entanglement entropy $S_\A$ is an ultraviolet effect, coming from modes of short wavelength supported near $\partial\A$, while
 in the bulk gravitational theory, the divergence in the RT formula for the entanglement entropy is an infrared effect, coming from the divergence of distances, areas,  and volumes
 as $z\to 0$.
 
 We have here considered only the leading divergence in the entanglement entropy.  But in fact the parallelism between the predictions of field theory on the boundary and the RT
 formula in the bulk continues
 for subleading divergences, which in general can be matched between the boundary and bulk descriptions. 
 For $d=2$, the entanglement entropy has  only the logarithmic divergence that we have analyzed, but for $d>2$, in general the expansion of $S_\A$ for small $\varepsilon$
 has a succession of less singular terms proportional to lower powers of $\frac{1}{\varepsilon}$,  with a $\log\frac{1}{\varepsilon}$ term at the end for even $d$.  (In some cases, depending
 on the dimensions of relevant operators in a particular theory, there can be a logarithm also for odd $d$.)
    When there is no logarithmic term, it is possible,
 by subtracting the power law divergences, to define a Weyl-invariant regularized version of the entanglement entropy.   When there is a logarithm, the finite part of the entanglement
 entropy has a Weyl anomaly, as explained for $d=2$ at the end of section \ref{sample}.  Divergences  and conformal anomalies in areas or volumes of extremal submanifolds asymptotic to
 the boundary of an asymptotically locally AdS manifold were originally studied with a different motivation \cite{KS,GW}.

\subsection{Derivation}\label{just}

Finally we will explain a path integral argument for  the RT formula, for a large class of states \cite{LM}.  See also section 2 of \cite{malda2} for another exposition, and \cite{Fursaev}  for an earlier attempt with  some of the ideas.

We return to the two-sided situation considered in section \ref{hm}, but now, instead of considering two copies of a CFT entangled in the thermofield double state,
we will consider two copies entangled in a more general state.  The aim is to argue that the RT formula computes the entanglement entropy between the two copies,
in this more general situation.  (After explaining this argument, we will briefly sketch a similar argument in the original setting of the RT formula 
where  one aims to compute the entropy of a region $\A$ in a single copy of the CFT.)
In these arguments, we will formulate the CFT on a $d$-manifold $Y$
that is static, meaning that it is a product $\S'\times \R$, where $\S'$ is a manifold of dimension $d-1$, $\R$ parametrizes the time, and the metric on $\S'\times \R$ 
is a simple product metric.   If $\S'=S^{d-1}$, then $\S'\times \R$ is the Einstein static universe, but we do not restrict to that case.

A first point to notice is that  there are  many CFT density matrices that can be conveniently studied via AdS/CFT duality.
The most familiar one is the thermal density matrix
\be\label{utt} \rho_\th =\frac{e^{-\beta H}}{Z_\th}, \ee
where $H$ is the Hamiltonian of the CFT formulated on the spatial manifold $\S'$.  As usual, $\beta$ is the inverse temperature and $Z_\th$ is the thermal partition function
computed in the boundary CFT.   Expectations of operators in the mixed state $\rho_\th$ can, of course, be computed directly in the CFT, but in AdS/CFT duality,
they can also be computed by  a Euclidean path integral over
manifolds $X$ that are asymptotically locally AdS with $\S'\times S^1_\beta$ for conformal boundary.   All such manifolds $X$ should be included, though frequently there is
one such manifold that dominates for small $G$.     Operator insertions in the CFT path integral are reflected in the gravitational description by the choice of boundary conditions
along $Y$.

\begin{figure}
 \begin{center}
   \includegraphics[width=6.1in]{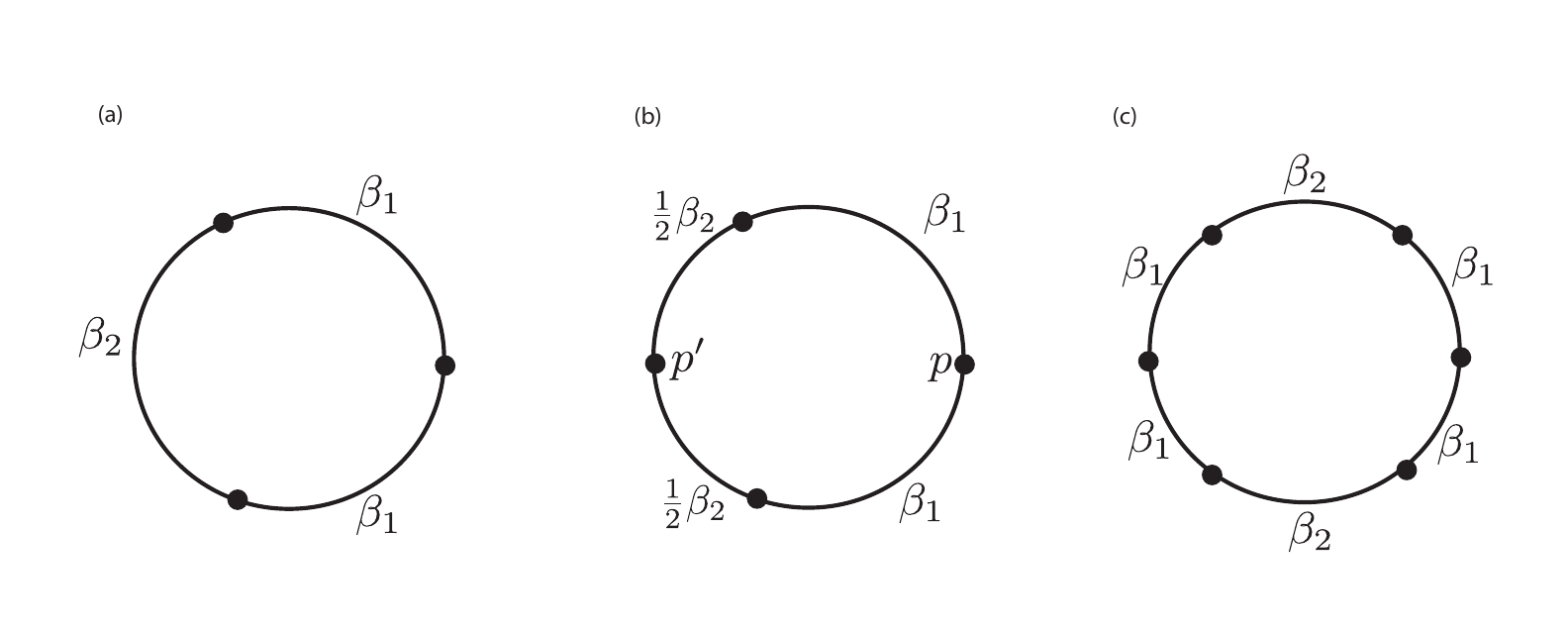}
 \end{center}
\caption{\footnotesize (a)  To study the density matrix $\rho=\frac{1}{Z} e^{-\beta_1 H} e^{-\beta_2 \t H} e^{-\beta_1H}$, one performs a path integral in the boundary CFT on
$\S'\times S^1_{2\beta_1+\beta_2}$   with boundary conditions that are not invariant under rotations of $S^1_{2\beta_1+\beta_2}$; rather 
$S^1_{2\beta_1+\beta_2}$ is divided into successive intervals of lengths $\beta_1$, $\beta_2$, and $\beta_1$  with boundary conditions set by $H$, $\t H$, and  again $H$.
In AdS/CFT duality, the  entropy is nonzero in order $1/G$ if this circle is contractible in the bulk manifold that dominates the gravitational description.
 (b)  If $H$ and $\t H$ are time-reversal invariant Hamiltonians, then the construction 
has a time-reversal symmetry that acts as a reflection on $S^1_{2\beta_1+\beta_2}$, leaving fixed two points  $p$ and $p'$, as sketched here.
(c) To compute $\Tr\,\rho^n$, we replace the circle $S^1_\beta$ by an $n$-fold connected cover of itself, repeating the boundary conditions $n$
times, as sketched here for $n=2$.  \label{deriv}}
\end{figure}

However, there are many other density matrices that in principle can be studied similarly in AdS/CFT duality.  
A typical example for illustration is the following.
  Make a relevant deformation of the CFT to get a quantum field theory that is not conformally invariant but is conformally invariant
at short distances.  Let $\t H$ be the Hamiltonian of this theory and define 
\be\label{mutt} \rho=\frac{1}{Z}e^{-\beta_1 H}e^{-\beta_2 \t H} e^{-\beta_1 H},~~~~~Z=\Tr\,e^{-\beta_1 H}e^{-\beta_2 \t H} e^{-\beta_1 H}. \ee
To show that $\rho$ is a density matrix, we need to show that it is positive and self-adjoint.   This is straightforward as
\be\label{ult}\rho=\V\V^\dagger, ~~\V=\frac{e^{-\beta_1 H}e^{-\beta_2 \t H/2}}{\sqrt{Z}}.\ee
AdS/CFT duality gives a gravitational recipe to compute expectation values of operators in the mixed state $\rho$.
The recipe involves a path integral over bulk manifolds $X$ whose conformal boundary is $\S'\times S^1_\beta$, with $\beta=2\beta_1+\beta_2$  (see fig. \ref{deriv}(a)), and with boundary conditions
that account for the modification of the Hamiltonian in the definition of $\rho$ and whatever operator insertions one wishes to make.
This construction has many obvious generalizations.   One can give the Hamiltonian a more general time-dependence, continuous or as in the preceding example
only piecewise continuous, as long as a condition like (\ref{ult}) is available to ensure that $\rho$ is a density matrix.  
Other generalizations are possible; for example, if $\O$ is an operator in the CFT -- possibly a product of local operators at different points in $\S'$ -- one can define
a density matrix
\be\label{zutt}\rho'=\frac{1}{Z'}e^{-\beta_1 H}\O e^{-\beta_2 H}\O^\dagger e^{-\beta_1 H},\ee
where $Z'$ is a normalizing factor.   Such a density matrix and many obvious combinations and extensions of these definitions can be conveniently studied in AdS/CFT duality.

The formula $\rho=\V\V^\dagger$, beyond proving that $\rho$ is a density matrix, means that we can view $\V$ as a purification of the density matrix $\rho$.
In fact, as  in the general discussion of eqns. (\ref{anytrans})-(\ref{nvector}),
the operator $\V$ 
on the CFT Hilbert space $\H$ can be associated
to a highly entangled vector $\Psi_{\V}$ in the tensor 
product $\H\otimes\H$ of two copies of  $\H$.    $\Psi_{V}$ is a  generalization  of the thermofield double state $\Psi_\TFD$: it is a pure state
of a doubled system with  density matrix $\rho$ on the original system.   In contrast to the thermofield double state, here no symmetry
is present, in general, between the two copies of $\H$.   The density matrix for the second copy is  $\t\rho=(\V^\dagger \V)^{\rm{tr}}$, where $\rm{tr}$ denotes the transpose.
This generically differs from $\rho$ though
$\rho$ and $\t\rho$ have the same entropies.

Let $\rho$ be any density matrix, such as the one in eqn. (\ref{mutt}),  that can be described in the CFT by a path integral on $\S'\times S^1_\beta$ for some $\beta$, though with operator insertions that may not be invariant under rotation of $S^1_\beta$.
  We would like to understand a Lorentz signature spacetime
related to $\rho$ in the same way that a thermal density matrix is related, if the temperature is high enough, to the extended Schwarzschild spacetime.
This is most straightforward if the density matrix $\rho$ has a time-reversal symmetry, acting on $\S'\times S^1_\beta$ as a reflection on $S^1_\beta$.  For example, if $H$ and $\t H$ are both time-reversal invariant Hamiltonians, then the density matrix $\rho$ has a time-reversal
symmetry that reverses the order of the three factors in eqn. (\ref{mutt}).   This symmetry acts by a reflection  of $S^1_\beta$ with two fixed points $p$ and $p'$, as illustrated in fig. \ref{deriv}(b).  
So the time-reversal operation acting on $\S'\times S^1_\beta$ has fixed point set $\S'\times (p\cup p')$.

Now, in AdS/CFT duality, consider the gravitational path integral in which we sum over bulk manifolds  that have $\S'\times S^1_\beta$
as their conformal boundary.  If there is a unique manifold $X$ that dominates this sum as  $G\to 0$, the time-reversal symmetry of the conformal boundary
must extend to a time-reversal symmetry of $X$.   Assuming that this is the case,\footnote{The alternative is that rather than a single manifold that dominates
for small $G$, there might be a pair of dominant manifolds exchanged by the time-reversal symmetry of the boundary conditions.  In that case, the time-reversal
symmetry of the density matrix is spontaneously broken as $G\to 0$.}  the time-reversal symmetry
of $X$ will have a fixed point set $\S$ whose conformal boundary will be the fixed point set $\S'\times (p\cup p')$ of the boundary.   Let $\tE$ be a Euclidean time coordinate
that vanishes along $\S$ and is odd under time-reversal.  The  time-reversal symmetry of the Euclidean manifold $X$ ensures that by
defining $t=-\i \tE$ and taking $t$ to be real, we get a real Lorentz signature spacetime $X_L$. (This analytic continuation leads to a real metric by the same reasoning
as  in section \ref{thermofield}, though now we are proceeding in 
reverse, starting in Euclidean signature and continuing to Lorentz signature.)
  We can view $X_L$ as the Lorentz signature spacetime that evolves from 
the initial value surface $\S$.

In the particular case of a thermal density matrix $\rho_\th =\frac{1}{Z}e^{-\beta H}$, the AdS/CFT correspondence gives a convenient bulk description of any power of $\rho_\th$,
such as $\rho_\th^n=\frac{1}{Z^n}e^{-n \beta H}$.   In particular, we can compute $\Tr\, e^{-n\beta H}$ by summing over Euclidean manifolds $X$ with conformal boundary
$\S'\times S^1_{n\beta}$.   It is then straightforward to compute the entropy by differentiating with respect to $n$, as we have done in section \ref{euclidean}.
The basic reason that this is possible is that the boundary conditions on $\S'\times S^1_\beta$ that are associated to the thermal density matrix $\rho_\th$ are local and are invariant
under rotations of $S^1_\beta$.   Hence one can study an arbitrary power of $\rho_\th$ by suitably changing the circumference of the circle.

A more generic density matrix, such as $\rho$ as defined in eqn. (\ref{mutt}),
 can still be described by a path integral on manifolds with conformal boundary
 $\S'\times S^1_\beta$, but now with  boundary conditions that are not invariant under rotations of $S^1_\beta$.  Accordingly, there  is no convenient way, either in the boundary CFT or in the
 bulk gravitational description, to access non-integer powers of $\rho$.   However, it is always straightforward to access positive integer powers of $\rho$.  To do so in the gravitational
 description,
 we just replace  $S^1_\beta$ by an $n$-fold unramified cover of itself.   This cover is a copy of $S^1_{n\beta}$, but with the boundary conditions, whatever they are, repeated
 $n$ times (fig. \ref{deriv}(c)).    A bulk path integral over manifolds asymptotic to $\S'\times S^1_{n\beta}$ gives a recipe in gravity to compute $\Tr\,\rho^n$ and therefore
 the R\'enyi entropy of order $n$.   As in the general discussion of the replica trick in section \ref{sample}, analytic continuation in $n$ then potentially enables one to compute
 the R\'enyi entropies of general $n$ and in particular the von Neumann entropy.

The implications  depend on the nature of the dominant bulk manifolds.
Not much is known in general about what sort of manifold $X$ dominates the gravitational description of a density matrix such as $\rho$.  The case
that is best understood is the case $\S'=S^{d-1}$ with $\rho$ a thermal density matrix.   In that case, as discussed in section \ref{negative}, there are two
different phases: at sufficiently large $\beta$, the dominant manifold is what we called $\AdS_\beta$ or thermal AdS, and at sufficiently small $\beta$, the
Euclidean black hole dominates.   They differ topologically in the following way.  $\AdS_\beta$ is the product of $S^1_\beta$ with another manifold $B$ (a $d$-dimensional ball).
By contrast,  in the Euclidean black hole solution,
the circle $S^1_\beta$ at infinity is contractible in the interior; it is the boundary of the cigar of fig. \ref{cigar}, and accordingly the topology of the Euclidean black hole solution is
different.   Each of these behaviors is possible for a density matrix that is
not thermal.   Indeed, in the case of a density matrix that is sufficiently close to a thermal one -- for instance, the density matrix defined in eqn. (\ref{mutt}) if the Hamiltonian
$\h H$ is a sufficiently small perturbation of $H$ -- we expect that the dominant manifold $X$ will be the same topologically as for a thermal density matrix, but with
a different metric.  It will take a sufficiently large perturbation of the density matrix to cause a jump in the topology of $X$.

Let us first see what result the replica trick gives in a case that is similar to the low temperature phase of a thermal density matrix.   That is, we will
assume that the Euclidean manifold that dominates the gravitational description of the density matrix $\rho$ is\footnote{A weaker topological condition would actually
suffice.   It is enough to know that the $X$ has an $n$-fold unramified cover that restricts on the conformal boundary to $\S'\times S^1_{n\beta}$.
The argument in the text will then show that this $n$-fold cover of $X$ is the manifold $X_n$ that dominates the computation of $\Tr\,\rho^n$.}    $X=B\times S^1_\beta$ for some
$B$.   In that case, there is an obvious candidate for  the manifold that dominates the computation of $\rho^n$, namely $X_n=B\times S^1_{n\beta}$, obtained by taking a connected  $n$-fold cover
of the circle at infinity.   It turns out that this is always the manifold that dominates the computation of $\Tr\,\rho^n$.   For this we can reason as follows.

Let $\rho$ be the density matrix  that is associated to $\S'\times S^1_\beta$ with some boundary conditions.
Then we define
\be\label{pillow}\rho=\frac{\h \rho}{\Tr\,\h\rho} \ee
where $\h\rho$ is defined via a gravitational path integral with the given boundary conditions and no normalizing factor, and  the normalizing
factor $1/\Tr\,\h\rho$   ensures that $\Tr\,\rho=1$.   
Hence
\be\label{illox}\Tr\,\rho^n =\frac{\Tr\,\h\rho^n}{\left(\Tr\,\h\rho\right)^n}. \ee
Suppose that $X=X_{[1]}$ is the bulk manifold that dominates the path integral over manifolds with conformal boundary $\S'\times S^1_\beta$ (with some given
boundary conditions associated to $\rho$), and $X_{[n]}$ dominates the path integral if $S^1_\beta$ is replaced by its $n$-fold cover $S^1_{n\beta}$ (with the boundary
conditions repeated $n$ times).   Then to leading order in $G$, denoting the action of the classical solution $X_{[n]}$ as  $I(X_{[n]})$, we have 
$\Tr\,\h\rho^n=e^{-I(X_{[n]})}$, and therefore from (\ref{illox})
\be\label{zillow} \Tr\,\rho^n=\exp(- I(X_{[n]})+n I(X_{[1]})). \ee
Equivalently, the R\'enyi entropy of order $n$ is
\be\label{illow}R_n(\rho)=\frac{1}{n-1}\left(I(X_{[n]})-n I(X_{[1]})  \right) +\O(1).\ee

Let us consider the $n$-fold cover $X_n$ as a candidate for what might be the dominant manifold $X_{[n]}$.
As $X_{n}$ is locally isomorphic
to $X$ but covers it $n$ times,  its classical action is $I(X_{n})=n I(X)$.   
Therefore, if $X_n$ is the dominant manifold in the computation of $\Tr\,\rho^n$, then eqn (\ref{illow}) tells us that $R_n(\rho)$ vanishes in order $1/G$.
If so, then by analytic continuation, all R\'enyi and von Neumann entropies vanish.
Could there be a more dominant contribution to $\Tr\,\rho^n$?   If there is another manifold that is more dominant than $X_n$, its action must be less
then $I(X_n)=n I(X)$, and in that case, eqn. (\ref{illow}) tells us that $R_n(\rho)<0 $ in order $1/G$.  For sufficiently small $G$, this contradicts the positivity 
of  R\'enyi entropies (eqn. (\ref{renyipos})).   So $X_n$ is the dominant manifold, and  $R_n(\rho)$ vanishes in order $1/G$.   By analytic continuation in $n$, this is true   
for all R\'enyi entropies $R_\alpha(\rho)$ and for the von Neumann entropy.\footnote{More explicitly, since $R_n(\rho)$ is non-increasing as a function of $n$ (eqn. (\ref{enyipos}))
and non-negative, its vanishing in order $1/G$ for any given $n$ implies vanishing for all $n'>n$.}

Now let us discuss the generalization of the high temperature thermal phase.   For sufficiently small $\beta$, as discussed in section \ref{negative}, 
the dominant manifold with $S^{d-1}\times S^1_\beta$ for its conformal boundary is the Euclidean black hole.\footnote{The analysis in section \ref{negative} was
written for $D=4$, $d=3$,  but the behavior is similar in any dimension.}   Topologically, the Euclidean black hole is  $S^{d-1}\times D$, where $D$, topologically
a disc, is the ``cigar'' in fig. \ref{cigar}.   If one replaces $S^{d-1}$ with any other $d-1$-manifold $\S'$, one  expects that for sufficiently small $\beta$, the behavior
is the same: the dominant Euclidean manifold is topologically $\S'\times D$.   The logic in expecting this is that in the limit of high temperatures, the thermal
ensemble of the CFT should not be sensitive to the spatial manifold on which the CFT is formulated.    One can also support this claim by studying the Einstein equations.

If $\rho$ is a density matrix that is not purely thermal, for example if we define $\rho$ by deforming away from a thermal density matrix as in eqn. (\ref{mutt}), the metric of the dominant
Euclidean manifold will change, but at least for a sufficiently
small perturbation, the topology will be the same.   Thus for a whole open set in the space of density matrices, we expect the dominant
manifold $X$ to be topologically $\S'\times D$.  Moreover, for a time-reversal invariant density matrix that is sufficiently close to a thermal one, we expect
the time-reversal symmetry to remain unbroken, so the dominant metric on $X=X_{[1]}$ will possess the time-reversal symmetry of the
boundary.    
   The fixed point set of this time-reversal symmetry will be topologically  $\S=\S'\times \R$, 
   as in the thermal case.  
   Here  $\S$ is the time-reversal invariant Cauchy hypersurface that was introduced earlier
   in constructing the Lorentz signature spacetime $X_L$ (but now we know the topology of $\S$ because the density matrix is close to a thermal one).
     The RT formula says that there should be  a minimum area surface $\Sigma\subset \S$ -- the RT surface -- whose area determines the entropy of $\rho$
in leading order:
\be\label{milgo|} S(\rho)=\frac{A(\Sigma)}{4G}+\O(1). \ee

For any given integer $n$, if $\beta$ is sufficiently small, then $n\beta$ is sufficiently small that the dominant manifold $X_{[n]}$ is still topologically $\S'\times D$, though with a different metric.  
The conformal boundary of $\S'\times D$ is $\S'\times S^1_{n\beta}$.   We recall that generically, the boundary conditions associated to the computation of $\Tr\,\rho^n$ are not
invariant under a general  rotation of $S^1_{n\beta}$, but they are invariant under the replica symmetry -- the group $\Z_n$ generated by a $2\pi/n$ rotation that cyclically permutes
the replicas.   We say that the replica symmetry is unbroken if the $\Z_n$ symmetry of the conformal boundary extends to a $\Z_n$ symmetry of the dominant manifold $X_{[n]}$.
This is certainly true for a thermal density matrix, so it is also true for a density matrix that is sufficiently close to a thermal one.   In the derivation of the RT formula, it is necessary
to assume that the replica symmetry is unbroken.

Now, consider the quotient 
 $X'_{[n]}=X_{[n]}/\Z_n$.  In general, the $\Z_n$ action on $X_{[n]}$ has fixed points.  In the example in which $X_{[n]} =\S'\times D$,   $\Z_n$ acts by a rotation of $D$, with a fixed
 point at the ``origin,''  so
 the fixed point set is a codimension two surface  $\Sigma_{[n]}\subset X_{[n]}$ that is a copy of $\S'$.  
In the quotient $X'_{[n]}$, $\Sigma_{[n]}$ becomes the locus of a conical singularity with opening angle $2\pi/n$.  Since $2\pi/n=2\pi+\varepsilon$ with
 \be\label{mino}\varepsilon=-2\pi\left(1-\frac{1}{n}\right), \ee the corresponding excess angle is $\varepsilon$.
 $X'_{[n]}$ is not a solution of Einstein's equations, because the equations fail along the locus $\Sigma_{[n]}$ of the conical singularity. 
 We can think of this conical singularity as the back-reaction on the geometry of a codimension two ``cosmic brane,'' whose tension is
 such that it produces a conical singularity with precisely  this cone angle.    The conical singularity contributes to the Einstein action,
 as analyzed in section \ref{another}.  Let $I_0(X'_{[n]})$ be the action of $X'_{[n]}$ integrating the Lagrangian density away from the singularity and ignoring the contribution
 of the conical singularity, and let $I(X'_{[n]})$ be the full action including the contribution of the singularity.   The relation between them, according to eqn. (\ref{dozo}) with
the value of $\varepsilon$ given in eqn. (\ref{mino})  and following the computation that led to eqn. (\ref{nofo}), is
 \be\label{dincox} I(X'_{[n]})=I_0(X'_{[n]})-\frac{A(\Sigma_{[n]})}{4G }\left(1-\frac{1}{n}\right) , \ee
 where as usual $A(\Sigma_{[n]})$ is the area of $\Sigma_{[n]}$.   
 On the other hand, the relation beween $I(X_{[n]})$ and $I_0(X'_{[n]})$ is
 \be\label{incox} I(X_{[n]})=n I_0(X'_{[n]}),\ee
 since away from $\Sigma_{[n]}$, $X_{[n]}$ is an unramified $n$-fold cover of $X'_{[n]}$.  
 So from (\ref{illow}), the leading order R\'enyi entropy is
 \be\label{hincox} R_n(\rho)=\frac{1}{n-1}\left( I(X_{[n]})-nI(X_{[1]})\right)=  \frac{n}{n-1}\left( I(X'_{[n]})-I(X_{[1]})\right) +\frac{A(\Sigma_{[n]}  ) }  {4G}. \ee
 
 In this form, analytic continuation in $n$ is possible, by simply varying the tension of the cosmic string or equivalently by varying the assumed cone angle $2\pi/n$.
There is no simple general formula for $R_n(\rho)$, because the metric of $X'_{[n]}$ depends on $n$ in a way that in general is difficult to control.   However,
there is a simple answer in the limit $n\to1$,
because in that limit, the excess angle $\varepsilon$ vanishes and $X'_{[n]}$ converges to the original manifold $X_{[1]}$. 
  As in section \ref{another}, since $X_{[1]}$ is a classical solution, its action is invariant to first order under any deformation
 that satisfies the boundary conditions.   To first order near $n=1$, we can view $X'_{[n]}$ as a first order deformation of $X_{[1]}$, so $I(X'_{[n]})-I(X_{[1]}) $ is of order $(n-1)^2$
 near $n=1$ and does not contribute to $S_\vN(\rho)=\lim_{n\to 1} R_n(\rho)$.   Hence
 \be\label{vnanswer} S_\vN(\rho)=\frac{A(\Sigma)}{4G},\ee
 where $\Sigma\subset X_{[1]}$ is the codimension two surface  that is the limit for $n\to 1$ of the surface $\Sigma_{[n]}\subset X'_{[n]}$.   
 
To recover the RT formula,  we must show that $\Sigma$ is a surface of minimal area in $X=X_{[1]}$.    This was explained in \cite{LM}, but rather than reproduce
 their reasoning in detail, we will explain an analogy that goes back to the work of Einstein, Infeld, and Hoffman (EIH) \cite{EIH}.  Einstein was dissatisfied with postulating that a test
 particle in a gravitational field propagates on a geodesic, and wanted to argue that this follows from the Einstein field equations.   In modern language, EIH considered a small
 black hole propagating in a gravitational field with a much larger radius of curvature.   The whole spacetime is governed by the Einstein field equations; it is not possible
 to make a separate postulate governing how the black hole propagates.   The gravitational field of the small black hole contributes to the geometry and the nonlinear Einstein equations
 determine the full solution including the trajectory of the black hole.   In the limit, however, that the black hole mass goes to zero, the black hole becomes a test particle that no longer
 influences the geometry and it is indeed necessary to give a condition that describes its trajectory.   EIH showed that in that limit,
 the Einstein equations for the spacetime reduce to the condition that the spacetime in which the black hole propagates satisfies the Einstein equations, and the black hole
 propagates on a geodesic in this spacetime.
 
 To possibly make this more intuitive, we can consider  a small body like the Moon (in a gravitational field of very large scale compared to the size of the Moon) rather than
 a black hole.
 Recall first that for a test particle, the geodesic equation is equivalent to the statement that the energy-momentum of the particle is covariantly 
 conserved.   If we consider not an infinitesimal test particle
 but a body like the Moon, its energy-momentum tensor $T_{\mu\nu}$ appears in the Einstein equations:
 \be\label{refo}R_{\mu\nu}-\frac{1}{2}g_{\mu\nu}R=8\pi G T_{\mu\nu}. \ee
 This results in  back-reaction of the Moon on the ambient spacetime geometry.
 Beyond the Einstein equations and the equation of state of the material making up the Moon, there is no need to impose any other equation.  
 The Einstein equations determine the spacetime, including the position of the Moon in it.    Since the Moon is a source for the geometry, there is no way to move the Moon without
 changing the geometry.
   Via the Bianchi identity, the Einstein equations imply
 that the stress tensor of the Moon is conserved, $D^\mu T_{\mu\nu}=0$.   So this need not be postulated separately.   However, in the limit in which the mass of the Moon
vanishes and we ignore its back-reaction on the geometry, we  drop the term $8\pi G T_{\mu\nu}$ from the right hand side of Einstein's equations.   Then Einstein's
 equations  no longer tell us that $D^\mu T_{\mu\nu}=0$ and we do need to impose this separately; that is, in this limit, 
 we need to impose the geodesic equation for the Moon as a separate condition.
 
 The analog here is the following.   As long as $n\not=1$, there is no meaningful condition  on the position of the  locus $\Sigma_{[n]}$ of a conical singularity in $X'_{[n]}$.
 The geometry of $X'_{[n]}$ away from the conical singularity is determined by the Einstein equations.   There is no way to move the singularity without changing the geometry
 and we are not free to change the geometry as it is determined by the Einstein equations.   If one considers the conical singularity to be sourced by a cosmic brane, one would
say that the Einstein equations including this source determine the spacetime geometry and imply that the energy-momentum tensor of the brane is conserved.
  In the limit that $n\to 1$, however, there is no conical singularity, the cosmic brane is not sourcing anything, and one does need to impose the condition that its energy-momentum
  tensor is conserved.   
 Generalizing the geodesic equation for a test particle, the condition of conservation of the stress tensor for a brane in an ambient spacetime is that its worldvolume has extremal area.
 Analogously to the EIH result and its counterpart for the Moon,
 it is shown in \cite{LM} that in the limit $n\to 1$, the Einstein equations for $X'_{[n]}$ go over to the Einstein equations for $X$ together
 with the condition that $\Sigma$ is a surface of extremal area.
 
 We also want to know that $\Sigma$ satisfies the homology constraint: it should be homologous to the conformal boundary of $X$.   This is not entirely clear in general,
 but it is clear in the case of a density matrix that is sufficiently
 close to a thermal one, in the sense that $X$ and $X_{[n]}$ are topologically what they would be in the thermal case.  The initial value
 surface $\S$ is then topologically a product $\S'\times \R$, and $\Sigma$ is topologically $\S'\times q$ (where $q$ is a point in $\R$), which is manifestly homologous to the conformal boundary.

 We have phrased these arguments for the case of a density matrix that is sufficiently close to being thermal that the topology of $X$ and $X_{[n]}$ is known.   However, much of this
 reasoning applies as long as $X_{[n]}$ is such that the replica symmetry is unbroken and the quotient $X'_{[n]}=X_{[n]}/\Z_n$ is topologically $X$ with a conical singularity at the $\Z_n$ fixed points.
 The codimension 2 fixed points comprise the RT surface.  It is not clear whether in general one should expect that $\Z_n$ has additional fixed points of codimension bigger than 2.
 A scaling argument indicates that they do not contribute to the Einstein-Hilbert action, so maybe such fixed points can occur and are not important.  It is not completely clear
 that in general the codimension 2 fixed point set satisfies the homology constraint of the RT formula, though this is certainly true in the almost thermal case.   
 
 This argument  also has an analog for the one-sided case that provided the original setting for the RT formula.   Here as in section \ref{original}, we consider 
  a $d$-dimensional CFT on a spatial manifold $Y$, with Cauchy hypersurface $\S'$, embedded in $Y$ at $t=0$.   Given a normalized quantum state $\Psi$ 
  of the CFT and a region $\A\subset \S'$ with boundary $\partial\A$, we let $\rho_\A$ be the density matrix for observations in the state $\Psi$ in the region $\A$.   
  We would like to compute the von Neumann
  entropy $S_\vN(\rho_\A)$.   As a first step, we consider the replica trick in the CFT to compute R\'enyi entropies $R_n(\rho_\A)$ for integer $n$.   For this, following the logic
  in section \ref{sample}, we replace $Y$ with $Y_{[n]}$, an $n$-fold cover of $Y$ branched over the codimension two manifold $\partial\A$.  The path integral on $Y_{[n]}$
  (with boundary conditions in the far past and far future appropriate to $n$ copies of $\Psi$) computes $\Tr\,\rho_\A^n$ and thus the $n^{th}$ R\'enyi entropy of $\rho_\A$.   
  On the other hand, we can apply AdS/CFT duality to the CFT formulated on $Y_{[n]}$ and thereby get a gravitational recipe to compute $R_n(\rho_\A)$.  In this recipe,
  we are supposed to sum over bulk manifolds  with conformal boundary $Y_{[n]}$. Among these manifolds, let $X_{[n]}$ have minimum action.
   Then AdS/CFT duality says that in leading order 
  \be\label{tracon} \Tr\,\rho_\A^n=\exp(-I(X_{[n]})).\ee
  (The normalizing factor $\exp(n I(X_{[1]})$ in eqn. (\ref{zillow}) is, according to AdS/CFT duality, equal to $(|\Psi|^2)^{-n}$, so it equals 1 if $\Psi$ is normalized.) 
  Now, as before, we assume that $X_{[n]}$ is invariant under the $\Z_{n}$ replica symmetry and we define $X'_{[n]}=X_{[n]}/\Z_n$.   $X'_{[n]}$ will have a conical
  singularity, with cone angle $2\pi/n$, on a codimension two manifold $\Sigma_{[n]}$ whose conformal boundary is $\partial\A$.  The conformal boundary of $\Sigma_{[n]}$ is $\partial \A$
   because 
 the conformal boundary of $X'_{[n]}$ is $Y'_{[n]}=Y_{[n]}/\Z_n$, and $Y'_{[n]}$ has a conical singularity along $\partial\A$.   As in the previous discussion, the formula (\ref{tracon})
 for $\Tr\,\rho_\A^n$ can be analytically continued in $n$ and leads to a simple answer in the limit $n\to 1$.   Following the same logic as before, one finds a formula
 for the von Neumann entropy of $\rho_\A$ in terms of the (renormalized) area of a surface $\Sigma$ that is the limit of $\Sigma_{[n]}$ for $n\to 1$:
 \be\label{zerdo} S_\vN(\rho_\A)=\frac{A(\Sigma)}{4G}. \ee
 The same argument as before shows that $\Sigma$ is a surface of extremal  area.   
 
 In both versions of the argument, in order to complete the derivation of the RT formula, we would like to know that $\Sigma$ has minimal area, not just extremal area.
 This question is vacuous for a density matrix of the full CFT that is sufficiently close to a thermal one, since the Euclidean black hole solution has a unique codimension two surface of
 extremal area (the bifurcation surface), and this extremal surface remains unique after any sufficiently small change in the spacetime. So in such a case an extremal surface  automatically
 has minimum area.   However, in general there can be multiple candidate
 RT surfaces (as we saw in some examples in section \ref{original}), and the extension of the derivation that we have explained to cover this case is subtle and not fully understood in general.  
   Let $X$ be the dominant bulk spacetime associated to the CFT density matrix $\rho$ (in the two-sided case) or to the CFT state $\Psi$ (in the
 one-sided version of the problem).   The candidate RT surfaces are the extremal area surfaces $\Sigma_\mu\subset X$, $\mu=1,\cdots,s$  that satisfy the homology constraint.
 If $\Sigma_\mu $ is any of these surfaces,  we can wrap a cosmic brane on $\Sigma_\mu$, and, after solving for back-reaction,
 we can find, at least for $n$ near 1,  a spacetime $X'_{\mu,n}$ with a conical singularity of cone angle $2\pi/n$ on a surface $\Sigma_{\mu,n}$ (which reduces to $\Sigma_\mu$ as $n\to 1$).   
 Hopefully, $X'_{\mu,n}$  can be continued to integers $n>1$ and
 at such values, hopefully $X'_{\mu,n}$ is a quotient $X_{\mu,n}/\Z_n$, where $X_{\mu,n}$ is a manifold that can contribute to the $n^{th}$ R\'enyi entropy of the relevant
 density matrix.   
 Among the $X_{\mu,n}$, the  one whose action $I(X_{\mu,n})$ is smallest is expected to  dominate  the  computation of the $n^{th}$ R\'enyi entropy.   
At $n=1$, $X_{\mu,n}$ reduces to the original manifold $X$, so its action is independent of $\mu$.  Moreover the derivative of $I(X_{\mu,n})$ with respect to $n$  at $n=1$
is $A(\Sigma_\mu)/4G$.
So for $n$ slightly greater than 1, the minimum of  $I(X_{\mu,n})$  is achieved by minimizing $A(\Sigma_\mu)$.
 Hence under the stated assumptions, the candidate RT surface of least area is indeed the one whose area determines the von Neumann entropy.

 In general, not much is known about constructing candidate manifolds $X_{\mu,n}$ for integer $n>1$ associated to candidate RT surfaces $\Sigma_\mu$.
In a few cases, this has been done.  For instance, in the example shown in fig, \ref{TwoIntervals} with two candidate RT surfaces $\Sigma_\mu$, $\mu=1,2$,
manifolds $X_{\mu,2}$ were explicitly described and studied in \cite{Headrick}, and this has been extended to $X_{\mu,n}$ \cite{BDHM}.

This discussion  of candidate RT surfaces
should alert us  to another subtlety in the derivation of the RT formula.   In general, in the limit $G\to 0$, the $n^{th}$ R\'enyi
entropy can have a phase transition as a function of $n$.   This will happen if the choice of $\mu$ that minimizes 
$I(X_{\mu,n})$ jumps as a function of $n$.   Therefore, it is oversimplified to present the derivation in terms of a family of manifolds $X_{[n]}$ that is analytic in $n$
and that dominate the $n^{th}$ R\'enyi entropy for each integer $n$. Rather, one has to think in terms of, roughly, a family of manifolds $X_{[n]}$ that depends analytically on $n$
 and that  minimizes the action, among all such families, if $n$ is sufficiently close to 1.    It was in part to avoid such issues that at the outset, we presented the derivation of the RT formula
  in terms of density
matrices that are sufficiently close to a thermal one.   For that class of density matrices, the extremal surface is unique and the topology of the manifolds $X_{[n]}$ is known,
so many tricky issues do not arise.

Even if we restrict to density matrices that are sufficiently close to a thermal one,
there is still an inconvenient issue.   In the case of a thermal density matrix,
it  is true that for any given $n>1$,
if $\beta$ is small enough, the $n^{th}$ R\'enyi entropy is dominated by a Euclidean black hole with inverse temperature $n\beta$.   
But for any given $\beta$, this is false if $n$ is too large.   Indeed, the Euclidean black hole solution with inverse temperature $n\beta$  does not
exist if  $n$ is too large, because as we saw in section \ref{negative}, there is a maximum value of the inverse temperature  for this solution to exist.
A possible point of view is to vary both $\beta$ and $n$, computing $I(X_{[n]})$ for those values  of $\beta$ and $n$ for which this makes sense
and then analytically continuing to $n=1$ and the desired value of $\beta$.    The last step will not be problematical, since if the Euclidean black
hole exists at inverse temperature $\beta$, it will also exist at inverse temperature $n\beta$ if $n$ is close enough to 1.

\subsection{Further Developments}\label{pageagain}

We will conclude by indicating, without detailed explanation, a few important further developments involving the RT formula.

\begin{figure}
 \begin{center}
   \includegraphics[width=3.1in]{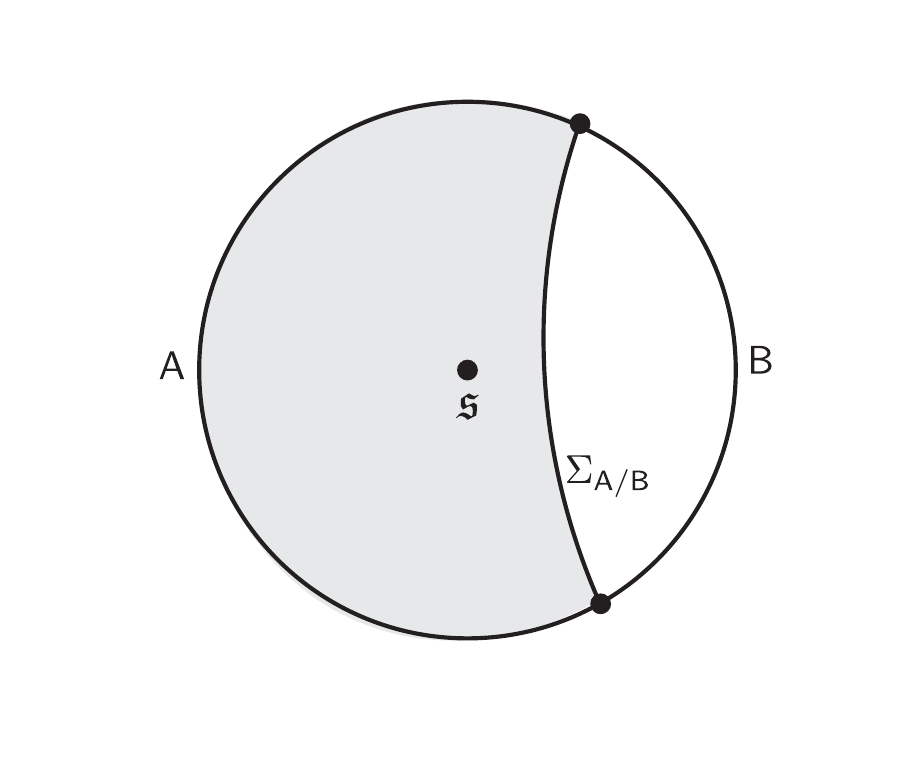}
 \end{center}
\caption{\footnotesize
 The shaded region $\S(\A)$ is bounded by $\A$ and its RT surface $\Sigma_{\A/\B}$; the unshaded region is the corresponding $\S(\B)$.   The entanglement wedge $\E(\A)$
is the domain of dependence of $\S(\A)$; similarly, the entanglement wedge $\E(\B)$ is the domain of dependence of $\S(\B)$. The spin $\mathfrak s$ at the center of the disc is contained in 
$\S(\A)$ but not in $\S(\B)$, so it can be measured by a boundary observer with access only to operators in region $\A$, but not by a boundary observer with access only to
operators in region $\B$.
\label{entanglementwedge}}
\end{figure}

A first question involves quantum corrections to the RT formula \cite{BDHM,FLM}.   Let $Y$ be a spacetime on which a holographic CFT is formulated, and suppose that (for some class of states in the
CFT) $X$ is the
manifold that dominates  the dual bulk description of the CFT on $Y$.   For some spatial region $\A\subset Y$, let $\Sigma_\A\subset X$ be the RT surface.   Let $\S$ be a Cauchy hypersurface
that contains both $\Sigma_\A$ and $\A$.  The homology constraint says that $\Sigma_\A$ and $\A$ are together the boundary of a region $\S_\A\subset \S$ (illustrated in an example in
fig. \ref{entanglementwedge}).  
Let $\rho_\A$ be the CFT density matrix for the region $\A$, and let $\rho_{\S_\A}$ be the density matrix of the bulk quantum fields for the region $\S_\A$.
A refinement of the RT formula with the one-loop quantum correction included is
\be\label{polly} S_\vN(\rho_\A)=\frac{A(\Sigma_\A)}{4G}+S_\vN(\rho_{\S_\A}). \ee
Here $S_\vN(\rho_{\S_\A})$ appears where $S_\out$ appeared  in Bekenstein's original definition (\ref{sgen}) of the generalized entropy, so one can think of the right hand side of eqn. (\ref{polly})
as the generalized entropy of the RT surface (while Bekenstein defined the generalized entropy of the black hole horizon, which in general is a time-dependent thermodynamic entropy, not
a von Neumann entropy, even if one interprets $S_\out$ as the von Neumann entropy of the fields outside the black hole).   
The proof of eqn. (\ref{polly}) is roughly as follows.   The derivation of the RT formula was based on interpreting the R\'enyi entropy $R_n(\rho_\A)$ in terms of the partition function $Z(X_n)$
of the bulk theory on a certain manifold $X_n$.  To get the RT formula, we approximated this partition function as the exponential of minus the classical action.  This led to a sort of classical
replica trick calculation that yielded the RT formula.   Suppose instead that we compute  $Z(X_n)$ more precisely, including the one-loop partition function of the bulk quantum fields.
Repeating the derivation of section \ref{just} but including the one-loop correction, we simply run into the replica trick calculation of $S_\vN(\rho_{\S_\A})$, accounting for the second
term on the right hand side of eqn. (\ref{polly}).

 AdS/CFT duality says that from a knowledge of the quantum state $\Psi$ of the boundary
CFT, one can extract a full knowledge of the bulk quantum state.  But suppose that one has access not to $\Psi$ but only to the density matrix $\rho_\A$ that describes CFT measurements
in the region $\A$ (and therefore also in its domain of dependence $D(\A)$).   What portion of the bulk spacetime is determined by a knowledge of $\rho_\A$?
The quantum-corrected RT formula (\ref{polly}) suggests an answer.   Consider a particle in the bulk spacetime such as a spin $\mathfrak s$  (fig. \ref{entanglementwedge}).   The quantum state of this  particle contributes to $S_\vN(\rho(\S_\A))$, and thereby,
via the formula (\ref{polly}), to $S_\vN(\A)$,  if its worldline passes
through $\S_\A$, and not otherwise.    This suggests that a knowledge of $\rho_\A$ determines the bulk quantum state in the region $\S_\A$ -- or equivalently in its domain of
dependence $D(\S_\A)$ --  and not outside.   The domain of dependence\footnote{The domain of dependence of a subset of an asymptotically Anti de Sitter spacetime was defined
in footnote \ref{dodads}.} of $\S_\A$ is called the {\it entanglement wedge} $\E(\A)$.  The upshot is that a knowledge of the boundary state
in region $\A$ determines the bulk state in the region $\E(\A)$, but not beyond.  Deducing the contents of the bulk region $\E(\A)$ from a knowledge of the density matrix $\rho_\A$ of the boundary 
region $\A$ is called
entanglement wedge reconstruction \cite{CKNR,HHLR,JLMS,DHW}.

We recall that a candidate RT surface for a given region $\A$ is defined as a surface of extremal area that satisfies certain topological conditions (it has the same boundary as $\A$ and
is homologous to $\A$).   The formula (\ref{polly}) and its
derivation, however, suggest that  instead of classical extremal surfaces that extremize
the area, we should consider {\it quantum extremal surfaces} that extremize the generalized entropy.   Here the generalized entropy of a surface $\Sigma_\A$ is defined
to be the right hand side of eqn. (\ref{polly}).   In this language, one states the quantum--corrected RT formula as follows.   A candidate quantum extremal surface associated to
some region $\A$ of the conformal boundary is a
surface $\Sigma$ that satisfies the appropriate topological conditions and  extremizes the generalized entropy.
An improved formula for $S_\vN(\rho_\A)$ is then proposed to be the generalized entropy of the quantum extremal surface that has the smallest generalized entropy \cite{NEW}.
The heuristic idea behind this is the following.  The derivation in section \ref{just} shows that the $A/4G$ term in the classical RT formula arises from minimizing the action of a certain
solution with replicated boundary conditions.  Instead of minimizing the action, it would be a better approximation to minimize the effective action, including the one-loop quantum correction.
But this has the effect of replacing $A/4G$ with the generalized entropy.  So instead of considering minimal surfaces that minimize $A/4G$, we should consider quantum minimal
surfaces that minimize the generalized entropy.   

At first sight, one might expect that to the extent that semiclassical reasoning is valid, the $S_\out$ term in the generalized entropy, which is of order $G^0$, would be negligible compared
to the area term, which is of order $G^{-1}$.   If so, it would not matter (except near an entanglement phase transition at which two candidate RT surfaces exchange dominance)
whether we minimize the area or the generalized entropy.   However, that is not the thole story.
It is possible for the correction term $S_\vN(\rho_{\S_\A})$ in the generalized entropy formula eqn. (\ref{polly}) to be large,
for example in the case of a black hole that has been emitting Hawking radiation for a long time.   Then the difference between classical and quantum versions of the RT formula can be
important.

Once one identifies the entropy of a boundary region $\A$ as the generalized entropy of a quantum extremal surface of minimum generalized entropy,
the possibility arises that a phase transition will occur in the entropy of a region if two quantum extremal surfaces exchange dominance.  In our discussion of the Page
curve, we encountered a phenomenon reminiscent of a phase transition: the entropy of an evaporating black hole is claimed to vary as a function of time in a way that
becomes nonanalytic at the Page time as $G\to 0$
 (see fig. \ref{pagecurve}).   This has been interpreted as resulting from a phase transition in the quantum extremal  surface \cite{P,AEMM}.  Roughly
speaking, prior to the Page time, the entropy of an evaporating black hole is described in terms of an empty quantum extremal surface, and after the Page time it
is described in terms of a quantum extremal surface located near the black hole horizon.   This is perhaps the most significant recent result about black hole thermodynamics, but  explaining it here would take us too far afield.   See \cite{malda,malda2} for
expositions.

Finally, we will briefly discuss the implications of entanglement wedge reconstruction for the nature of the holographic map from bulk degrees of freedom to boundary degrees of freedom.
As a motivating example, we return to the case of  $\AdS_3$ with an initial value surface $\S$ that is a hyperbolic disc.    Consider a state that is fairly close to the $\AdS_3$ ground state, but with a spin $1/2$ particle $\mathfrak s$ in the center of $\S$, and  possibly some other particles distributed throughout  $\S$. 
Let $\O$ be the operator that measures the spin of particle $\mathfrak s$ in some chosen direction.   This operator is not defined on all states of the theory -- the particle $\mathfrak s$
may be absent, or the metric or topology of $\S$ may be so different from what is assumed in fig. \ref{RTpic2} that $\S$  may have no well-defined ``center.''  But on a suitable subspace $\H_0$ of bulk states, the operator $\O$ is meaningful, at least semiclassically
(that is, for small $G$, or equivalently for  large $N$ in the boundary CFT).
 AdS/CFT duality says that the CFT Hilbert space $\H_\CFT$ has a subspace $\H_{0,\CFT}$ that describes the bulk states $\H_0$, and there is a CFT operator $\O_\CFT$
such that a CFT measurement of $\O_\CFT$ is equivalent to a bulk measurement of $\O$.   (It does not matter how we define $\O_\CFT$ on states orthogonal to $\H_{0,\CFT}$;
we can simply define it to annihilate those states.)   Thus, conditional on knowing that the system is in the subspace $\H_0$ in which the question makes sense, an observer with
access to the CFT on the conformal boundary of spacetime can measure the bulk spin $\mathfrak s$ by measuring the operator $\O_\CFT$.

But what about an observer who has access to only part of the conformal boundary of spacetime?   For example, in fig. \ref{entanglementwedge}, consider an observer who has access only
to operators that can be defined in  region $\A$ (or equivalently in the domain of dependence $D(\A)$). 
Entanglement wedge reconstruction asserts that such an observer can measure the spin of particle $\mathfrak s$ if and only if it passes through the corresponding bulk region $\S(\A)$ or
equivalently
its domain of dependence $\E(\A)$.     From the figure, we see that, with $\A$ comprising more than half of the boundary of the disc and $\B$ comprising
less than half, the center of the disc is contained in $\S(\A)$ and not in $\S(\B)$.   So in this particular example, the prediction is that an observer with access only to region $\A$
of the conformal boundary can measure the spin of particle $\mathfrak s$, and an observer with access only to region $\B$ cannot.   Obviously, a rotation of the diagram  would
change nothing essential.   The general statement, for a single interval, is that the CFT restricted to an interval $I$
encodes the information needed to measure spin $\mathfrak s$ if and only if $I$ comprises more than half of the boundary of the disc.

    \begin{figure}
 \begin{center}
   \includegraphics[width=3.61in]{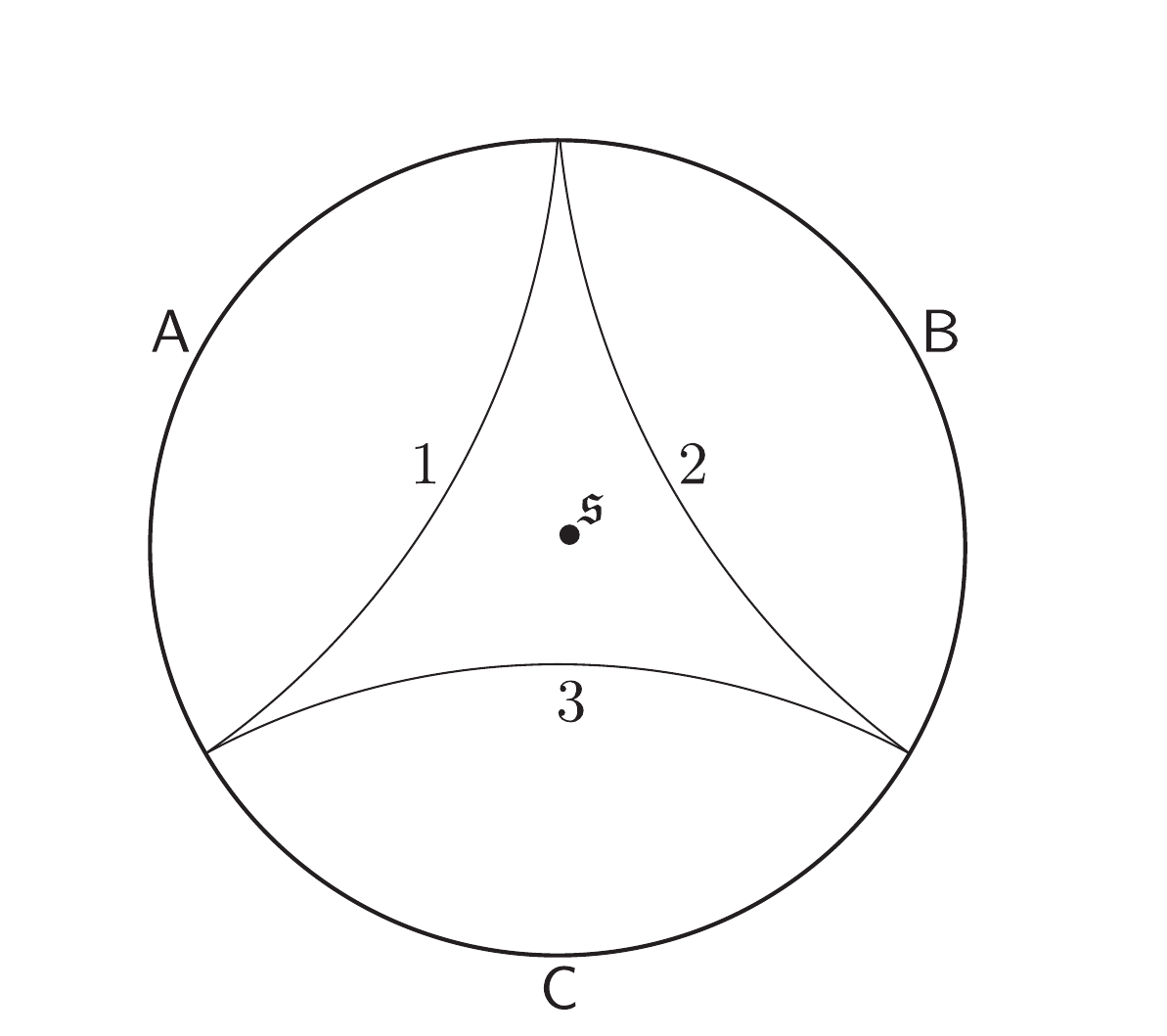}
 \end{center}
\caption{\footnotesize The boundary  of the Cauchy hypersurface $\S$ is divided in three equal parts $\A$, $\B$, $\CC$, disjoint except for endpoints (or with very small overlaps).  The
entanglement wedges of $\A$, $\B$, and $\CC$ are bounded respectively by the geodesics labeled $1,2,$ and $3$.  The resulting entanglement wedges do not
contain the center of the disc and a CFT  observer with access to only one of $\A$, $\B$, or $\CC$ cannot measure the spin $\mathfrak s$ that is located  there.   
However, the entanglement wedges of the joint regions $\A\B$, $\B\CC$, and $\CC\A$ are bounded respectively by the geodesics $3$, $1$, and $2$.  Each of these
entanglement wedges contain the center of the disc, so an observer with access to any  two of the three regions can measure the spin in question.\label{TripleDivision}}
\end{figure} 

This has fascinating implications for the way bulk information is encoded in the AdS/CFT correspondence \cite{ADH,PYHP}.  To illustrate why, consider a simple
case in which the boundary of the disc $\S$ is divided as the union of three intervals $\A,$ $\B$, and $\CC$ of equal size, disjoint except for their boundaries (fig. \ref{TripleDivision}).
Any one of the regions has an entanglement wedge that does not contain the center of the disc, so an observer with access to only one of the three regions cannot measure the
spin of particle $\mathfrak s$.   However, the entanglement wedge of the union of any two of the three regions does contain the center of the disc.   This tells us that bulk information
is stored in the boundary in a way that is {\it distributed} (because two regions together contain information that either region separately would not) and {\it redundant} (losing access
to any one of the three regions does not cause the information to become inaccessible as it is fully encoded in the other two regions).

Classically, such distributed, redundant encoding can be achieved as follows.   Suppose for example that one wishes to store in $\partial \S$ a knowledge of the value of a classical 
$\Z_2$-valued spin $s$.
One way to do this is to pick three elements $a,b,c\in\Z_2$ that are  random except for a condition $a+b+c=s$ mod 2.   (For example, one can pick $a$ and $b$ at random and
pick $c$ so that $a+b+c=s$ mod 2.)   Then one can, say,  store the values of  $a$ and $b$ in region $\A$, the values of $b$ and $c$ in region $\B$, and the values of $c$ and $a$ in
region $\CC$.   Clearly, any one of regions $\A$, $\B$, and $\CC$ does not contain any information about the value of $s$, but any two regions taken together do contain this
information.

Encoding the quantum spin $\mathfrak s$ in the boundary CFT is a more subtle problem than encoding a classical spin $s$.   The reason is that the operators that can be used
to measure $\mathfrak s$ in the bulk do not commute with each other, and the encoding in the boundary CFT must be made in such a way that {\it any} bulk measurement 
is equivalent to some boundary measurement.  This is loosely described by saying that what must be encoded is quantum information, not just classical information.
Distributed, redundant encoding of quantum information is possible  \cite{Shor}, though this is much more subtle than in the classical case.   
In fact, distributed, redundant encoding of information
 would be essential to the functioning
of a hypothetical large scale quantum computer.   The basic reason for this is that any computer, classical or quantum, makes errors, since the components from which it is built are
not perfect.  To do any large scale computation reliably therefore
requires a method to correct errors.     Quantum error correction depends on distributed, redundant encoding of a quantum state.   See for example \cite{PYHP} for a basic
introduction.

\section{What is a White Hole?}\label{whitehole}

    \begin{figure}
 \begin{center}
   \includegraphics[width=5.9in]{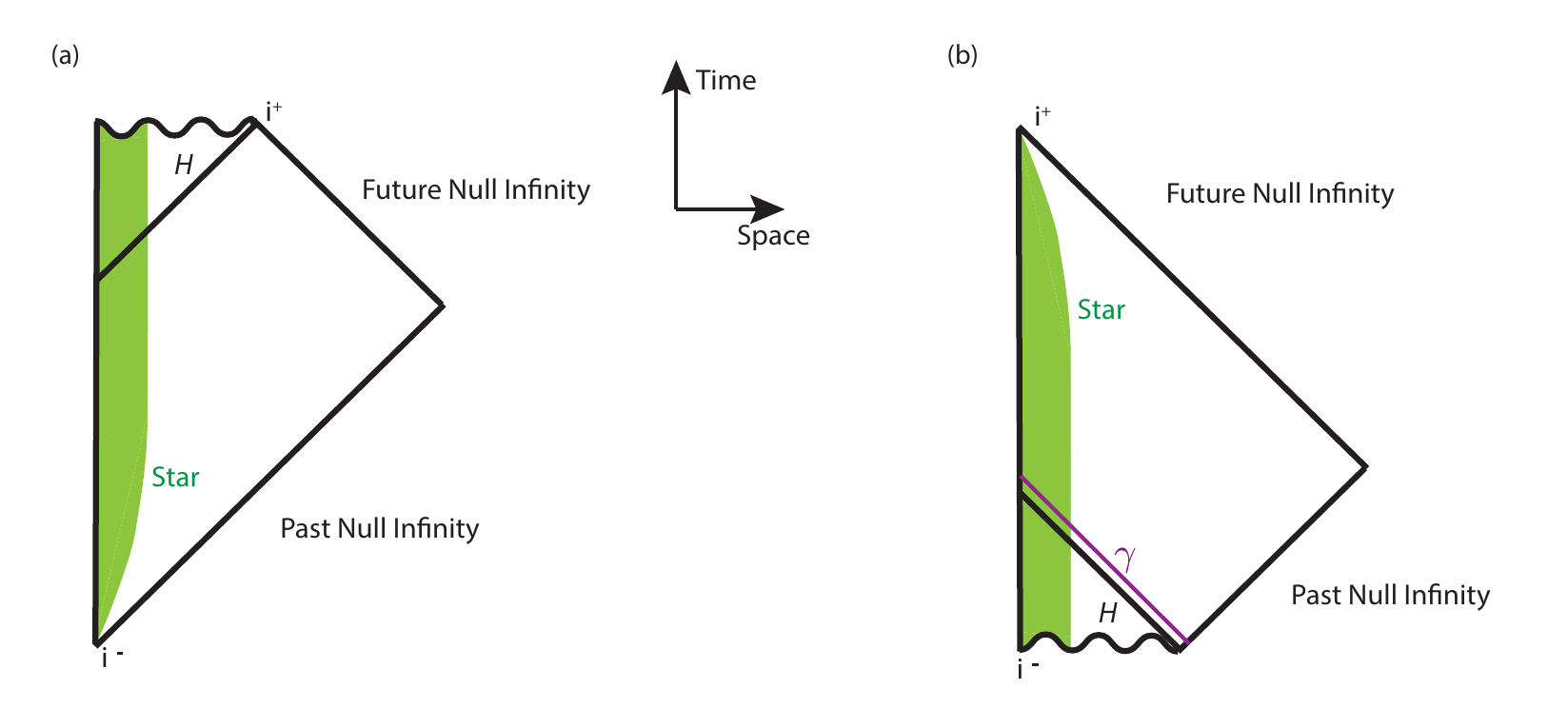}
 \end{center}
\caption{\footnotesize  In (a), a star collapses to form a black hole.  The diagonal black line labeled $H$ is the future horizon of an outside observer.  The wiggly line at the top
represents the black hole singularity. $\i^-$ and $\i^+$ are past and future timelike  infinity for an observer outside the horizon.
Sketched in (b) is the time-reversed spacetime: a white hole spontaneously emits a star.  The diagonal black line labeled $H$ is the past horizon of an outside observer
-- bounding the region that the observer cannot enter or influence. $\i^-$ and $\i^+$ are past and future timelike infinity for an observer outside the horizon.  The diagonal 
line $\gamma$ is the worldline of an infalling massless particle that was injected from a great distance far in the past.   Initially, its energy was of of order the Hawking temperature;
by the time it collides with the star that is emerging from the white hole, its energy is exponentially large.  This exponential blueshift is the time-reversal  of the exponential
redshift on which  Hawking's derivation of black hole evaporation is based.   \label{blackandwhite}}
\end{figure} 

A white hole is the time-reversal of a black hole.    A black hole has a future horizon with the property that from outside the future horizon one cannot see what is behind it.
A white hole has a past horizon with the property that from outside the past horizon one cannot influence what is behind it.   Anything can fall into a black hole, and conversely,
anything can emerge spontaneously from a white hole.    Fig. \ref{blackandwhite}(a) is a Penrose diagram describing the collapse of a star to form a black hole.  In the time-reversed Penrose diagram
of fig.  \ref{blackandwhite}(b), a white hole spontaneously emits a star.  In fact, one might prefer to describe fig. \ref{blackandwhite}(b) by saying that the white hole spontaneously converts
into a star since on any spacelike surface that is sufficiently far in the future, an ordinary star is all that there is.   

A white hole singularity is certainly an example of a naked singularity that can be seen by an observer outside the white hole.   
Penrose's cosmic censorship conjecture states roughly  
that in Einstein's theory a naked singularity never forms to the future of a Cauchy hypersurface.
  Even if this conjecture is true, it has no bearing on the spacetime of fig. \ref{blackandwhite}(b), in which the naked singularity is  to the past of any Cauchy hypersurface.   Just like the black hole spacetime, we can consider the white hole spacetime to be the result of evolving Einstein's equations forwards and backwards in time, starting from
a smooth, complete Cauchy hypersurface.    It is a perfectly valid solution of Einstein's equations.

What does black hole thermodynamics say about a white hole?   The collapse of a star to form a black hole is a highly irreversible phenomenon in which the
entropy sharply increases.   So its time-reversal  is a phenomenon that strongly violates the Second Law of Thermodynamics, with a sharp reduction in the thermodynamic entropy
when the white hole converts into a star.

As we discussed in the introduction, in ordinary physics, processes that involve a macroscopic violation of the Second Law can happen at least in theory, but they are prohibitively unlikely
(and  not observed in practice) because they require extreme fine-tuning of the initial conditions.   An ice cube in a cup of hot tea will melt, so applying time-reversal symmetry,
one can envisage initial conditions for the cup of hot tea such that an ice cube will spontaneously form, say near time $t=0$.  But with slightly different initial conditions, this will not happen at all.   Starting with special initial conditions that
will lead to a violation of the Second Law, how big  must a perturbation be to produce initial conditions in which the Second Law will not be violated?  An important part of the answer
to this question is that the farther in the past the perturbation acts, the smaller it can be.    That is because only a very tiny fraction of the phase space volume -- or in quantum mechanical
terms, only a very small fraction of possible quantum states -- correspond to initial conditions that will lead to a macroscopic violation of the Second Law near $t=0$.
The presumably highly non-integrable dynamics of a vast number of molecules
making up a cup of hot tea will scramble the special initial conditions that lead to violation of the second law with far more numerous 
generic ones that do not.   In a generic non-integrable many-body system,
there is some ``scrambling time'' $t_\sc$ that suffices to mix the phase space so thoroughly that 
 a generic perturbation of just one or a few molecules at or prior to time $-t_\sc$ would be expected to prevent violation of the Second Law near $t=0$.

Black hole thermodynamics  suggests that a similar statement will be true for a white hole spacetime, since such a spacetime represents a violation of the Second Law.
A very small perturbation, acting sufficiently far in the past, should have a drastic effect
on the white hole spacetime and in particular it should be possible for a small perturbation sufficiently far in the past to prevent the evolution of this spacetime from violating
the Second Law.  

This reasoning alone does not tell us how far in the past a perturbation should act in order to radically affect the black hole spacetime.   However, known results about
black holes give a clue.   Black holes
are believed to be the fastest scramblers allowed by quantum mechanics \cite{SuSe}, with a scrambling time of order
\be\label{wono} t_\sc = \frac{1}{T_\sH}\log S \ee
where $S\sim GM^2$ is the entropy.
This observation is also related to the chaotic nature of black hole dynamics, which was found in  \cite{SS},  partly presaged in  earlier ideas \cite{DT}.
(In general, among systems of a given temperature and entropy, a black hole has the fastest possible  growth of chaos \cite{StMC}.)   Converting from black holes
to white holes, 
it is natural to guess that $t_\sc$ is also the answer to the question, ``How far in the past should a perturbation be made in order to radically affect the white hole spacetime?''

It is not hard to see why that is the case, by adapting the arguments that have been used for Hawking radiation and for chaos.\footnote{Essentially  the following argument
was first presented by Eardley \cite{Eardley}.}
In the derivation of Hawking radiation in section \ref{bhevaporation}, we observed the following.   Suppose that an outgoing massless quantum with an energy of
order the Hawking temperature $T_\sH$ is observed by a distant observer at a late retarded time $t_\ret$.  We can trace that particle back to an outgoing particle just outside the horizon
at an early time with a huge energy of order $T_\sH e^{t_\ret/4GM}$ (times a factor that depends on how far away is the distant observer in question, and how early is the ``early time''
that we consider, but not on $t_\ret$).    Thus by waiting for a while, the observer can see particles of modest energy coming out that at an early time were near the horizon
with an exponentially large energy.

The time-reversal of the retarded time $t_\ret=t-r$ is $-t-r$, which is the negative of the advanced time $t_\adv=t+r$.   By time-reversing the process described in the last paragraph, we learn
 the following.   Suppose that far in the past, at advanced time $t_\adv = - \tau$, an observer at a great distance from the white hole sends a massless particle with energy of order $T_\sH$  
inwards towards the white hole, as depicted in fig \ref{blackandwhite}(b).  With a probability of order one, this particle might be reflected back by the potential barrier discussed in section \ref{graybody}, but with a probability that
is also of order one, it will continue in to the origin and eventually encounter the material that is emerging from the white hole.   If so, by the time the incident particle reaches the star
that is emerging from the white hole, it will have an energy of order $T_\sH e^{\tau/4GM}$ (times a factor that depends on how far away is the distant observer in question, but not
on $\tau$).  With $\tau$ of order $t_\sc$, this will exceed the ADM mass of the white hole spacetime.   What happens next is hard to say, as our knowledge of physics
does not suffice to describe the collision of
 such an ultrarelativistic particle
 with the star that is emerging from the white hole.    But presumably whatever happens
will represent a substantial perturbation of the white hole spacetime.  The production of a black hole seems likely, especially if several such particles are incident on the white hole
from different directions.  The black hole will have a large entropy, so its formation will ensure that the Second Law is not violated.

So far, we have considered the white hole spacetime as a purely classical background.   Quantum mechanically,
we have to take the Hawking radiation into account, and fine-tuning of the white hole spacetime is even more glaring.
  Quantum mechanically, in fig. \ref{blackandwhite}(a), there is in the far future outgoing Hawking radiation, with a thermal spectrum modulated by gray body factors.
   The time-reversed state in fig. \ref{blackandwhite}(b) has incoming Hawking radiation in the far past with the same thermal spectrum  modulated by 
gray body factors.  
The incoming radiation is entangled with modes behind or near the past horizon of the white hole
in just such a way as to build up the quantum correlations
 that make a quantum state that is smooth along the horizon.    
This imploding shell of Hawking radiation converging on the white hole with just the right spectrum to make that work
would look highly fine-tuned to a distant observer.   And if the distant observer disturbs
this fine-tuning by adding or subtracting an additional incoming particle  at a very early time, that will lead to a drastic change in the spacetime evolution such as we already discussed.

So as suggested by  black hole thermodynamics, the white hole spacetime is fine-tuned and unstable to a perturbation acting far enough in the past.

\vskip1cm
 \noindent {\it {Acknowledgements}}   
 I thank Z. Komargodsky, J. Maldacena, and D. Stanford for discussions, and   A. Herdershee, T. Jacobson,   B. Zhao, and especially R. Wald  for comments on the manuscript.
  Research supported in part by NSF Grant PHY-2207584.  No data were generated in the course of this work.
 \bibliographystyle{unsrt}

\begin{thebibliography}{99}

\bibitem{bekenstein}J. D. Bekenstein, ``Black Holes and Entropy,'' Phys. Rev. {\bf D7} (1973) 2333-2346.

\bibitem{hawking}S. W. Hawking, ``Particle Creation By Black Holes,''
Commun. Math. Phys. {\bf 43} (1975) 199-220.

\bibitem{Christodoulou}
D. Christodoulou, ``Reversible and Irreversible Transformations in Black-Hole Physics,'' Phys. Rev. Lett. {\bf 25} (1970) 1596.

\bibitem{BCH}
J. M. Bardeen,   B. Carter, and S. W. Hawking, ``The Four Laws of Black Hole Mechanics,'' 
Commun. Math. Phys. {\bf 31} (1973) 161-70.

\bibitem{wald}R. M. Wald, {\it General Relativity} (University of Chicago Press, 1984).

\bibitem{HI}
S. W. Hawking and W. Israel, eds., {\it General Relativity: an Einstein Centenary Survey} (Cambridge University Press, 1979).



\bibitem{malda} A. Almheiri, T. Hartman, J. Maldacena, E. Shaghoulian,
and A. Tajdini, ``The Entropy of Hawking Radiation,'' Rev. Mod. Phys. {\bf 93} (2021) 35002, arXiv:2006.06872.

\bibitem{malda2}A. Almheiri, T. Hartman, J. Maldacena, E. Shaghoulian,
and A. Tajdini, ``Replica Wormholes and The Entropy of Hawking Radiation,''   JHEP {\bf 05} (2020) 013, arXiv:1911.12333 

\bibitem{Witten}
E. Witten, ``Light Rays, Singularities, and All That,'' Rev. Mod. Phys. {\bf 92} (2020) 045004, arXiv:1901.03928.

\bibitem{AreaTheorem}S. W. Hawking, ``Black Holes In General Relativity,'' Commun. Math. Phys. {\bf 25} (1972) 152-66.

\bibitem{StromingerVafa}
A. Strominger and C. Vafa, ``Microscopic Origin of the Bekenstein-Hawking Entropy,'' 
Phys. Lett. {\bf B379} (1996), 99-104, hep-th/9601029.

\bibitem{nohair}
W. Israel, ``Event Horizons in Static Vacuum Spacetimes,'' Phys. Rev. {\bf 164} (1967) 1776-9.

\bibitem{nohair2}
B. Carter, ``Axisymmetric Black Hole Has Only Two  Degrees of Freedom,''
Phys. Rev. Lett. {\bf 26} (1971) 331-333..

\bibitem{IW}V. Iyer and R. Wald, ``Some Properties of the Noether Charge and a Proposal for Dynamical Black Hole
Entropy,'' Phys. Rev. {\bf D50}  (1994) 846-64.

\bibitem{Page}D. Page, ``Particle Emission From a Black Hole: Massless Particles From an
Uncharged, Nonrotating Hole,'' Phys. Rev. {\bf D13} (1976) 198-206.

\bibitem{FH}
K. Fredenhagen and R. Haag, ``On The Derivation of the Hawking
Radiation Associated With the Black Hole,''
Commun. Math. Phys. {\bf 127} (1990) 273.

\bibitem{ItFromBit}
J. A. Wheeler, ``Information, Physics, Quantum: The Search For Links,'' in S. Kobayashi, ed., {\it 3rd International Symposium on Foundations of Quantum Mechanics in Light} (Physical
Society of Japan, 1990), available at \url{https://philpapers.org/archive/WHEIPQ.pdf}.

\bibitem{Malda}
J. Maldacena, ``Comments on Magnetically Charged Black Holes,'' JHEP {\bf 04} (2021) 079, arXiv:2004.06084.

\bibitem{Equil1}  G. W. Gibbons and M. J. Perry, ``Black Holes in Thermal Equilibrium,'' 
Phys. Rev. Lett. {\bf 36} (1976) 965-7.

\bibitem{Equil2}G. W. Gibbons and M. J. Perry, ``Black Holes and Thermal Green Functions,''
Proc. R. Sol. Lond. {\bf A358} (1978) 467-94.

\bibitem{U} W. G. Unruh, ``Notes on Black-Hole Evaporation,'' Phys. Rev. {\bf D14} (1976) 870-92.


\bibitem{GPY}
D. J. Gross, M. J. Perry, and L. Yaffe, ``Instability of Flat Space at Finite Temperature,''
Phys. Rev. {\bf D25} (1982) 330-55.

\bibitem{HawkingPage}S. W. Hawking and D. Page, ``Thermodynamics of Black Holes in Anti de Sitter Space,'' Commun. Math. Phys. {\bf 87}  (1983) 577.

\bibitem{UW} W. G. Unruh and N. Weiss, ``Acceleration Radiation in Interacting Field Theories,'' Phys. Rev. {\bf D29} (1984) 1656-62.


\bibitem{BW}J. Bisognano and E. Wichmann, ``On The Duality Condition For Quantum Fields,'' J. Math.
Phys. 17 (1976) 303-21.

\bibitem{Sewell}G. L. Sewell,
``Quantum Fields On Manifolds: PCT and Gravitationally Induced Thermal
States,'' Ann. Phys. {\bf  141} (1982) 201-24.

\bibitem{Notes}E. Witten, ``Notes on Some Entanglement Properties of Quantum Field Theory,'' Rev. Mod. Phys. {\bf 90} (2018) 045003,
arXiv:1803.04993.

\bibitem{GH}G. W. Gibbons and S. W. Hawking, ``Action Integrals and Partition Functions in Quantum Gravity,''
Phys. Rev. {\bf D15} (1977) 2752-6.

\bibitem{HH}
S. W. Hawking and J. Hartle, ``Path Integral Derivation of Black Hole Radiance,'' Phys. Rev. {\bf D13} (1976) 2188-2203.

\bibitem{Y}
J. W. York, ``Role of Conformal Three-Geometry in the Dynamics of Gravitation,'' Phys. Rev, Lett. {\bf 28} (1972) 1082.

\bibitem{CT}
S. Carlip and C. Teitelboim, ``The Off-Shell Black Hole,'' arXiv:gr-qc/9312002.

\bibitem{hp}
S. W. Hawking and D. N. Page, ``Thermodynamics of Black Holes in Anti-de Sitter Space,''
Commun. Math. Phys. {\bf 87} (1983) 577-88. 

\bibitem{MaldaDual}J. M. Maldacena, ``The Large $N$ Limit
Of Superconformal Field Theories and Supergravity,''
Adv. Theor. Math. Phys. {\bf 2} (1998) 231-52.

\bibitem{GKP}S. S. Gubser, I. R. Klebanov, and A. M. Polyakov,
``Gauge Theory Correlators From Noncritical String Theory,'' Phys. Lett. {\bf B428} (1998) 105-14.

\bibitem{EW}E. Witten, ``Anti-de Sitter Space and Holography,'' Adv. Theor. Math. Phys.
{\bf 2} (1998) 253-91.

\bibitem{BF} P. Breitenlohner and D. Z. Freedman, ``Stability in Gauged Extended Supergravity,''
Annals Phys. {\bf 144} (1982) 249.

\bibitem{FG}C. Fefferman and C. R. Graham, ``Conformal Invariants,'' in {\it Elie Cartan et les Math\'ematiques d'Aujourdhui}
(Asterisque, 1985) 95.

\bibitem{VH}
V. Hubeny, ``The AdS/CFT Correspondence,'' Class. Quant. Grav. {\bf 32} (2015) 12, arXiv:1501.00007.

\bibitem{MaldaNew}
J. M. Maldacena, ``The AdS/CFT Correspondence,'' available at \url{https://link.springer.com/referenceworkentry/10.1007/978-981-19-3079-9_65-1#Sec14}.

\bibitem{SW}
N. Seiberg and E. Witten, ``The D1/D5 System and Singular CFT,''
JHEP {\bf 04} (1999) 017, hep-th/9903224.

\bibitem{GH2}
G. W. Gibbons and S. W. Hawking, ``Cosmological Event Horizons, Thermodynamics,
and Particle Creation,'' Phys. Rev. {\bf D15} (1977) 2738-2751.

\bibitem{FHN}R. Figari, R. Hoegh-Krohn, and C. R. Nappi, ``Interacting
Relativistic Boson Fields in the De Sitter Universe With Two Space-Time
Dimensions,''
Commun. Math. Phys. {\bf 44} (1975) 265-278.

\bibitem{TJ}
T. Jacobson, ``A Note On Hartle-Hawking Vacua,''   Phys. Rev. {\bf D50} (1994) R6031-R6032, gr-qc/9407022.

\bibitem{KW} B. S. Kay and R. M. Wald, ``Theorems on the Uniqueness and Thermal
Properties of Stationary, Nonsingular, Quasifree States On Space-Times with a Bifurcate Horizon,'''
Phys. Rept. {\bf 207}  (1991) 49-136. 


\bibitem{CT2} N. A. Chernikov and E. A. Tagirov, ``Quantum theory of scalar field in de Sitter space-time,''
Annales de l'Institut Henri Poincar\'{e} A IX (1968) 109.

\bibitem{SS2}C. Schomblond and P. Spindel,  ``Conditions d'unicit\'{e}
pour le propagateur $\Delta_1(x;y)$  du champ scalaire dans l'univers de de Sitter,''  Annales de l'Institut Henri Poincar\'{e} A XXV (1976) 67.


\bibitem{BD}T. S. Bunch and P. Davies,
``Quantum Field Theory in de Sitter Space:
Renormalization by Point Splitting,'' Proc. Roy. Soc. London {\bf A360} (1978) 117-34.

\bibitem{Mo}
E. Mottola, ``Particle Creation in de Sitter Space,'' Phys. Rev. {\bf D31} (1985) 754.

\bibitem{Al}
B. Allen, ``Vacuum States in de Sitter Space,'' Phys. Rev. {\bf D32} (1985) 3136.



\bibitem{bc}
R. L. Bishop and R. L. Crittenden, {\it Geometry of Manifolds} (Academic Press, 1964). 


\bibitem{I}
W. Israel, ``Thermo-field Dynamics of Black Holes,'' Phys. Lett. {\bf 57} (1976) 107-10.

\bibitem{EinsteinRosen}
A. Einstein and N. Rosen, ``The Particle Problem in The General Theory of Relativity,'' Phys. Rev. {\bf 48} (1935) 73-7.

\bibitem{topocensor} J. L. Friedman, K. Schleich, and D. M. Witt, ``Topological Censorship,'' Phys. Rev. Lett. {\bf 71} (1993)
1486-9, arXiv:gr-qc/9305017.


\bibitem{topotwo} G. J. Galloway, K. Schleich, D. Witt, and E. Woolgar, ``The AdS/CFT Correspondence And Topological Censorship,''  Phys. Lett. {\bf B505} (2001) 255-62, hep-th/9912119.

\bibitem{MaldaSuss} J. Maldacena and L. Susskind, ``Cool Horizons for Entangled Black Holes,'' Fortschritte fur Physik {\bf 61} (2013) 781-811.

\bibitem{ETH}J. M. Deutsch, ``Quantum Statistical Mechanics in a Closed System,'' Phys. Rev {\bf A43} (1991) 2046-9.

\bibitem{ETH2}
M. Srednicki, ``Chaos and Quantum Thermalization,'' Phys. Rev. {\bf E50} (1994) 888-901.

\bibitem{Preskill}J, Preskill, ``Chapter 10: Quantum Shannon Theory,''
available at \url{http://theory.caltech.edu/~preskill/ph219/chap10_6A_2022.pdf}.

\bibitem{WittenQI}E. Witten, ``A Mini-Introduction to Information Theory,'' La Rivista del Nuovo Cimento {\bf 43} (2020) 187,
arXiv:1805.11965.

\bibitem{LR}
 E. H. Lieb and M. B. Ruskai, ``Proof Of The Strong Subadditivity Of Quantum Mechanical Entropy,''  J. Math. Phys. {\bf 14} (1973) 1938.

 \bibitem{PN} 
 M. A. Nielsen and D. Petz, ``A Simple Proof of the Strong Subadditivity Inequality,'' Quantum
Information and Computation {\bf 5} (2005) 507-13, arXiv:quant-ph/0408130.

\bibitem{Wall}A. C. Wall, ``A Proof of the Generalized Second Law for Rapidly Changing
Fields and Arbitrary Horizon Slices,'' Phys. Rev. {\bf D85} (2012) 104049, arXiv:1105.3445.

\bibitem{PageCurvePaper}D. Page, ``Information in Black Hole Radiation,''  Phys. Rev. Lett. {\bf 71} (1993) 3743-46,
hep-th/9306083.

\bibitem{P}G. Penington, ``Entanglement Wedge Reconstruction and the Information Paradox,''
JHEP {\bf 09} (2020) 002, arXiv:1905.08255.

\bibitem{AEMM}A. Almheiri,  N. Engelhardt, D. Marolf,  and H. Maxfield, ``The Entropy of Bulk Quantum
Fields and the Entanglement Wedge of an Evaporating Black Hole,'' JHEP {\bf 12} (2019) 063, arXiv:1905.08762.

\bibitem{SorkinA}
R.D. Sorkin, ``On The Entropy of a Vacuum Outside a Horizon,'' in B. Bertotti, F. de Fellice, and A. Pascolini, eds., {\it General Relativity and Gravitation, proceedings of
the GR10 Conference, Padova 1983}  (Consiglio Nazionale della Ricerche, Roma, 1983) Vol.
2, available at arXiv:1402.3589.

\bibitem{SorkinB}
L. Bombelli, R.K. Koul, J. Lee and R.D. Sorkin, ``Quantum Source of Entropy for Black Holes,'' Phys. Rev. {\bf D34}
(1986) 373.

\bibitem{Thooft}
G. 't Hooft, ``On The Quantum Structure Of A Black Hole,'' Nucl. Phys. {\bf B256} (1985) 727.


\bibitem{Srednicki}M. Srednicki, ``Entropy and Area,'' Phys. Rev. Lett. {\bf 71} (1993) 666-9, arXiv:hep-th/9303048.

\bibitem{SU}
L. Susskind and J. Uglum, ``Black Hole Entropy in
Canonical Quantum Gravity and Superstring Theory,'' hep-th/9401070.

\bibitem{Jacobson}
T. Jacobson, ``Black Hole Entropy and Induced Gravity,''
arXiv:gr-qc/9404039.

\bibitem{Sakharov}
A. D. Sakharov, ``Vacuum Quantum Fluctuations In Curved Space And The Theory
Of Gravitation,''  Sov. Phys. Dokl. {\bf 12} (1968) 1040 [Dokl. Akad. Nauk Ser. Fiz. 177
(1968) 70], 
reprinted in Gen. Rel. Grav. {\bf 32} (2000) 365-367.


\bibitem{CW}C. Callan and F. Wilczek, ``On Geometric Entropy,''  Phys. Lett. {\bf B333} (1994) 55-61, arXiv:hep-th/9401072.

\bibitem{HLW}
C. Holzhey, F. Larsen, and F. Wilczek, ``Geometric and Renormalized Entropy in Conformal Field Theory,'' Nucl. Phys. {\bf B424}
(1994) 443-67,
arXiv:hep-th/9403108.

\bibitem{AE}
S. F. Edwards and P. W. Anderson, ``Theory of Spin Glasses,'' J. Phys. {\bf F5} (1975) 965.

\bibitem{CC}
P. Calabrese and J. Cardy, ``Entanglement Entropy and Quantum Field Theory,'' J.Stat.Mech. {\bf 0406} (2004) P06002, arXiv:hep-th/0405152.

 
\bibitem{Boas}
R. P. Boas, Jr., ``Entire Functions'' (Academic Press, New York, 1954).

\bibitem{DFMS}
L. J. Dixon, D. Friedan, E. Martinec, and S. H. Shenker,
``The Conformal Field Theory of Orbifolds,'' Nucl. Phys. {\bf B282} (1987) 13-73.

\bibitem{orbifold}L. J. Dixon, J. A. Harvey, C. Vafa, and E. Witten, ``Strings on Orbifolds,''
Nucl. Phys. {\bf B261} (1985) 678-86.

\bibitem{Bek2}J. D. Bekenstein, ``Universal Upper Bound on the Entropy-to-Energy
Ratio for Bounded Systems,'' Phys. Rev. {\bf D23} (1981) 287-98.

\bibitem{Casini}H. Casini, ``Relative Entropy and the Bekenstein Bound,'' Class. Quant. Grav. {\bf 25} (2008) 205021,
arXiv:0804.2182.

\bibitem{MMR}
D. Marolf, D. Minic, and S. F. Ross, ``Notes on Spacetime Thermodynamics
and the Oberver-dependence of Entropy,'' Phys. Rev. {\bf D69} (2004) 064006, arXiv:hep-th/03120022.

\bibitem{Araki}H. Araki, ``Relative Entropy of States of Von Neumann Algebras,'' Publ. RIMS, Kyoto Univ. {\bf 11} (1976) 809-33.

\bibitem{KLS}J. Kudler-Flam, S. Leutheusser, A. A. Rahman,
G. Satishchandran, and A. J. Speranza, ``A Covariant Regulator
for Entanglement Entropy: Proofs of the Bekenstein Bound and
QNEC,'' arXiv:2312.07646.

\bibitem{RT}S. Ryu and T. Takayanagi, ``Holographic Derivation of Entanglement Entropy from AdS/CFT,''
Phys. Rev. Lett. {\bf 96} (2006) 181602,  arXiv:hep-th/0603001.

\bibitem{HRT}V. E. Hubeny, M. Rangamani and T. Takayanagi, ``A Covariant Holographic Entanglement
Entropy Proposal, JHEP {\bf 07} (2007) 062, arXiv:0705.0016.

\bibitem{LM}
A. Lewkowycz and J. Maldacena, ``Generalized Gravitational Entropy,'' JHEP {\bf 08} (2013) 090,
arXiv:1304.4926.

\bibitem{BDHM} T. Barrella, X. Dong, S. A. Hartnoll and V. L. Martin, ``Holographic Entanglement Beyond
Classical Gravity,''  JHEP {\bf 09} (2013) 109, arXiv:1306.4682.


\bibitem{FLM}  T. Faulkner, A. Lewkowycz and J. Maldacena, ``Quantum Corrections to Holographic
Entanglement Entropy,''  JHEP {\bf 11} (2013) 074, arXiv:1307.2892.

\bibitem{NEW}
N. Engelhardt and A. C. Wall, ``Quantum Extremal Surfaces: Holographic Entanglement Beyond the Classical Regime,''
JHEP {\bf 01} (2015) 073.

\bibitem{SS}
S. H. Shenker and D. Stanford, ``Black Holes and the  Butterfly Effect,'' JHEP {\bf 03} (2014) 067,
arXiv:1306.0622.

\bibitem{SSagain}S. H. Shenker and D. Stanford, ``Multiple Shocks,''
JHEP {\bf 12} (2014) 046, arXiv:1312.3296.



\bibitem{DT}
T. Dray and  G. 't Hooft, ``The Effect of Spherical Shells of Matter on the Schwarzschild Black Hole,''
Commun. Math. Phys. {\bf 99} (1985) 613.


\bibitem{Twice} N. Engelhardt, G. Penington, and A. Shahbazi-Moghaddam, ``Twice Upon A Time: Timelike Separated
Quantum Extremal Surfaces,'' JHEP {\bf 01} (2024) 033, arXiv:2308.16226.


\bibitem{BTZ}M. Ba\~{n}ados, C. Teitelboim, and J.Zanelli, ``The Black Hole In Three Dimensional Spacetime,'''  Phys. Rev. Lett. {\bf 69} (1992) 1849-51,
  arXiv:hep-th/9204099. 


\bibitem{MWW}
D Marolf, S. Wang, and Z. Wang, ``Probing Phase Transitions of Holographic Entanglement Entropy With Fixed Area States,''
arXiv:2006.10089.

\bibitem{TakHead}M. Headrick and T. Takayanagi, ``A Holographic Proof of the Strong Subadditivity of Entanglement Entropy,''
Phys. Rev. {\bf D76} (2007) 106013.

\bibitem{Wall2}A. Wall,  ``Maximin Surfaces, and the Strong Subadditivity of the Covaeriant Holographic
Entanglement Entropy,'' Class. Quant. Grav. {\bf 31} (2014) 225007, arXiv:1211.3494.

\bibitem{KS}M. Henningson and K. Skenderis, ``Weyl Anomaly For Wilson Surfaces,'' JHEP {\bf 9906} (1999) 012, 
arXiv:hep-th/9905163.

\bibitem{GW} R. Graham and E. Witten, ``Conformal Anomaly of Submanifold
Observables in AdS/CFT Correspondence,''
Nucl. Phys. {\bf B546} (1999) 52-64.



\bibitem{Fursaev}
D. V. Fursaev, ``Proof of the Holographic Formula for Entanglement Entropy,'' arXiv:hep-th/0606184.




\bibitem{EIH}A. Einstein, L. Infeld, and B. Hoffman, ``The Gravitational Equations and the Problem of Motion,'' Ann. Math. {\bf 39} (1938) 65-100.



\bibitem{Headrick}
M. Headrick, ``Entanglement R\'enyi Entropies in Holographic Theories,''   Phys. Rev. {\bf D82} (2010) 126010,   arXiv:1006.00473.




\bibitem{CKNR} B.  Czech, J. L Karczmarek, F.  Nogueira, and M. Van Raamsdonk,
``The Gravity Dual of a Density Matrix,''  Class.  Qtm.  Grav., {\bf 29} (2012) 155009.

\bibitem{HHLR}
 M.  Headrick, V. E. Hubeny, A. Lawrence, and M.  Rangamani. ``Causality \& Holographic Entanglement Entropy,''  JHEP {\bf 12} (2014) 162.
 
\bibitem{JLMS}
 D.  L Jafferis, A.  Lewkowycz, J.  Maldacena, and S Josephine Suh, ``Relative Entropy
Equals Bulk Relative Entropy,''  JHEP  {\bf 4} (2016).


\bibitem{DHW} X. Dong, D. Harlow, and A.  C Wall, ``Reconstruction of Bulk Operators Within the
Entanglement Wedge in Gauge-Gravity Duality,''  Phys.  Rev. Lett. {\bf  117} (2016) 021601.










\bibitem{ADH}
A. Almheiri, X. Dong, and D. Harlow, ``Bulk
Locality and Quantum Error Correction in AdS/CFT,''
JHEP {\bf 04}  (2015) 163 , arXiv:1411.7041.

\bibitem{PYHP}
 F.Pastawski, B. Yoshida, D. Harlow, and
J. Preskill, ``Holographic Quantum Error-Correcting
Codes: Toy Models for the Bulk/Boundary Correspondence,'' JHEP {\bf 06} (2015) 149, arXiv:1503.06237.


\bibitem{Shor}
P. W.  Shor, ``Fault-Tolerant Quantum Computation,''  Proceedings of 37th Conference on Foundations of Computer Science,  IEEE Comput. Soc. Press. (1996),  pp. 56-65..



\bibitem{SuSe}
Y. Sekino and L. Susskind, ``Fast Scramblers,'' JHEP {\bf 10} (2008) 065, arXiv:0808.2096.





\bibitem{StMC}J. Maldacena, S. H. Shenker, and D. Stanford, ``A Bound on Chaos,''
JHEP {\bf 08} (2016) 106, arXiv:1503.01409.

\bibitem{Eardley}
D. M. Eardley, ``Death of White Holes in the Early Universe,''  Phys. Rev. Lett. {\bf 33} (1974) 442-4.

\end{thebibliography}

\end{document}